\documentclass[10pt,a4paper]{book}

\usepackage{fancyhdr}
\fancypagestyle{plain}{%
\fancyhead{} 
}

\usepackage{graphicx} 
\usepackage{dcolumn}  
\usepackage{bm}       
\usepackage{amsmath} 
\usepackage{amssymb}

\usepackage{graphicx}
\usepackage{hyperref}
\usepackage{latexsym}
\usepackage{dsfont}

\renewcommand{\maketitle}{
\thispagestyle{empty}
\begin{changemargin}{+ 2 cm}{-.1 cm}{-1.9 cm}{-.5 cm}{-.5 cm} 
\begin{minipage}[t]{0.4\linewidth}
\begin{center}
\includegraphics[width=2.8 cm]{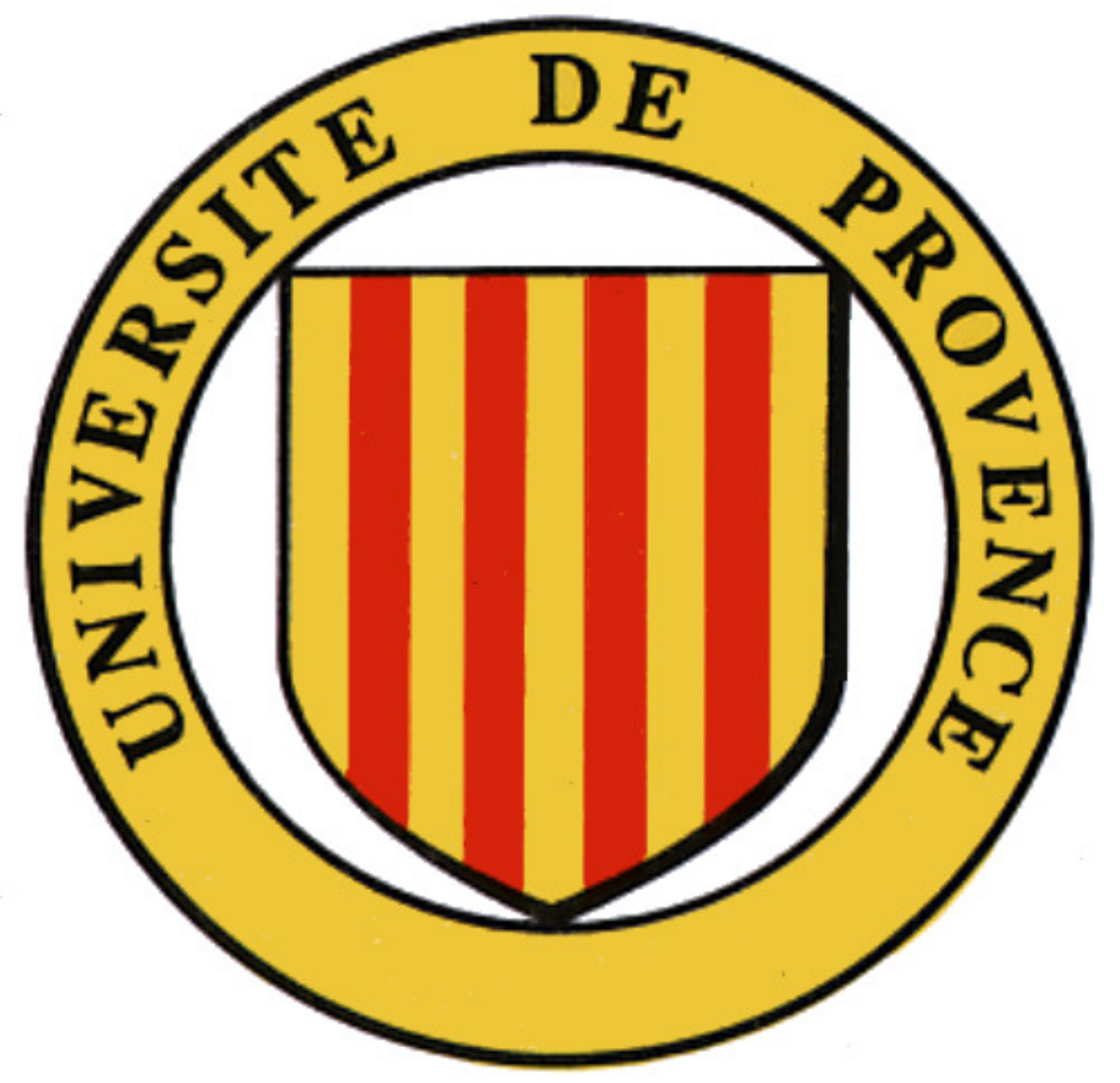}\\
\footnotesize \sf
Physique des Interactions Ioniques et Mol\'eculaires\\ 
Centre de St. J\'er\^ome - UMR CNRS 6633\\
Avenue Escadrille Normandie-Ni\'emen\\
F-13397 Marseille, France\\
\end{center}
\end{minipage}
\hfill
\begin{minipage}[t]{0.4\linewidth}
\begin{center}
\includegraphics[width=2.8 cm]{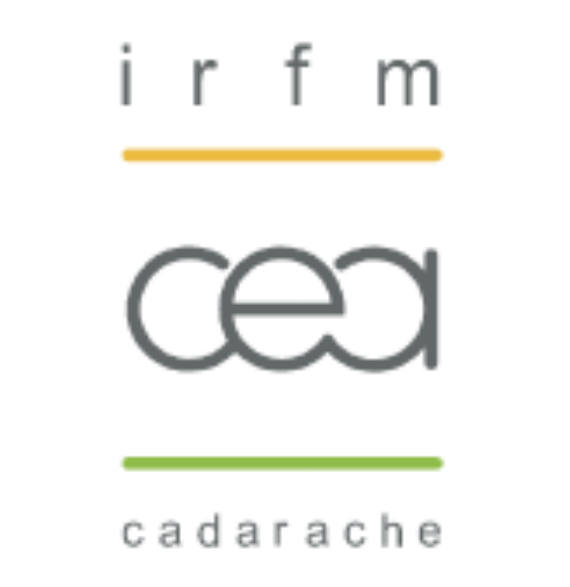}\\
\footnotesize \sf
Institut de Recherche sur la Fusion Magn\'etique\\
Association Euratom-CEA\\
CEA Cadarache\\
F-13108 Saint Paul-Lez-Durance, France\\
\end{center}
\end{minipage}
\vspace{1.6cm}
\begin{center}
{ 
\huge
A quasi-linear gyrokinetic transport model \\ for tokamak plasmas \\} 
\vspace{1.2cm}
{\rmfamily\large  by}\\
\vspace{0.5cm}
{\rmfamily\Large\bfseries Alessandro Casati}\\
\vspace{1.6cm}
{\rmfamily \large A thesis in candidacy for the degree of}\\
\vspace{.8cm}
{\rmfamily \bfseries \large Doctor of Philosophy}\\
\vspace{0.2cm}
{\rmfamily \bfseries \large Universit\'e de Provence (Aix-Marseille I), France}\\
\vspace{0.2cm}
{\rmfamily \large Speciality: \textit{Plasma Physics}}\\
\vspace{1.8 cm}
\begin{tabular}{lll}
\hline
  { Dr. Clarisse Bourdelle} & {CEA, Research scientist} &{CEA supervisor}\\
  { Dr. Ambrogio Fasoli}    & {EPFL, Professor} &{Referee}\\
  { Dr. Pascale Hennequin}  & {CNRS, Research director} &{Chair}\\ 
  { Dr. Fr\'ed\'eric Imbeaux} & {CEA, Research scientist} &{CEA supervisor}\\
  { Dr. Frank Jenko}        & {IPP, Research director} &{Invited member}\\ 
  { Dr. Alexander Shekochihin} & {University of Oxford, Lecturer} &{Referee}\\
  { Dr. Roland Stamm}      & {Universit\'e de Provence, Professor} &{Thesis director}\\ 
 \hline
\end{tabular}
\end{center}
\end{changemargin}
}

\begin{document}



\cleardoublepage

\maketitle
\thispagestyle{empty}

\frontmatter

\tableofcontents

\chapter{Introduction} 
\label{Introduction}

\indent The long way of trying to reproduce on the Earth the processes which make the stars brighten, is the more than 50 years long history of nuclear fusion. Despite the last half century has seen an extraordinary improvement of the scientific knowledge and the associated human technological capabilities, the goal of employing the nuclear fusion reactions as profitable and peaceful energy source, has still to be achieved. This large amount of efforts and time that have been spent may suggest that achieving nuclear fusion on the Earth is not an easy task. Indeed, it is not.\\
\indent Fusion relies on the nuclear interaction between light elements which are combined to form heavier ones, releasing an amount of energy. On the other hand, such interactions are possible if the original components have an high free energy, thus implying that the reacting medium is locally far from the thermodynamical equilibrium. Practically, that means that the mixture of the reacting components, typically deuterium and tritium, have enough energy to be completely ionized, thus forming a plasma: external constraints have then to balance its internal pressure. At the very beginning of this work, it is important to highlight that the actual crucial issue of the nuclear fusion can be extremely summarized by a single word: \textsl{confinement}. From the problem of the energy confinement in fact, stems a wide galaxy of both scientific and technological topics which are subjects of the living history of the nuclear fusion research. Even if this thesis will only deal with a very specific issue in this long list, the main concern that has inspired this work was to clearly locate the research activity within this global framework. Understanding, predicting and controlling the \textsl{confinement} mechanisms are the main keys to thrive over the scientific and technological challenges of fusion. \\ 
\indent A bit more quantitatively, a characteristic time scale can be defined as the ratio between the energetic content of the plasma and the power losses: this defines the so called energy confinement time $\tau_E$. In the case of the tokamak configuration, where a particular magnetic topology of nested flux surfaces provides a pressure that counterbalances the internal plasma one, the losses are mainly due to conductive effects. The energy confinement time can then be associated to a diffusive time $a/\chi_{th}$, where $a$ is the typical macroscopic scale of the plasma column and $\chi_{th}$ is the effective thermal diffusivity. Leaving aside any technological implication, which \textit{de facto} represents a wide and very active domain of research within the fusion community, this study will treat the issue of confinement only in terms of a problem of transport of energy and matter. \\
\indent The transport in tokamak plasmas is usually referred as anomalous. The anomaly makes reference to the experimental evidence, verified on many devices since the 1980s, that the measured thermal diffusivities are larger by one or two order of magnitudes, respectively for the ions and the electrons, than the expectations based only on the collisional processes (neoclassical theory). Effectively, the anomalous tokamak transport is largely due to the micro-turbulence that affects the plasma dynamics. The main drives of these turbulent micro-instabilities were already known in the 1960s-70s: these are namely the temperature, density and velocity gradients transverse with respect to the magnetic flux surfaces. The resulting instabilities originate fluctuations in the plasma pressure as well in the electromagnetic field associated to the spatial charge distribution. The final macroscopic result is the presence of turbulent fluxes of particle, energy and momentum.\\
\indent The quest for comprehensive and reliable models of the tokamak anomalous transport lasts from several decades. From one side, the combined efforts of sharing the increasingly large amount of experimental results collected from many fusion facilities across the world, allowed the formulation of semi-empirical scaling laws. The underlying idea of this approach is to try to identify a statistically coherent correlation, which is therefore transferred into a predictive capability, between a number of physically relevant dimensionless parameters (related for ex. to the size of the device, the plasma pressure, etc.) and the critical fusion quantities, typically the energy confinement time. On the opposite side, the ambition of the first principle modelling is to elaborate a theoretically based understanding of the plasma dynamics. Because of the complexity of the underlying processes, which often makes that the problem cannot be analytically solved, the first principle quantitative information is more and more deferred to the numerical simulations, strongly pushed by the recent advances of the computational capabilities. These two different approaches, the experimentally driven formulation of scaling laws and the coupled theoretical-numerical effort for first principle models, refer respectively to an inductive rather than a deductive knowledge process, or in other words, to a top-down rather than a bottom-up strategy. Not pretending to epistemology, the scientific understanding does not rely on a purely inductive neither deductive reasoning, but on a balanced hypothetical-deductive method. The work here proposed aims developing a reduced transport model for the core of tokamak plasmas which stems from the combined contributions of theoretical, numerical and experimental insights.\\
\indent A rigorous first principle modelling of the tokamak turbulence should deal with the self-consistent dynamics between the electromagnetic fields and the plasma response with respect to the field perturbations. The resulting problem is of great complexity, since it involves a nonlinear dynamics in both the spatial and the velocity phase spaces. The actual state of the art is represented by the coupling of the Maxwell equations with a reduced 5-dimensional description of the plasma response: this is the so called gyro-kinetic approximation, which in fact results from the average over the fast particle gyro-motion. Moreover, a number of additional simplifications are usually adopted, provided the fact that the plasma dynamics is intrinsically extended over disparate temporal and spatial scales. Still, the quantitative information that can be gained from the numerical solution of such an electromagnetic nonlinear system requires huge computational resources and it is hardly applicable for inferring predictions over the experimental macroscopic time scales of a typical tokamak discharge.\\
\indent The approach of this thesis work adopts a different strategy. The turbulent transport model here proposed follows from a gyro-kinetic quasi-linear approximation. The relevance of such a solution goes well beyond the motivations of simplifying the nonlinear nature of the problem and therefore lowering the computational cost: the reduction of complexity under a minimal set of hypotheses is a physically challenging issue. This logic of modeling allows identifying whether the tokamak turbulence is reminiscent or not of the linear dynamics, resulting in a great progress of the physical understanding.\\
\indent A linear response that keeps the fundamental kinetic wave-particle resonance mechanism in the velocity space is chosen. On the other hand, the nonlinear self-consistency of the system is broken and split into two basic parts: (1) the linear response of the plasma and (2) the saturated fluctuating potential. The hypotheses underlying the linear response, and therefore its limits, have to be accurately verified against fully nonlinear simulations. The saturation regime can instead be studied by mean of both turbulence measurements in the plasma core, and nonlinear simulations. At any level of the whole process of the model formulation, simplified analytical models can greatly enhance our understanding of both the experimental, hence real, and the numerical, hence artificial, observations.\\
The final goal of such a reduced quasi-linear gyro-kinetic model is the solution of the time-dependent transport problem. Finally, the detail of the energy and particle turbulent transport at the microscopic scales predicted by the model, is recast into an upper-level solver which integrates all the information concerning the equilibrium, the sources and the transport of the plasma: the time evolution of the thermodynamically relevant quantities is therefore predicted.\\
\indent In conclusion, this thesis work tackles the issue of the energy and particle confinement in the core of tokamak plasmas through the formulation of a reduced physical model: the related understanding, validity and limits stem from the concurrence of the theoretical, experimental and numerical analysis.



\chapter{R\'esum\'e} 
\label{Resume}

\indent Le long chemin d'essayer de reproduire sur Terre les processus qui font briller les \'etoiles, est l'histoire, longue de plus que 50 ans, de la fusion nucl\'eaire. Bien que la derni\`ere moiti\'e du si\`ecle ait connu une extraordinaire am\'elioration de la connaissance scientifique et des capacit\'es technologiques associ\'ees, l'objectif d'utiliser les r\'eactions de fusion nucléaire en tant que source d'\'energie rentable et pacifique, n'a pas encore \'et\'e atteint. Cette grande quantit\'e d'efforts et de temps qui a \'et\'e d\'epens\'ee sugg\`ere que la r\'ealisation de la fusion nucl\'eaire sur Terre n'est pas une t\^ache facile.\\
\indent La fusion repose sur l'interaction \`a l'\'echelle nucl\'eaire entre des ions l\'egers qui se combinent pour former des ions plus lourds, en lib\'erant de l'\'energie. D'autre part, cette fusion est possible si les composants d'origine ont une \'energie libre \'elev\'ee, ce qui implique que ce moyen r\'eactif est au niveau local loin de l'\'equilibre thermodynamique. En pratique, cela signifie que les composants qui interviennent dans les r\'eactions, g\'en\'eralement le deut\'erium et le tritium, ont suffisamment d'\'energie pour \^etre compl\`etement ionis\'ees, formant ainsi un plasma: des contraintes externes doivent donc pourvoir \^a \'equilibrer la pression interne. Il est important de souligner que la question cruciale de la fusion nucl\'eaire peut \^etre synth\'etis\'ee par le mot \textsl{confinement}. \`A partir de la question du confinement de l'\'energie en fait, r\'esulte une grande vari\'et\'e de probl\`emes \`a la fois scientifiques et technologiques qui constitue les sujets de l'histoire de la recherche sur la fusion nucl\'eaire. M\^eme si cette th\`ese ne portera que sur une question tr\`es pr\'ecise dans cette longue liste, la principale pr\'eoccupation qui a inspir\'e ce travail a \'et\'e de bien localiser l'activit\'e de recherche dans ce cadre global. Comprendre, pr\'evoir et contr\^oler les m\'ecanismes du confinement sont les cl\'es principales pour r\'eussir dans les d\'efis scientifiques et technologiques de la fusion. \\ 
\indent De fa\c con un peu plus quantitative, une \'echelle de temps caract\'eristique du confinement peut \^etre d\'efinie comme le rapport entre le contenu \'energ\'etique du plasma et la puissance perdue: cela est la d\'efinition du temps de confinement de l'\'energie  $\tau_E$. Dans le cas de la configuration tokamak, o\`u une topologie magn\'etique des surfaces de flux imbriqu\'ees fournit une pression qui contrebalance celle int\'erieure du plasma, les pertes dans le coeur sont principalement dues \`a des effets de conduction. Le temps de confinement de l'\'energie peut alors \^etre associ\'e \`a un temps de diffusion $a/\chi_{th}$, o\`u $a$ est la grandeur macroscopique de la colonne de plasma et $\chi_{th}$ est la diffusivit\'e thermique efficace. Laissant de c\^ot\'e toute implication technologique, qui repr\'esente \textit{de facto} un domaine large et actif de la recherche dans la fusion, cette \'etude traite la question du confinement seulement en termes d'un probl\`eme de transport de l'\'energie et de la mati\`ere. \\
\indent Le transport dans les plasmas de tokamak est g\'en\'eralement qualifi\'e comme anormal. Cette anomalie fait r\'ef\'erence \`a l'\'evidence, v\'erifi\'ee sur des nombreuses installations depuis les ann\'ees 1980, que les diffusivit\'es thermiques sont plus grandes de un ou deux ordres de grandeur, respectivement pour les ions et les \'electrons, des pr\'edictions bas\'ees uniquement sur les processus de collision (th\'eorie n\'eoclassique). En effet, le transport anormal dans les tokamaks est largement d\^u \`a la micro-turbulence qui affecte la dynamique du plasma. Les m\'ecanismes principaux \`a la base de ces micro-instabilités \'etaient d\'ej\`a connus dans les ann\'ees 1960-70: ce sont notamment les gradients de temp\'erature, densit\'e et vitesse perpendiculaires aux surfaces de flux magn\'etique. Ces instabilit\'es sont \`a l'origine de fluctuations de la pression du plasma et du champ \'electromagn\'etique associ\'e \`a la distribution spatiale de charge. Le r\'esultat macroscopique est la pr\'esence des flux turbulents de particules, d'\'energie et de moment angulaire.\\
\indent La qu\^ete de mod\`eles exhaustifs et fiables du transport anormal dans les tokamaks dure depuis plusieurs d\'ecennies. D'un c\^ot\'e, les efforts combin\'es de partage d'une grande quantit\'e de r\'esultats exp\'erimentaux obtenus aupr\`es des nombreuses installations de fusion \`a travers le monde, a permis l'\'elaboration des lois d'\'echelle semi-empiriques. Cette approche permet de trouver une corr\'elation statistique coh\'erente, qui est transf\'er\'ee en capacit\'e de pr\'ediction, entre un certain nombre des param\`etres physiques adimensionnels (li\'es par ex. \`a la taille de l'appareil, la pression du plasma, etc.) et des quantit\'es critiques pour la fusion, g\'en\'eralement le temps de confinement d'\'energie. De l'autre c\^ot\'e, l'ambition de la mod\'elisation de la turbulence consiste \`a acqu\'erir une compr\'ehension de la dynamique du plasma bas\'ee sur la th\'eorie. En raison de la complexit\'e des processus non-lin\'eaires sous-jacents, qui fait souvent que le probl\`eme ne peut pas \^etre r\'esolu analytiquement, seule la simulation num\'erique, fortement pouss\'ee par les r\'ecents progr\`es des capacit\'es de calcul, permet de donner des r\'eponses quantitatives. Ces deux approches, la formulation des lois d'\'echelle et les efforts th\'eoriques et num\'eriques, sont respectivement des processus de connaissance inductive et d\'eductive, soit des strat\'egies top-down plut\^ot que bottom-up. Sans pr\'etention d'\'epist\'emologie, la connaissance scientifique ne se fonde pas sur un raisonnement purement inductif ni d\'eductif, mais sur une m\'ethode hypoth\'etique-d\'eductive. Le travail propos\'e ici vise \`a l'\'elaboration d'un mod\`ele de transport r\'eduit pour le coeur des plasmas de tokamak, qui int\`egre des contributions \`a caract\`ere th\'eorique, num\'erique et exp\'erimental.\\
\indent Une rigoureuse mod\'elisation aux premiers principes de la turbulence dans les tokamaks devrait traiter la dynamique auto-consistante entre les champs \'electromagn\'etiques et la r\'eponse du plasma \`a l'\'egard des perturbations du champ. Le probl\`eme qui en r\'esulte est d'une grande complexit\'e, car il implique une dynamique non-lin\'eaire \`a la fois dans les espaces de phase en espace et en vitesse. L'\'etat de l'art actuel est repr\'esent\'e par le couplage des \'equations de Maxwell avec une description r\'eduite \`a 5 dimensions de la r\'eponse du plasma: c'est l'approximation dite gyro-cin\'etique, qui r\'esulte de la moyenne sur le mouvement de giration rapide des particules le long du champ magn\'etique. En outre, un certain nombre de simplifications suppl\'ementaires est g\'en\'eralement adopt\'e, utilisant le fait que la dynamique du plasma est intrins\`equement \'etendue sur des \'echelles temporelles et spatiales disparates. Toutefois, l'information quantitative qui peut \^etre retir\'ee par la solution num\'erique d'un tel syst\`eme \'electromagn\'etique non-lin\'eaire exige d'\'enormes ressources de calcul, et elle est actuellement difficilement applicable pour obtenir des pr\'edictions sur les \'echelles de temps macroscopiques d'une d\'echarge typique de tokamak.\\
\indent L'approche de ce travail de th\`ese adopte une strat\'egie diff\'erente. Le mod\`ele de transport turbulent ici propos\'e part d'une approximation gyro-cin\'etique quasi-lin\'eaire. L'importance d'une telle solution va au-del\`a des raisons de simplification de la nature non-lin\'eaire du probl\`eme afin de r\'eduire le co\^ut de calcul. En effet, cette logique de mod\'elisation permet d'identifier dans quelle mesure la turbulence des tokamaks montre des r\'eminiscences  de la dynamique lin\'eaire, entra\^inant une progression sensible de la compr\'ehension physique.\\
\indent Une r\'eponse lin\'eaire qui conserve le m\'ecanisme fondamental de r\'esonance cin\'etique onde-particule dans l'espace des vitesses est choisie. D'autre part, l'auto-consistance non-lin\'eaire du système, ici d\'efectueuse, est scind\'ee en deux termes : (1) la r\'eponse lin\'eaire du plasma et (2) le potentiel \'electrostatique satur\'e fluctuant. Les hypoth\`eses qui sous-tendent la r\'eponse lin\'eaire, et donc ses limites, doivent \^etre attentivement confront\'ees \`a des simulations enti\`erement non-lin\'eaires. Le r\'egime de saturation peut \^etre \'etudi\'e \`a la fois par moyenne des mesures de turbulence dans le coeur du plasma, et par des simulations non-lin\'eaires. A tout niveau de l'ensemble du processus de la formulation du mod\`ele, des mod\`eles analytiques simplifi\'es peuvent grandement am\'eliorer notre compr\'ehension de la complexité\'e \`a la fois exp\'erimentale, donc r\'eelle, et num\'erique, donc artificielle, des observations.\\
\indent En conclusion, ce travail de th\`ese aborde la question du confinement de l'\'energie et des particules dans le coeur des plasmas de tokamak \`a travers la formulation d'un mod\`ele physique r\'eduit: la validit\'e et les limites de ce mod\`ele d\'ecoulent de la concurrence entre l'analyse th\'eorique, exp\'erimentale et num\'erique. \\

\noindent \textbf{Contenu de la th\`ese}\\

\noindent Le contenu de ce travail de th\`ese est le suivant. \\
\indent Dans le Chapitre 2, la strat\'egie de confinement adopt\'ee par les tokamaks est introduite. Les principales instabilit\'es du plasma, responsables du transport turbulent de l'\'energie et la mati\`ere dans un tel syst\`eme, sont trait\'ees. Les deux repr\'esentations fondamentales du plasma, celle fluide et celle cin\'etique, sont ici bri\`evement pr\'esent\'ees. Une attention particuli\`ere est consacr\'ee aux raisons pour lesquelles une approche gyro-cin\'etique a \'et\'e pr\'ef\'er\'ee dans la mod\'elisation quasi-lin\'eaire. Un exemple pertinent pour le tokamak est pr\'esent\'e afin de souligner l'importance de retenir correctement la r\'esonance cin\'etique onde-particule: il s'agit de la d\'ependance du seuil d'instabilit\'e lin\'eaire en fonction du rapport des temp\'eratures $T_i/T_e$ des modes \'electroniques et ioniques. \\
\indent Le Chapitre 3 traite la question de la r\'eponse quasi-lin\'eaire. Premi\`erement, la d\'erivation du mod\`ele, appel\'e QuaLiKiz, et ses hypoth\`eses sous-jacentes permettant d'obtenir des flux turbulents d'\'energie et de particules, sont pr\'esent\'ees. Deuxi\`emement, la validit\'e de la r\'eponse quasi-lin\'eaire est confront\'ee \`a des simulations gyro-cin\'etiques non-lin\'eaires afin de: (a) identifier les temps caract\'eristiques dominant la dynamique turbulente, (b) comparer les relations de phase entre les champs fluctuants, (c) examiner l'intensit\'e globale du transport, normalis\'e \`a l'intensit\'e du potentiel satur\'e.  \\
\indent Le Chapitre 4 porte sur le mod\`ele de la saturation non-lin\'eaire. Dans la premi\`ere partie, les simulations gyro-cin\'etiques non-lin\'eaires sont valid\'ees quantitativement par rapport aux mesures de turbulence dans le tokamak Tore Supra. Les spectres des fluctuations de densit\'e tant dans l'espace, $k$, que en fr\'equence, $\omega$, sont \'etudi\'es aussi en termes de mod\`eles analytiques simplifi\'es. Sur la base de cette validation non-lin\'eaire avec les mesures, le mod\`ele de saturation qui est introduit dans QuaLiKiz est pr\'esent\'e et discut\'e. \\
\indent Le Chapitre 5 est consacr\'e \`a qualifier les r\'esultats de QuaLiKiz. Les flux quasi-lin\'eaires d'\'energie et des particules sont compar\'es aux pr\'edictions des simulations non-lin\'eaires pour une large gamme des param\`etres. Finalement, le couplage de QuaLiKiz dans le solveur de transport int\'egr\'e CRONOS est pr\'esent\'e. Cette proc\'edure permet de r\'esoudre la d\'ependance temporelle dans le probl\`eme de transport, et donc une application directe du mod\`ele \`a l'exp\'erience. Des r\'esultats pr\'eliminaires concernant l'analyse exp\'erimentale sont enfin examin\'es. \\

\mainmatter

\chapter{Foreword}

\section{Basics around nuclear fusion}

\begin{figure}[!htbp]
  \begin{center}
    \leavevmode
      \includegraphics[width=6 cm]{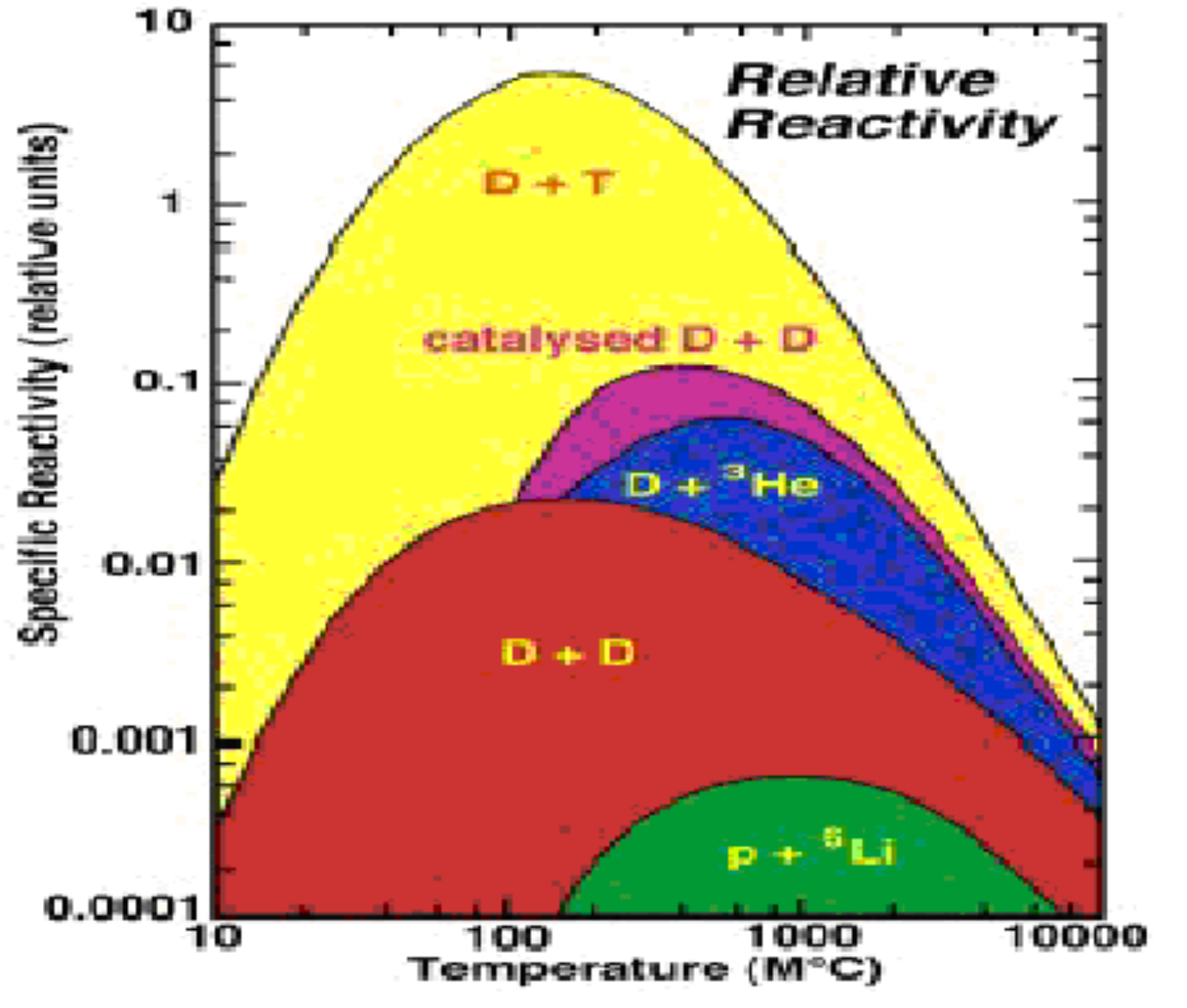}
    \caption{Reactivity (reaction cross-section) for different nuclear fusion processes.}
    \label{DTreaction}
  \end{center}
\end{figure}

\indent The nuclear fusion reactions rely on the interaction between light nuclei on the scale of characteristic range of the nuclear strong force: the overcome of the electrostatic Coulomb barrier gives rise to the fusion of the nuclei to form heavier elements and releasing the difference of the binding energy of the original components. Because of the presence of the Coulomb repulsion, the nuclei involved in the fusion reaction must have a kinetic energy of the order of energy barrier: the elements will therefore form a completely ionized plasma.\\
\indent The nuclear fusion processes are extremely common in the universe as they represent the source of energy of the stars. Inside the sun, for example, a series of nuclear reactions converts mainly hydrogen nuclei into helium ones through the so called proton cycle, whose net result is:
\begin{eqnarray}
	4p^+\qquad\Longrightarrow\qquad ^{4}He+2e^++2\nu+27~\rm{MeV} \label{cap1eq1}
\end{eqnarray}
These are processes that have a particularly high energy efficiency, but the drawback of an extremely low reaction rate. This is not a problem in the stars, because the extreme pressure of the core provides a gravitational confinement of the particles for a sufficiently long time, so that a balance is established between the radiation pressure due to the thermonuclear reactions and the gravitational compression.\\
\indent In order to reproduce on the earth controlled nuclear fusion reactions, it is necessary to adopt appropriate solutions both in terms of the nuclear reactivity and the plasma confinement. Although from the theoretical point of view several fusion reactions can be explored, as shown in Fig. \ref{DTreaction},
the way that nowadays appears as the most accessible and promising one, relies on the fusion between between deuterium D and tritium T nuclei:
\begin{eqnarray} 
	^{2}D + \:^{3}T \qquad\Longrightarrow\qquad n +\:^{4}He+17.6~\rm{MeV} \label{cap1eq2}
\end{eqnarray}
As shown in Fig. \ref{DTreaction}, the D-T reaction maximizes the reactivity at the lowest ion temperature, with a cross-section which is on average higher by two orders of magnitude compared to the other reaction channels. It is interesting to note that for the most part of the elementary fusion reaction processes, the overcome of the Coulomb barrier between the nuclei occurs through a quantum tunneling effect.\\
\indent The D-T reactions are also interesting when considering the coupling with the following one:
\begin{eqnarray}
	n + \:^{6}Li \qquad\Longrightarrow\qquad ^{3}T + \:^{4}He+4.8~\rm{MeV} \label{cap1eq3}
\end{eqnarray}
The nuclear reactions \ref{cap1eq1} and \ref{cap1eq2} suggest in fact that the energetic neutrons produced by D-T reaction \ref{cap1eq1}, could interact, within an appropriate structure that is usually referred to as blanket, with lithium atoms, thus regenerating the tritium. This would end in a twofold advantage: the first one relates to the possibility of adopting a closed cycle for the tritium in a future fusion reactor, while the second one refers to the creation of a thermal energy source associated with the neutron-heated lithium, hence usable for electricity production purposes.\\
\indent The internal total energy of a thermonuclear plasma can be defined as:
\begin{eqnarray}
	W=\frac{3}{2} \int dr^3\left(n_iT_i+n_eT_e\right) \label{cap1eq4}
\end{eqnarray}
The famous energy confinement time can then be formally expressed as:
\begin{eqnarray}
	\tau_e=\frac{W}{P_{in}-dW/dt}  \label{cap1eq5}
\end{eqnarray}
where $P_{in}$ indicates the external input power. The break-even condition simply foresees that the nuclear fusion power overcomes the external input power: this is usually expressed by the so-called Q-factor, i.e. $Q=\frac{P_{fus}}{P_{in}}>1$. On the other hand, the ignition corresponds to a self-sustaining reaction, i.e. $Q\rightarrow\infty$, which physically means that the nuclear fusion power associated to the $\alpha$-particles balances the power losses. Historically, this criterion has been translated into an operational condition, the famous Lawson's triple product:
\begin{eqnarray}
	n_iT_i\tau_e>3\cdot10^{21}~\rm{m^{-3}\:keV\:s}  \label{cap1eq6}
\end{eqnarray}
The ion temperatures which are required are typically of the order of $T_i\approx10-30~\rm{keV}$. Seeking the maximization of the product of the plasma density and the energy confinement time leads to two different strategies, that are at the present day the two ways the scientific community follows for achieving the nuclear fusion on the Earth.\\
\indent The way of the inertial confinement plans to maximize the ion density, at the detriment of the energy confinement time. Practically, the inertial fusion most probably consists of a pulsed regime of micro-esplosion of small D-T targets; the ignition is triggered, according to several different schemes, by compressional waves induced with powerful lasers or accelerated particle beams.\\
The second approach is the magnetic fusion. In this case, the ion density is much smaller with respect to a gas at the atmospheric pressure, but the energy confinement time aims to reach macroscopic scales of the order of the s. An appropriate magnetic configuration, which is not necessarily the tokamak one, is responsible to spatially confine the hot temperature plasma.



\section{Outline of this work}

The outline of this thesis work is the following. In Chapter \ref{cap2-plasma-instabilities}, the framework of the tokamak strategy to deal with confinement, hence the main plasma instabilities which are responsible for turbulent transport of energy and matter in such a system are introduced. The two principal plasma representations, the fluid and the kinetic ones, are here briefly introduced. Particular attention is devoted to the reasons why a gyro-kinetic approach has been preferred in the quasi-linear modelling. A tokamak relevant case is presented in order to highlight the relevance of a correct accounting of the kinetic wave-particle resonance: the example deals with the $T_i/T_e$ dependence of the linear instability threshold for the ion and electron modes. The Chapter \ref{cap3-quasilinear-approximation} discusses the issue of the quasi-linear response. Firstly, the derivation of the model, called QuaLiKiz, and its underlying hypotheses to get the energy and the particle turbulent flux are presented. Secondly, the validity of the quasi-linear response is verified against the nonlinear gyrokinetic simulations in order to: (a) identify the dominant characteristic times of the turbulent dynamics, (b) compare the phase relations between the fluctuating fields between the linear and the nonlinear phase, (c) examine the overall transport intensity, normalized to the intensity of the saturated potential, between the linear and the nonlinear regime. The Chapter \ref{cap4-improving-QL} deals with the model of the nonlinear saturation. In the first part, the nonlinear gyrokinetic simulations are quantitatively validated against turbulence measurements on the Tore Supra tokamak. Both the spatial $k$ and the frequency $\omega$ spectra of the density fluctuations are investigated also in terms of simplified analytical models. Consequently, the saturation model that is assumed in QuaLiKiz is presented and discussed. The Chapter \ref{cap-model-operation} is devoted to qualify the global outcomes of QuaLiKiz. Both the quasi-linear energy and the particle flux are compared to the expectations from the nonlinear simulations, across a wide scan of tokamak relevant parameters. Therefore, the coupling of QuaLiKiz within the integrated transport solver CRONOS is presented: this procedure allows to solve the time-dependent transport problem, hence the direct application of the model to the experiment. The first preliminary results regarding the experimental analysis are finally discussed.  

\chapter{Description of the plasma instabilities}
\label{cap2-plasma-instabilities}

\section{General framework}
\label{sec-general-framework} 

\indent The purpose of this first section is to give a general overview of the main instabilities which can drive anomalous transport in magnetically confined plasmas. This framework is intended to introduce the basic mechanisms which are modeled in this thesis work and will be extensively discussed in the following sections. Even if specific attention will be devoted to the toroidal magnetic configurations, this description should not suffer of a loss of generality. In fact, the two main instabilities detailed here below, namely the interchange and drift-waves, can commonly characterize the turbulence of electrically charged media in the presence of a eventual background magnetic field. \\
\indent In the first paragraph, a brief description of the charged particles trajectories in toroidal magnetic geometries is given; their characteristic drift velocities and the invariants of motion are introduced. \\
In the second paragraph the main physical features of the interchange and drift-wave instabilities are presented. The first mechanism is formally analogous the hydrodynamic Rayleigh-B\'ernard instability, and can take place even in two dimensions thanks to the non-homogeneity of the magnetic field. The second one arises from both the single particle drifts and the collective behavior of the plasma. This is essentially a three dimensional effect, relying on the plasma response to the perturbations in the direction parallel to the magnetic field. \\

\subsection{Geometry and particle motion}
\label{sec-geometry}

\begin{figure}[!htbp]
  \begin{center}
    \leavevmode
      \includegraphics[width=11 cm]{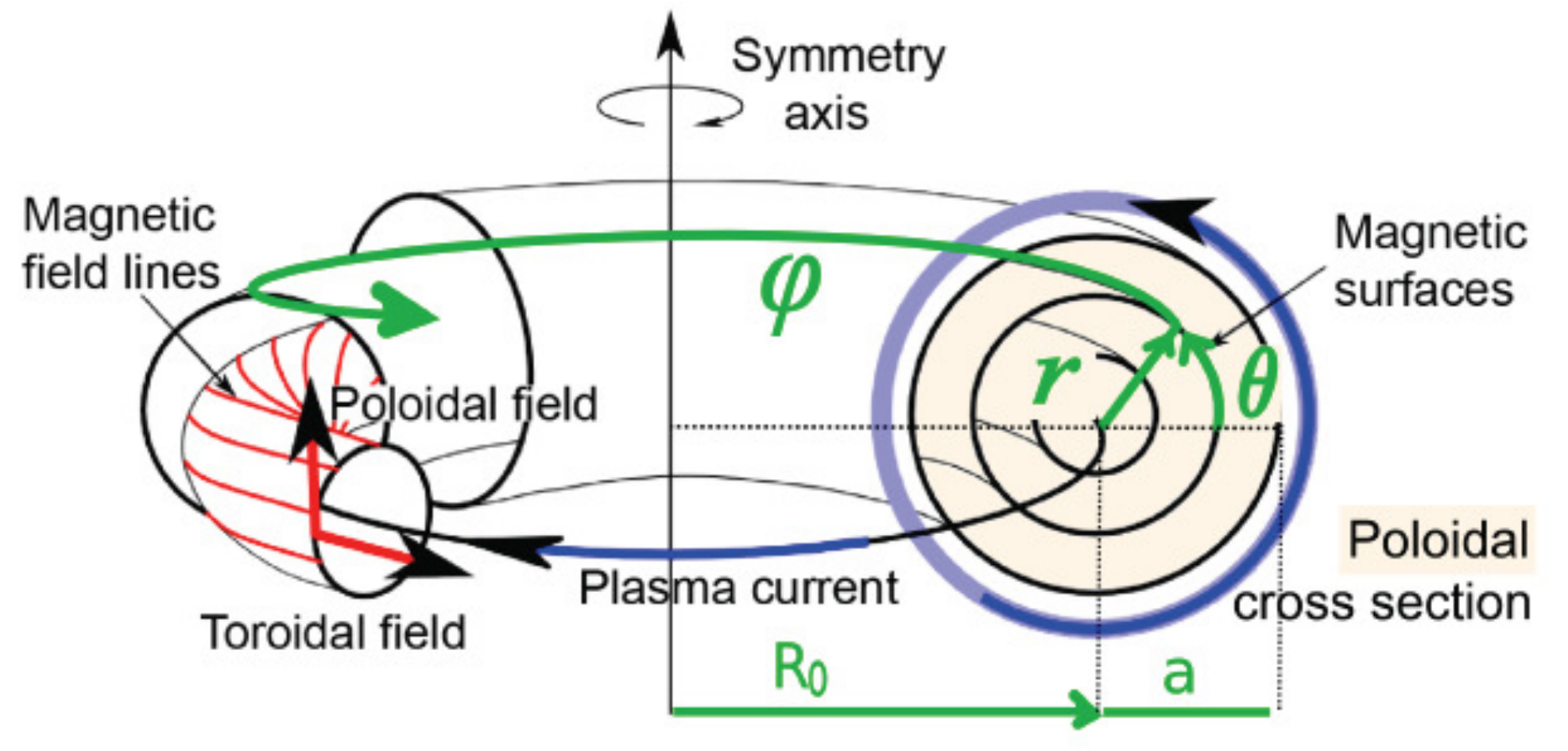}
    \caption{Scheme of the simplified magnetic geometries, characterized by circular concentric flux surfaces}
    \label{Figgeomagn}
  \end{center}
\end{figure}

\indent The tokamak magnetic geometry is axis-symmetric and consists of a series of closed nested surfaces. The toroidal component of the magnetic field is produced by external coils, while the poloidal field is originated by the current induced in the plasma. The resulting magnetic field is usually expressed by the relation:
\begin{eqnarray}
	\mathbf{B}=I(\psi)\nabla\varphi + \nabla\psi\times\nabla\varphi \label{Bfield}
\end{eqnarray}
In Eq. \eqref{Bfield} $\psi$ is the magnetic poloidal flux normalized to $2\pi$, while $\varphi$ stands for the toroidal angle and $I(\psi)$ is a flux function. The field lines are winded around the magnetic flux surfaces. An important parameter for tokamak plasmas defines the winding rate of these field lines, constant on a given flux surface, in the following way:
\begin{eqnarray}
	q(\psi)=\frac{1}{2\pi} \int^{2\pi}_{0} \frac{\mathbf{B}\cdot\nabla\varphi}{\mathbf{B}\cdot\nabla\theta} \rm{d}\theta \label{qdef}
\end{eqnarray}
where $\theta$ is the poloidal angle. $q(\psi)$ is the so called safety factor. Introducing the radial coordinate $r$ that labels a given magnetic flux surface through a simple dependence of the flux function $I=I(r)$, all the quantities of interest can the be decomposed on the orthonormal basis $(\hat{\mathbf{e}}_r,\hat{\mathbf{e}}_{\theta},\hat{\mathbf{e}}_{\varphi})$.\\
\indent The most recent tokamak configurations adopt a plasma cross-section that presents finite elongation and triangularity, with the typical D shaping, leading to improved stability. Nevertheless, a more simple circular geometry is still of great interest. Within this approximation, exemplified in Figure \ref{Figgeomagn}, the flux surfaces are assumed circular and concentric, and the magnetic field can be simply written as $\mathbf{B}=B_{\theta}\hat{\mathbf{e}}_{\theta} + B_{\varphi}\hat{\mathbf{e}}_{\varphi}$, giving:
\begin{eqnarray}
	B_{\varphi}=\frac{B_0 R_0}{R_0+r cos\theta} \approx B_0(1-\frac{r}{R_0}cos\theta) \label{Bphicomp}
\end{eqnarray}
where the $B_0$ is the reference field value on the magnetic axis and $R_0$ is the major radius of the plasma. 
In this context, it is worth noting that the magnetic field $\mathbf{B}$ can be as well expressed through the so called Clebsch representation \cite{kruskal58}: the latter is characterized by the definition of the new variable $\zeta=\varphi-q(\psi)\theta$. This advantage of this procedure follows from the fact that the magnetic field can now be written as:
\begin{eqnarray}
	\mathbf{B}=\nabla\eta \times \nabla\psi \label{Bfalign}
\end{eqnarray}
which satisfies $ \mathbf{B} \cdot \nabla\zeta = \mathbf{B} \cdot \nabla\psi=0$. The new coordinates system $(\psi,\theta,\zeta)$ appears then to be a field-aligned representation, where the variable $\theta$ refers simultaneously to (a) an angle in the poloidal plane (at fixed $\varphi$), or (b) a parametrization of distance along the field line (at fixed $\zeta$). The field-aligned coordinates systems appears particularly useful when applied to numerical simulations of tokamak microturbulence: examples will be provided in the following of this thesis work.\\

\noindent \textbf{Particle motion}\\

\indent Under the effect of the just described magnetic field, a charged particle with mass $m$ and electric charge $e$ undergoes a cyclotron gyromotion, which is characterized by typical spatial and temporal scales: these are respectively the Larmor cyclotron gyroradius $\rho_c=mv_{\perp}/eB$ and the gyrofrequency $\omega_c=eB/m$, where $v_{\perp}$ is the amplitude of the particle velocity transverse to $\mathbf{B}$. The motion of this charged particle in a arbitrary time-dependent electromagnetic field cannot be analytically solved. On the other hand, these trajectories are found to be integrable under the hypothesis of scale separation. The key concept relies on the decoupling of the fast cyclotron dynamics in the plane transverse to $\mathbf{B}$, from the slow motion of the particle guiding center. In magnetic fusion applications, this limit appears to be a very good approximation, since the external field $\mathbf{B}$ can be assumed quasi-static in terms of spatial and time variations\footnote{An important exception that will not be here treated is related to the plasma heating through RF power}. In particular, the following ordering is established by the adiabatic theory \cite{Nicholson}:
\begin{itemize}
	\item The typical spatial variations of the magnetic field are negligible compared to the particle gyroradius
	  \begin{eqnarray} \rho_s \frac{\nabla B}{B} \ll 1 \end{eqnarray}
	\item The typical time variation of the magnetic field can be neglect in comparison to the particle gyrofrequency
	  \begin{eqnarray} 
	  \frac{\rm{d} B}{\rm{d} t} \approx \mathbf{v}\cdot\nabla\mathbf{B} \approx \frac{v_{th}}{R_0}B \ll    \omega_cB 
	  \end{eqnarray}
\end{itemize}
where $v_{th}$ is the thermal speed of the particle.\\
\indent Within this framework, i.e. with the hypothesis of a quasi-static electromagnetic field, three fundamental adiabatic invariants are conserved along the particles trajectories; these are:
\begin{itemize}
	\item The total energy of the particle $E$
	  \begin{eqnarray} E \equiv \frac{1}{2}mv_{G\parallel}^2 + \mu B + e\phi \end{eqnarray}
	where the presence of $\phi$ refers to the eventual presence of stationary electric potential and the index G recall the guiding center framework.
	\item The magnetic moment $\mu$
	  \begin{eqnarray} \mu \equiv \frac {mv_{\perp}^2}{2B_G} \end{eqnarray}
	In other words $\mu$ expresses the magnetic flux across the particle gyromotion; a direct consequence of the invariance of this quantity is the variation of the Larmor gyroradius along the field line.
	\item The kinetic toroidal moment $M$
	  \begin{eqnarray} M \equiv e\psi + mRv_{G\phi} \end{eqnarray}
	It is possible to show that this invariant is a direct consequence of the axis-symmetry of the system. Within the Hamiltonian mechanics formulation $M$ is in fact the conjugate momentum of the toroidal angle $\varphi$, $M \equiv \partial L/ \partial \dot{\varphi}$, where the Lagrangian of the particle $L=1/2mv^2 -e\phi -e \mathbf{A} \cdot \mathbf{v}$ ($\mathbf{A}$ is the potential vector) does not depend on $\varphi$. This is an idealization of the real tokamak plasmas, where inhomogeneities in the toroidal magnetic field are present due to the finite numbers of coils.
\end{itemize}
\indent At this point it is particularly useful to use the just described invariants in order to introduce the normalized velocity-space coordinates, defined as
\begin{eqnarray} \mathcal{E}=\frac{E}{T_s} \end{eqnarray}
\begin{eqnarray} \lambda=\frac{m_s \mu}{E} \end{eqnarray}
\begin{eqnarray} \varsigma=sgn(v_{s\parallel}) \end{eqnarray}
such that $(\mathcal{E},\lambda)$ are again unperturbed constants of motion, while $\varsigma$ is the sign of the parallel velocity. This notation helps in highlighting a general feature of the dynamics of charged particles in non-uniform magnetic fields, i.e. the magnetic mirror phenomenon. Due to the radial dependence of $B$ in tokamak plasmas, some particles have not enough energy in the parallel direction to undergo a complete turn in the poloidal direction. Writing the parallel velocity for a given particle with the new notation, we have
\begin{eqnarray} 
  v_{\parallel}=\sqrt{\frac{2T_s}{m_s} } \sqrt{\mathcal{E}} \sqrt{1-\lambda b(r,\theta)} \varsigma
\end{eqnarray}
where $b(r,\theta)=B(r,\theta)/B(r,0)$. It appears that the particles satisfying the condition $\lambda=b(r,\theta_0)^{-1}$ have a bouncing motion along the field line between the poloidal angles $\theta=-\theta_0$ and $\theta=+\theta_0$. Two main classes can then be distinguished, passing and trapped particles.\\
\indent The fundamental equation governing the guiding center dynamics can be derived from the non-relativistic Lorentz equation of motion: $m \rm{d} \mathbf{v}/ \rm{d}t = e(\mathbf{E} + \mathbf{v}\times\mathbf{B})$. The latter can be adopted to the evolution of the guiding center with an extra term which embeds the effect of the fluctuations of $\mathbf{B}$ at the Larmor scale, giving:
\begin{eqnarray} 
  m \frac{\rm{d} \mathbf{v_G}}{\rm{d}t} = e(\mathbf{E}+\mathbf{v_G}\times\mathbf{B})-\mu\nabla\mathbf{B} 
  \label{eqgcenter}
\end{eqnarray}
The projection of the Eq. \eqref{eqgcenter} on the transverse and parallel directions allows the understanding of some essential features of the particles dynamics in tokamak plasmas. Starting with the expansion on the perpendicular plane, four basic drifts mechanisms for the guiding center motion are identified:
\begin{itemize}
	\item The $\mathbf{E}\times \mathbf{B}$ drift
	  \begin{eqnarray} \mathbf{v_{E\times B}}=\frac{\mathbf{E}\times\mathbf{B}}{B^2} \end{eqnarray}
	\item The $\nabla B$ drift
	\begin{eqnarray} 
	  \mathbf{v}_{\nabla B} = \frac{\mathbf{B}}{eB^2} \times \left(\mu B\frac{\nabla B}{B} \right)
	\end{eqnarray}
	\item The curvature drift
	\begin{eqnarray} 
	  \mathbf{v}_c = \frac{\mathbf{B}}{eB^2}\times \left( mv_{\parallel}^2\frac{\mathbf{N}}{R} \right) 
	  \qquad \text{where } \frac{\mathbf{N}}{R} = \frac{\nabla_{\perp}B}{B} + \frac{\nabla_{\perp}p}{B^2/\mu_0}
	\end{eqnarray}
	Introducing the dimensionless ratio between the kinetic and magnetic plasma pressure $\beta \equiv p/\left(B^2/2\mu_0 \right)$, usually of the order of 0.01 in tokamaks, the second term of the quantity $\mathbf{N}/R$ is negligible, so that $\mathbf{N}/R \approx \nabla_{\perp}B / B$.
	\item The polarization drift (higher order)
	\begin{eqnarray} 
	  \mathbf{v}_p = \frac{m \mathbf{B}}{eB^2} \times \left[\frac {\rm{d}}{\rm{d}t} \left(\mathbf{v_{E\times B}}+\mathbf{v}_{\nabla B} + \mathbf{v}_c \right) \right] 
	\end{eqnarray}
\end{itemize}

The appearance of these drift motions define some fundamental properties of the tokamak plasma instabilities described in the next paragraph.

\subsection{Drift waves and interchange instabilities}
\label{sec-drift-interchange}

\indent 
The purpose of this paragraph is only to show the main physical mechanisms at the origin of the turbulence object of the modeling presented in this work. In tokamak plasmas there are in fact generally many free energy sources, originating a very wide spectrum of micro/macro-instabilities, whose exhaustive discussion is far beyond the scope of the present work. A general distinction can be made on the pressure-driven modes or the current-driven modes. The latter are typically driven by the current flowing in the parallel direction and they are usually described within the framework of fluid MHD (\textit{Magneto Hydro Dynamics}) models. Even if this class of instabilities has important consequences in tokamak plasmas, setting for example intrinsic limits on the total plasma current and pressure, these modes will not be treated here. Indeed, the quasi-linear turbulence modeling presented in this work presupposes a fixed magnetic equilibrium, where the pressure driven micro-instabilities are completely decoupled from the evolution of the macroscopic magnetic field. The self-consistent interplay between the MHD and the pressure driven instabilities has just started to be explored, and represents one of the next research challenges in the progress of understanding magnetic fusion. \\
\indent The drift-wave instabilities are a key mechanism in tokamak plasma turbulence. In the case of electrostatic turbulence, the transport is set by the fluctuations in the $E\times B$ drift velocity
\begin{eqnarray}  \delta \mathbf{v_E} = \frac{\mathbf{B} \times \nabla \delta \phi}{B^2}  \end{eqnarray}
The thermal velocities of electrons differs from the ions one by a factor $\sqrt{m_i/m_e}$; in first approximation then, the electrons can be supposed to instantaneously respond to the potential fluctuations characterized by a smaller frequency with respect to the their parallel dynamics, i.e. $\omega \ll k_{\parallel}v_{th,e}$. In the parallel force balance equation for electrons
\begin{eqnarray} 
	  m_en_e\frac{\rm{d} \mathbf{v_e}}{\rm{d} t} \cdot \frac{\mathbf{B}}{B} = en_e\nabla_{\parallel}\phi - \nabla_{\parallel}p_e + \frac{\nu_em_ej_{\parallel}}{e}
	\end{eqnarray}
the dominant terms are the electric field and the pressure gradient. This relation reduces then to $\nabla_{\parallel}p_e = en_e\nabla_{\parallel}\phi$, when neglecting the temperature fluctuations and in the collisionless limit. A linear Boltzmann response directly follows as:
\begin{eqnarray}  \frac{\delta n_e}{n_e} = \frac{e \delta \phi}{T_e} \label{adiabeq}
	\end{eqnarray}
Eq. \eqref{adiabeq} is usually referred as the hypothesis of adiabatic electron response. Following a fluid approach \cite{braginskii65}, the fluid dynamics equation is:
\begin{eqnarray}
	mn\left(\partial_t+\mathbf{u}\cdot\nabla\right)\mathbf{u}=nq\left(-\nabla\phi+\mathbf{u}\times\mathbf{B}\right)-\nabla p -\nabla\cdot\bar{\bar{\Pi}}
	\label{adddeqq1}
\end{eqnarray}
where $\mathbf{u}$ and $\mathbf{v}$ are respectively the fluid and kinetic velocity while the tensor $\bar{\bar{\Pi}}$ contains the non-diagonal terms of the pressure tensor $\bar{\bar{P}}=p\bar{\bar{I}}+\bar{\bar{\Pi}}=\frac{1}{m}\int d^3v\left(\mathbf{v}-\mathbf{u}\right)\otimes\left(\mathbf{v}-\mathbf{u}\right)f$. At the first order of the expansion in the parameter $\epsilon\equiv\frac{\omega}{\omega_c}\sim\frac{u_\perp k_\perp}{\omega_c}$, the perpendicular projection of the Eq. \eqref{adddeqq1} reduces to $nq\left(-\nabla\phi+\mathbf{u}_\perp\times\mathbf{B}\right)-\nabla_\perp p=0$. Hence, the perpendicular fluid velocity at the first order reads:
\begin{eqnarray}
	\mathbf{u}_\perp^{1} = \mathbf{u}_E + \mathbf{u}^* = \frac{\mathbf{B}\times\nabla\phi}{B^2} + \frac{\mathbf{B}}{B^2}\times\frac{\nabla p}{en}
	\label{dimagnvel}
\end{eqnarray}
where $\mathbf{u}^*$ is called the diamagnetic drift velocity.\\
\begin{figure}[!htbp]
  \begin{center}
    \leavevmode
      \includegraphics[width=11 cm]{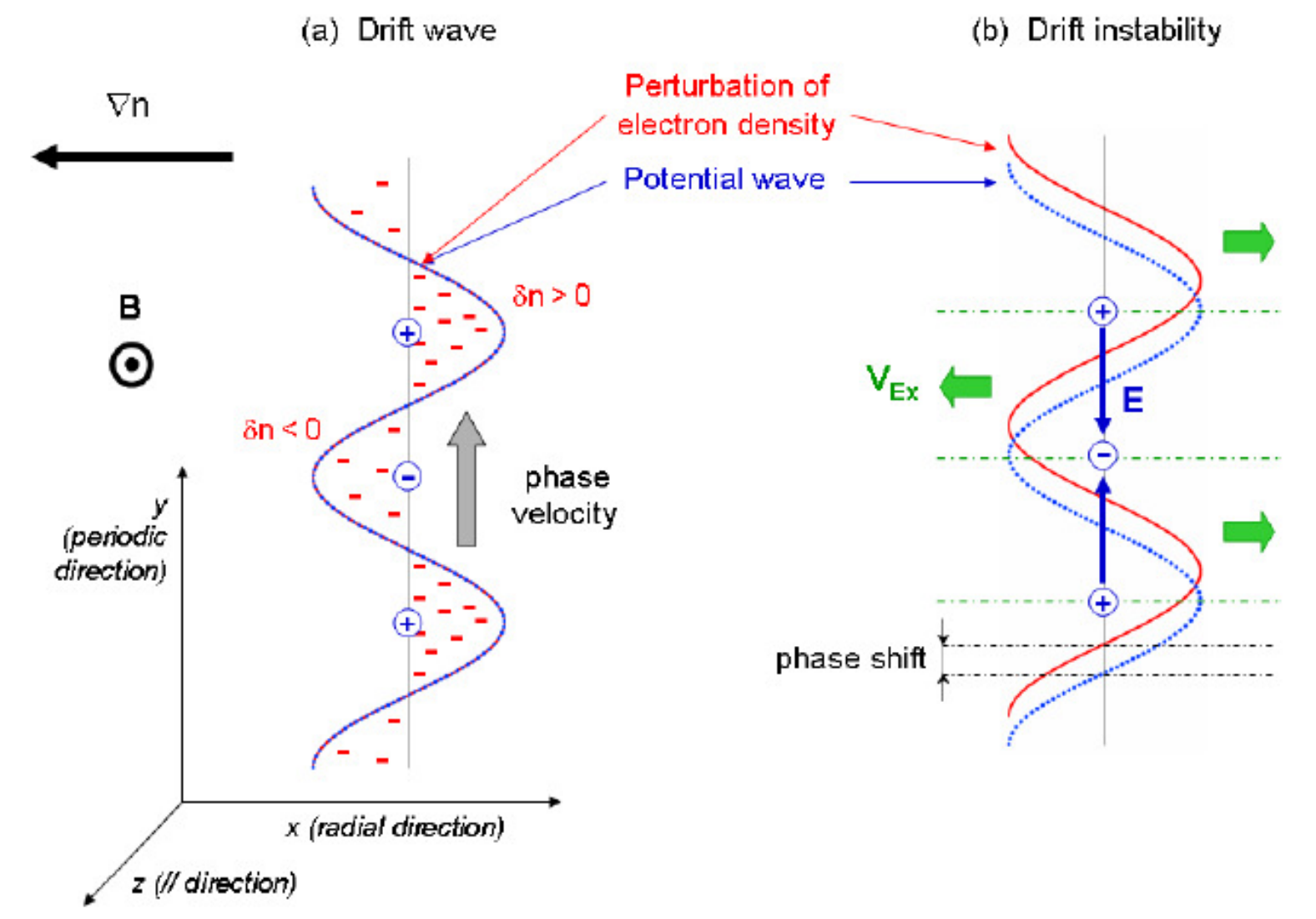}
    \caption{Exemplification of the mechanism of the drift wave (a) and the drift wave instability (b).}
    \label{driftwave}
  \end{center}
\end{figure}
Under the hypothesis of adiabatic electron response described by Eq. \eqref{adiabeq}, the electric potential and the electron density fluctuations are in phase. In the case of perturbations of the electric potential, an oscillation of the wave is produced, propagating along the perpendicular y-direction with a phase velocity. The drift wave frequency is actually of the order of the electron diamagnetic frequency, following the dispersion relation: 
\begin{eqnarray}  \omega=\omega_e^* =- \frac{k_y \rho_i}{v_{th,i}} \frac{\nabla_r n_e}{n_e}
	\end{eqnarray}
Retaining a possible phase shift between the density and potential fluctuations, turns into assuming a response that can be written as $\delta n_k/n=\left(1-i\delta_k\right)e\delta\phi_k/T_e$ ( $k$ is the wave vector in the direction transverse both to the density gradient and to the magnetic field  ). This situation is exemplified in Figure \ref{driftwave}: the radial component of the $E \times B$ drift results in a net average motion due to the non-zero phase shift between $\delta n_e$ and $\delta\phi$, in a way that an initial perturbation will be sustained and amplified. The same representation illustrates that the modes characterized by a positive phase shift, i.e. with $\delta_k < 0$, will be damped. In tokamak plasmas, there are mainly two possible mechanisms responsible for a non-vanishing phase shift: the wave-particle kinetic resonances and the breaking of the hypothesis of electrons adiabadicity (cf Eq. \eqref{adiabeq}). Both of them will be treated in the following of this work. The drift-waves represent then an instability mechanism that does not depend on the toroidal plasma geometry: for this reason the drift-wave instability defines the so called `slab' branch of several tokamak unstable modes. \\
\indent The other big class of instabilities present in the core of tokamak plasmas, are the pressure-driven family of interchange modes. The physical mechanism relies on the amount of free energy that is released, under certain conditions, by the exchange of two flux tubes around a given field line. Formally, this kind of plasma instability is analogous to the hydrodynamic Rayleigh-B\'enard instability, whose origin derives from the fact that the fluid temperature gradient is aligned with the gravitational force.  In tokamak plasmas, the interchange instability is due to both the inhomogeneity of the magnetic field (analogy with the gravity) and on the departures from the thermodynamical equilibrium through the presence of large pressure gradients (analogy with the temperature gradients). An interchange mode is then unstable only when the magnetic curvature, i.e. $\nabla B$, is aligned with the pressure gradient $\nabla p$. This condition is verified only on the low-field side of the toroidal plasma geometry, while the high-field side is stable with respect to the interchange drive: the stability condition may in fact be written as $\nabla p \cdot \nabla B<0$. The electric potential is the analogue of the hydrodynamic stream function for the Rayleigh-B\'enard problem. When the condition $\nabla p \cdot \nabla B>0$ is verified, the $E \times B$, the $\nabla B$ and the curvature drifts combine resulting into a destabilization of the convective cells, i.e. iso-contours of the electric potential perturbations.
\begin{figure}[!htbp]
  \begin{center}
    \leavevmode
      \includegraphics[width=9 cm]{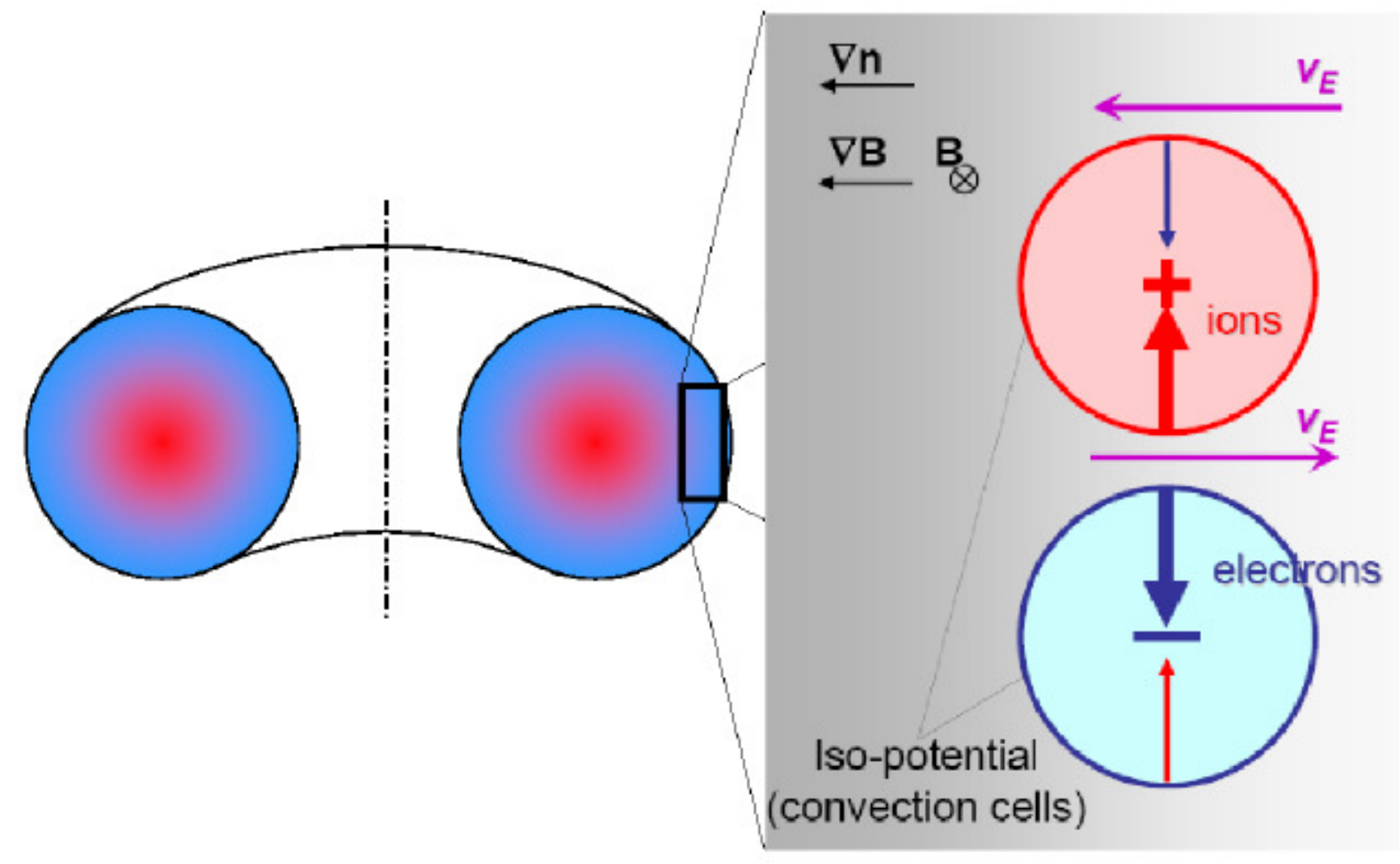}
    \caption{Exemplification of the mechanism of the interchange instability on the tokamak low-field side.}
    \label{interchange}
  \end{center}
\end{figure}
Referring to scheme of Figure \ref{interchange}, it appears that the electric drift $v_E$ changes of sign between the two cells, while the curvature and the $\nabla B$ drifts are vertical, but in opposite directions for the ions and the electrons. Globally, the particle motion leads to a positive charge accumulation in the positive cell and vice-versa, thus sustaining the electric potential and the convective cells instability.\\
The analogy between the tokamak interchange and the hydrodynamic Rayleigh-B\'enard instability is nevertheless limited by a relevant factor. This is the intrinsic anisotropy of tokamak micro-turbulence: the unstable modes tend to be aligned to the magnetic field, with a typical transverse size remarkably smaller than the parallel one. The tokamak turbulence is then often considered as quasi-2D. The fast particle motion along the magnetic field lines make that they experience both unstable and stable regions with respect to the interchange drive, respectively on low and high field sides: the parallel current appears to be stabilizing in this sense. Since the interchange instability can be excited only in presence of a non-vanishing magnetic curvature, intrinsic in the tokamak configuration, the related unstable modes define a so called `toroidal' branch.\\

\begin{figure}[!htbp]
  \begin{center}
    \leavevmode
      \includegraphics[width=8 cm]{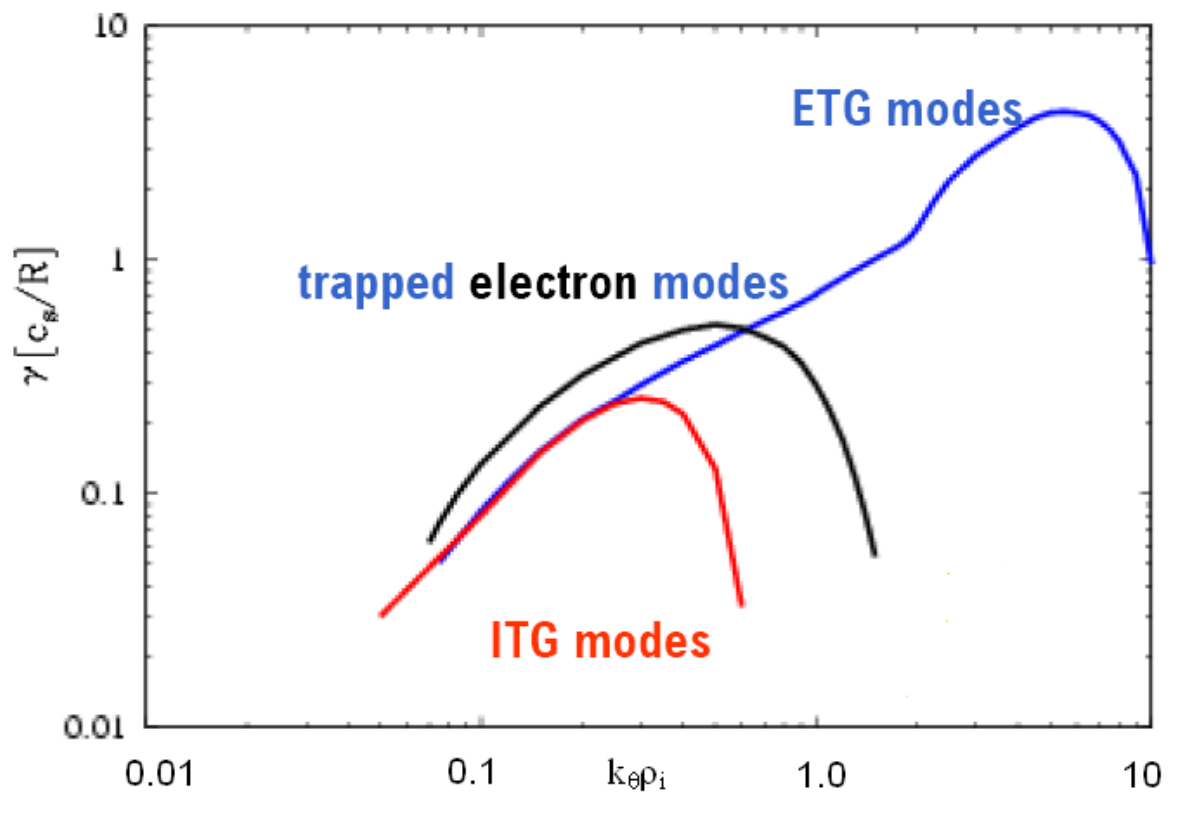}
    \caption{A typical spectrum of the linear growth rates for the ITG, TEM and ETG tokamak plasma instabilities.}
    \label{sample-ITG-TEM-ETG-sp}
  \end{center}
\end{figure}
\indent At this point it is useful to introduce the main classes of tokamak instabilities which are believed to be the main responsible for the anomalous transport of energy and particles in the plasma core. Most of them can be ascribed to the drift-wave or the interchange mechanisms. During the motion along the magnetic field lines, the trapped particles undergo the vertical drift which is at the origin of the banana shape of their trajectories. The radial extension of the banana width $\delta_{b}$ can be estimated using the conservation property of the toroidal kinetic moment $M=mRv_\phi+e\psi$, giving $\delta_b\cong\frac{2q\rho_c}{\sqrt{r/R}}$, where $\rho_c$ is the Larmor gyro-radius. For typical values of tokamak safety factor $q$ and local aspect ratio $r/R$ the following ordering is obtained:
\begin{eqnarray}
	\delta_{b,i} < \rho_{c,i} \lesssim \delta_{b,e} < \rho_{c,e}
	\label{adddeqq3}
\end{eqnarray}
Relation \eqref{adddeqq3} can be used to infer the characteristic lengths of the corresponding tokamak instabilities, i.e. in the order, TIM (trapped ion modes), ITG (ion temperature gradient), TEM (trapped electron modes) and ETG (electron temperature gradient).\\
\begin{itemize}
	\item ITG, Ion Temperature Gradient: these are electrostatic ion modes. They are also often referred to as $\eta_i$ modes, since the most relevant parameter for their turbulent drive is the ratio $\eta_i=\mathrm{d}(\log T_i)/\mathrm{d}(\log n)$. ITG modes include both interchange and drift-wave instabilities, namely (1) slab modes, (2) toroidal modes and (3) trapped ion modes. Their characteristic wavelength is bigger or comparable than the ion Larmor radius, such that $k_{\theta}\rho_i\lesssim1.0$.
	\item TEM, Trapped Electron Mode: these are electrostatic modes due to trapped electrons active at ion spatial scales. Their free energy source can be $\nabla T_e$ as well $\nabla n_e$ and they are usually distinguished between (1) collisionless and (2) dissipative (due to collisions) trapped electron modes. The precise limit on their characteristic wavelength is somewhat arbitrary, since they largely overlap with ITG modes, but a reasonable criterion can be expressed as $k_{\theta}\delta_e \lesssim 1.0$ (where $\delta_e$ is the typical radial width of the trajectories of the trapped electrons).
	\item ETG, Electron Temperature Gradient: these are electrostatic electron modes analogous to the ionic $\eta_i$ ones. Their free energy sources are again both $\nabla T_e$ and $\nabla n_e$, i.e. $\eta_e=\mathrm{d}(\log T_e)/\mathrm{d}(\log n)$, but their typical wavelengths are on the electron rather than the ion scales, then $k_{\theta}\rho_e\lesssim1.0$.
	\item Electromagnetic modes: these can be drift-wave or interchanges instabilities driven by the fluctuations of the full electromagnetic potential or micro tearing modes.
	\item Fluid-like modes: these can unstable modes active in the plasma periphery and driven by $\nabla p$, as the resistive balloning modes.
\end{itemize}

\section{Frameworks for the tokamak micro-turbulence}
\label{sec-kinetic-approach}

\indent In the last section, we have introduced the general framework and the basic mechanisms which are at the origin of the anomalous transport of energy and particles in the core of tokamak plasmas. In order to progress towards a formulation of a transport model able to gain reliable predictions, it is important to detail the plasma representation. 
Historically, the plasma dynamics has been described either within a fluid or a kinetic representation. The aim of the present section is to justify the relevant reasons for which the quasi-linear model here proposed, adopts a kinetic description. The latter is in fact a unique feature among all the other actual quasi-linear tokamak transport models, which are instead mainly based on a fluid description.

\subsection{Kinetic and fluid approaches}
\label{sec-kinetic-fluid}

\indent One of the major concern in the representation of the plasma dynamics is the issue of coherence: this kind of constraint is at the origin of relevant increase of complexity of the plasma turbulence with respect to more common neutral fluids turbulence. The motion of the charged particles in fact, induce electromagnetic fields which back-reacts on the charge and current particle densities according to the Maxwell equations.\\
A first level of description could try to directly deal with the time evolution of every single $i$ particle trajectory, according to the dynamics equation $\mathbf{\dot{v}}_i(t)=q_s/m_s\left(\mathbf{E}^m+\mathbf{v}_i(t)\times \mathbf{B}^m\right)$. Here the microscopic electric $\mathbf{E}^m$ and magnetic fields $\mathbf{B}^m$ have to be simultaneously coherent to the positions and the velocity of the particles themselves through the Maxwell equations. The treatment of this kind of problem, which is at the origin of the Klimontovich equation \cite{Nicholson}, appears as completely untractable since it would end up in tracking typical numbers of $\rm{10^{23}}$ particles.\\
\indent The starting point of the kinetic representation is the introduction of a phase space which contains a significant statistical information of the given system. The six-dimensional phase space takes the form $(\mathbf{q},\mathbf{p})$,where $\mathbf{q}=(\mathbf{q}_1,\mathbf{q}_2,\mathbf{q}_3)$ is a three-dimensional spatial basis, and $\mathbf{p}=(\mathbf{p}_1,\mathbf{p}_2,\mathbf{p}_3)$ is a momentum basis. A distribution function $f$ is then introduced in oder to represent the probability density for the system to be found in the elementary phase space volume $\rm{d} \mathbf{q} \rm{d} \mathbf{p}$, i.e. $f\left(\mathbf{q},\mathbf{p},t\right) \equiv \left\langle N\left(\mathbf{q},\mathbf{p},t\right) \right\rangle$, where the brackets represent an ensemble average.
Within the Hamiltonian mechanics formulation, it can be shown \cite{Nicholson} that the particle approach following from the Klimontovich equation, lead to a master Boltzmann kinetic equation in the form:
\begin{eqnarray}  \frac{\mathrm{d} f}{\mathrm{d} t} = \frac{\partial f}{\partial t} \left( \mathbf{\dot{q}}\cdot \partial_{\mathbf{q}} + \mathbf{\dot{p}}\cdot \partial_{\mathbf{p}} \right)f = \frac{\partial f}{\partial t} - \left[ H,f \right] = \mathcal{C}\left( f \right) 
  \label{kinmaster}
	\end{eqnarray}
where the right term $\mathcal{C}\left( f \right)$ is a generic collision operator, and the general expression for the Hamiltonian is:
\begin{eqnarray}
	H = \frac{1}{2}mv_\parallel^2+e\phi+\mu B
\end{eqnarray}
Depending on the form of this collision operator, Eq. \eqref{kinmaster} is also referred to as the Fokker-Plank equation.
Concerning the application of Eq. \eqref{kinmaster} to tokamak plasmas, two relevant issues has to be stated:
\begin{itemize}
	\item When dealing with hot thermonuclear grade plasmas, the mean free path of the particles is usually very large and it can be found that the collision rate is very small. A widely used approximation of Eq. \eqref{kinmaster} relies then on considering the collisionless problem $\mathrm{d} f/\mathrm{d} t = \mathcal{C}\left( f \right)=0$: this hypothesis defines the so called Vlasov equation
	\begin{eqnarray}  \frac{\mathrm{d} f}{\mathrm{d} t} = \frac{\partial f}{\partial t} - \left[ H,f \right] = 0 
  \label{Vlasoveq}
	\end{eqnarray}
	\item The system defined though the full kinetic equation \eqref{kinmaster} is still of almost untractable complexity, since it involves a 6$\times$M dimensional space (where M is the number of species in the plasma; often M=3, i.e. ions, electrons and one impurity). In the case of strongly magnetized plasma, a further significant simplification of the kinetic \eqref{kinmaster} or Vlasov \eqref{Vlasoveq} equations is the gyrokinetic approximation. Essentially this relies on a scale separation argument, averaging the fast gyromotion of the charged particles along the field lines, allowing to pass from a 6$\times$M to a 5$\times$M dimensional space.
\end{itemize}
\indent Even if the kinetic, and more in particular the gyrokinetic approximation, is of great interest in magnetic fusion plasmas and will be extensively used in this work, this approach is still far to be trivial both from the analytical and also from the numerical point of view. This complexity has been at the origin of another major step in the simplification process, defining a fluid approach. Here, only the hierarchy of moments of the kinetic equation \eqref{kinmaster} is considered, through a projection of the latter expression on the velocity basis $\left( 1,v,v^2...v^k \right)$. It immediately appears that, thanks to the integration over the velocity space $\mathrm{d}v^3$, the advantage of this kind of representation is the further dimensional space reduction from 6$\times$M to 3$\times$M. Consequently, this kind of approach allows a more tractable formulation from both the analytical and numerical point of view, allowing to more easily manage the equations and to understand the physical mechanisms at play. Historically this is the reason why the most part of the actual turbulent transport models deal with a more or less advanced fluid description. Nevertheless, two main drawbacks are intrinsic in the fluid approach: 
\begin{itemize}
	\item Due to the velocity space integration, the fluid moments can hardly account for the interaction between waves and particles, as long as resonances in the velocity space are present: the most relevant example is the mechanism of the linear Landau damping. This is particularly true for the hot and nearly collisionless thermonuclear plasmas. For the same reason, the fluid approach has difficulties to distinguish between passing, trapped and suprathermal particles that characterize tokamak plasmas, as well to treat the finite Larmor radius effects.
	\item The hierarchy of fluid equations obtained by higher order moments on the kinetic equation is potentially infinite; practically, a truncation of this hierarchy at a certain order in the $v^k$ moments represents a closure assumption. The latter is a crucial point for any fluid model: a large research activity in tokamak plasmas has been focused on the improvement of these closures, aiming to recover the most relevant effect of the kinetic approach. Even if this area will not be deepened in this work, the most relevant class of advanced fluid closure is the so called Gyro-Landau-fluid set of equations, which is widely used for a number of both quasi-linear transport model and also nonlinear numerical codes. The problem of the fluid closure for nearly collisionless plasmas remains nowadays an open issue and a subject of active research.
\end{itemize}

\subsection{The gyro-kinetic approximation}
\label{sec-gyrokinetic}

\indent Thanks to quasi-periodicity of the charged particles trajectories in tokamak plasmas, the Hamiltonian $H$ of the system can be described in terms of the angle-actions variables $(\boldsymbol{\alpha},\mathbf{J})$ such that the Hamilton equations are respected:
\begin{eqnarray}  \dot{\boldsymbol{\alpha}} = \frac{\partial H}{\partial \mathbf{J}} = \mathbf{\Omega_J}
	\end{eqnarray}
\begin{eqnarray}  \dot{\mathbf{J}} = \frac{\partial H}{\partial \boldsymbol{\alpha}} = 0
	\end{eqnarray}
The angle-action variables system $\boldsymbol{\alpha},\mathbf{J}$ is here formally defined. The first pair refers to cyclotron gyro-motion:
\begin{align}
	\alpha_1 &\equiv \Omega_1t \\
	J_1 &\equiv -\frac{m}{e}\mu \\
	\Omega_1 &= -\Omega_2\oint\frac{d\theta}{2\pi Jv_\parallel}\frac{eB\left(\mathbf{x}\right)}{m} \qquad 
	J\left(\psi,\theta\right) \equiv \frac{\mathbf{B\cdot\nabla\theta}}{B}
\end{align}
The second pair is written as:
\begin{align}
	\alpha_2 &\equiv \Omega_2t \\
	J_2 &\equiv \bar{\epsilon}_{pass}e\phi_T\frac{J_3}{e} + \oint\frac{d\theta}{2\pi J}mv_\parallel \\
	\Omega_2 &= \left(  \frac{1}{2\pi}\oint\frac{d\theta}{J\sqrt{\frac{2}{m}\left(E-mB\right)}}  \right)^{-1}
\end{align}
where $\bar{\epsilon}_{pass}$ is a constant whose value is $1$ in the case of passing particles and $0$ otherwise, $\phi_T$ is the toroidal flux normalized to $2\pi$. Moreover here $\oint\equiv\int_0^{2\pi}$ for the passing particles, while for the trapped particles $\oint\equiv2\int_{-\theta_0}^{\theta_0}$, where $\theta_0$ is the angle defining the banana bouncing motion.\\
Finally the third pair is:
\begin{align}
	\alpha_3 &\equiv \Omega_3t \\
	J_3 &\equiv M = e\psi+mRv_{gc,\varphi} \\
	\Omega_3 &= \left\langle \Lambda \right\rangle_{\alpha_2} + \bar{\epsilon}_{pass}q\left(\psi\right)\Omega_2
\end{align}
where $gc$ stands for guiding center, $\psi$ for the reference magnetic surface around which the guiding center is evolved and  $\Lambda\equiv\mathbf{v}_{gc,\perp}\cdot\nabla\varphi-q\left(\psi\right)\mathbf{v}_{gc,\perp}\cdot\nabla\theta+J v_\parallel\psi\partial_\psi q$.\\	
This formulation allows to easily recover the integrability of the system and the existence of three invariants along the $J_i$ trajectories. According to a perturbative approach, it is then possible to rewrite the Hamiltonian $H$ and the distribution function $f$ as:
\begin{eqnarray}  H\left( \boldsymbol{\alpha},\mathbf{J},t \right) = H_0\left( \mathbf{J},t \right) + \sum_{\mathbf{n},\omega} \delta h_{\mathbf{n},\omega}\left( \mathbf{J} \right)e^{i\left(\mathbf{n}\cdot \boldsymbol{\alpha}-\omega t \right)}
	\label{HFourier}
	\end{eqnarray}
\begin{eqnarray}  f\left( \boldsymbol{\alpha},\mathbf{J},t \right) = f_0\left( \mathbf{J},t \right) + \sum_{\mathbf{n},\omega} \delta f_{\mathbf{n},\omega}\left( \mathbf{J} \right)e^{i\left(\mathbf{n}\cdot \boldsymbol{\alpha}-\omega t \right)}
	\label{fFourier}
	\end{eqnarray}
where the time evolution of the equilibrium quantities is supposed much slower than the perturbations one. At a reference equilibrium, the Vlasov problem $\partial f_0-\left[H_0,f_0\right]=0$ will define a certain distribution function $f_0$, whose form is a priori fixed also by Coulomb collisions: this configures the neoclassical theory problems, which describes the collisional transport apart from any turbulent fluctuation and will not deepened here. Within this framework, a local Maxwellian will be adopted:
\begin{eqnarray}  f_0\left( \mathbf{J},t \right) = \frac{n\left(t\right)}{\left( 2\pi mT \left(t\right) \right)^{3/2}}e^{-H_0\left(\mathbf{J}\right)/T\left(t\right)}
	\label{Boltzmannf}
	\end{eqnarray}
where $n$ and $m$ are respectively the particle density and mass.\\
\indent It is of interest to derive a linearised expression for the response of the perturbation $\delta f$ to small fluctuations. At the first order, the Vlasov equation \eqref{Vlasoveq} can be then be rewritten as:
\begin{eqnarray}  \partial_t \delta f - \left[H_0,\delta f\right] - \left[\delta h,f_0\right] = 0
	\end{eqnarray}
It is important to note that this results follows from the scale separation on the temporal dynamics, i.e. using the quasi-stationary condition for the equilibrium distribution function $\partial_t f_0 = \left[ H_0,f_0 \right]=0$. Using the Fourier spectral decomposition of Eq. \eqref{HFourier} and \eqref{fFourier}, the following important relation for the linear response of $\delta f$ is derived:
\begin{eqnarray}  \delta f_{\mathbf{n},\omega}\left( \mathbf{J} \right) = -\frac{\mathbf{n} \cdot \partial_{\mathbf{J}}f_0}{\omega - \mathbf{n} \cdot \partial_{\mathbf{J}}H_0 + i0^+} \delta h_{\mathbf{n},\omega}\left( \mathbf{J} \right)
	\label{linfresp}
	\end{eqnarray}
where a resonance appears at the frequency $\omega=\mathbf{n} \cdot \partial_{\mathbf{J}}H_0=\mathbf{n} \cdot \boldsymbol{\Omega}_{0\mathbf{J}}$. The presence of this singularity is at the origin of the Landau damping, a crucial kinetic mechanism which implies an energy transfer between the waves and the particles. Formally, the expression does still have a validity in the mathematic sense of distributions, when eliminating the indetermination on the position of the pole with respect to the real axis. Physically, the solution corresponds to a causality constraint, imposing null perturbations for $t\rightarrow-\infty$; mathematically, this is done through the addition of the imaginary term $i0^+$, where $0^+$ is small but strictly positive. The presence of the Landau resonance in the linear kinetic response will be deepened also in the following of this work, about the formulation of the quasi-linear transport model.\\
A further step with respect to Eq. \eqref{linfresp} can be done within the hypothesis of an equilibrium Boltzmann distribution $f_0$ of the type of \eqref{Boltzmannf}, leading to:
\begin{eqnarray}  \delta f_{\mathbf{n},\omega}\left( \mathbf{J} \right) = -\frac{f_0\left( \mathbf{J} \right)}{T} \left(1- \frac{\omega - \mathbf{n} \cdot \boldsymbol{\Omega}^*}{\omega - \mathbf{n} \cdot \boldsymbol{\Omega}_{0\mathbf{J}} + i0^+} \right) \delta h_{\mathbf{n},\omega}\left( \mathbf{J} \right)
	\label{linfrespB}
	\end{eqnarray}
where
\begin{eqnarray}  \boldsymbol{\Omega}^* = T \left[ \frac{\mathrm{d}}{\mathrm{d}\mathbf{J}}log\left(n\right) + \left(\frac{H_0}{T}-\frac{3}{2}\right)\frac{\mathrm{d}}{\mathrm{d}\mathbf{J}}log\left(T\right) \right]
	\label{gendiamagnv}
	\end{eqnarray}
Eq. \eqref{gendiamagnv} is the expression for the generalized diamagnetic frequency, in analogy with the definition found by Eq. \eqref{dimagnvel}. The linear response \eqref{linfrespB} is particularly meaningful from the physical point of view: the first term corresponds to the adiabatic response of the distribution function, while the second one represents the non-adiabatic component. It will be shown that only the latter term originates a turbulent transport via the fluctuations.\\

\indent The gyrokinetic theory was introduced in the attempt of describing the strongly magnetized plasma dynamics over time scales that set apart from the fast gyromotion. In the case of tokamak plasmas, where the spatial and temporal variations of the external magnetic field are weak, the charged particles undergo 3 types of quasi-periodic motions: (1) the fast gyromotion along the magnetic field lines, (2) an intermediate bounce motion along the parallel direction due to the parallel gradients and (3) a slow motion across the field lines driven by magnetic curvature and the transverse gradients. Formally, the gyrokinetic theory is an improvement with respect to the gyro-center coordinates, introducing a new set of gyrocenter coordinates which account for a gyro-averaged perturbed dynamics. A complete overview of the modern formulation of the gyrokinetic theory can be found in Ref. \cite{brizard07}.\\
\indent The traditional derivation of the gyrokinetic approach follows from a number of fundamental assumptions on the spatial and temporal ordering, comparable to the ones leading to the adiabatic theory. The parameter $\epsilon$, ratio between the particle (ion) Larmor radius $\rho_c$ and a macroscopic length L (plasma radius or the density gradient length $n/\left|\nabla_r n\right|$) is defined:
\begin{eqnarray}
	\epsilon = \frac{\rho_c}{L} \ll 1
\end{eqnarray}
Three basic frequency scales exist: the fast cyclotron motion at the frequency $\omega_c$, a medium frequency which is the typical one for the turbulent fluctuations $\omega_{fl}\sim\dfrac{v_{th}}{L}\sim\epsilon\omega_c$ and the slow frequency of the macroscopic transport $\omega_{tr}\sim\dfrac{v_{th}}{L}\epsilon\sim\epsilon^3\omega_c$.\\
The distribution function and the fields are split into a slowly varying (in time and space) equilibrium part and a fast varying fluctuating part, such that
\begin{align}
   f\left(\mathbf{x},\mathbf{v},t\right)=f_0\left(\mathbf{x},\mathbf{v},t\right)+\delta f\left(\mathbf{x},\mathbf{v},t\right) 
   \qquad &\mathbf{B}\left(\mathbf{x},t\right)=\mathbf{B}_0\left(\mathbf{x},t\right)+\delta \mathbf{B}\left(\mathbf{x},t\right)
   \\ &\mathbf{E}\left(\mathbf{x},t\right)=\delta \mathbf{E}\left(\mathbf{x},t\right)
\end{align}
The gyrokinetic ordering foresees:
\begin{enumerate}
	\item Small fluctuations, of the order of $\epsilon$ 
	\begin{eqnarray}
   \frac{\delta f}{f_0} \sim \frac{\left|\delta\mathbf{B}\right|}{\left|\mathbf{B}_0\right|} \sim
   \frac{\left|\delta\mathbf{E}\right|}{\left|v_{th}\mathbf{B}_0\right|} \sim \epsilon
\end{eqnarray}
  \item Slowly varying equilibrium, on the macroscopic length and time scales ($\tau_{tr}\sim\frac{L}{\epsilon^2v_{th}}$ is the macroscopic transport time scale)
  \begin{eqnarray}
   \nabla f_0\sim\frac{f_0}{L} \qquad \nabla \mathbf{B}_0\sim\frac{\mathbf{B}_0}{L} \\
   \frac{\partial f_0}{\partial t} \sim \frac{f_0}{\tau_{tr}} \qquad \frac{\partial \mathbf{B}_0}{\partial t} \sim \frac{\mathbf{B}_0}{\tau_{tr}}
\end{eqnarray}
  \item Fast spatial fluctuations across $\mathbf{B}_0$, on the microscopic length $\rho_c$
   \begin{eqnarray}
   \left|\frac{\mathbf{B}_0}{B_0}\times\nabla\delta f\right| \sim \frac{\delta f}{\rho_c} \qquad
   \left|\frac{\mathbf{B}_0}{B_0}\times\nabla\right|\delta\mathbf{B} \sim \frac{\delta\mathbf{B}}{\rho_c} \qquad
   \left|\frac{\mathbf{B}_0}{B_0}\times\nabla\right|\delta\mathbf{E} \sim \frac{\delta\mathbf{E}}{\rho_c}
\end{eqnarray}
  \item Slowly varying fluctuations along $\mathbf{B}_0$, on the macroscopic length $L$
   \begin{eqnarray}
   \frac{\mathbf{B}_0}{B_0}\cdot\nabla\delta f \sim \frac{\delta f}{L} \qquad
   \frac{\mathbf{B}_0}{B_0}\cdot\nabla\delta\mathbf{B} \sim \frac{\delta\mathbf{B}}{L} \qquad
   \frac{\mathbf{B}_0}{B_0}\cdot\nabla\delta\mathbf{E} \sim \frac{\delta\mathbf{E}}{L}
\end{eqnarray}
  \item Fluctuations varying on a medium time scale (the fluctuations time scale is $\tau_{fl}\sim\frac{L}{v_{th}}$)
  \begin{eqnarray}
   \frac{\partial \delta f}{\partial t} \sim \frac{\delta f}{\tau_{fl}} \qquad
   \frac{\partial \delta \mathbf{B}}{\partial t} \sim \frac{\delta \mathbf{B}}{\tau_{fl}} \qquad
   \frac{\partial \delta \mathbf{E}}{\partial t} \sim \frac{\delta \mathbf{E}}{\tau_{fl}}
\end{eqnarray}
   \item Collisions act on the medium time scale of the turbulent fluctuations
   \begin{eqnarray}
   \nu\sim\omega_{fl} \qquad \rm{hence} \qquad \frac{\nu}{\omega_c}\ll1
\end{eqnarray}
\end{enumerate}
These ordering are then defined splitting the distribution function and the fields as a slowly varying (in time and space) equilibrium part, characterizing the background quantities, and the fast varying fluctuating parts. Since the information on the fast gyromotion dynamics is not relevant in this framwork, it is of interest to define an averaging procedure in the Fourier space:
\begin{eqnarray}  \int_0^{2\pi}\frac{\mathrm{d}\theta'}{2\pi}e^{\mathbf{k}\cdot \mathbf{\rho}_s} = J_0\left(k_{\perp}\rho_s\right)
	\end{eqnarray}
where $J_0$ is the Bessel function. Finally, within the gyrokinetic framework, the general Vlasov equation \eqref{Vlasoveq} takes the following form for a given plasma species $s$:
\begin{eqnarray}  \frac{\partial \bar{f}_s}{\partial t} + \left( \mathbf{v_{E\times B}} + \mathbf{v}_{\nabla B} + \mathbf{v_c} \right) \cdot \nabla_{\perp}\bar{f}_s + v_{\parallel}\nabla_{\parallel}\bar{f}_s + \dot{v_{\parallel}}\partial_{v_{\parallel}}\bar{f}_s = 0
	\label{GKeq}
	\end{eqnarray}
where the hat corresponds to a gyro-averaged quantity. As already mentioned, in order to describe a coherent problem, the plasma dynamics is constrained also by the electromagnetic Maxwell equations. The Debye length is much smaller than the scale lengths of the fluctuations which are described: for this reason, a local electro-neutrality condition can be imposed. Within this limit, Eq. \eqref{GKeq} has to be consistently solved with the Poisson-Amp\`ere equations
\begin{eqnarray}  \epsilon_0\nabla^2\phi = \sum_s e_s \int\mathrm{d}^3vf_s \qquad 
									\nabla^2_{\perp}A_{\parallel} = -\mu_0 \sum_s e_s \int\mathrm{d}^3v v_{\parallel}f_s
	\label{PAeq}
	\end{eqnarray}
\indent The solution of the time-evolving nonlinear system described by the set of equations \eqref{GKeq}-\eqref{PAeq} (mapped on a realistic toroidal magnetic geometry), could in principle provide an accurate information about the turbulence dynamics in tokamak plasmas, according to the degree of approximation adopted by the gyrokinetic formulation. Nevertheless, the complexity of the problem is not only analytically largely untractable, but nowadays, and at least for the coming 10 years, also too costly from the numerical point of view. \\

\noindent \textbf{Solving the linear gyrokinetic dispersion relation} \\
\label{Kinezero-descript}

\indent Solving the linear gyrokinetic plasma dispersion relation, i.e. the linearized gyrokinetic equation coupled to the quasi-neutrality condition, is a much easier task with respect to the nonlinear system defined by Eqs. \eqref{GKeq}-\eqref{PAeq}. Still this approach can provide a great amount of information on the tokamak plasma turbulence, hence this is the strategy which has been followed by this work.\\
This paragraph is dedicated firstly to illustrate the general structure of the gyrokinetic electrostatic linear problem, and therefore a particular formulation that is used to compute numerical solutions. Moreover, the following discussion will provide the basic framework which is employed by the quasi-linear transport model QuaLiKiz. QuaLiKiz in fact is entirely based on a linear gyrokinetic eigenvalue code, Kinezero \cite{bourdelle02}. The hypotheses and the approximations underlying Kinezero are then completely shared with QuaLiKiz.\\
\indent The electro-neutrality constraint appearing in the first of Eqs. \eqref{PAeq} can be rewritten according to a variational approach \cite{garbet90}, giving:
\begin{eqnarray}  \sum_s\mathcal{L}_s\left(\omega\right)=0 \qquad \mathrm{with} \qquad \mathcal{L}_s\left(\omega\right)=-\sum_{\mathbf{n},\omega}\int d^3x\:e_s\delta n_{\mathbf{n},\omega}^s\left(\mathbf{x}\right) \delta\phi_{\mathbf{n},\omega}^*\left(\mathbf{x}\right)   \label{kinez-eq1}
	\end{eqnarray}
The linear gyrokinetic plasma dispersion relation is obtained combining the linearized response for the distribution function Eq. \eqref{linfrespB}, with the quasi-neutrality condition Eq. \eqref{kinez-eq1}, leading to:
\begin{eqnarray}  
  \sum_s\frac{e_s^2f_0^s}{T_s}\left[\sum_{\mathbf{n}}\left\langle\ \delta\phi_{\mathbf{n},\omega} \delta\phi_{\mathbf{n},\omega}^* \right\rangle_{\boldsymbol{\alpha},\mathbf{J}} - \sum_{\mathbf{n}}\left\langle\ \frac{\omega - \mathbf{n} \cdot \boldsymbol{\Omega}^*}{\omega - \mathbf{n} \cdot \boldsymbol{\Omega}_{0\mathbf{J}} + i0^+}  \delta\phi_{\mathbf{n},\omega} \delta\phi_{\mathbf{n},\omega}^* \right\rangle_{\boldsymbol{\alpha},\mathbf{J}} \right]=0   \label{kinez-eq2}
	\end{eqnarray}
Formally, the solution of the Eq. \eqref{kinez-eq2} provides for each $\mathbf{n}$ one or more complex eigenvalues $\omega=\omega_r+i\gamma$. If $\gamma>0$, then the eigenvalue is associated with a linear unstable eigenmode which is exponentially growing in time with the finite real frequency $\omega_r$.\\
The solution of the eigenvalue problem defined by Eq. \eqref{kinez-eq2} is still not trivial even from the numerical point of view, because of the high order of the matrix which are involved. Recently, accurate eigenvalue numerical solvers for the Eq. \eqref{kinez-eq2} have been developed \cite{kammerer08,waltz09}: even if less expensive than the full nonlinear simulations, presently these codes still require a significant amount of computational resources, that make them not immediately applicable within a time-evolving integrated transport solver. Consequently, the quasi-linear transport model here proposed is based on an further approximation of the Eq. \eqref{kinez-eq2}, which is numerically solved by the code Kinezero. One of the most relevant feature of this code is in fact its significantly lower computational requirement with respect to the other gyrokinetic linear solvers. The approximations that are employed by Kinezero are then here below briefly recalled.\\
\begin{enumerate}
	\item \indent \textsl{Lowest order of the ballooning representation}. In the coordinate system $\left(r,\theta,\varphi\right)$, the hypothesis of perfect toroidal axisymmetry implies that $e^{in\varphi}$ is an eigenvector. Moreover, the tokamak micro-instabilities are characterized by the strong anisotropy such that $k_\parallel\ll k_\perp$. The ballooning representation uses these arguments in order to reduce the spatial dimensionality of the problem \cite{connor93}. In particular, the eigenvector $\delta\phi\left(r,\theta,\varphi,t\right)$ is re-written as:
\begin{eqnarray}  
   \delta\phi\left(r,\theta,\varphi,t\right)=\sum_{n,\omega}\int\frac{d\theta_0}{2\pi}\sum_{l=-\infty}^{+\infty} \delta\hat{\phi}_{n\omega\theta_0}\left(\theta+2\pi l\right)e^{i\left[n\left(\varphi-q\left(r\right)\left(\theta+2\pi l-\theta_0\right)\right)-\omega t\right]}     \label{kinez-eq4}
	\end{eqnarray}
The ballooning angle $\theta_0$ is here taken to be $\theta_0=0$, since in the most cases the unstable modes are ballooned on the low-field side. Within the lowest order of the ballooning representation, the second radial derivatives of the equilibrium quantities are neglected, i.e. $\left|d\right|<\left(\nabla_r\log A\right)^{-1}$, where $A$ stands for a generic equilibrium quantity and $d$ is the distance between two resonant surfaces $d=-\left(n\nabla_r q\right)^{-1}$. This lowest order limit is formally no longer valid when the first radial derivative of an equilibrium quantity is approaching to zero: this applies in particular when the magnetic shear $\left|s\right|\rightarrow0$.\\
  \item \indent \textsl{Trial Gaussian eigenfunction}. In order to quickly find the eigenvalues of the linear dispersion relation, Kinezero adopts a trial $\delta\phi$ eigenfunction. The latter one is chosen to be the most unstable analytical solution in the fluid limit for strongly ballooned modes. It can be shown that $\delta\phi\left(k_r\right)$ is a Gaussian in the form:
\begin{eqnarray}  
   \delta\phi\left(k_r\right)=\delta\phi_0\frac{\sqrt{w}}{\pi^{1/4}}e^{-k_r^2w^2/2}     \label{kinez-eq5}
	\end{eqnarray}
The mode width $w$ is an important parameter which is as well calculated in the fluid limit, assuming the interchange as the dominant instability. The complete expression for $w$ can be found in Ref. \cite{bourdelle02}.\\
  \item \indent \textsl{Functionals for trapped and passing particles}. The complex zeros of the Eq. \eqref{kinez-eq2}, re-written as $\mathcal{D}\left(\omega\right)=\sum_s\frac{n_sZ_s^2}{T_s}\left(1-\left\langle \mathcal{L}\right\rangle_{tr,s}-\left\langle \mathcal{L}\right\rangle_{ps,s}\right)=0$ are numerically solved, where the ion, electron and one impurity species are accounted in both their trapped \textit{tr} and passing \textit{ps} domains. The frequencies are:
\begin{eqnarray}  
   n\Omega^*_s=-\frac{k_{\theta}T_s}{e_sBR}\left[\frac{\nabla_r n_s}{n_s}+\frac{\nabla_r T_s}{T_s}\left(\mathcal{E}-\frac{3}{2}\right)\right] \label{kinez-eq6} \\
   n\Omega_{0Js} = -\frac{k_{\theta}T_s}{e_sBR}\mathcal{E}\left(2-\lambda b\right)f\left(\lambda\right) + k_{\parallel}v_{\parallel}     \label{kinez-eq6b}
	\end{eqnarray}
where $k_\theta=nq\left(r\right)/r$ is the toroidal wave-number and $f\left(\lambda\right)$ a function of $\lambda$ depending on the magnetic geometry and the MHD parameter $\alpha$, differing for trapped and passing particles \cite{bourdelle02}. In the resonance frequency both the vertical drift and the parallel motion appear, while the terms eventually coming from $\nabla v_\parallel$ and $\nabla v_\perp$ are neglected.\\
Firstly focusing on the trapped particles, the corresponding velocity integration domain, using the $\mathcal{E},\lambda$ variables, is:
\begin{align}  
   &\left. \left\langle \ldots \right\rangle_{\mathcal{E},\lambda} \right|_{\textrm{trapped}} = 
    \int_{0}^{+\infty}\frac{2}{\sqrt{\pi}}d\mathcal{E} \int_{\lambda_c}^{1}d\lambda \frac{1}{4\bar{\omega}_2}J_0^2\left(k_\perp\rho_{cs}\right)J_0^2\left(k_r\delta_s\right) \nonumber \\
    & \bar{\omega}_2 = \frac{1}{\oint\frac{d\theta}{2\pi}\frac{1}{\sqrt{1-\lambda\left(r,\theta\right)b}}}
    \qquad \lambda_c = \frac{\mu B\left(r,\theta=0\right)}{\mu B\left(r,\theta=\pi\right)}
        \label{kinez-eq7}
	\end{align}
$J_0^2\left(k_\perp\rho_{cs}\right)$ and $J_0^2\left(k_r\delta_s\right)$ are Bessel functions standing respectively for cyclotron gyroaverage and the bounce average. As the trapped particles are mainly efficient at low wave-numbers, the hypothesis that $n\ll \frac{k_r}{\pi\nabla_r q}$, therefore $k_r d\gg\pi$ is used. Moreover, because of the assumption of a local Maxwellian equilibrium, the energy integration of the Bessel functions are separated from the rest. Finally, the collisions on the trapped electrons are also retained through a Krook operator \cite{romanelli07}.\\
Concerning the passing particles velocity integration, this can be written as:
\begin{eqnarray}  
   \left. \left\langle \ldots \right\rangle_{\mathcal{E},\lambda} \right|_{\textrm{passing}} =
    \int_{0}^{+\infty}\frac{2}{\sqrt{\pi}}d\mathcal{E} \int_{0}^{\lambda_c}d\lambda \frac{1}{4\bar{\omega}_2} \frac{1}{2} \sum_{\varsigma}\varsigma J_0^2\left(k_\perp\rho_{cs}\right)    \label{kinez-eq8}
	\end{eqnarray}
where the hypothesis of $n\gg \frac{k_r}{\pi\nabla_r q}$, therefore $k_r d\ll\pi$, is used. Also the $\lambda$ integration is approximated, so that:
\begin{eqnarray}  
   n\Omega_s\approx\left\langle n\Omega_s\right\rangle_\lambda \qquad k_{\parallel}v_{\parallel} \approx \left\langle k_{\parallel}v_{\parallel}\right\rangle_\lambda = \varsigma \frac{s}{q} \frac{v_{th,s}}{R}w \sqrt{\mathcal{E}}     \label{kinez-eq9}
	\end{eqnarray}
A wide documentation on the benchmark efforts made comparing Kinezero against other linear gyrokinetic codes is reported in Refs. \cite{bourdelle02,romanelli07,bourdelle07}.\\
\end{enumerate}
	
\subsection{Choosing the kinetic approach}
\label{sec-kinvsfluidappr}

\indent In the previous paragraphs we have briefly introduced the two most common frameworks used in the description of the tokamak plasma turbulence, namely the kinetic and the fluid approach. A unique feature of the quasi-linear model proposed in this work is the use of a gyrokinetic formulation, while most of the actual quasi-linear transport model are fluid \cite{weiland00} or gyro-fluid \cite{waltz97,garbet96}. This paragraph and the next one will be devoted to provide demonstrations that supports the preference in favor of the kinetic approach. In particular it will be shown that:
\begin{itemize}
	\item The fluid approach, which intrinsically requires an arbitrary closure, typically overestimates the values of the linear threshold of the tokamak micro-instabilities.
	\item The kinetic framework consistently recovers the fluid limit.
	\item Several mechanisms which are at play in tokamak plasmas are a direct consequence of the presence of the kinetic resonances.
\end{itemize}



\indent A comprehensive understanding of the fluid approximation is based on the derivation following from the kinetic approach. The main interest of a kinetic treatment for deriving the linear unstable modes thresholds concerns the accurate treatment of the wave-particle resonance at the frequency $\omega_R$ such that:
\begin{eqnarray}  \omega_R=\partial_{\mathbf{J}}H_0\approx\omega_D\left(v_{\parallel}^2,v_{\perp}^2\right)+k_{\parallel}v_{\parallel}
	\end{eqnarray}
where $\omega_D$ is the magnetic (curvature and $\nabla \mathbf{B}$) drift frequency, $k_{\parallel}$
 is the parallel wave vector, while $v_{\parallel}$ and $v_{\perp}$ are, respectively, the parallel and perpendicular velocities. We can assume that the magnetic drift component of the resonance will be dominant on the instability growth rates, since the parallel dynamics is subdominant for $s/q \ll 2$ \footnote{$s$ is the so called magnetic shear, defined as $s=r \nabla_r q/q$}. Neglecting the $k_{\parallel}v_{\parallel}$ term in the resonance and the finite Larmor radius (FLR) effects, the linear kinetic density response for the $s$ species $\delta n_s$ directly follows from the linear response \eqref{linfrespB} and can be written in the following way:
\begin{eqnarray}  \frac{\delta n_s}{n_s}=-\frac{e_s \delta \phi}{T_s}\left[1-\left\langle \frac{\omega-\omega_s^*}{\omega-\omega_{Ds}} \right\rangle_{\mathcal{E},\lambda} \right]
	\label{gykinrespapp} \end{eqnarray}
where the characteristic frequencies are expressed as $\omega_s^*=\omega_{ns}^*+\omega_{Ts}^*\left(\mathcal{E}-3/2\right)$, $\omega_{Ds}=\lambda_s\omega_{DTs}\mathcal{E}$ and $\lambda_s=\cos \chi+s\chi\sin \chi$ ($\chi$ is a coordinate along the magnetic field line). The resonance and the diamagnetic frequencies are analogous to the ones already introduced by Eqs. \eqref{kinez-eq6}-\eqref{kinez-eq6}; here the vertical drift and the diamagnetic terms driven by the density and the temperature gradients are distinguished, such that:
\begin{eqnarray}  \omega_{DTs}=2k_{\theta}\frac{T_s}{e_s B}\frac{\nabla_r B}{B} \end{eqnarray}
\begin{eqnarray}  \omega_{ns}^*=k_{\theta}\frac{T_s}{e_s B}\frac{\nabla_r n_s}{n_s} \qquad \omega_{Ts}^*=k_{\theta}\frac{T_s}{e_s B}\frac{\nabla_r T_s}{T_s} \end{eqnarray}
where $k_{\theta}$ is the poloidal wave-number.\\
\indent Dealing firstly with ion ITG modes, a very crude fluid limit (without any closure) can be derived from the linear kinetic density response through a second order expansion of the Eq. \eqref{gykinrespapp} based on the condition $\omega_{Di}/\omega \ll 1$, i.e. considering that the mode frequency is far from the resonance. The resulting ion response $\delta n_i/n_i$ has then to be coupled to the quasi-neutrallity condition $\delta n_i=\delta n_e$ and assuming adiabatic electrons ($\delta n_e/n_e=e\delta \phi/T_e$). The following second order dispersion relation is then derived:
\begin{eqnarray}  \frac{T_i}{T_e}\omega_{DTi}^2\left(\frac{\omega}{\omega_{DTi}}\right)^2 + \omega_{DTi}\left(\omega_{ni}^*-\frac{3}{2}\omega_{DTi}\right) \left(\frac{\omega}{\omega_{DTi}}\right) + \nonumber \\
+ \frac{3}{2}\omega_{DTi}\left(\omega_{ni}^*+\omega_{Ti}^*\right) - \frac{15}{4}\omega_{DTi}^2=0
  \label{ITGfluidisp} \end{eqnarray}
Since at the threshold the imaginary part of $\omega$ is vanishing, the following expression for the critical ITG $R/L_{Ti}$ can be obtained:
\begin{eqnarray}  \left.\frac{R}{L_{Ti}}\right|_{th}^{\mathrm{crude\:fluid}} = \frac{R}{L_{n}}\left(-1-\frac{1}{2}\frac{T_e}{T_i}+\frac{1}{12}\frac{T_e}{T_i}\frac{R}{L_{n}}\right)+\frac{3}{4}\frac{T_e}{T_i}+5  \label{ITGcrfluithr} \end{eqnarray}
\indent It is interesting to compare the latter expression for the ITG threshold obtained by a primitive fluid expansion, with a more sophisticated fluid model which accounts for a particular closure. Leaving the derivation details in the appendix \ref{appx1-TiTe}, the advanced fluid Weiland model \cite{weiland00} operates through this strategy, providing the following expression for the ITG linear threshold:
\begin{eqnarray}  \left.\frac{R}{L_{Ti}}\right|_{th}^{\mathrm{Weiland}} = \frac{R}{L_{n}}\left(\frac{2}{3}-\frac{1}{2}\frac{T_e}{T_i}+\frac{1}{8}\frac{T_e}{T_i}\frac{R}{L_{n}}\right)+\frac{1}{2}\frac{T_e}{T_i}+\frac{20}{9}\frac{T_i}{T_e}  \label{ITGadWfluithr} \end{eqnarray}
When comparing Eq. \eqref{ITGcrfluithr} and Eq. \eqref{ITGadWfluithr}, it appears that a relevant difference is carried by the parametric dependence of the thresholds on the ratio $T_i/T_e$: in particular, while the Weiland ITG threshold increases for higher $T_i/T_e$, the opposite behavior is shown by the crude fluid limit. In fact, as it will be demonstrated in the following, for conditions close to the flat density limit, the hypothesis of mode frequencies far from the kinetic resonance (crude fluid limit) completely fails in reproducing the correct $T_i/T_e$ dependence of the ITG threshold, which is instead linked to kinetic resonance effects. The closure introduced by the advanced fluid models has specifically the aim of recovering some of the essential physics carried by the presence of the kinetic resonance. The linear threshold derived from the Weiland advanced fluid model appears then to be a good candidate to be compared with completely kinetic approach. Before this direct comparison, a brief discussion on the linear threshold of the trapped electron modes is here given.\\
\indent The problem of the $\nabla T_e$ linear instability threshold for trapped electron modes has been quite extensively addressed through both analytical efforts and numerical simulations. The TEM threshold problem can be again addressed starting from the linear kinetic response given by Eq. \eqref{gykinrespapp} in the presence of a nonzero fraction of trapped electrons $f_t$ \footnote{In this case the quasi-neutrality condition will be modified as $\delta n_i=\delta n_{e,pass}+\delta n_{e,trap}$ where the non-adiabatic trapped electron response is included, while passing electrons are still treated as adiabatic}. In this case, the hypothesis of retaining only the magnetic drift contribution of the kinetic resonance is even more justified, since passing electrons are considered adiabatic. On the other hand, a more subtle question is arising when coupling the electron kinetic response with the ion one. In fact, if for ITG modes the hypothesis of electron adiabaticity can be valid at high collisionality, the symmetric choice of adiabatic ions is not feasible, since $\nu_{ei} \gg \nu_{ii}$ (where $\nu_{ij}$ is the collision frequency between species $i$ and $j$). Hence, the non-adiabatic response of both trapped electrons and ions through quasi-neutrality is retained, giving:
\begin{eqnarray}  1-f_t \left\langle \frac{\omega-\omega_e^*}{\omega-\omega_{De}} \right\rangle_{\mathcal{E},\lambda} = 
-\frac{T_e}{T_i}\left[1-\left\langle \frac{\omega-\omega_i^*}{\omega-\omega_{Di}} \right\rangle_{\mathcal{E},\lambda} \right] = 
-\aleph \frac{T_e}{T_i}
	\label{kineqtrape} \end{eqnarray}
The choice for the ion response is put in evidence through the parameter $\aleph$ appearing in the last term of Eq. \eqref{kineqtrape}. Even if $\aleph$ will depend primarily on the modes frequency $\omega$ and on both $\nabla_r n$ and $\nabla_r T$, two extreme cases can be drawn: keeping the condition $\aleph=1$ means adopting ion adiabaticity, while on
the other hand $\aleph=0$ implies entirely neglecting the ion response.\\
\indent Even in this case, it is initially possible to adopt a crude fluid limit based on Eq. \eqref{kineqtrape}, considering conditions far from the electron resonance, i.e. $\omega_{De}/\omega \ll 1$; a second order expansion for the non-adiabatic
electron response is then allowed. According to this approximation, the following $R/L_{Te}$ TEM threshold expression can
be derived:
\begin{eqnarray}  \left.\frac{R}{L_{Te}}\right|_{th}^{\mathrm{crude\:fluid}} = \frac{K_t}{3\left(1+\aleph \mathcal{F}\right)} \left( \frac{3}{2}-\frac{R}{2L_n} \right)^2-\frac{R}{L_n}+5
  \label{TEMfluidthr} \end{eqnarray}
where $K_t=\frac{f_t}{1-f_t}$ and $\mathcal{F}=\frac{T_e}{T_i}\frac{1}{1-f_t}$. Within this simple fluid approximation, the two opposite limits $\aleph=1$ and $\aleph=0$ lead to discrepancies in the temperature ratio behavior of the TEM threshold. In fact, when choosing $\aleph=1$ (adiabatic ions), a TEM $R/L_{Te}$ threshold raising with higher $T_i/T_e$ is found. Within the second limit $\aleph=0$ instead, the TEM threshold exhibits no temperature ratio dependence, as often found in literature \cite{weiland00,peeters05}. In fact, as done for the ITG threshold, Eq. \eqref{TEMfluidthr} can be compared with the prediction of the Weiland advanced fluid model that gives:
\begin{eqnarray}  \left.\frac{R}{L_{Te}}\right|_{th}^{\mathrm{Weiland}} = \frac{K_t}{2} \left( 1-\frac{R}{2L_n}^2 \right)+\frac{2}{3}\frac{R}{L_n}+\frac{20}{9K_t}
  \label{TEMWadthr} \end{eqnarray}
\begin{figure}[!htbp]
  \begin{center}
    \leavevmode
      \includegraphics[width=10 cm]{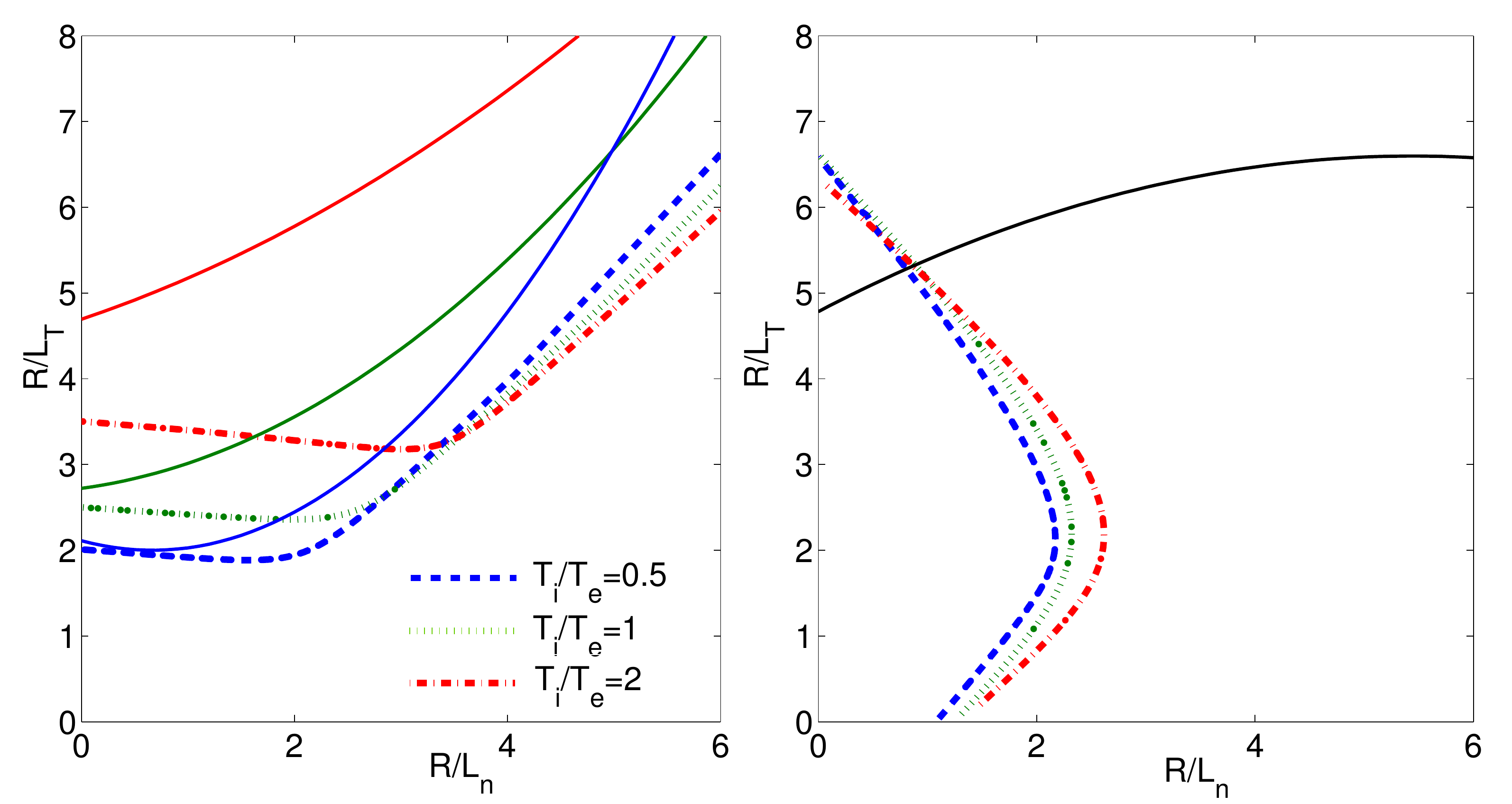}
    \caption{(a) ITG threshold according to the Weiland advanced fluid model (full lines) and computed by the linear gyrokinetic code Kinezero (dotted lines) in the plane $R/L_n,R/L_T$ for different values of the ratio $T_i/T_e$. (b) TEM threshold according to the Weiland advanced fluid model (full line) and computed by the linear gyrokinetic code Kinezero (dotted lines) in the plane $R/L_n,R/L_T$ for different values of the ratio $T_i/T_e$.}
    \label{Fluidvskin-thr}
  \end{center}
\end{figure}
\indent Finally, it is relevant to quantitatively compare the expectations for the linear ITG and TEM thresholds from the kinetic and the advanced fluid approaches. In Fig. \ref{Fluidvskin-thr}, the parametric space referred to the normalized density and temperature gradients $\left(R/L_n,R/L_T\right)$ is explored, distinguishing between linearly stable and unstable regions according to the Weiland model and to the numerical results obtained using the linear gyrokinetic code Kinezero. It clearly appears that even the advanced fluid approach overestimates the linear threshold with respect to the fully kinetic result. Even if the closure is able of reproducing the overall $T_i/T_e$ dependence of the ITG threshold, the critical gradients significantly differ between the two formulations. Moreover, no temperatures ratio dependence for the TEM threshold is expected according to the Weiland fluid  model, while the linear gyrokinetic simulations reveal that also the TEM threshold is affected by the $T_i/T_e$ variations. The fluid overestimates would then imply non-negligible errors when applied into a quasi-linear transport model, since no turbulent transport can be predicted in the linearly stable region. Therefore, there are strong indications to prefer the kinetic framework if one hopes to realistically describe the tokamak turbulent transport using a quasi-linear theory.\\


\subsection{The $T_i/T_e$ dependence of the linear ITG-TEM thresholds}
\label{sec-TiTeexample}

\indent In order to reinforce the argument in favor of the use of the kinetic approach with respect to the fluid one, a particular issue which is relevant for realistic tokamak applications, will be here below discussed in detail. As anticipated in the previous paragraph, this example deals with the temperatures ratio $T_i/T_e$ dependence for the linear threshold of the ITG-TEM unstable modes. Firstly, a simple analytical derivation will be used to highlight that the correct temperatures ratio dependence of both the ITG and TEM linear thresholds is directly linked to the presence of the kinetic resonance, providing expectations in agreement with the linear gyrokinetic simulations. Secondly, it will be shown that the fluid expectations naturally appear as a particular limit of the kinetic framework. The whole analysis help moreover in understanding the impact of the relevant parameter $T_i/T_e$ on the tokamak micro-turbulence.\\
\indent From the experimental point of view, a large number of evidences have revealed that the plasma performance is significantly improved when the ratio between ion and electron temperature $T_i/T_e$ is increased \cite{weiland05,asp05,wolf03,manini04,petty99,deboo99}. The hot ion high confinement mode (H-mode) has allowed achieving the highest fusion yields in the Joint European Torus (JET) \cite{jetteam99,thompson93}; similar results have been reported also in DIII-D \cite{burrell91}, with the 'supershot' in the Tokamak Fusion Test Reactor (TFTR) \cite{strachan87} and with the advanced scenarios in ASDEX-Upgrade \cite{sips02}. However in the next step devices, the dominant electron heating by fusion $\alpha$-particles in the center coupled with the thermal equilibration between ions and electrons at high densities, will lead to a ratio $T_i/T_e$ slightly below unity. Hence, it is of interest to have a theoretically based understanding about the impact of the temperatures ratio on micro-turbulence, which strongly affects the plasma performance.\\
\indent Some dedicated transport analysis have already assessed the $T_i/T_e$ dependence on the energy confinement time \cite{asp05,petty99} and on the ion heat transport together with the role of the radial gradient of toroidal velocity $\nabla v_{tor}$ \cite{manini06}. Nevertheless few systematic analytical or numerical study of the temperatures ratio impact on the instability thresholds of ion temperature gradient ITG modes and trapped electron modes TEM has been carried out. Experimentally, the existence of such a threshold on the electron temperature gradient length has been proved \cite{hoang01,ryter05}. Several formulations for the critical normalized temperature gradient length $R/L_{Ti,e}=R|\nabla_r T_{i,e}|/T_{i,e}$ of these instabilities have been proposed according to the fluid or the kinetic approaches. In the case of pure ion modes, which have been firstly thoroughly studied, a threshold increasing with $T_i/T_e$ is well supported inside all the present formulations \cite{hahm89,weiland00}. Conversely, no impact coming from temperatures ratio variations was foreseen for the electron modes threshold \cite{weiland00,peeters05}. This study has systematically assessed for the first time also the $T_i/T_e$ dependence of the TEM threshold \cite{casati08}.\\

\noindent \textbf{TEM threshold within a kinetic approach}\\

\begin{figure}[!htbp]
  \begin{center}
    \leavevmode
      \includegraphics[width=12 cm]{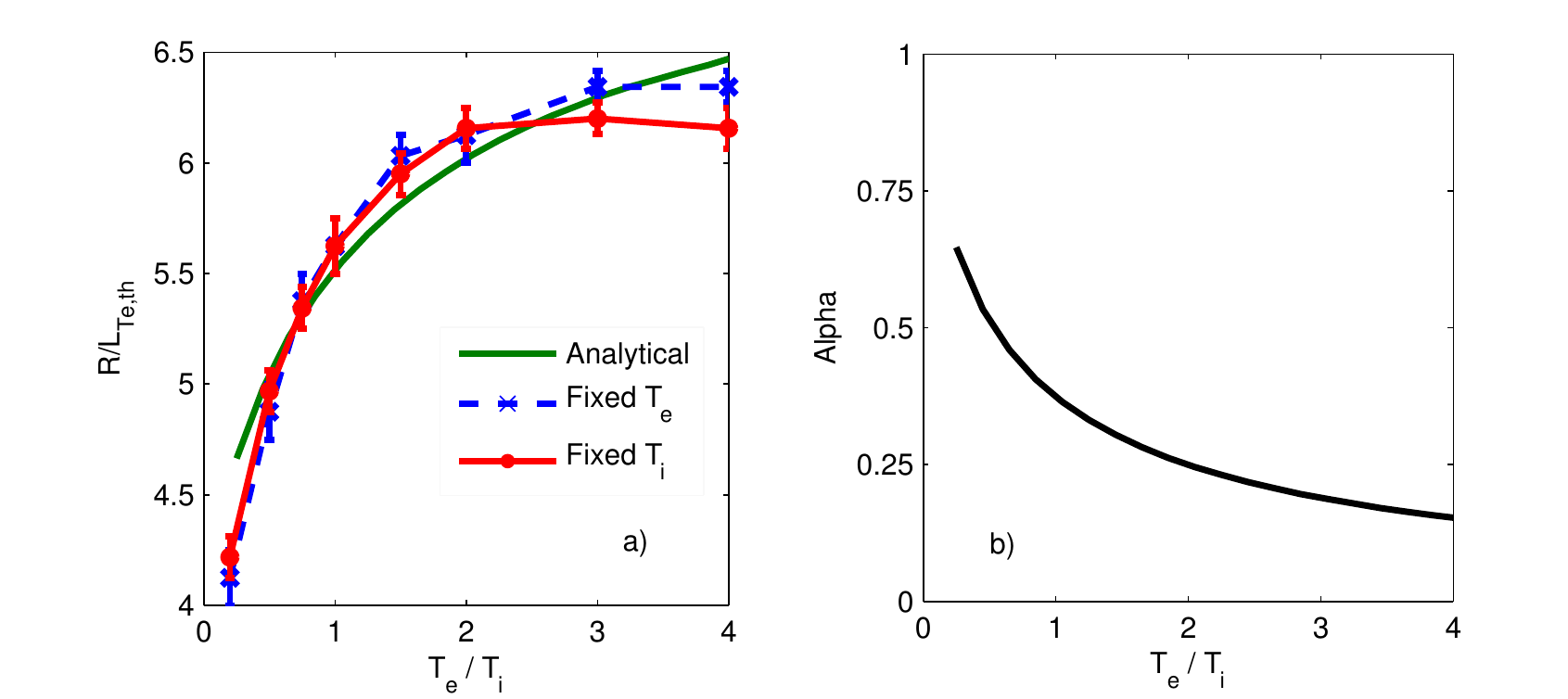}
    \caption{(a) $T_e/T_i$ dependence of the TEM instability threshold calculated with KINEZERO in the flat density limit at fixed $T_i$ and $T_e$. The numerical thresholds present intrinsic error bars, whose amplitude simply depends on the chosen step
size for the $R/L_{Ti}$ scan. Analytical predictions following from Eq. \eqref{kintrape3} are also plotted considering $\lambda_e=0.66$. (b) Analytical estimates for the $\aleph$ parameter appearing in Eq. \eqref{kintrape3}.}
    \label{figTiTe2}
  \end{center}
\end{figure}

Firstly, for reasons of simplicity, pure trapped electron modes in the strict flat density limit will be considered, thus
implying $R/L_{Ti}=R/L_n=0$; these assumptions will be later relaxed. The problem is addressed starting from the kinetic
dispersion relation \eqref{kineqtrape} considering a mode in the electron diamagnetic direction. The threshold for the mode instability is derived isolating the imaginary contribution coming from the resonance, using the relation $\lim_{\epsilon \rightarrow 0} \frac{1}{x \pm i\epsilon}=\mathrm{PP}\left(\frac{1}{x}\right)\mp i\pi \delta\left(x\right)$, thus giving:
\begin{align} &\qquad 1-f_t \left\langle \left[ \omega-\omega_{ne}^*-\omega_{Te}^*\left(\mathcal{E}-\frac{3}{2}\right) \right] \right. \nonumber \\ &\left.
\left\{ \mathrm{PP} \left(\frac{1}{\omega-\omega_{DTe}\mathcal{E}}\right) - i \pi \delta\left( \omega-\omega_{DTe}\mathcal{E}\right) \right\}  
 \right\rangle_{\mathcal{E},\lambda} + \aleph \frac{T_e}{T_i}=0
	\label{kintrape1}  \end{align}
At the threshold the imaginary part of Eq. \eqref{kintrape1} is vanishing, leading to:
\begin{eqnarray}  \omega = \omega_{ne}^*+\omega_{Te}^*\left(\frac{\omega}{\omega_{DTe}}-\frac{3}{2}\right)
  \label{kintrape1b} \end{eqnarray}
When using the latter relation into the real part of Eq. \eqref{kintrape1} (note that the dependence on the normalized energy $\mathcal{E}$ will not be present anymore), the following relation can be obtained:
\begin{eqnarray}  1+\aleph\frac{T_e}{T_i}-f_t\left\langle \frac{\omega_{Te}^*}{\omega_{DTe}} \right\rangle_{\lambda}=0
  \label{kintrape2} \end{eqnarray}
A last approximation regards neglecting the integration over $\lambda$. Even if $\lambda$ accounts for the mode structure through a $\chi$ coordinate along the field lines, if the eigenfunction has a strongly ballooned structure, $\omega_{DTe}$ can be replaced with its value at $\chi \approx 0$ \cite{romanelli89,guo93}. Within these hypotheses, Eq. \eqref{kintrape2} reduces to a basic expression for the TEM instability threshold:
\begin{eqnarray}  \left. \frac{R}{L_{Te}} \right|_{th} = \frac{2\lambda_e}{f_t}\left(1+\aleph\frac{T_e}{T_i} \right)
  \label{kintrape3} \end{eqnarray}
where the dependence on the parameter $\lambda_e$ has been left explicit. In the case of trapped electron modes, the localization of the eigenfunction can be affected by the minor role of parallel dynamics; for this reason the value of $\lambda_e$ is not expected equal to $1$ and it will be treated as a free parameter including a dependence on the magnetic shear $s$ \cite{garbet05}.\\
Equation \eqref{kintrape3} highlights that explicit $T_e/T_i$ dependence for the $R/L_{Te}$ TEM threshold can be potentially achieved for $\aleph \ne 0$. An implicit expression for $\aleph$ has been analytically obtained and numerically solved making use of the resonant condition \eqref{kintrape1b}: the details on this derivation are discussed in Appendix A. Equation \eqref{kintrape3} then provides us a new reference for studying the $R/L_{Te}$ TEM threshold scaling with $T_e/T_i$ in the flat density limit.\\
\indent This analytical approach has been tested against linear gyrokinetic simulations with the code Kinezero. In the framework of the $T_i/T_e$ study here presented, the set of plasma parameters reported in the Table \ref{TableTiTe} is defined in order to test the analytical predictions.
\begin{table}[!h]
	\begin{center} \begin{tabular}{|c|c|c|c|c|c|c|c|} \hline
			$B_0 [T]$ & $R_0/a$ & $r/a$ & $q$ & $s$ & $\alpha_{MHD}$ & $Z_{eff}$ & $\nu_{ei}$ \\
			\hline
		  2.8 & 3.0 & 0.4 & 1.4 & 0.8 & 0.0 & 1.0 & 0.0 \\ \hline
	\end{tabular} \end{center}
	\caption{\label{TableTiTe}Plasma parameters adopted for the linear gyrokinetic simulations here presented and performed with the code Kinezero.}
\end{table} \\
The numerical simulations consider 30 toroidal wave numbers in the range $0.1<k_{\theta}\rho_s<2.0$. Collisionless trapped electron modes in the flat density limit have been simulated imposing $R/L_{Ti}=0$ and $R/L_n=0$; a scan on the ratio $T_e/T_i$ from $0.25$ to $4$ has been performed fixing alternately $T_{e,i}=4~\rm{keV}$. The instability threshold has been identified within the interval of $R/L_{Te}$ values where the linear growth rates become nonzero. Results are plotted in Fig. \ref{figTiTe2} together with the analytical predictions for the TEM threshold and for $\aleph$. With the present set of plasma parameters the value of $\lambda_e=0.66$ has been considered; the latter choice is reasonable if compared to the
form proposed in Ref. \cite{garbet05} giving $\lambda_e=1/4+2s/3=0.783$.\\
\indent Several new and interesting features emerge from results shown in Fig. \ref{figTiTe2}. First of all, an effective dependence of the $R/L_{Te}$ TEM threshold on the electron to ion temperature ratio is both numerically and analytically recognized in the flat density limit. This evidence has been observed here for the first time and it is particularly remarkable for tokamak relevant conditions. Indeed a linear fit for $0.5<T_e/T_i<1.5$ gives in fact $R/L_{Te,th}=1.1\frac{T_e}{T_i}+4.4$. The $T_e/T_i$ dependence weakens for $T_e \gg T_i$ where the ion response becomes negligible. Moreover this feature is clearly not dependent on fixing $T_e$ or $T_i$ and numerical results confirm that having neglected FLR effects does not turn into a too severe simplification.\\
In the flat density limit, the TEM threshold is therefore largely dominated by the kinetic resonant effects. Moreover, consistent evaluation of the non-adiabatic ion response is crucial for successfully reproducing the linear increase of TEM $R/L_{Te,th}$ with $T_e/T_i$ at $\nabla_r n=0$, especially for $T_e \approx T_i$. A final statement has to be specified concerning the range of validity of the present approach. Equation \eqref{kintrape3} has been in fact derived retaining only the resonant contribution of the electron response; this means that the validity of this choice is limited to conditions of temperature and density gradients satisfying $R/L_n <\approx 3/2R/L_{T,th}$ and $R/L_{T,th} >\approx 2$ (see Appendix \ref{appx1-TiTe} for details). The problem of the temperature's ratio threshold dependence beyond this limit requires then a different approach.\\

\begin{figure}[!htbp]
  \begin{center}
    \leavevmode
      \includegraphics[width=12 cm]{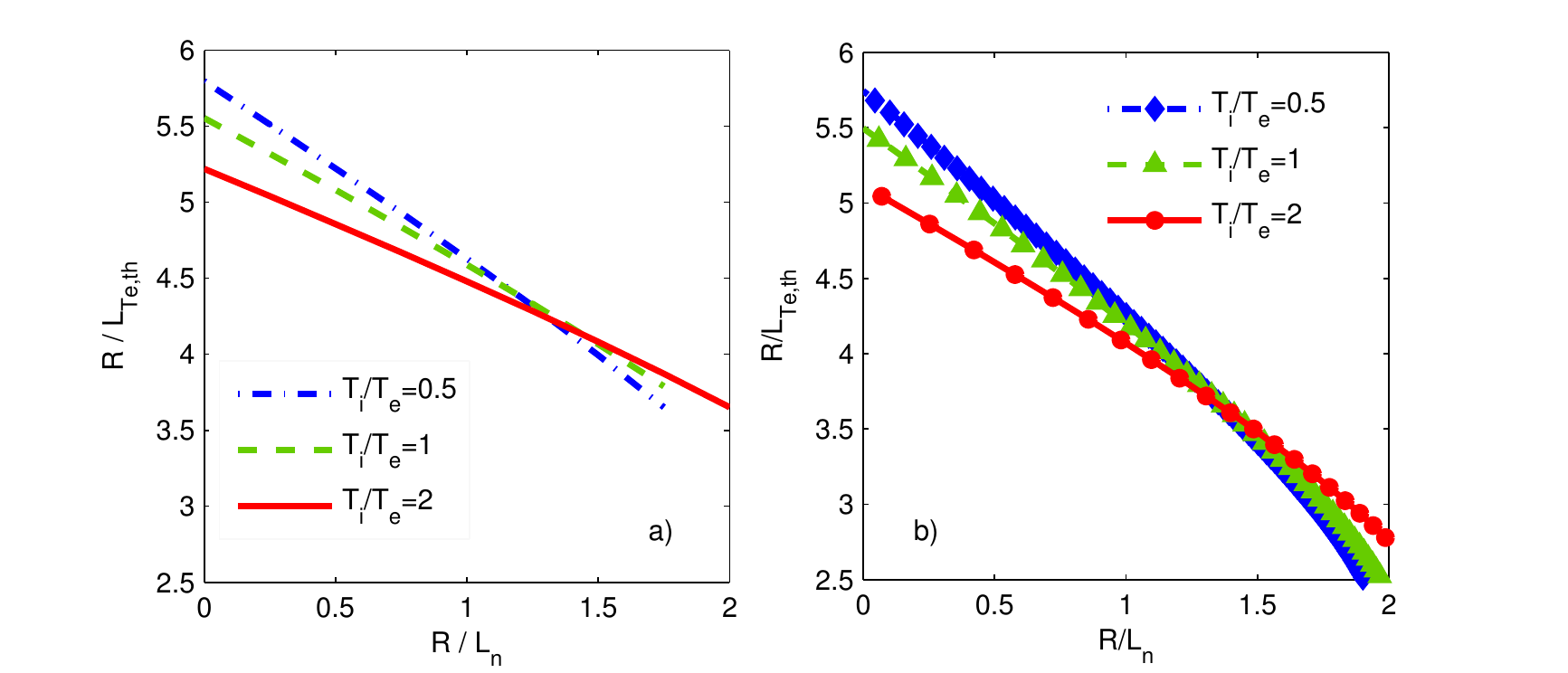}
    \caption{(a) Analytical predictions for the $R/L_{Te}$ TEM threshold vs $R/L_n$, imposing $R/L_{Ti}=2.5$ for different ratios $T_i/T_e$. (b) $R/L_{Te}$ TEM instability threshold calculated with Kinezero on a single wave number (maximum of the linear spectrum) considering $R/L_{Ti}=0.65R/L_{Te}$ and for the same ratios $T_i/T_e$.}
    \label{figTiTe3}
  \end{center}
\end{figure}

\indent Conditions of both $\nabla_r n=0$ and $\nabla_r T_i=0$ adopted until here are not commonly compatible with realistic
plasma scenarios. In the following, we will discuss the temperature ratio dependence of the TEM $R/L_{Te,th}$ in the presence
of non-adiabatic ion response driven by nonzero $\nabla_r n$ and $\nabla_r T_i$.\\
Modes rotating in the electron diamagnetic direction obeying the linear kinetic dispersion relation \eqref{kintrape1} are considered. Nonzero $R/L_{Ti}$ will be simply treated as a constant, playing the role of driving additional contribution in the non-adiabatic ion response. In other words, the possible contemporary presence of two unstable solutions (electron and ion
branches) is not retained. Within our analytical approach based on the electron resonance condition, the ion non-resonant response has been evaluated including the additional contributions of both $\omega_{ni}^*,\omega_{Ti}^*\neq0$, as detailed in Appendix A. In this case the value of $\aleph$ appearing in Eq. \eqref{kintrape1} depends on the parameters $T_i/T_e$, $R/L_n$, $R/L_{Ti}$, and $f_t$. The effect of magnetic shear is incorporated in the free parameter $\lambda_e$.\\
\indent This approach has been tested against linear gyrokinetic simulations with Kinezero, performed on a single wave number corresponding to the maximum of the linear TEM spectrum ($k_{\theta}\rho_s \approx 0.5$) and looking for the critical normalized
gradients. Conditions of dominant TEM turbulence have been achieved in the simulations imposing $R/L_{Ti}=0.65R/L_{Te}$, while the analytical calculations have been performed fixing $R/L_{Ti}=2.5$; finally the values $T_i/T_e=0.5, 1, 2$ have been considered. Both simulation results and analytical expectations for the TEM threshold are presented in Fig. \ref{figTiTe3}.\\
\indent A good agreement between analytical and numerical results is observed; small discrepancies have to be ascribed to the role of $R/L_{Ti}$, which is kept constant within the analytical approach while it is multiplied by a factor of $0.65$ with respect to $R/L_{Te,th}$ in the simulations. At zero density gradients, but in presence of non-negligible $\nabla_r T_i$, numerical results and analytical expectations agree in recognizing an increase of the $R/L_{Te}$ TEM threshold with higher $T_e/T_i$. This behavior has been already found with $R/L_{Ti}=0$ in Fig. \ref{figTiTe2}, meaning that a modest ion temperature gradient is not sufficient to significantly affect the $T_e/T_i$ TEM threshold scaling.\\
For higher values of density gradients, above $R/L_n \approx 1.2$ for this set of plasma parameters, an inversion
of the temperature ratio dependence of $R/L_{Te,th}$ is observed. In this case, the $\nabla_r T_e$ TEM threshold is slightly increasing with higher $T_i/T_e$. From the analytical point of view, the peculiar reversal in the temperature ratio dependence can be explained only if accounting for the ion response dynamics within the kinetic modes dispersion relation. This effect is clearly due to the role of nonzero $\nabla_r n$ in the non-adiabatic ion response.\\

\noindent \textbf{ITG threshold within a kinetic approach}\\

\begin{figure}[!htbp]
  \begin{center}
    \leavevmode
      \includegraphics[width=7 cm]{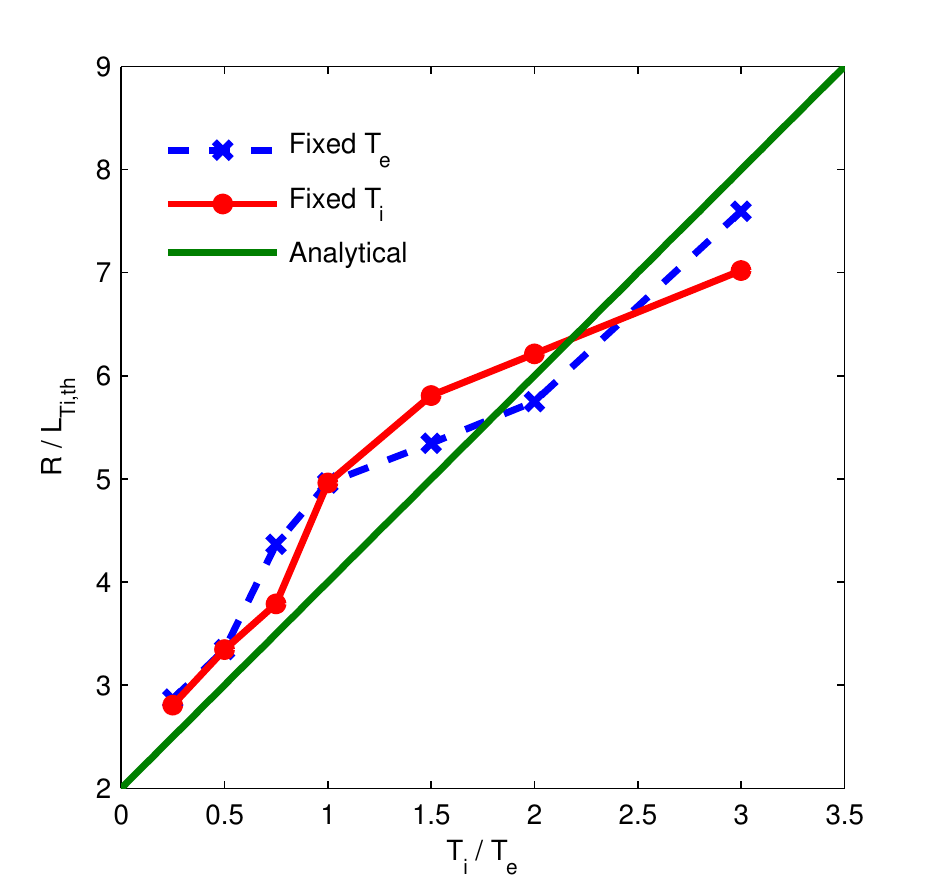}
    \caption{$T_i/T_e$ dependence of the ITG instability threshold calculated with KINEZERO in the flat density limit at fixed $T_i$ and $T_e$. Analytical predictions following from Eq. \eqref{kinitglin2} are also plotted.}
    \label{figTiTe5}
  \end{center}
\end{figure}

\indent We have anticipated that the correct temperature ratio dependence of the ITG threshold in the flat density limit is due to the role of the kinetic resonance. This result was already known by previous works \cite{romanelli89} and it can be also recovered by advanced fluid model like the Weiland model \cite{weiland00}; in Appendix A the derivation of such a model is briefly recalled, highlighting the presence of a resonant denominator acting within the fluid dispersion relation. Here instead, the $T_i/T_e$ dependence of the ITG threshold in the absence of density gradients is derived directly from the ion linear kinetic equation \eqref{gykinrespapp}, coupled with the quasi-neutrality condition and adiabatic electrons. Similarly to the procedure adopted for the electron modes, isolating the contribution coming from the ion kinetic resonance (imaginary part) provides a condition on the mode frequency:
\begin{eqnarray}  \omega = \omega_{ni}^*+\omega_{Ti}^*\left(\frac{\omega}{\omega_{DTi}}-\frac{3}{2}\right)
  \label{kinitglin1} \end{eqnarray}
The latter relation can be used for rewriting the real part of the linear ion kinetic response \eqref{gykinrespapp} coupled to adiabatic electrons, resulting in the following threshold expression:
\begin{eqnarray}  \left. \frac{R}{L_{Ti}} \right|_{th} = 2\left(1+\frac{T_i}{T_e}\right)
  \label{kinitglin2} \end{eqnarray}
\indent This simple relation has been tested against numerical simulations using KINEZERO with $R/L_{Te}=0$ and $R/L_n=0$. A scan on the ratio $T_i/T_e$ from $0.25$ to $3.0$ has been performed alternately fixing $T_{e,i}=4~\rm{keV}$. The resulting ITG threshold versus $T_i/T_e$ is presented in Fig. \ref{figTiTe5}. 
Numerical simulations highlight that the ITG instability threshold is clearly raising with increasing $T_i/T_e$ on the whole interval. Despite having neglected the parallel ion dynamics and simplified the $\lambda$-integration, the agreement with the analytical form by Eq. \eqref{kinitglin2} is quite satisfactory. Some discrepancies are nevertheless observed when fixing the ion rather than the electron temperature; these differences are most probably ascribed to the role of passing ions. In the flat density limit, the ITG instability threshold is largely dominated by the role of kinetic resonance (magnetic drift contribution), which is responsible for its clear increase with higher $T_i/T_e$.\\

\begin{figure}[!htbp]
  \begin{center}
    \leavevmode
      \includegraphics[width=12 cm]{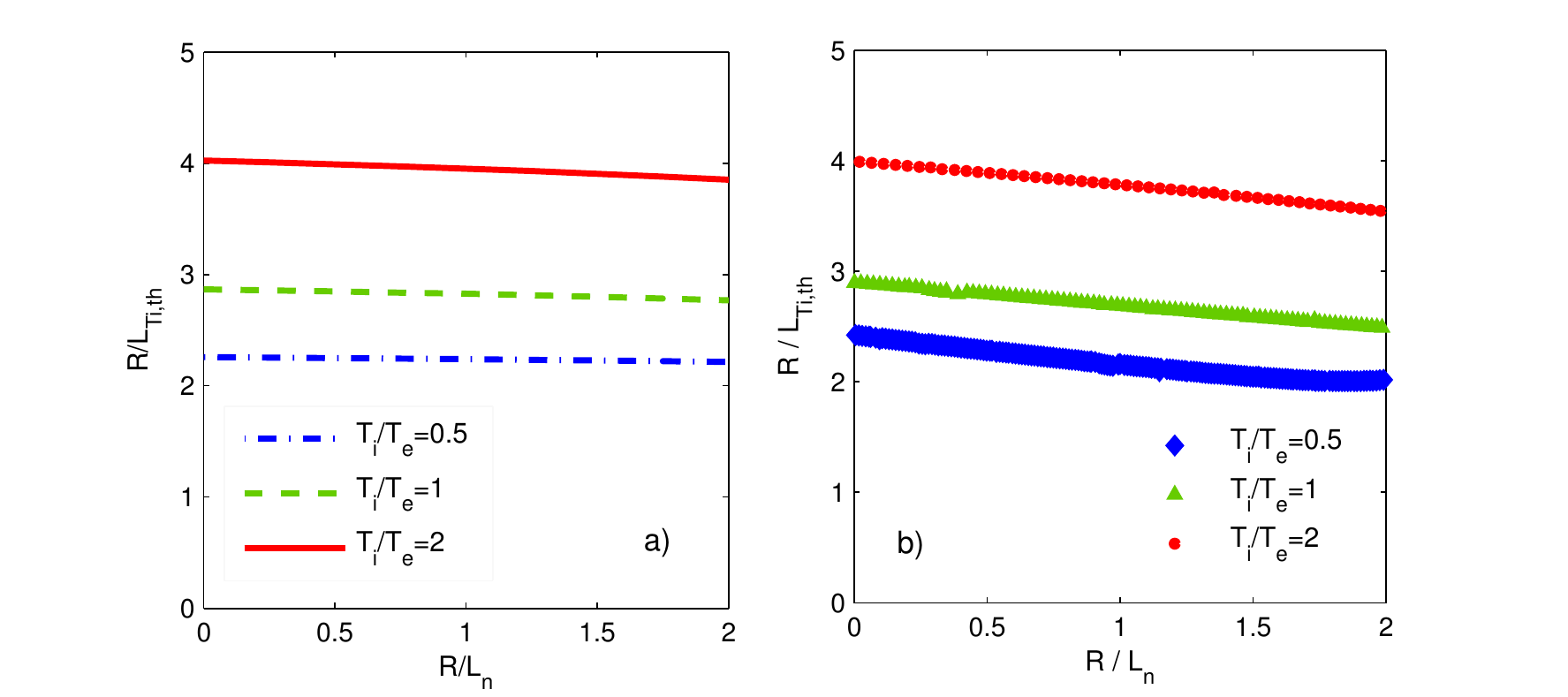}
    \caption{(a) Analytical predictions for the ITG threshold vs $R/L_n$ imposing $R/L_{Te}=3$ for different values of $T_i/T_e$. (b) ITG instability threshold calculated with KINEZERO on a single wave number (maximum of the linear spectrum) and considering $R/L_{Te}=0.4R/L_{Ti}$ for the same ratios $T_i/T_e$.}
    \label{figTiTe6}
  \end{center}
\end{figure}

\indent Electrons are usually retained adiabatically when deriving the ITG threshold; nevertheless including the non-adiabatic response due to trapped electrons (hence nonzero $\nabla_r n$ and $\nabla_r T_e$) leads to a more accurate treatment. As detailed in Appendix A, an analytical procedure very close to the one already applied for TEM has been adopted. Analytical predictions have been tested against numerical results. The ion modes thresholds have been calculated with Kinezero on a single wave number (maximum of the linear ITG spectrum) for ion turbulence dominated plasma. In the simulations, $R/L_{Te}=0.4R/L_{Ti}$ has been imposed, while $R/L_{Te}=3$ has been considered within the analytical calculations. Fig. \ref{figTiTe6} summarizes the results for $T_i/T_e=0.5, 1, 2$.\\
\indent The ion modes threshold does not exhibit any inversion in the temperature ratio dependence, which conserves the same $T_i/T_e$ scaling found in the flat density limit. Including the electron non-resonant response leads to a slight decrease of the ITG threshold at higher density gradients, as confirmed by the simulations; the latter effect is not captured by the advanced fluid models.\\
Briefly, kinetic resonant effects playing in the temperature ratio ITG threshold dependence have been verified beyond the usual hypothesis of adiabatic electrons. The increase of ITG threshold with higher $T_i/T_e$ is not significantly affected by nonzero $\nabla_r n$ and $\nabla_r T_e$. If in the strict flat density limit, ion and electron modes thresholds have opposite temperature ratio dependence, a critical value in $R/L_n$ exists, above which an analogous scaling of both the critical $R/L_{Ti,e}$ increasing with higher $T_i/T_e$ is expected.\\

\noindent \textbf{Linear stability diagrams and $T_i/T_e$ impact on ITG-TEM growth rates}\\

\begin{figure}[!htbp]
  \begin{center}
    \leavevmode
      \includegraphics[width=7 cm]{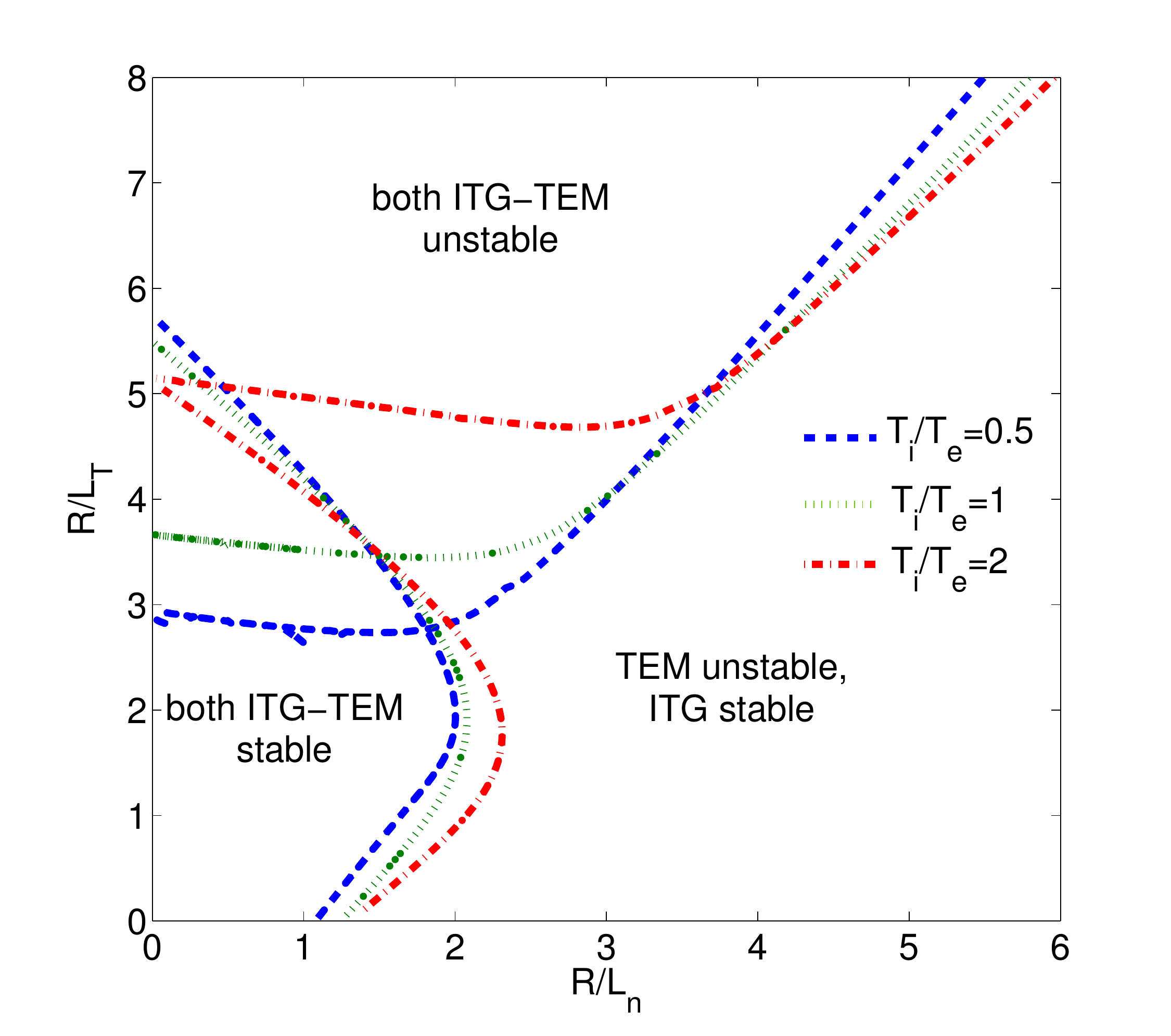}
    \caption{Stability diagrams of both ion and electron modes at different ratios $T_i/T_e$ calculated with KINEZERO on a single wave number (maximum of the linear spectrum=, (a) imposing $R/L_{Ti}=0.6R/L_{Te}$, (b) imposing $R/L_{Ti}=R/L_{Te}$.}
    \label{figTiTe7}
  \end{center}
\end{figure}

\indent 
The mode frequency relations (respectively Eq. \eqref{kinitglin1} for ion and Eq. \eqref{kintrape1b} for electron modes) are consistent with the hypothesis of modes propagating in the ion (electron respectively) diamagnetic direction; the latter condition turns into the following limits: $R/L_n<\approx 3/2R/L_{T,th}$ and $R/L_{T,th}>\approx2$ (see Appendix A). These criteria define the region where resonant kinetic effects are dominant; until here, the $T_i/T_e$ dependence of ITG and TEM temperature gradient thresholds have been studied within these limits. Here instead we extend the study of temperature ratio dependence of instability thresholds when considering frequencies away from the magnetic drift ones, where the non-resonant effect will impact the thresholds: in other words, the latter conditions correspond to what has been previously referred as the crude fluid limit.\\
\indent Stability diagrams for ITG and TE modes in the plane $\left(R/L_n ,R/L_T\right)$ have been numerically obtained with the code Kinezero. Simulations have been performed considering $R/L_{Ti}=0.6R/L_{Te}$ (Fig. \ref{figTiTe7}) on a single wave number corresponding to the maximum of the linear ITG-TEM spectrum. Two unstable branches coexist, corresponding to modes in the ion and in the electron diamagnetic directions. Concerning the ion branch, the resonant kinetic effects are dominant at low density gradients and the ITG threshold clearly raises with the ratio $T_i/T_e$. At higher $\nabla_r n/n$ and $\nabla_r T_i/T_i$, non-resonant terms (and therefore the crude fluid limit) become more and more relevant; this results in a reversed $T_e/T_i$ scaling of the ITG threshold, as already predicted by the analytical fluid limit of Eq. \eqref{ITGcrfluithr}.\\
\indent For a wide region of the plane $\left(R/L_n ,R/L_T\right)$, resonant kinetic effects are indeed at play for coupled ITG-TEM modes, significantly affecting their instability thresholds. In this case, for $R/L_n\gtrsim1.2$ these results highlight that raising the ratio $T_i/T_e$ with $R/L_{Ti}=R/L_{Te}$ is simultaneously widening the stability region of both ion and electron modes through a direct increase of their instability thresholds. This observation could eventually be regarded as a theoretical insight for the large amount of experimental evidence that recognizes the beneficial effect of high $T_i/T_e$ on the plasma confinement. Nevertheless, one of the main points raised by this work highlights that the $T_i/T_e$ thresholds scalings are far from being universally valid, since they can be reversed at higher values of density gradients. Moreover, the results presented in this work are referred to a specific set of plasma parameters (namely safety factor $q$, magnetic shear $s$, fraction of trapped particles, etc.) and the mode thresholds are sensible to all of them; a lot of care is then required when comparing these predictions with the experiments. More realistic self-consistency between different transport channels (ion/electron particle and heat transport) has not been addressed here and it is beyond the scope of this present analysis.\\

\begin{figure}[!htbp]
  \begin{center}
    \leavevmode
      \includegraphics[width=12 cm]{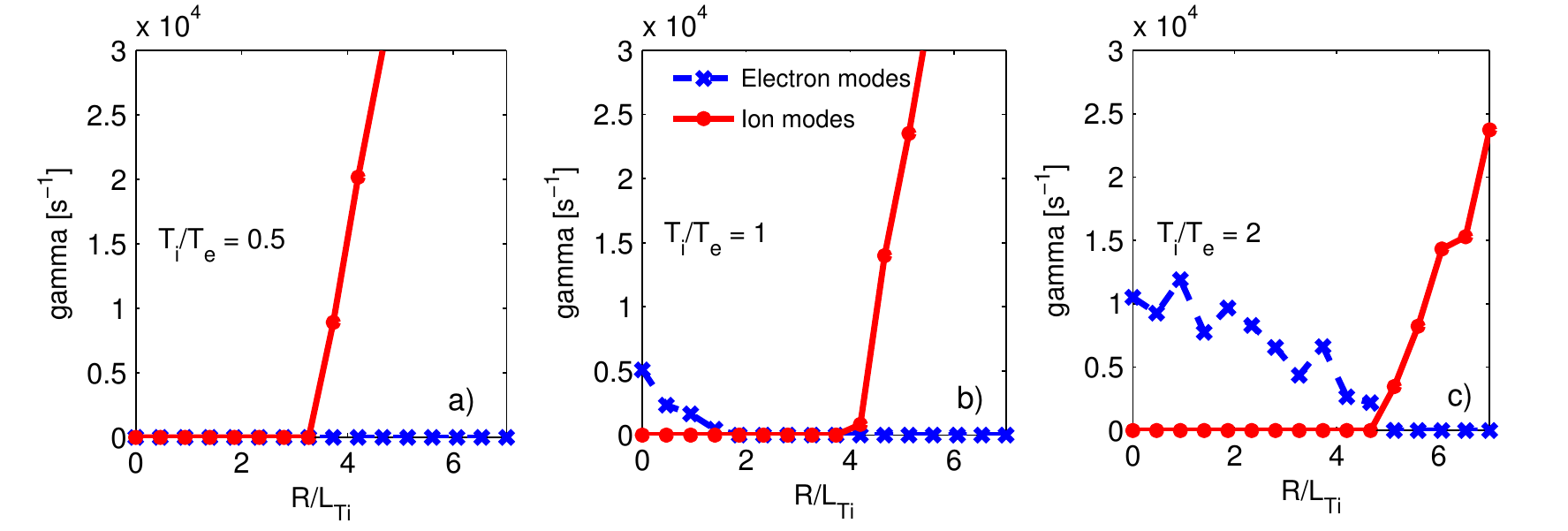}
    \caption{Linear growth rates vs $R/L_{Ti}$ considering $R/L_n=0$ and $R/L_{Te}=6$; both the first and the second most unstable solution (in the ion and electron dimagnetic directions) are plotted for (a) $T_i/T_e=0.5$, (b) $T_i/T_e=1.0$, (c) $T_i/T_e=2.0$.}
    \label{figTiTe8}
  \end{center}
\end{figure}

\begin{figure}[!htbp]
  \begin{center}
    \leavevmode
      \includegraphics[width=12 cm]{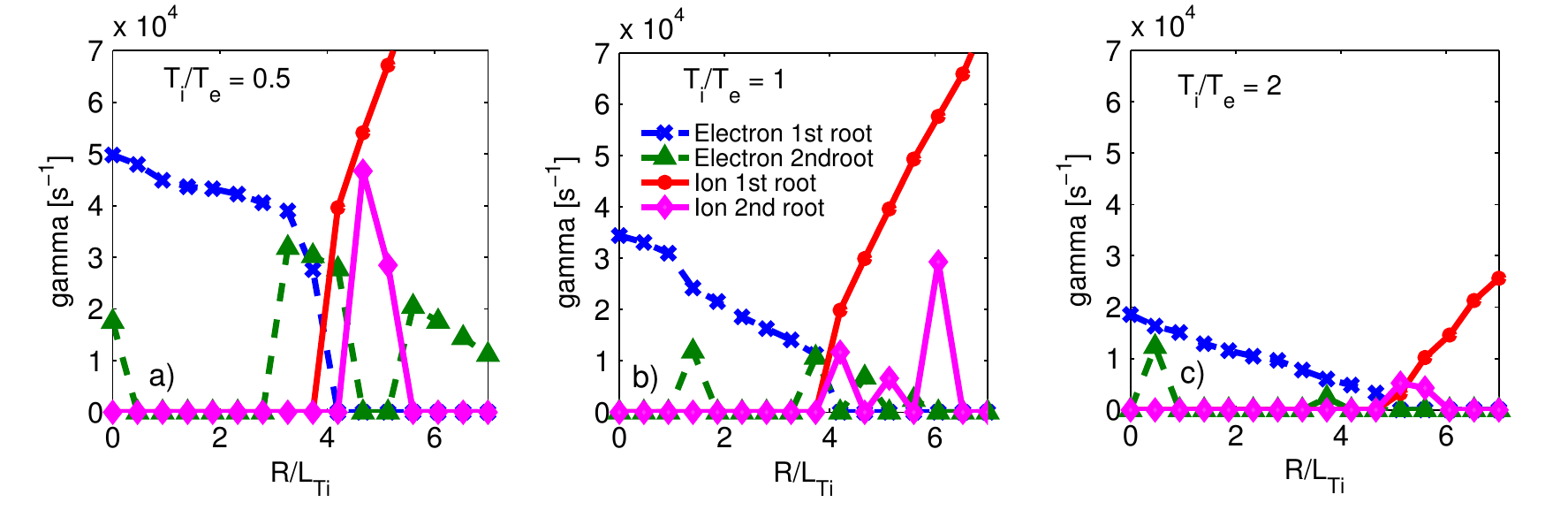}
    \caption{Linear growth rates vs $R/L_{Ti}$ considering $R/L_n=3$ and $R/L_{Te}=5$; both the first and the second most unstable solution (in the ion and electron dimagnetic directions) are plotted for (a) $T_i/T_e=0.5$, (b) $T_i/T_e=1.0$, (c) $T_i/T_e=2.0$.}
    \label{figTiTe9}
  \end{center}
\end{figure}

\indent A final relevant question could be: do these revised temperature ratio thresholds dependences actually turn into appreciable variations of the modes linear growth rates?\\ 
The growth rates are in fact expected to strongly affect turbulent heat and particle fluxes. For this reason the transition from TEM to ITG turbulence has been studied with Kinezero using the set of plasma parameters of Table \ref{TableTiTe}. The electron temperature gradient has been fixed sufficiently high to have TEM turbulence active, while a scan over $R/L_{Ti}$ has allowed moving from electron to ion turbulence at $T_i/T_e=0.5, 1, 2$. Moreover the role of normalized density gradients has been taken into account considering two different cases: $R/L_n=0$ (Fig. \ref{figTiTe8}) and $R/L_n=3$ (Fig. \ref{figTiTe9}). In these conditions, ITG and TE modes are often coupled; the following plots distinguish the growth rates of modes propagating in the ion ($\gamma_{0i}$) and in the electron ($\gamma_{0e}$) diamagnetic direction.\\
\indent Results of Fig. \ref{figTiTe8} in the flat density limit exhibit consistent behavior with both the previously found analytical and numerical thresholds scalings. The net effect of raising the ratio $T_i/T_e$ has opposite effects on the two branches, i.e., stabilizing for ion modes but destabilizing for electron ones; moreover the transition from electron to ion modes shifts towards higher $R/L_{Ti}$ when increasing $T_i/T_e$. Hence when moving from TEM to ITG dominated plasmas at negligible $\nabla_r n/n$, one expects a reversal of the $T_i/T_e$ impact on the confinement.\\
In the case of Figs. \ref{figTiTe9}, the role of nonzero density gradients $R/L_n=3$ reverses the temperature ratio dependence
of electron linear growth rates, while leaving unaffected the impact on the ion ones. Here in fact the increase of $T_i/T_e$ lowers not only the ion but also the electron growth rates, breaking the opposite $T_i/T_e$ scaling observed in the flat density limit. As the thresholds, the linear growth rate behavior of both ITG and TEM modes due to the role of the temperature ratio can undergo deep changes depending on the values of the density gradients.\\

\noindent \textbf{In summary}\\

\begin{enumerate}
	\item An analytical approach based on the kinetic resonance contribution for deriving estimates on the instability thresholds is able of successfully reproduce the results from linear gyrokinetic simulations. When considering modes in the electron (ion resp.) magnetic drift direction, a relevant point is also the self-consistent treatment of the ion (electron resp.) response in both its adiabatic and non-resonant contributions. 
	\item At low values of normalized density gradients, the ITG $R/L_{Ti,th}$ increasing with higher $T_i/T_e$ is due to the drift magnetic resonant contribution acting within the kinetic mode dispersion relation. Regarding pure trapped electron modes at $\nabla_r n/n=0$, significant raising of $R/L_{Te,th}$ with higher $T_e/T_i$ has been analytically and numerically found for $T_i\approx T_e$. Hence linearly, at low $\nabla_r n/n$ the plasma confinement with higher $T_i/T_e$ is expected to degrade for TEM turbulence and to improve for ITG turbulence.
	\item When instead increasing the normalized density gradient at $\nabla_r T_i/T_i\ne0$, the $T_e/T_i$ scaling of the $\nabla_r T_e/T_e$ TEM threshold reverses, due to the non-adiabatic ion response: in these conditions, increasing the ratio $T_i/T_e$ turns in raising both TEM and ITG thresholds. Similar inversion is not found for the ion modes threshold. The numerical linear gyrokinetic stability diagrams in the plane $\left(R/L_n ,R/L_T\right)$ allow to identify the conditions where resonant effects are overcome by non-resonant terms.
	\item A major step is still required in order to apply the present understanding to the tokamak discharges. The self-consistency between particle and heat transport channels for both species and the sensibility to several plasma parameters have to be accurately accounted for within a physically comprehensive gyrokinetic quasi-linear transport model.
	\item A significant region for feasible tokamak plasmas scenarios in the plane $\left(R/L_n ,R/L_T\right)$ has then been
studied and identified as dominated by the physics of magnetic drift kinetic resonance. A comprehensive kinetic approach is essential to correctly capture both the resonant and the non-resonant components of the plasma response, without the need of any additional arbitrary closure. Finally, the gyro-kinetic framework adopted in the quasi-linear transport model presented in this work, is believed to be a necessary feature for advances in first principle tokamak transport modeling.
\end{enumerate}
\chapter{The quasi-linear approximation}
\label{cap3-quasilinear-approximation}

\indent In this chapter the new quasi-linear gyro-kinetic transport model QuaLiKiz is presented. More in particular the attention will be focused on the quasi-linear approximation used for evaluating the plasma response for the transported quantities. In fact, the structure of the quasi-linear turbulent flux results composed by two main parts, according to the scheme:
\begin{eqnarray}  \mathrm{QL\:flux}\propto \mathrm{QL\:response} \otimes \mathrm{saturated\:potential}
	\label{cap3eq1r1} \end{eqnarray}
The first term represents the quasi-linear plasma response while the second one is the intensity of the saturated fluctuating potential.\\
\indent In the first section, the equations allowing to derive a quasi-linear response for the transported quantities, are introduced. Particular attention will be devoted to the ordering of the characteristic times and to the presence of kinetic resonances. On one hand, the ordering of the typical times governing the plasma dynamics, defines the limits where the quasi-linear approximation is expected to be valid. On the other hand, the contribution from both resonant and non-resonant terms appears to be crucial in the quasi-linear expression for the tokamak turbulent flux.\\
Once the quasi-linear formulation has been established, the hypothesis of a linear response for the transported quantities (particles and energy) to the fluctuating potential has to be validated against comprehensive nonlinear simulations. This challenging aspect has been here treated according to different levels of analysis, namely the analysis on: (1) the characteristic quasi-linear and nonlinear evolution times, (2) the phase relation between the fluctuating potential and the transported quantity, (3) the overall transport weights (independent of the fluctuating potential intensity) in both nonlinear and quasi-linear regimes. 

\section{The quasi-linear energy and particle fluxes}
\label{sec-quasilinear energy and particle}

\subsection{Quasi-linear ordering and hypotheses}
\label{sec-quasilinear-ordering}

\indent After more than 40 years from the first pioneering papers \cite{vedenov62,drummond62}, quasi-linear theory (\textit{QLT}) remains still nowadays an open subject of research that can provide a very powerful instrument for plasma physics understanding. Very large amount of literature often accompanied by controversies on the formulation and on the limits of QLT has been produced. Significant reviews can be found for example in \cite{laval99,krommes02,escande02,balescu05}.
Even if most part of the theoretical efforts in QLT has been applied to 1D plasma turbulence, several QL transport models have been proposed for the tokamak relevant 3D drift wave turbulence, providing feasible and commonly used predictive tools for the evolution of the thermodynamic quantities in tokamak plasmas. Among them we recall here GLF23 \cite{waltz97}, TGLF \cite{staebler07}, IFS-PPPL \cite{kot95}, MMM95 \cite{bateman98}, Weiland model \cite{weiland00}. Such efforts are nowadays more and more supported because, despite the apparently crude approximations adopted, QLT has revealed for a relevant number of cases a good agreement with both experimental results \cite{kot95,waltz97} and nonlinear gyrokinetic simulations \cite{jenko05,staebler05,staebler07,lin07}.\\
Inferring from the linear structure of the turbulence significant information on the nonlinear regime represents a great challenge; hence extremely accurate care has to be used in understanding the hypotheses of the underlying physics.\\

\indent The general framework of QLT is a mean field theory for the evolution of a time averaged distribution function $f_0=\left\langle f \right\rangle_{\tau}$ of the plasma particles population, governed by a Vlasov equation \eqref{Vlasoveq}. A first time ordering concerning the time scale used for the distribution function averaging is the following:
\begin{itemize}
	\item $\tau > 1/\gamma$, where $\gamma$ is the dominant unstable mode growth rate. In other words, the characteristic time scale $\tau$ is larger than the typical time of the microscopic fluctuations.
	\item $\tau < T_0$, where $T_0$ is the equilibrium evolution time. This means that the characteristic time scale $\tau$ is smaller than the macroscopic equilibrium evolution time.
\end{itemize}
A Fourier perturbative approach on both the Hamiltonian $H=H_0+\delta h$ and the distribution function $f=f_0+\delta f$ (introduced by Eq. \eqref{HFourier}-\eqref{fFourier}), is applied assuming:
\begin{eqnarray}  \frac{\delta f}{f_0} \approx \frac{\delta h}{H_0} \ll 1
	\end{eqnarray}
Averaging the Vlasov equation over the time $\tau$ leads to:
\begin{eqnarray}  \frac{\partial f_0}{\partial t} - \left\langle \left[\delta h,\delta f\right] \right\rangle = 0
	\label{QLeq1} \end{eqnarray}
while in the $\delta f$ equation, when neglecting the terms that are quadratic in the fluctuations amplitude, the following expression is obtained:
\begin{eqnarray}
   \frac{\partial\delta f}{\partial t}-\left[H_0,\delta f\right]-\left[\delta h,f_0\right]=0
\end{eqnarray}
Using the angle-action variables $(\boldsymbol{\alpha},\mathbf{J})$, the nonlinear terms of \eqref{QLeq1} can be written as:
\begin{align}  \left\langle \left[\delta h,\delta f\right] \right\rangle &=
   \int_0^{\tau}\frac{dt}{\tau}\int_0^{2\pi}\frac{d^3\alpha}{\left(2\pi\right)^3} \left( \partial_{\boldsymbol{\alpha}} \delta h \partial_{\mathbf{J}}\delta f - \partial_{\mathbf{J}}\delta h\partial_{\boldsymbol{\alpha}} \delta f \right) \nonumber \\
   &= \partial_{\mathbf{J}} \int_0^{\tau}\frac{dt}{\tau}\int_0^{2\pi}\frac{d^3\alpha}{\left(2\pi\right)^3} \delta f \partial_{\boldsymbol{\alpha}} \delta h
	\label{QLeq2} \end{align}
\indent QLT means that the fluctuating part of the distribution function appearing in Eq. \eqref{QLeq2} is approximated by a linear coherent response, i.e. using the important Eq. \eqref{linfresp}. Hence, a fundamental quasi-linear diffusion equation can be derived:
\begin{eqnarray}  \frac{\partial f_0}{\partial t} = \nabla_{\mathbf{J}} \cdot \left\{ \sum_{\mathbf{n},\omega} \mathbf{n}{ }\textrm{ Im}\left( \frac{\mathbf{n} \cdot \partial_{\mathbf{J}}f_0} {\omega - \mathbf{n} \cdot \boldsymbol{\Omega}_{0\mathbf{J}} + i0^+} \right) \right\} \left|\delta h_{\mathbf{n},\omega}\right|^2  
  \label{QLeq3} \end{eqnarray}
In the Eq. \eqref{QLeq3}, the single-particle propagator $G_{\mathbf{n},\omega}=1/\left(\bar{\omega}-\mathbf{n} \cdot \boldsymbol{\Omega}_{0\mathbf{J}}\right)$, dependent on both the wave number and the frequency, is of central importance. The appearing frequency $\omega$ is in principle real, nevertheless $\bar{\omega}=\omega+i0^+$ is introduced in the particle propagator. The strictly positive additional term $+i0^+$ represents the imaginary quantity following the Landau prescription on the resonance, needed for granting causality in the process. With this aim, since the temporal structure of the fluctuations is in the form $e^{i\omega t}$, it is sufficient adding to the real frequency $\omega$ a strictly positive but infinitesimal growth rate $+i0^+$, such that fluctuations cancel out for $t\rightarrow -\infty$. On the other side, in the limit of $0^+\rightarrow 0$, the term $\textrm{Im}\left(G_{\mathbf{n},\omega}\right)$ gives origin to a singular Dirac function; thus meaning that in presence of a discrete spectrum in $\mathbf{n}$ and $\omega$ the quasi-linear diffusion coefficient of Eq. \eqref{QLeq3} is not properly defined. This aspect is often arbitrarily solved passing from a discrete summation over $\mathbf{n}$ to an integral over a continuous space, thus recovering a proper mathematical expression. Nevertheless, since diffusion results from an intrinsically irreversible process, the appearance of a diffusion coefficient from the QL equations (which are a priori symmetric by time-reversal) is not a trivial point and deserves additional care. \\
Substituting the real frequency $\omega$ with $\bar{\omega}$ in the particle propagator does not simply fulfill causality through a vanishing imaginary contribution. Conversely, the diffusion coefficient of Eq. \eqref{QLeq3} has to be necessarily linked to intrinsic nonlinear effects leading to irreversibility through stochastic mixing of the particles orbits in the phase space. The hypothesis
\begin{eqnarray}  \bar{\omega}=\omega+i0^+\rightarrow\omega+i\nu \qquad \nu>0
	\label{QLeq4} \end{eqnarray}
where $\nu$ is definitely non negligible and positive, is actually the key point for passing from a conventional resonance localized QLT to a renormalized QLT.\\
\indent The stochastic character of this renormalized QLT, which results into a turbulent particles orbit diffusion, can be made clearer in terms of a Chirikov criterion \cite{chirikov79} for a chaotic state; that is simply expressed by:
\begin{eqnarray}  \sigma=\frac{\delta}{\Delta}>1
	\label{eqchirikov} \end{eqnarray}
where $\delta$ and $\Delta$ are respectively the width of the resonant islands and the distance between them in the phase space. In terms of the Chirikov criterion \eqref{eqchirikov}, the non-resonant contributions to QL diffusivity arise from the distortions of Kolmogorov-Arnold-Moser (KAM) torus in the phase space: a regime where the particles trajectories are not anymore integrable is gained when condition \eqref{eqchirikov} is fulfilled and the pathological resonance localized character of diffusivity is broken in favor of a large scale diffusive behavior. Since $\delta\propto\sqrt{\left|\delta h\right|}$, the resonance overlapping can take place either when resonances are getting closer, or when the amplitude of the fluctuating potential is raising. One would however expect that when moving towards strong turbulence with increasing $\left|\delta h\right|$, a limit in the amplitude of $\left|\delta h\right|$ exists beyond which QLT is not valid anymore, because of the previous assumption of $\delta h/H_0\ll1$ and of the linearization of the particles trajectories. Nevertheless, this seems still to remain an open point of QLT, in particular for 3D drift wave turbulence, which could be investigated by future full-$f$ gyrokinetic simulations. A proof of the validity of QL equations in the strongly nonlinear regime has been indeed given within the framework of Hamiltonian dynamics in \cite{escande03} for 1D Langmuir turbulence.\\
\indent Within a renormalized QLT then, accounting for $i\nu$ inside the particle propagator $G_{\mathbf{n},\omega}$ effectively corresponds to a stochastic renormalization to nonlinear effects. Historically, this has been at the origin of the so called resonance broadening theory \textit{RBT}, firstly initiated by Dupree in \cite{dupree66} and followed by several other works, leading also to more elaborate theories like the direct interaction approximation \textit{DIA} \cite{orszag67,dupree68,weinstock70}. The reason of the name \textit{resonance broadening} can be clearly understood since the term $\textrm{Im}\left(G_{\mathbf{n},\omega}\right)$ of Eq. \eqref{QLeq3} produces a broader resonance function with non negligible $i\nu$, in contrast with the singular resonance localized expression found for $i0^+\rightarrow0$:
\begin{eqnarray}  \textrm{Im}\left(G_{\mathbf{n},\omega}\right) = \frac{\nu}{\left(\omega-\mathbf{n} \cdot \boldsymbol{\Omega}_{0\mathbf{J}}\right)^2+\nu^2} \stackrel{\nu \rightarrow 0}{\longrightarrow} \pi\delta\left(\omega-\mathbf{n} \cdot \boldsymbol{\Omega}_{0\mathbf{J}}\right)   
	\label{QLeq5} \end{eqnarray}
where the limit passage from a Lorentzian broadening around the resonance to a local $\delta$-function is shown. In other words, the RBT accounts for both the resonant and non-resonant contributions to QL diffusivity, while the limit $i0^+\rightarrow0$ includes only the resonant terms.\\
\indent A fundamental question regards the value of such nonlinear resonance broadening $\nu$. In the current framework of renormalized QLT, a rather crude nonlinear dispersion relation is considered by:
\begin{eqnarray}  \omega=\omega_{\mathbf{n}}^{lin}+i\left(\gamma_{\mathbf{n}}^{lin}-\nu_{\mathbf{n}}\right)
	\label{QLeq6} \end{eqnarray}
Eq. \eqref{QLeq6} describes a simple dynamics, where, for each mode, the linear growth rate $\gamma_{\mathbf{n}}^{lin}$ competes with the nonlinear damping $\nu_{\mathbf{n}}$. Within this picture, saturation is then linked to the resonance broadening mechanism, when the nonlinear dissipation balances the linear instability drive, giving
\begin{eqnarray}  \nu_{QL,\mathbf{n}}=\gamma_{\mathbf{n}}^{lin}
	\label{QLeq7} \end{eqnarray}	 
Relation \eqref{QLeq7} actually represents the renormalized QL practical recipe for accounting the resonance broadening in Eq. \eqref{QLeq5}. This kind of assumption is in fact widely used inside several quasi-linear transport models such as GLF23, TGLF, Weiland model, IFS-PPPL, MMM95.\\
\indent The stochastic nature of the present QLT can be regarded in terms of a random walk phenomenology. Within this framework, one can estimate the perpendicular diffusion (cross field transport in the case of tokamak plasmas) as:
\begin{eqnarray}  D_{\perp} \approx \frac{1}{2}\left\langle \left|\delta v_{\perp}\right|^2\right\rangle\tau_{wp}
	\label{QLeq8} \end{eqnarray}
where $\delta v_{\perp}$ are the velocity fluctuations in the perpendicular direction (the $E\times B$ drifts in tokamak plasmas); $\tau_{wp}$ is a characteristic wave-particle interaction time, or more intuitively the effective lifetime of the field pattern. In order to test the validity of such QL model, one should compare this time scale $\tau_{wp}$ with a characteristic nonlinear time $\tau_{NL}$. QLT applies in fact when the particles motion perturbations remain small during $\tau_{wp}$, while one expects that for times $\tau>\tau_{NL}$ the linearization of the trajectories and the QL diffusivities will not be valid anymore. This means that a safe condition of validity for this QL random walk representation results from the following ordering of these time scales: 
\begin{eqnarray}  \tau_{wp}<\tau_{NL}
	\label{QLeq9} \end{eqnarray}
The latter criterion reflects the need of avoiding that particles undergo to a \textsl{trapping condition} in the resulting electric field pattern, which is incompatible within the present QL formulation \footnote{The use of the word \textsl{trapping}, originally introduced by Dupree, could however be misleading, since coherent phase space islands are destroyed when considering a spectrum of waves.}. It is however important to notice that also different approaches accounting for \textsl{turbulent trapping} have been developed for example with the so called \textit{clump theory} \cite{dupree72,adam79,krommes02}.\\
\indent The characteristic nonlinear time can be estimated starting from a general definition of the diffusion coefficient in the velocity space \cite{krommes02,krommes84}:
\begin{eqnarray}  D=\lim_{t\rightarrow\infty}\frac{\left\langle \left| \mathbf{v}\left(t\right)-\mathbf{v}_0 \right|^2\right\rangle}{2t}=\int_{t_0}^{\infty}C\left(\tau\right)d\tau
	\label{QLeq10} \end{eqnarray} 
where $\mathbf{v}_0$ is the velocity initial condition and the average $\left\langle \ldots \right\rangle$ is to be taken over an ensemble of particles. The appearing correlation function is defined as:
\begin{eqnarray}  C\left(\tau\right) = \frac{q^2}{m^2}\left\langle \delta\mathbf{E}\left[\mathbf{x}\left(t_0+\tau\right),t_0+\tau\right]\delta\mathbf{E}\left[\mathbf{x}\left(t_0\right),t_0\right] \right\rangle
	\label{QLeq11} \end{eqnarray} 
Relation \eqref{QLeq11} actually defines a Lagrangian, taken along the orbits, correlation function for the fluctuating electric field $\delta\mathbf{E}$. These correlations will decay on a nonlinear time scale $\tau_{L}$, measuring the decorrelation time of the turbulent structures experienced in a framework moving with the particles velocity. Indeed, one can demonstrate \cite{krommes84} that the time integration according to \eqref{QLeq10} lead to a velocity diffusion coefficient similar to the one embedded in Eq. \eqref{QLeq3}, with a resonance broadened function for the single particle propagator (in general not necessarily with a Lorentzian shape). The principal effect of this kind of approach is then the introduction of a characteristic resonance broadening width $\nu_{NL}=\tau_{L}^-1$ in the single particle propagator $G_{\mathbf{n},\omega}$.\\
\indent The former discussion about stochastic QLT can then be reviewed in the light of the latter implications. In particular, a useful criterion for the validity of a simple QL random walk model can be gained if testing the QL ordering \eqref{QLeq9}, where the nonlinear time coincides with the just defined Lagrangian one: $\tau_{NL}=\tau_{L}$. In the case of tokamak microturbulence, this kind of approach has been used for example in \cite{lin07}, by mean of massive nonlinear gyrokinetic particle in cell (\textit{PIC}) simulations of ETG turbulence. The same argument will be also deepened later in this work.\\ 
Another kind of approach can rely on the comparison between the resonance broadening width adopted in the renormalized QL model according to Eq. \eqref{QLeq7} with the nonlinear $\nu_{NL}$. That would imply that a nonlinear Lagrangian time is expected to be close to the inverse of a linear growth rate: 
\begin{eqnarray}  \tau_{NL,\mathbf{n}}\approx \frac{1}{\gamma_{\mathbf{n}}^{lin}}
	\label{QLeq12} \end{eqnarray} 	  
However this point should deserve great additional care for a further improvement of QL transport models, and represents an open issue also discussed in the following.\\

\noindent \textbf{Heuristic derivation of a saturation rule}\\

\indent Until here the main hypothesis of the present QL formulation has been the stochasticity of the particles trajectories in the phase space, thus allowing a random walk model. The second main critical issue for gaining QL predictions for the turbulent fluxes is the saturation level of the fluctuating potential. This argument will be the subject examined in Chapter 4. Nevertheless, in the present framework, it is worth noting that a heuristic derivation of a commonly adopted saturation model, namely the mixing length rule, can be coherently derived from the latter considerations.\\
Following from the former discussion on the QL hypothesis of resonance broadening driven saturation Eq. \eqref{QLeq6}-\eqref{QLeq7}, one expect that also the saturation level is linked with the QL resonance broadening width $\nu_{QL,\mathbf{n}}$. At this regard a significant conclusion can be gained if supposing, without a rigorous justification, the equality between $\nu_{QL,\mathbf{n}}$ and the inverse of nonlinear decay time of the Lagrangian correlation function \eqref{QLeq11}. An estimation of this time scale can be obtained making use of a quite strong hypothesis of random Gaussian statistics for the saturated fluctuating electric field \cite{krommes02} :
\begin{eqnarray}  \frac{1}{\tau_{NL,\mathbf{k}}}\approx\left\langle k_{\perp}^2 \right\rangle D_{\perp}
	\label{QLeq13} \end{eqnarray}  
where the nonlinear decay time scales like the inverse of a Dupree-Kolmogorov-like time \cite{chirikov79,krommes02}. It has to be stressed that the assumption of Gaussian statistics leading to relation \eqref{QLeq13} constitutes a strong additional hypothesis, consistent with a Markovian limit of a diffusion equation. A priori, this kind of choice has not any rigorous justification, since for developed plasma turbulence the fluctuating potential will be self-consistent with the particle distribution function through the Poisson-Amp\`ere equation. Finally it is just sufficient of using the already discussed equality $\nu_{QL,\mathbf{n}}\approx 1/\tau_{NL,\mathbf{n}}$ to obtain:
\begin{eqnarray}  D_{\perp}\approx\frac{\gamma_{\mathbf{k}}^{lin}}{\left\langle k_{\perp}^2 \right\rangle}
	\label{QLeq14} \end{eqnarray}  	 
Eq. \eqref{QLeq14} is usually referred as a mixing length saturation rule: within this renormalized QLT, it arises from the resonance broadening. The rule \eqref{QLeq14} is a very important practical relation adopted in several QL transport model; despite it is not rigorous, there are several indications showing that this kind of hypothesis can at least qualitatively  reproduce nonlinear fluxes. Recently, a remarkable confirmation of this kind of behavior has been identified also by mean of comprehensive nonlinear gyrokinetic simulations of tokamak TEM turbulence \cite{merz08}.\\
From expression \eqref{QLeq14} it also clearly appears that the hypotheses adopted in deriving the estimate \eqref{QLeq13} are crucial. Non-Gaussian deviations of the saturated field statistics could in fact result in a perpendicular diffusion coefficient $D_{\perp}$ which deviates from the linear scaling with $\gamma_{\mathbf{n}}^{lin}$, as prescribed by the usual mixing length estimate \eqref{QLeq14}. Finally one should notice that the arbitrary dependence of the QLT on the wave-vector $\mathbf{n}$ is here appearing in expressions \eqref{QLeq13}-\eqref{QLeq14}, but this argument will be deepened in the following.

\subsection{A new quasi-linear gyrokinetic model, QuaLiKiz}
\label{sec-quasilinear-new-model}

\indent The quasi-linear framework just described has been applied to the calculation of the turbulent fluxes of energy and particle in tokamak plasmas. The gyrokinetic linear solver Kinzero \cite{bourdelle02}, introduced in the previous paragraph \ref{Kinezero-descript}, has been upgraded to the quasi-linear transport model named QuaLiKiz \cite{casati09b,bourdelle07}.\\
\indent The gyrokinetic quasi-linear formulation of the particle $\Gamma_s$ and energy $Q_s$ flux for each plasma species $s$, follows from the velocity moments integration of the diffusion coefficient embedded in Eq. \eqref{QLeq3}. The derived expressions are:
\begin{align} & \qquad \Gamma_s = \textrm{Re} \left\langle \delta n_s \frac{ik_{\theta}\delta\phi}{B} \right\rangle = \nonumber \\ 
&- \frac{n_s}{R} \left(\frac{q}{rB}\right)^2 \sum_{n} \int\frac{d\omega}{\pi}n^2 
\left\langle \sqrt{\mathcal{E}} e^{-\mathcal{E}} \left( \frac{R\nabla_rn_s}{n_s}+\left(\mathcal{E}-\frac{3}{2}\right)\frac{R\nabla_rT_s}{T_s}+\frac{\omega}{n\omega_{ds}} \right) \right. \nonumber \\
&\left. \qquad \textrm{Im} \left( \frac{1}{\omega -n\Omega_s\left(\mathcal{E},\lambda\right) + i0^+} \right) \mathcal{J}_0^2\left(k_{\perp}\rho_s\right) \right\rangle_{\mathcal{E},\lambda} 
\left|\delta\phi_{n,\omega}\right|^2
	\label{QLpartflux} \end{align}
\begin{align} & \qquad Q_s = \textrm{Re} \left\langle \frac{3}{2}\delta P_s \frac{ik_{\theta}\delta\phi}{B} \right\rangle = \nonumber \\ 
&- \frac{n_s T_s}{R} \left(\frac{q}{rB}\right)^2 \sum_{n} \int\frac{d\omega}{\pi}n^2 
\left\langle \mathcal{E}^{3/2} e^{-\mathcal{E}} \left( \frac{R\nabla_rn_s}{n_s}+\left(\mathcal{E}-\frac{3}{2}\right)\frac{R\nabla_rT_s}{T_s}+\frac{\omega}{n\omega_{ds}} \right) \right. \nonumber \\
&\left. \qquad \textrm{Im} \left( \frac{1}{\omega -n\Omega_s\left(\mathcal{E},\lambda\right) + i0^+} \right) \mathcal{J}_0^2\left(k_{\perp}\rho_s\right) \right\rangle_{\mathcal{E},\lambda} 
\left|\delta\phi_{n,\omega}\right|^2
	\label{QLenergyflux} \end{align}
The sum is over the toroidal wave-numbers $n$,while the frequencies are:
\begin{align}  & n\omega_{ds}=-\frac{k_{\theta}T_s}{e_sBR} \qquad 
   n\Omega_s = -\frac{k_{\theta}T_s}{e_sBR}\mathcal{E}\left(2-\lambda b\right)f\left(\lambda\right) + k_{\parallel}v_{\parallel} \nonumber \\
   &\qquad k_{\parallel}v_{\parallel} \approx \varsigma \frac{s}{q} \frac{v_{th,s}}{R}w \sqrt{\mathcal{E}}
   \qquad k_{\theta}=\frac{nq}{r}
	\label{QLeq15} \end{align}
The details on the notation and the integration over the passing and trapped domains are identical to the ones introduced for the description of Kinezero in the paragraph \ref{Kinezero-descript}. Thanks to the relation $k_{\theta}=nq/r$, there is a correspondence between the toroidal wave-numbers labeled by $n$ and $k$, which will be from here equivalently used in the text.

\indent A first relevant consideration stems from the unphysical breakdown of the particle flux ambipolarity when using a strict resonance localized quasi-linear diffusion. Considering for simplicity a plasma of ions and electrons only, the quasi-neutrality condition implies $\delta n_{e,k}=\delta n_{i,k}$ for a given wave-number $k$: thus automatically turns into the ambipolar fluxes $\Gamma_{e,k}=\Gamma_{i,k}$. Making use of the useful resonance limit \eqref{QLeq5}, for $0^+\rightarrow 0$, the quasi-linear expression \eqref{QLpartflux} implies for a given $k$:
\begin{align} &\Gamma_{k,s} = - \frac{n_s}{R} \left(\frac{q}{rB}\right)^2 n^2 
\left\langle \sqrt{\mathcal{E}} e^{-\mathcal{E}} \left( \frac{R\nabla_rn_s}{n_s}+\left(\mathcal{E}-\frac{3}{2}\right)\frac{R\nabla_rT_s}{T_s} + \right. \right. \nonumber \\ 
&\qquad \left. \left. +\frac{n\Omega_s\left(\mathcal{E},\lambda\right)}{n\omega_{ds}} \right) \mathcal{J}_0^2\left(k_{\perp}\rho_s\right) \right\rangle_{\mathcal{E},\lambda} 
\left|\delta\phi_{n,n\Omega_s}\right|^2
	\label{QLeq17} \end{align}
Since the ion and electron resonances do not normally overlap, it appears that the particle flow is not automatically ambipolar, i.e. $\Gamma_{e,k} \ne \Gamma_{i,k}$. This breakdown following from Eq. \eqref{QLeq17}, makes that quasi-linear resonance localized formulation is usually rejected in the tokamak transport modeling in favor of a renormalized QLT, introduced in paragraph \ref{sec-quasilinear-ordering}.\\
\indent In principle, two kinds of broadening exist in the quasi-linear fluxes expressed by Eq. \eqref{QLpartflux}-\eqref{QLenergyflux}, whose physical origin is distinct and has to be specified. The first one actually coincides with the just mentioned RBT (see Eq. \eqref{QLeq5}): as already explained, it corresponds to a addition of a non-negligible term $+i0^+\rightarrow+i\nu$ in the particle propagator $G_{n,\omega}$. The second one is instead related to an intrinsic $\omega$-spectral shape of the fluctuating potential $\left|\delta\phi_{n,\omega}\right|^2$. Since the properties of $\left|\delta\phi_{n,\omega}\right|^2$ derive from the nonlinear saturation, it is reasonable to assume that this spectral quantity can deviate from a singular Dirac function localized at the frequency of the linearly unstable mode; conversely, it can exhibit a finite broadening around a given frequency. Here we refer to this second broadening mechanism as frequency broadening, in order to discriminate it from the previous resonance broadening. General models on the $\left|\delta\phi_{n,\omega}\right|^2$ frequency spectrum are commonly considered by statistical plasma turbulence theories \cite{krommes02,waltz83}: a relevant example is the so called pole approximation of the Lorentz line broadening model used by the DIA theory \cite{kadomtsev65,waltz83}. Letting $\omega_{0k}$ be the linear $k$-mode frequency, this model takes the following form:
\begin{eqnarray}  \left|\delta\phi_{k,\omega}\right|^2 = \left|\delta\phi_{k}\right|^2 
  \frac{\alpha_k}{\left[\omega-\left(\omega_{0k}+\delta\omega_k\right)\right]^2+\alpha_k^2}
	\label{QLeq18} \end{eqnarray}
where $\omega_{0k}+\delta\omega_k$ presents a nonlinear shift with respect to the corresponding linear frequency, while $\alpha_k$ is the frequency broadening or the nonlinear decorrelation rate.\\
\indent A rather general approach that can be adopted in tokamak transport quasi-linear models, is describing the $\left|\delta\phi_{k,\omega}\right|^2$ spectrum through a sum of $\omega$-broadenings around the different linear eigenmodes labeled by $j$:
\begin{eqnarray}  \left|\delta\phi_{k,\omega}\right|^2 = \left|\delta\phi_{k}\right|^2 \sum_j\mathcal{S}_{k,j}\left(\omega\right) 
\qquad \mathcal{S}_{k,j}\left(\omega\right) = \frac{\alpha_{k,j}}{\left(\omega-\omega_{0k,j}\right)^2+\alpha_{k,j}^2}
	\label{QLeq19} \end{eqnarray}
It is worth noting that with respect to the DIA model (Eq. \eqref{QLeq18}), the nonlinear shift $\delta\omega_k$ has been skipped. Coupling in the quasi-linear particle flux \eqref{QLpartflux} both the frequency (Eq. \eqref{QLeq19}) and the resonance broadenings (Eq. \eqref{QLeq5}) would end up in the following expression:
\begin{align} \Gamma_s &\propto - \sum_{n} \int\frac{d\omega}{\pi}n^2 
\left\langle \sqrt{\mathcal{E}} e^{-\mathcal{E}} 
 \textrm{Im} \left( \frac{\omega -n\omega_s^*}{\omega -n\Omega_s\left(\mathcal{E},\lambda\right) + i0^+} \right) \mathcal{J}_0^2\left(k_{\perp}\rho_s\right) \right\rangle_{\mathcal{E},\lambda} 
\left|\delta\phi_{n,\omega}\right|^2 \nonumber \\
 &= - \sum_{n,j} \int\frac{d\omega}{\pi} n^2
 \left\langle \sqrt{\mathcal{E}} e^{-\mathcal{E}} 
 \textrm{Im} \left( \mathcal{S}_{k,j}\left(\omega\right) \frac{\omega -n\omega_s^*}{\omega -n\Omega_s\left(\mathcal{E},\lambda\right) + i\nu_{kj}} \right) \mathcal{J}_0^2\left(k_{\perp}\rho_s\right) \right\rangle_{\mathcal{E},\lambda} 
\left|\delta\phi_{n,j}\right|^2
	\label{QLeq20} \end{align}
where for simplicity we have used $n\omega_s^*=-\frac{k_{\theta}T_s}{e_s B}\left[\frac{1}{L_{ns}}+\frac{1}{L_{Ts}}\left(\mathcal{E}-\frac{3}{2}\right)\right]$; the integration over $d\omega$ can in principle be analytically performed by residues.\\
\indent At this regard, most QL tokamak transport models assume more or less implicitly that for each wave number $k$ exists a well defined frequency $\omega$ such that $\omega\rightarrow\omega_{0k}$. In other words this choice corresponds to:
\begin{eqnarray}  \mathcal{S}_{k,j}\left(\omega\right) \rightarrow \delta\left(\omega-\omega_{0k,j}\right)
	\label{QLeq21} \end{eqnarray}
On the contrary, QuaLiKiz explicitly assumes the Lorentzian frequency broadening expressed by Eq. \eqref{QLeq19}, while keeping the localized resonance limit $i0^+\rightarrow0$. The combination of these choices results in a formulation completely equivalent to the more common renormalized quasi-linear theory, i.e. properly accounting for both the resonant and non-resonant contributions to the quasi-linear flux. Indeed, it is possible to write:
\begin{align} \Gamma_s &\propto - \sum_{n,j} \int\frac{d\omega}{\pi} n^2
 \left\langle \sqrt{\mathcal{E}} e^{-\mathcal{E}} 
 \textrm{Im} \left( \mathcal{S}_{k,j}\left(\omega\right) \frac{\omega -n\omega_s^*}{\omega -n\Omega_s + i0^+} \right) \mathcal{J}_0^2\left(k_{\perp}\rho_s\right) \right\rangle_{\mathcal{E},\lambda} 
\left|\delta\phi_{n,j}\right|^2 \nonumber \\
 &\stackrel{0^+ \rightarrow 0}{\longrightarrow} - \sum_{n,j} n^2
 \left\langle \sqrt{\mathcal{E}} e^{-\mathcal{E}} 
 \frac{\alpha_{k,j}}{\left(n\Omega_s-\omega_{0k,j}\right)^2+\alpha_{k,j}^2} \left( n\Omega_s -n\omega_s^* \right) \mathcal{J}_0^2\left(k_{\perp}\rho_s\right) \right\rangle_{\mathcal{E},\lambda} 
\left|\delta\phi_{n,j}\right|^2 \label{QLeq22}\\
&= - \sum_{n,j} n^2
 \left\langle \sqrt{\mathcal{E}} e^{-\mathcal{E}} 
 \left( n\Omega_s -n\omega_s^* \right) \textrm{Im} \left( \frac{1}{\omega_{0k,j} + i\alpha_{k,j} -n\Omega_s} \right) \mathcal{J}_0^2\left(k_{\perp}\rho_s\right) \right\rangle_{\mathcal{E},\lambda} 
\left|\delta\phi_{n,j}\right|^2
	\label{QLeq23} \end{align}
The latter demonstration proves finally that the frequency broadening model proposed in QuaLiKiz (Eq. \eqref{QLeq22}), coincides with the renormalized QLT (Eq. \eqref{QLeq23}) used in several other tokamak quasi-linear models like GLF23, TGLF, Weiland model, etc., under the following conditions:
\begin{enumerate}
	\item $\mathcal{S}_{k,j}\left(\omega\right)$ is chosen with a Lorentzian shape. Any different functional dependence for $\mathcal{S}_{k,j}\left(\omega\right)$ would have in fact lead to a discrepancy between the two approaches.
	\item $\alpha_{kj}=\nu_{kj}$. The broadening introduced in the renormalized QLT through the prescription $\omega\rightarrow\omega_{0kj}+i\nu_{kj}$ is identical to the Lorentzian frequency spectrum width $\alpha_{kj}$.
\end{enumerate}
\indent At the actual time, neither the formulation proposed in QuaLiKiz, nor the more common renormalized QL models, simultaneously accounts for both these broadenings, as suggested by Eq. \eqref{QLeq20}. In principle both these widths, namely the resonance and the frequency broadenings, could be object of study through dedicated nonlinear gyrokinetic simulations. In practice, separating these two effects is not an easy task, because the fluctuations of the potential and of the particle distribution function are back-reacting one on the other through the Poisson-Amp\`ere equation.\\
\indent A crucial point for all the actual quasi-linear transport models is the choice for the broadening $\alpha_{kj}$. The practical solution involves the balancing of the linear growth rate, relying on considerations expressed by Eq. \eqref{QLeq7}, then giving:
\begin{eqnarray}  \alpha_{kj} = \gamma_{kj}
	\label{QLeq24} \end{eqnarray}
Eq. \eqref{QLeq24} is presently adopted by QuaLiKiz as well by all the other tokamak quasi-linear transport models. Nevertheless, this hypothesis represents an assumption that is not rigorously justified. A remarkable example where the rule \eqref{QLeq24} breaks down can be easily found in the case of a linearly stable mode at a given wave-number $\bar{k}$, i.e. $\gamma_{\bar{k}}=0$. Whereas in this case quasi-linear models find $\Gamma_{\bar{k}}^{QL}=0$ following from \eqref{QLeq24}, nonlinear simulations \cite{waltz07} show instead a finite value of the turbulent flux $\Gamma_{\bar{k}}^{NL}\ne0$. One way to account for these kind of non-local couplings in the $k$ space would be to replace the rule \eqref{QLeq24} by the non-local one $\alpha_{kj} = \gamma_{kj} + k^2D$. In that way, even if $\gamma_{\bar{k}}=0$, a finite contribution to the quasi-linear flux will be retained. The additional term $+k^2D$, where $D$ is a general total diffusivity independent of $k$, can still be heuristically justified through statistical plasma theories \cite{krommes02}; more specifically the nonlinear time derivative becomes $d/dt\rightarrow\partial/\partial t-k^2D$, resulting in the modified linear growth rate $\gamma_k^{'}=\gamma_k+k^2D$.\\
On the other hand, also turbulence measurements can provide useful insights about this issue, and the argument will be deepened in Chapter \ref{cap4-improving-QL}.\\

\section{Validating the quasi-linear response}

\indent The purpose of this section is the validation of the assumption of the linear response of the transported quantities to the fluctuating potential. The only way to quantify the goodness of the quasi-linear response for the turbulent energy and particle transport is an accurate comparison with comprehensive nonlinear simulations.\\
\indent Despite the large number of existing tokamak quasi-linear transport models, most of them are commonly applied, more or less successfully, for predictive simulations of plasma the discharges (i.e. calculating the time evolution of the $T_e$, $T_i$ and $n_e$ profiles), without a clear picture of the capabilities and the limits of this kind of modeling.\\
Only very recently, for example in the case of the TGLF model, an exhaustive comparison of the total turbulent fluxes has been performed between the quasi-linear (TGLF) predictions and the full nonlinear gyrokinetic expectations \cite{kinsey08}. Even if this work represents a great advance, the relevance of this validation mostly relies on the specific choices adopted in the quasi-linear model that contribute to the final estimation of the total flux. In fact, when a theory based model fails to match the nonlinear gyrokinetic simulated transport (or the experimental transport for that matter), it is generally not possible to know if there is a failure in the accuracy of the underlying physics (e.g. gyrofluid versus gyrokinetic description), of the nonlinear saturation rule, or of the quasi-linear response.\\
\indent In this thesis work, the validity of a tokamak relevant quasi-linear approximation is systematically studied apart from any hypothesis on the nonlinear saturation mechanism. By using both linear and nonlinear gyrokinetic simulations, the aim is then to isolate and quantify the success or failure of the quasi-linear approximation itself. This analysis is believed to be a crucial point to get insight about the effective predictive capabilities of the model. The validation of the quasi-linear response is structured according to the following arguments:
\begin{itemize}
	\item In order to verify if the quasi-linear expected time ordering is respected, the characteristic nonlinear and linear times are defined and compared in nonlinear gyrokinetic simulations of tokamak microturbulence .
	\item The cross-phase between the fluctuating transported quantity (energy and particle) and potential are compared in the nonlinear and in the linear regime.
	\item An overall transport weight is defined and a relevant ratio between the quasi-linear and the nonlinear response is quantified.
\end{itemize}

\noindent \textbf{The reference case for the nonlinear simulations}\\

\indent The most part of the nonlinear gyrokinetic simulations presented in this work are performed using the code GYRO \cite{candy03}. Further details about the GYRO code and the nonlinear gyrokinetic simulations are reported in Appendix \ref{appx2-GYRO-GYSELA}. A standard reference case of study is defined in Table \ref{GA-std-table} and it will be later referred as \textit{GA-standard case}; here below a summary of the physical parameters which are adopted:
\begin{table}[!h]
	\begin{center} \begin{tabular}{|c|c|c|c|c|c|c|c|c|c|c|} \hline
			$R_0/a$ & $r/a$ & $R/L_{Ti}$ & $R/L_{Te}$ & $R/L_{n}$ & $q$ & $s$ & $T_i/T_e$ & $\rho_*$ & $\nu_{ei}$ & $\beta$ \\
			\hline
		  3.0 & 0.5 & 9.0 & 9.0 & 3.0 & 2.0 & 1.0 & 1.0 & 0.0025 & 0.0 & 0.0 \\ \hline
	\end{tabular} \end{center}
	\caption{\label{GA-std-table}Plasma parameters defining the \textit{GA-standard case}. Unless otherwise specified, this set refers to electrostatic turbulence in circular $s-\alpha$ magnetic geometry. Moreover, $\alpha_{MHD}=0.0$ and $Z_{eff}=1.0$.}
\end{table}
The numerical grid and the model assumptions for the GYRO nonlinear simulations are here briefly summarized (otherwise specified, these parameters apply to all the GYRO simulations presented in this chapter):
\begin{itemize}
  \item Local (flux-tube) simulations with periodic radial boundary conditions
	\item Drift-kinetic electrons (electron FLR are neglected) with real mass ratio $\sqrt{m_i/m_e}=60$, electrostatic, collisionless
	\item Box size in the perpendicular directions $\left[L_x/\rho_s,L_y/\rho_s\right]=\left[126,126\right]$, radial resolution $\Delta x/\rho_s=0.75$
	\item 16 Complex toroidal harmonics, covering $0.0<k_y\rho_s<0.75$, 12 grid points in the parallel direction, 15 in the gyroaverage and 5 in the radial derivative
	\item 128-point velocity space discretization per spatial cell, 8 pitch angles, 8 energies and 2 signs of velocity
	\item Statistical averages refer to typical time intervals $100<t<1000$, where time is expressed in $a/c_s$ units
\end{itemize}
In the GYRO code, upwind dissipative advection schemes are used in order to provide the dissipation and time irreversibilty required for the achievement of statistically steady states of turbulence; this numerical dissipation occurs only in the real space and arises from the upwind operators. Hence GYRO does not use any velocity-space dissipation other than possibly the collision operator. It has been shown \cite{candy03jcp} that adding upwind dissipation to radial advection terms is required to smooth over sub-grid-scale numerical disturbances associated with electron Landau layer physics. The radial upwind differencing in the drift-advection terms is then necessary when solving the electron equations on a $\rho_i$-scale grid. In cases where electron Landau layer effects do not need to be resolved, upwind dissipation eliminates unwanted numerical effects which would occur in a non-dissipative schemes.

\subsection{Characteristic turbulence times}
\label{sec-turbulence-times}

\begin{figure}[!htbp]
  \begin{center}
    \leavevmode
      \includegraphics[width=9 cm]{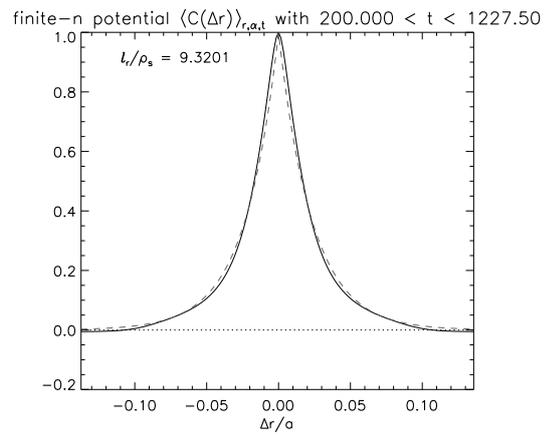}
    \caption{Example of a radial correlation function $C_r\left(\Delta r\right)$ obtained from a nonlinear GYRO simulation using kinetic electrons and collisions. The dotted line is the exponential fit.}
    \label{ExampleCrGYRO}
  \end{center}
\end{figure}

\begin{figure}[!htbp]
  \begin{center}
    \leavevmode
      \includegraphics[width=9 cm]{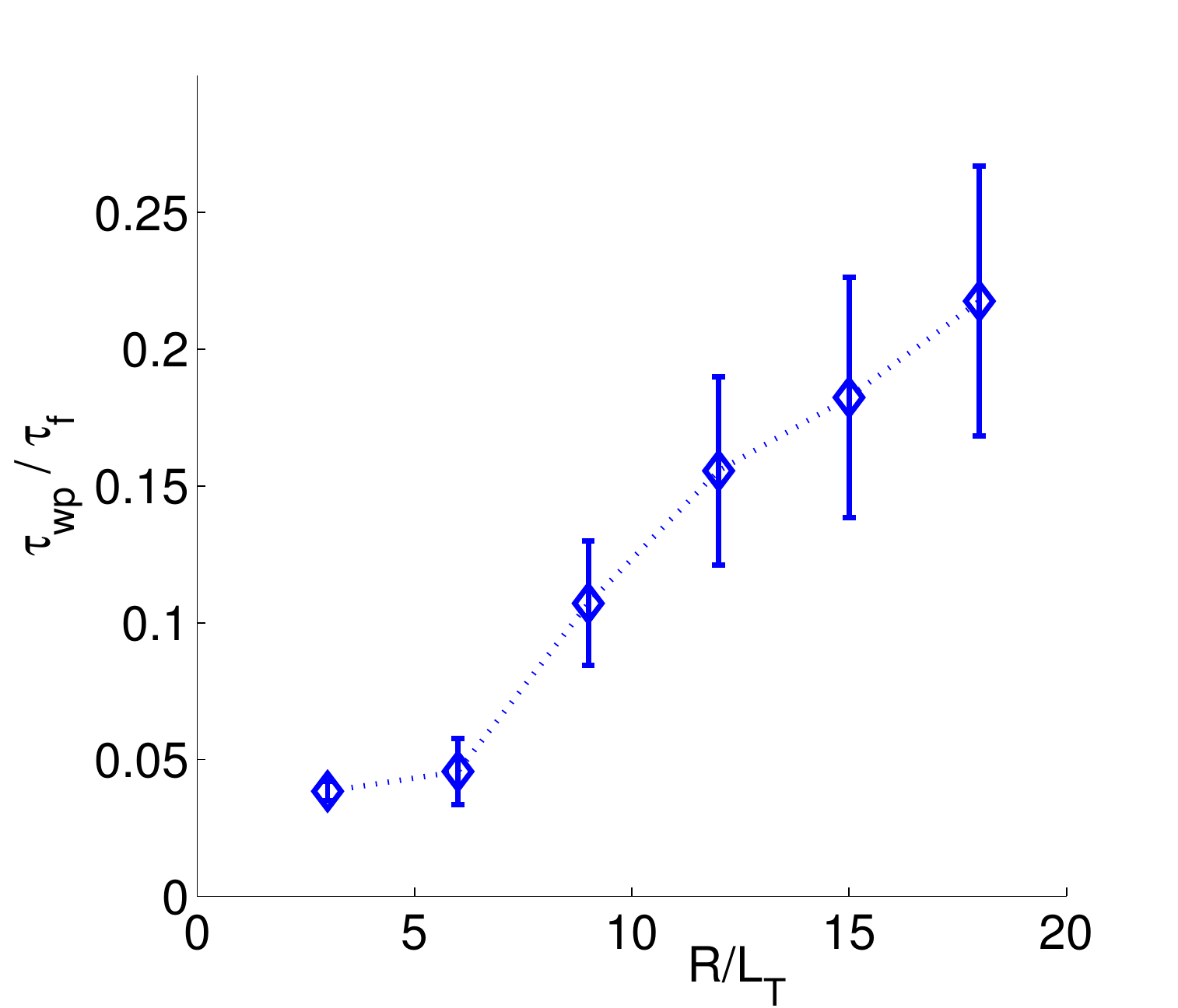}
    \caption{Kubo-like number $\mathcal{K}$ (Eq. \eqref{QLval3}) computed from GYRO nonlinear simulations versus a $R/L_T$ scan on the GA-standard case; vertical bars in the graph refer to the statistical evaluation of the intrinsic turbulence intermittency predicted by the nonlinear simulations.}
    \label{Gyro-Kubo}
  \end{center}
\end{figure}

\indent Historically, the quasi-linear theory has been elaborated for test particles \cite{krommes02}. As already pointed in paragraph \ref{sec-quasilinear-ordering}, this point can be easily understood, since the quasi-linear theory does not provide a self-consistent treatment; this is due to the fact that there is no back-reaction of the perturbed quantities on the fluctuating potential.\\
A powerful tool derives from fluid turbulence theories, the so called Kubo number $\mathcal{K}$ \cite{kubo63}, a property that can be defined for an advecting velocity field. The Kubo number is the ratio between (1) the wave-particle interaction time $\tau_{wp}$ characterizing the lifetime of the pattern the particle senses, and (2) a flight time $\tau_f$(or eddy turnover time), characterizing the time a particle would spend around the field structure. $\tau_f$ can also be interpreted as the time for the macroscopic flow to advect a perturbation across the system. Then,
\begin{eqnarray}  \mathcal{K}\equiv\frac{\mathrm{wave-particle\:time}}{\mathrm{eddy}\: \mathrm{turnover}\: \mathrm{time}} = \frac{\tau_{wp}}{\tau_f}
	\label{QLval1} \end{eqnarray}
The main difficulty comes from the fact that usually the advecting field is a function of both the 3D-position and time, so that $\tau_{wp}$ should be a Lagrangian time, which is typically hard to compute. According to the statistical plasma turbulence theories, a condition of validity for this quasi-linear framework is that the particles should not be trapped in the field (see paragraph \ref{sec-quasilinear-ordering}, Eq. \eqref{eqchirikov} and Eq. \eqref{QLeq9}). In terms of the Kubo number, this condition turns in:
\begin{eqnarray}  \mathcal{K}<1
	\label{QLval2} \end{eqnarray}
Note that quasi-linear ordering set by Eq. \eqref{QLval2} through the Kubo number is completely analogous to the condition expressed by Eq. \eqref{QLeq9}. It is worth noting that there are examples where quasi-linear theory has been shown to work up to $\mathcal{K}\approx1$ \cite{ottaviani92}. Ideally one would compute this Kubo number from nonlinear gyrokinetic simulations.\\
\indent Since the tokamak nonlinear gyrokinetic simulations consistently compute the coupled Vlasov-Maxwell equations, the definitions of these characteristic times have to be reviewed. $\tau_f$ can be calculated as the ratio between an auto-correlation length $L_c$, and the velocity at which the fluctuating quantities are radially transported $\delta v_r$:
\begin{eqnarray}  \tau_f=\frac{L_c}{\delta v_r} \qquad \Rightarrow \qquad \mathcal{K}=\frac{\tau_{wp}\delta v_r}{L_c}
	\label{QLval3} \end{eqnarray}
Referring to the tokamak geometry, $L_c$ is evaluated as a radial correlation decay length. The definition of the latter one requires the introduction of the 2D correlation function:
\begin{eqnarray}  C_{r,\varphi}\left(\Delta r,\Delta\varphi\right) = 
  \frac{\left\langle \delta\phi\left(r,\varphi,t\right) \delta\phi^*\left(r+\Delta r,\varphi+\Delta\varphi,t\right) \right\rangle_{r,\varphi,t}} { \left\langle \left|\delta\phi\left(r,\varphi,t\right)\right|^2 \right\rangle_{r,\varphi,t} }
	\label{QLval4} \end{eqnarray}
The radial correlation function $C_r\left(\Delta r\right)$ is calculated by taking the maximal value along the ridge of $C_{r,\varphi,t}\left(\Delta r,\Delta\varphi\right)$. The function $C_r\left(\Delta r\right)$ typically presents an exponential decay over the radial separation $\Delta r$, as shown in Fig. \ref{ExampleCrGYRO}.\\
Finally, the radial correlation length is defined as $L_c=\frac{\left\langle \Delta r C_r \right\rangle_r}{C_r}$.\\
$\tau_{wp}$, the effective lifetime of the field pattern, is estimated according to the arguments leading to Eq. \eqref{QLeq8}: together with the expression of the radial velocity fluctuations at a given toroidal wave-number $k$, $\tau_{wp}$ can be written as:
\begin{eqnarray}  \tau_{wp}=\frac{2\left\langle D \right\rangle_{r,\varphi,t}}{\left\langle  \left|\delta v_r\right|^2 \right\rangle_{r,\varphi,t} } \qquad  \delta v_{r,k}=i\frac{k_{\theta}\delta\phi_k}{B}
	\label{QLval5} \end{eqnarray}
where $D$ is the particle diffusivity, $D=-\Gamma_{e,i}/\nabla_r n$. This approach has also been applied in Ref. \cite{lin07}, in the case of nonlinear simulations of ETG tokamak turbulence.\\
\indent In the present work, this Kubo-like number defined by Eq. \eqref{QLval3} is computed from nonlinear gyrokinetic simulations of coupled ITG-TEM turbulence using GYRO. With reference to the GA-standard case, a wide scan on the normalized temperature gradient $R/L_T$ has been explored, varying simultaneously $R/L_{Ti}=R/L_{Te}$. The Kubo-like number $\mathcal{K}$ computed from each GYRO simulation is shown in Fig. \ref{Gyro-Kubo}.
For these parameters, $\mathcal{K}$ exhibit values well below 1.0 even for highest gradients in the scan. These estimations provide then a useful information on the expected quasi-linearity character of tokamak relevant microturbulence, involving both ion and electron unstable modes. Interestingly, the quasi-linear criterion $\mathcal{K}<1$ (Eq. \eqref{QLval2}) appears here to be largely respected. The numerical solution of the full nonlinear dynamics within the gyrokinetic framework has then revealed that the particles do not undergo to a trapping condition in the fluctuating electric field pattern. Hence, in the parametric region here explored, these results justify further attempts towards the quasi-linear modeling of the tokamak turbulent fluxes.

\subsection{Nonlinear versus linear phase relations}
\label{sec-cross-phases}

\begin{figure}[!htbp]
  \begin{center}
    \leavevmode
      \includegraphics[width=14 cm]{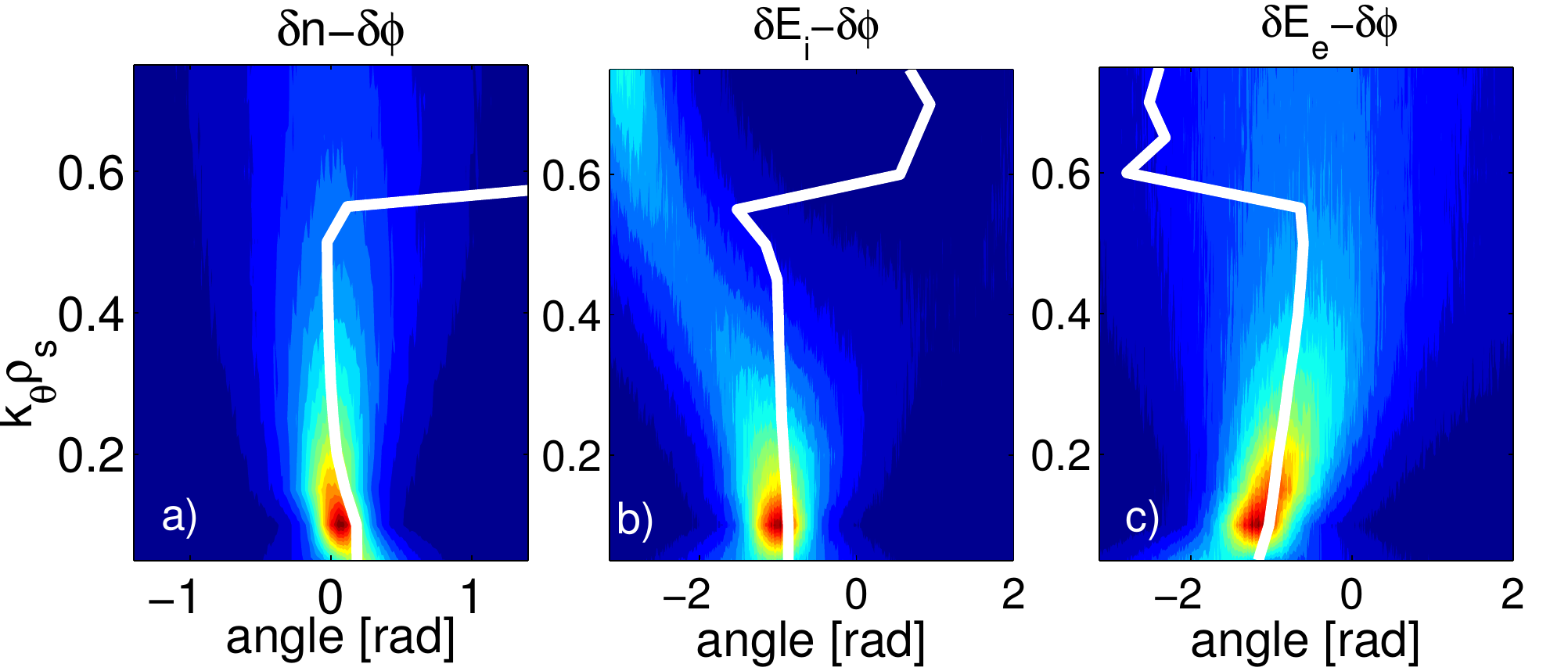}
    \caption{PDF of the nonlinear cross-phases (color contour plot) and the linear cross-phase of the most unstable mode (white line): a) $\delta n-\delta\phi$, b) $\delta E_i-\delta\phi$ and c) $\delta E_e-\delta\phi$ from a local GYRO simulation on the GA-ITG-TEM standard case.}
    \label{crossph-1}
  \end{center}
\end{figure}

\indent Apart from the criterion based on the ordering of the turbulence characteristic times just described, another relevant test for verifying the hypothesis of the quasi-linear response relies on the phase relations between the fluctuating fields. Referring for example to the particle flux, the general nonlinear expression for a given wave-number $k$ can be written as:
\begin{eqnarray}  \Gamma_k=\textrm{Re}\left\langle \delta n_k \frac{ik_{\theta}\delta\phi_k^*}{B}\right\rangle = 
\left\langle \left|\delta n_k\right| \frac{k_{\theta}\left|\delta\phi_k\right|}{B}\sin\Delta\Phi_k^{\delta n-\delta\phi}\right\rangle
	\label{QLval11} \end{eqnarray}
where $\Delta\Phi_k^{\delta n-\delta\phi}$ denotes the cross-phase between the fluctuating particle and potential fields. Expressions analogous to Eq. \eqref{QLval11} can be written for both the energy fluxes, defining the ion and electron energy cross-phases, respectively $\Delta\Phi_k^{\delta E_i-\delta\phi}$ and $\Delta\Phi_k^{\delta E_e-\delta\phi}$. The representation of Eq. \eqref{QLval11} is particularly physically meaningful, since it highlights that the turbulent transport can only arise in presence of phase shift between the fluctuating potential and the transported quantities. Phase shifts (or cross-phases equivalently) close to $0$ or $\pi$ are then indicative of a small flux \footnote{It immediately appears that within the approximation of adiabatic electrons, where the density response is simply taken as $\delta n/n=e\delta\phi/T_e$ without any phase shift, no particle flux can be retained.}. \\
\indent The quasi-linear theory can not give any insight on the product of the fluctuations absolute values $\left|\delta n_k\right|\left|\delta\phi_k\right|$, while it provides well defined phase relations between the fluctuating fields (see Eqs. \eqref{QLpartflux}-\eqref{QLenergyflux})\footnote{The linear dispersion relation used to derive the quasi-linear fluxes naturally implies the definition of a complex quantity, with a well defined phase relation between $\delta\phi$ and $\left[\delta n,\delta E_i,\delta E_e\right]$}. The comparison of the cross-phases in the linear and the nonlinear regimes allows to verify the validity of the quasi-linear response hypothesis. Nevertheless, as it will be deepened in the following, the linear cross-phases represent a relevant, but still not the whole information carried by the quasi-linear approximation.\\
\indent Practically, this aspect is investigated through both nonlinear and linear gyrokinetic simulations using the GYRO code. In the nonlinear regime, the cross-phase is not a constant quantity, but it presents oscillations reflecting the statistical intermittency of the turbulence. For this reason, the nonlinear cross-phases $\Delta\Phi^{nl}_k$ will be presented as probability density functions (PDF) of the phase shift between the fluctuating quantities computed by the simulation, i.e. between $\left[\delta n_k\left(r,\bar{\theta},t\right),\delta E_{i,k}\left(r,\bar{\theta},t\right),\delta E_{e,k}\left(r,\bar{\theta},t\right)\right]$ and $\delta \phi^*_k\left(r,\bar{\theta},t\right)$; where $\bar{\theta}=0$, $r$ spans over the whole radial domain and a nonlinear saturation $t$-interval is considered.
Conversely, only the linear most unstable mode is used for the estimation of the quasi-linear cross-phases, which will be referred as $\Delta\Phi_{k,1}^{ql}$.\\
\indent Fig. \ref{crossph-1} presents the results from the GYRO simulations on the GA-ITG-TEM standard case, which nonlinearly finds dominant (at low $k_{\theta}\rho_s<0.4$) saturated modes in the ion diamagnetic direction. Previous works in the plasma literature have already analyzed this issue for pure electron TEM turbulence \cite{dannert05,jenko05}. It appears that the linear phase shifts, i.e. the white lines in Fig. \ref{crossph-1}, remain close to the maximal PDF values of the nonlinear cross-phases. A very good agreement is therefore simultaneously observed for coupled ITG-TEM turbulence for all the particle and energy transport channels. It has to be however noticed that the quasi-linear phase shifts significantly breakdown at high $k$. Nevertheless, this failure is expected to have a small impact on the goodness of the quasi-linear approximation. As quantitatively detailed in the following in fact, most part of the transport is driven at scales corresponding to $k_{\perp}\rho_s\approx 0.2$, making that the cross-phases for $k_{\perp}\rho_s> 0.5$ contribute less to the total turbulent flux.\\
\begin{figure}[!htbp]
  \begin{center}
    \leavevmode
      \includegraphics[width=14 cm]{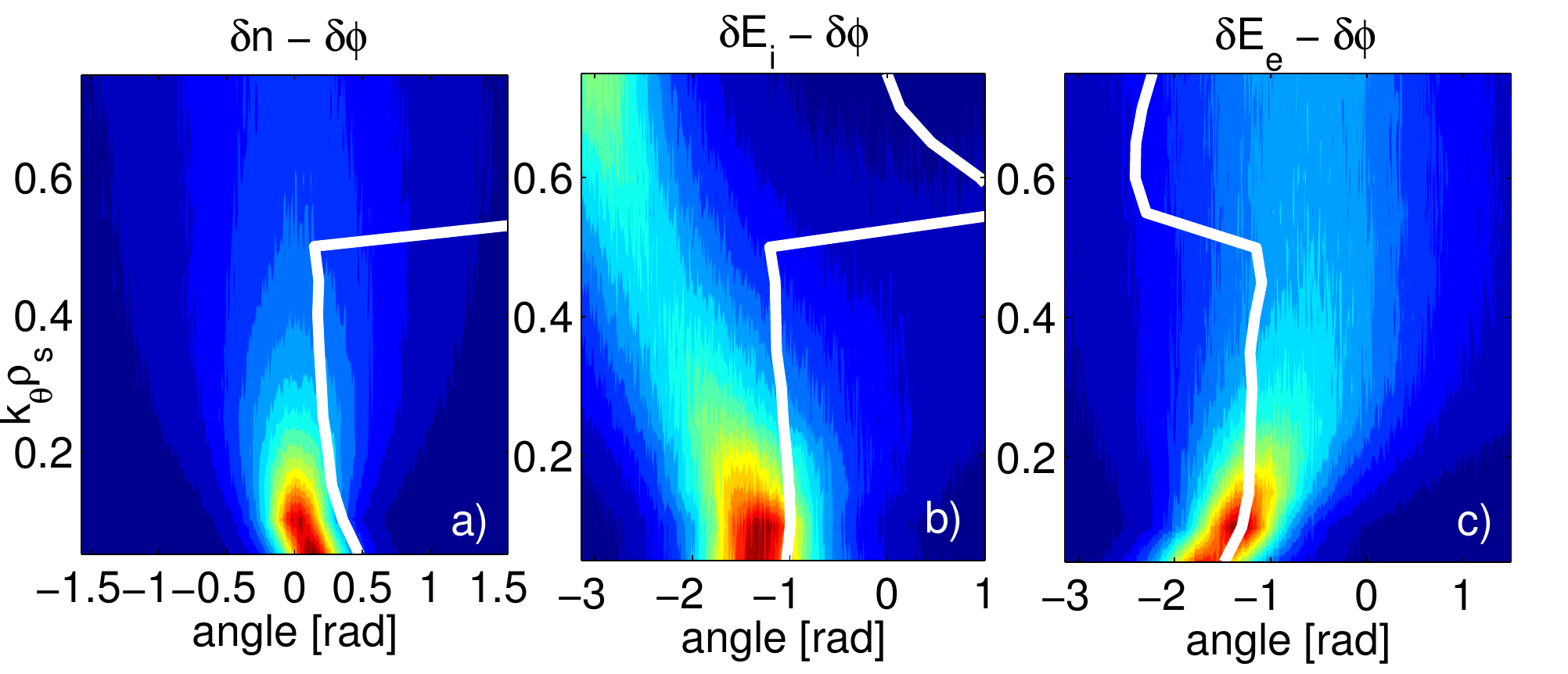}
    \caption{PDF of the nonlinear cross-phases (color contour plot) and the linear cross-phase of the most unstable mode (white line): a) $\delta n-\delta\phi$, b) $\delta E_i-\delta\phi$ and c) $\delta E_e-\delta\phi$ from a local GYRO simulation on the modified GA-ITG-TEM case with $R/L_{Ti,e}=9.0\Rightarrow18.0$.}
    \label{crossph-3}
  \end{center}
\end{figure}
\begin{figure}[!htbp]
  \begin{center}
    \leavevmode
      \includegraphics[width=14 cm]{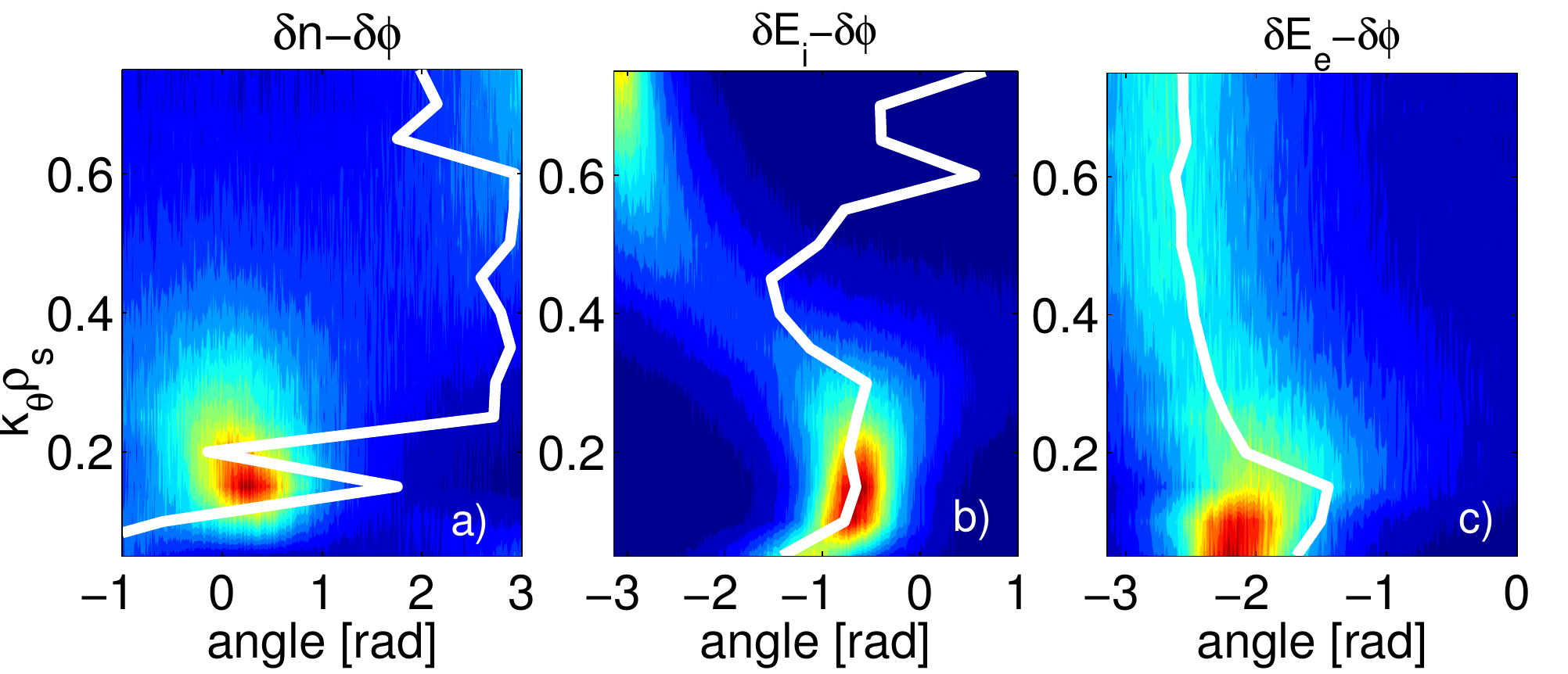}
    \caption{PDF of the nonlinear cross-phases and the linear cross-phase of the most unstable mode (white line): a) $\delta n-\delta\phi$, b) $\delta E_i-\delta\phi$ and c) $\delta E_e-\delta\phi$ from a local GYRO simulation on the modified GA-ITG-TEM case with $R/L_{Ti}=9.0\Rightarrow6.0$.}
    \label{crossph-2}
  \end{center}
\end{figure}
\indent It is important to check that the agreement on the cross-phase relations between the linear and the nonlinear regimes is respected across a variation of the plasma parameters. Hence, two modified GA-ITG-TEM standard cases are analyzed. The first one, reported in Fig. \ref{crossph-3}, is intended to test the quasi-linear response when enhancing the turbulent drives by increasing both the ion and electron temperature gradients, i.e. $R/L_{Ti}=R/L_{Te}=9.0\Rightarrow18.0$. As shown in Fig. \ref{crossph-3}, the linear phase shifts relative to the electron and ion energy transport appear still reasonably close to the nonlinear values, while a more pronounced failure is found for the particle transport. With these parameters, the GYRO nonlinear simulation predict a strong inward flow, mostly due the dominant low-$k$ ITG turbulence. On the other hand, since the values of the linear $\delta n-\delta\phi$ cross-phases are more positive than the nonlinear phase shifts, the quasi-linear estimate for the particle flux is possibly expected to disagree with the nonlinear flux for a non negligible over-prediction. The quasi-linear particle flux, especially for strong turbulence cases and inward flows, deserves then a careful treatment: a more exhaustive and quantitative approach will be introduced in the following paragraph.\\
A second case explore turbulence conditions where the ion ITG turbulence drive is lowered. This is realized changing only the ion temperature gradient with respect to the GA-ITG-TEM standard case, i.e. $R/L_{Ti}=9.0\Rightarrow6.0$, where ITG and TEM unstable branches are in competition with similar linear growth rates: results are shown in Fig. \ref{crossph-2}. Here the breakdown of the linear cross-phases with respect to the nonlinear phase shifts is even more evident. The most relevant failure is again on the particle transport, where the linear result significantly departs from the nonlinear PDF. The jumps observed in the linear $\delta n-\delta\phi$ cross-phase are due to the variations of the linear most unstable mode between ion and electron diamagnetic directions across the $k_{\theta}$ wave-numbers. Moreover, also the phase shifts relative to the electron transport exhibit a relevant disagreement, with the values of the linear cross-phases well below to the nonlinear ones. Also in this case then, these results could suggest that the quasi-linear estimates on the turbulent fluxes deviate from the nonlinear predictions. Nevertheless, it is important to remind that the linear cross-phases shown in Figs. \ref{crossph-1}, \ref{crossph-3} and \ref{crossph-2} refer only to the linear leading mode, while in general several unstable mode in both the electron and the ion diamagnetic direction are active. These sub-dominant modes are expected to drive a non-negligible contribution to the total turbulent fluxes, especially in conditions like the ones reported in Fig. \ref{crossph-2}, where ITG and TEM are competing at similar $k$ scales.\\
\indent In summary, the analysis of the linear versus nonlinear cross-phases appears as a powerful tool for validating the hypothesis of the quasi-linear response. Here we have shown that a good agreement on the phase shifts is obtained for conditions of coupled ITG-TEM turbulence, simultaneously on the particle and energy transport channels (Fig. \ref{crossph-1}). The breakdown of the linear cross-phases generally observed at high $k$, is not expected to have a drastic impact on the total turbulent fluxes which are mainly driven at $k_{\perp}\rho_s<0.5$.\\
Nevertheless, two main drawbacks of the quasi-linear validation approach based on the analysis of the cross-phases emerged; these are:
\begin{itemize}
	\item The quasi-linear cross-phases shown in Figs. \ref{crossph-1}, \ref{crossph-3} and \ref{crossph-2} are derived from the linear most unstable mode, while it is reasonable to expect that also sub-dominant modes play a role in the estimation of the total flux. Even if in principle the cross-phases relative to these sub-dominant linear eigenmodes can be numerically solved, that would not bring a definitive information on the final goodness of the quasi-linear response. The latter one in fact, will result from the weighted contributions of several unstable branches. The cross-phases of the linear unstable modes do not carry any insight on these relative weights. A more clever way to account for the role of sub-dominant modes have then to be elaborated.
	\item As suggested by the previous point, the cross-phases only partially represent the information provided by the quasi-linear theory (at least within the formulation presented in Sec. \ref{sec-quasilinear energy and particle}). In order to quantitatively address the validation of the quasi-linear response, another strategy has to be adopted, able to capture the whole outcome of the quasi-linear approximation.
\end{itemize}
Both these points motivate dedicated approaches to the validation of quasi-linear modeling and they will be originally treated in the next paragraph.

\subsection{Quasi-linear transport weights}
\label{sec-transport-weights}

\indent The nonlinear particle as well ion and electron energy fluxes $\left[\Gamma_e,Q_i,Q_e\right]$ and their respective effective diffusivities $\left[D_e,\chi_i,\chi_e\right]$ are given by the correlation of the particle and energy moment fluctuations per wave-number $k$ $\left[\delta n_{e,k},\delta E_{i,k},\delta E_{e,k}\right]$ with the radial $E\times B$ velocity  fluctuations $\delta v_{r,k}$:
\begin{align}  \left[\Gamma_e,Q_i,Q_e\right] &\equiv \left[-D\nabla_r n,-n_i\chi_i\nabla_r T_i,-n_e\chi_e\nabla_r T_e\right] \nonumber \\ 
  &= \textrm{Re} \sum_k \left\langle \left[\delta n_{e,k}^{nl},\delta E_{i,k}^{nl},\delta E_{e,k}^{nl}\right] \delta v_{r,k}^{*nl}\right\rangle
	\label{QLval6} \end{align}
where the superscript \textit{nl} refer to actual fluctuations that can be computed for example from a nonlinear simulation. Conversely, the quasi-linear turbulent fluxes are obtained convolving a quasi-linear response with a saturated potential spectral intensity (Eq. \eqref{QLpartflux}-\eqref{QLenergyflux}). The quasi-linear convolution takes the general following form:
\begin{eqnarray}  \textrm{QL-flux} \propto \sum_{k,j} \textrm{QL-weight}_{k,j} \otimes \textrm{Spectral-intensity}_{k,j}
	\label{QLval7} \end{eqnarray}
The first part derives from the linear correlation of the density (or energy) perturbations and the radial $E\times B$ velocity perturbations. The spectral intensity captures instead the saturated strength of the turbulence, which depends on the nonlinear coupling of the wave numbers. The index $j$ spans over a discrete number of unstable modes across the continuous frequency $\omega$.\\
\indent It is important at this point to detail the relation that links the quasi-linear transport weights appearing in Eq. \eqref{QLval7} and the cross-phases just previously defined:
\begin{eqnarray}  \textrm{QL-weight}_{k,j} = \left\langle \frac{k_{\theta}}{B}\frac{\left|\left[\delta n_{k,j},\delta E_{i,k,j}\delta E_{e,k,j}\right]\right|}{\left|\delta\phi_{k,j}\right|} \sin\Delta\Phi_{k,j}^{\left[\delta n,\delta E_i,\delta E_e\right]-\delta\phi} \right\rangle 
	\label{QLval12} \end{eqnarray}
where the here brackets refer to a flux-surface average. Eq. \eqref{QLval12} points out the reason why the linear cross-phases alone do not carry the whole information on the quasi-linear response, as mentioned in paragraph \ref{sec-cross-phases}. In fact, the amplitude ratios $\left|\left[\delta n_{k,j},\delta E_{i,k,j}\delta E_{e,k,j}\right]\right|/\left|\delta\phi_{k,j}\right|$ appearing in Eq. \eqref{QLval12} are effectively part of the quasi-linear transport weights, defining linear relative amplitude weightings for each unstable mode.\\
\indent There are three main strategies that can be adopted when practically calculating these transport weights, and therefore the quasi-linear turbulent fluxes, according to expression \eqref{QLval7}. These are here below summarized.
\begin{enumerate}
	\item \textbf{Eigenvalue code approach}. A discrete number of linear eigenmodes are found by a numerical solver; for each wave-number $k$ there are $j$ different solutions. $\textrm{QL-weight}_{k,j}$ is then unambiguously referred to the corresponding complex linear eigenvalue $\omega_{k,j}=\omega_{0k,j}+i\gamma_{0k,j}$. 
Defining the following linear complex functional (that immediately derives from Eqs. \eqref{QLpartflux} and \eqref{QLenergyflux}):
\begin{eqnarray}  \mathcal{F}_{k,j} = \int\frac{d\omega}{\pi} \left\langle \mathcal{E}^p e^{-\mathcal{E}} 
 \mathcal{S}_{k,j}\left(\omega\right) \frac{\omega -n\omega_s^*}{\omega -n\Omega_s\left(\mathcal{E},\lambda\right) + i\nu_{kj}} \mathcal{J}_0^2\left(k_{\perp}\rho_s\right) \right\rangle_{\mathcal{E},\lambda}
	\label{QLval7b} \end{eqnarray}
where $p=1/2$ when dealing with particle transport and $p=3/2$ for energy transport, the general expression for the quasi-linear weight is straightforward:
\begin{eqnarray}  \textrm{QL-weight}_{k,j} = \textrm{Im} \left( \mathcal{F}_{k,j} \right)
	\label{QLval7c} \end{eqnarray}
In this framework, the expression of the quasi-linear cross-phase is instead:
\begin{eqnarray}  \Delta\Phi_{k,j}^{ql} = \arctan \frac{\textrm{Im} \left( \mathcal{F}_{k,j} \right)}{\textrm{Re} \left( \mathcal{F}_{k,j} \right)}
	\label{QLval7d} \end{eqnarray}
From Eqs. \eqref{QLval7b}, \eqref{QLval7c} and \eqref{QLval7d}, it appears that the actual value of the quasi-linear weights, as well of the cross-phases, here depend on the choice on the resonance and frequency broadenings, respectively $\nu_{kj}$ and $\mathcal{S}_{k,j}$ retained within the linear dispersion relation, as discussed in paragraph \ref{sec-quasilinear-new-model}. The QuaLiKiz model for example, calculates Eq. \eqref{QLval7b} in the limit of $\nu_{kj}\rightarrow0^+$ and $\mathcal{S}_{k,j}$ with a Lorentzian shape. For this reason, a systematic verification of the quasi-linear response, apart from any further hypothesis, is investigated through the following two points rather than with the eigenvalue approach.\\
  \item \textbf{mQL, Leading mode approach}. In the case of initial value codes (such as the linear version of GYRO, see Appendix \ref{appx2-GYRO-GYSELA} for details), only the linear leading mode can be numerically solved; the index $j$ is then here fixed to $j=1$. In this case, the quasi-linear weight can be written as:
\begin{eqnarray}  \textrm{QL-weight}_{k,1} = \frac{\textrm{Re} \left\langle \left[\delta n_{e,k,1}^{lin},\delta E_{i,k,1}^{lin},\delta E_{e,k,1}^{lin}\right] \delta v_{r,k,1}^{*lin}\right\rangle}{\left\langle \left|\delta\phi_{k,1}\right|^2 \right\rangle}
	\label{QLval8} \end{eqnarray}
while the cross-phase takes the form:
\begin{eqnarray}  \Delta\Phi_{k,1}^{ql} = \arctan \frac{\textrm{Im} \left( \left[\delta n_{k,1}^{lin},\delta E_{i,k,1}^{lin},\delta E_{e,k,1}^{lin}\right] \right)}{\textrm{Re} \left( \left[\delta n_{k,1}^{lin},\delta E_{i,k,1}^{lin},\delta E_{e,k,1}^{lin}\right] \right)} - 
\arctan \frac{\textrm{Im} \left( \delta\phi_{k,1}^{lin*} \right)}{\textrm{Re} \left( \delta\phi_{k,1}^{lin*} \right)}
	\label{QLval8b} \end{eqnarray}
where the superscript \textit{lin} refers to a linear simulation. It is worth noting that both the eigenmodes appearing at the numerator and at the denominator of Eq. \eqref{QLval8} are exponentially diverging in time\footnote{The linear eigenmode temporal dependence follows $\delta\phi_k\left(t\right)=\delta\phi_k e^{-i\omega_{k,1}t+\gamma_{k,1}t}$}, since they are linearly unstable: the ratio between these two quantities is instead properly defined. Moreover, the advantage of this approach is that the quasi-linear weight of Eq. \eqref{QLval8} does not depend on any additional choice on the broadenings, contrarily to the eigenvalue approach. The time evolution of a normal linear mode in the initial value codes is in fact solved using the quasi-neutrality constraint $\delta n_{e,k,1}=\delta n_{i,k,1}$, so that the quasi-linear particle flows are automatically ambipolar, i.e. $\Gamma_{e,k,1}^{QL}=\Gamma_{i,k,1}^{QL}$.\\
\indent Nevertheless, according to this formulation, one should be able to identify and separate a nonlinear spectral intensity for each unstable mode $j$ at the wave-number $k$, i.e.
\begin{eqnarray}  \textrm{Spectral-intensity}_{k,j} = \left\langle \left|\delta\phi_{k,j}^{nl}\right| \right\rangle
	\label{QLval9} \end{eqnarray}
Unfortunately, this kind of information can not be inferred from nonlinear simulations, since the distinguishable structure of the linear eigenmodes does not survive in the nonlinear saturation regime. Therefore, only one $k$-resolved nonlinear saturation spectrum can be computed, i.e. $\left\langle \left|\delta\phi_{k}^{nl}\right| \right\rangle$. In order to gain an estimation of the quasi-linear turbulent flux according to Eq. \eqref{QLval7}, the mQL approach operates through the approximation $\left\langle \left|\delta\phi_{k,1}^{nl}\right| \right\rangle \approx \left\langle \left|\delta\phi_{k}^{nl}\right| \right\rangle$. It is important to detail the main consequences of this approximation.\\
Since the quasilinear weight of the leading mode is expected to be larger than that for the sub-dominant modes, the full spectral intensity applied only to the leading mode, will tend to produce a larger quasi-linear transport than it would be obtained if correctly distributed over both leading and sub-dominant modes. In fact, the leading mode in each toroidal wave-number typically refers to an ion or electron directed unstable branch. Moreover, within each branch, both the initial value mQL and the eigenvalue code approaches keep only the most outward ballooning mode, which is typically the most unstable. Effectively, there is a continuum of lesser ballooning and less unstable modes with smaller quasi-linear weight and lesser but not zero spectral weight.  These are left uncounted, while in principle they can contribute to the quasi-linear transport.\\
\indent Practically, the accuracy of the mQL approach is tested by two-step gyrokinetic simulations: firstly, a linear run is performed to get the quasi-linear weight on the leading mode, then a nonlinear simulation on the same parameters allows to get the nonlinear spectral intensity as well as the actual nonlinear transport. The quasi-linear fluxes $\left[\Gamma_e^{mQL},Q_i^{mQL},Q_e^{mQL}\right]$ obtained according to Eq. \eqref{QLval8} can then be compared to the nonlinear transport in each channel and wave-number. The ratios $\left[\Gamma_e^{mQL}/\Gamma_e^{nl},Q_i^{mQL}/Q_i^{nl},Q_e^{mQL}/Q_e^{nl}\right]$ (otherwise called overages) are physically meaningful, since they are independent of the structure of the saturation spectrum that is fixed by the nonlinear simulation.\\
	\item \textbf{fQL, Full frequency spectrum approach}. In contrast to the practical interest of the previous two approaches, this latter one, referred as fQL, is mainly of theoretical interest. The aim in this case is to suggest what might be ultimately captured from the quasi-linear theory if we could accurately model the full frequency spectrum of the nonlinear saturation, and in particular the portion not captured by the leading normal mode (or modes) in each wave-number. The fQL can be interpreted as the limit passage from the discrete mode $j$ summation of Eq. \eqref{QLval7} to an appropriate integration over the continuous frequency $\omega$, i.e. $\sum_j\rightarrow\int d\omega$.\\
\indent As already mentioned, there is no practical way to directly test the full frequency spectrum approach, i.e. first finding the quasi-linear weights over a wide range of frequencies, then secondly capturing the corresponding frequency dependent nonlinear spectral intensities for each wave-number. The method here developed to study the fQL approach relies instead on the simultaneous treatment of both nonlinear plasma species and linear tracers \cite{waltz09b}. The simulations retain the main species, i.e. ions and electrons at full densities $n_{e,main}$ and $n_{i,main}$, and identical tracer ion and electron species at negligible densities $n_{e,tr}$ and $n_{i,tr}$, such that $\frac{n_{e,tr}}{n_{e,main}}=\frac{n_{i,tr}}{n_{i,main}}\ll 1$. Hence, the main plasma species alone drive the potential fluctuations and the true (not external or artificial) saturation spectral intensity. Due to their negligible densities, the tracer species have no feedback on the potential fluctuations through the Poisson equation. If both the nonlinear $E\times B$ as well the linear terms are retained in the gyrokinetic equations governing the motion of the tracer species, then the turbulent fluxes of the main species and of the tracers are identical. However, if the tracers have only linear motion, by artificially deleting the nonlinear terms, the resulting tracer turbulent fluxes $\left[\Gamma_{e,tr}^{fQL},Q_{i,tr}^{fQL},Q_{e,tr}^{fQL}\right]$ can be truly taken as reliable quasi-linear transport estimates, here referred as the fQL fluxes. Further details on this method, as it has been implemented in the GYRO code, are reported in Appendix \ref{appx2-GYRO-GYSELA}. The quasi-linear weight according to the fQL approach can then be defined only for each toroidal wave-number $k$ (while the sum over the index $j$ of Eq. \eqref{QLval7} has no sense anymore here):
\begin{eqnarray}  \textrm{QL-weight}_{k} = \frac{\left[\Gamma_{e,tr,k}^{fQL},Q_{i,tr,k}^{fQL},Q_{e,tr,k}^{fQL}\right]}
  {\left\langle \left|\delta\phi_k\right|^2 \right\rangle}
	\label{QLval10} \end{eqnarray}
\indent The relevance of this method relies on the information carried by the transport ratios (overages) between the linear tracers and the nonlinear main species. The overages $\left[\Gamma_{e,tr}^{fQL}/\Gamma_e^{nl},Q_{i,tr}^{fQL}/Q_i^{nl},Q_{e,tr}^{fQL}/Q_e^{nl}\right]$ should in fact quantify the goodness of the best quasi-linear approximation, since no additional hypothesis has been introduced apart from the suppression of the nonlinear dynamics in the gyrokinetic equation. It is reasonable to expect the overages from the fQL approach to be somewhat smaller than those from the mQL approximation. The reason is due to the fact that the fQL approach does not discount any less unstable mode, while properly accounting for the whole structure of every sub-dominant mode. Conversely, the fQL approach exhibits an intrinsic failure of the ambipolarity condition, i.e. $\Gamma_{e,tr}^{fQL}\ne\Gamma_{i,tr}^{fQL}$. The tracer species dynamics is in fact not self-consistently evolved through the Poisson equation; this point is a natural consequence of the nature of the quasi-linear problem, where the feedback of the linear particle motions on the fluctuating fields is neglected. Nevertheless, since the nearly adiabatic electrons are expected to control the particle transport, the tracer fQL ion particle transport is usually neglected.
\end{enumerate}
\noindent The validity of the quasi-linear response can then be systematically tested by mean of both the mQL and fQL approaches. The key measures of success for the quasi-linear response approximation are then:
\begin{itemize}
	\item The quasi-linear over nonlinear overages have to be nearly the same in each transport channel (particle, ion and electron energy) and, less importantly, in each wave-number.
	\item The overages should be nearly constant across a wide variety of physical case parameters.
\end{itemize}

\begin{figure}[!htbp]
  \begin{center}
    \leavevmode
      \includegraphics[width=9 cm]{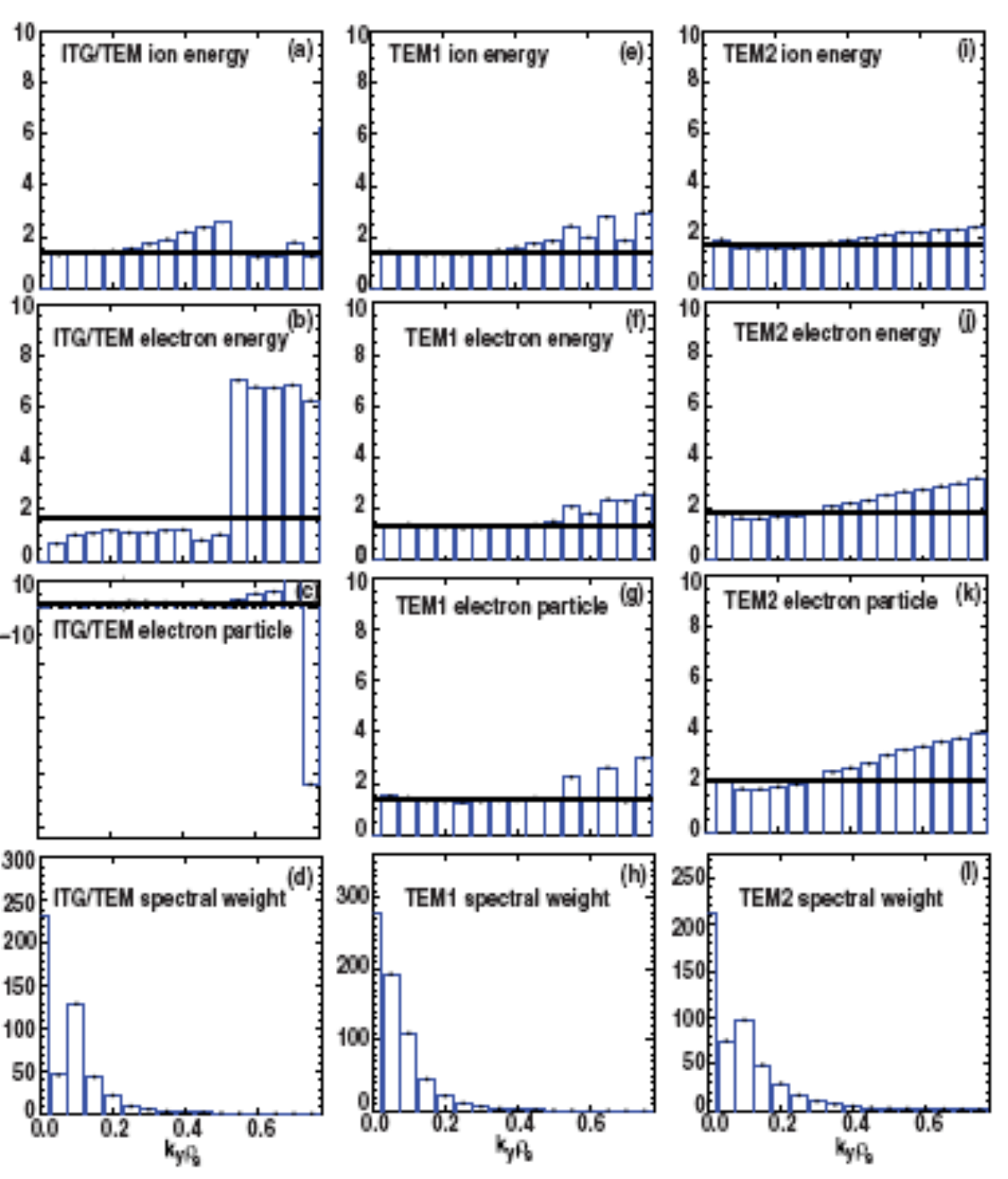}
    \caption{Wave-number spectrum of overage for ion energy (a, e, i), electron energy (b, f, j), and particle (c, g, k) transport, and nonlinear spectral intensity (d, h, l) for ITG-TEM, TEM1 and TEM2 cases. The horizontal lines indicate the net overage in each channel as given in Table \ref{mQL-table}.}
    \label{mQL-spectra}
  \end{center}
\end{figure}

\indent The validity of the quasi-linear response is tested using the gyrokinetic code GYRO. In addition to the GA-standard case already defined in Table \ref{GA-std-table}, two other sets of plasma parameters are defined: while keeping fixed all the other parameters, the normalized density and temperatures gradients are summarized in Table \ref{GA-ITG-TEMtable}.
\begin{table}[!h]
	\begin{center} \begin{tabular}{|c|c|c|c|}
			\hline
			{ } & GA-ITG-TEM & GA-TEM1 & GA-TEM2 \\
			\hline $R/L_{Ti}$ & 9.0 & 6.0 & 3.0 \\ 
			\hline $R/L_{Te}$ & 9.0 & 6.0 & 3.0 \\
			\hline $R/L_{n} $ & 3.0 & 6.0 & 9.0 \\ \hline
	\end{tabular} \end{center}
	\caption{\label{GA-ITG-TEMtable}Density and temperatures gradients lengths characterizing the three sets of plasma parameters.}
\end{table}
In the GA-standard case (ITG-TEM), the ITG mode linearly dominates at low $k$, while TEM leads for $0.5<k_{\theta}\rho_s<0.75$. For GA-TEM1, one half of the linear modes in the spectrum are ITG and the other half are TEM with similar growth rates. Finally in GA-TEM2, TEM linearly dominates at all $k_{\theta}\rho_s<0.75$.
\begin{table}[!h]
	\begin{center} \begin{tabular}{|c|c|c|c|}
			\hline
			{ } & $\chi_i^{mQL}/\chi_i^{nl}$ & $\chi_e^{mQL}/\chi_e^{nl}$ & $D_e^{mQL}/D_e^{nl}$ \\
			\hline GA-ITG-TEM & 17.9/12.1=\textbf{1.47} & 5.83/3.42=\textbf{1.70} & -3.21/-2.01=\textbf{1.79} \\ 
			\hline GA-TEM1 & 21.8/15.0=\textbf{1.45} & 21.0/15.3=\textbf{1.37} & 7.86/5.50=\textbf{1.42} \\
			\hline GA-TEM2 & 43.8/25.9=\textbf{1.69} & 56.3/30.3=\textbf{1.85} & 12.2/5.9=\textbf{2.07} \\ \hline
	\end{tabular} \end{center}
	\caption{\label{mQL-table}Overages for the two-step tests of the mQL approach as ratios of quasi-linear to nonlinear diffusivities for different sets of plasma parameters; diffusivities are in Gyro-Bohm units, where $\chi_{GB}=c_s/a\rho_s^2$.}
\end{table}
Table \ref{mQL-table} shows the quasi-linear/nonlinear overages for the two-step tests according to the mQL approach. The average across the results presented in Table \ref{mQL-table} gives an overage of 1.64, with an RMS of the normalized deviations of 13.5\%, which has to be considered a successful result. Fig. \ref{mQL-spectra} reports the corresponding overages across wave-numbers as well as the nonlinear saturation spectral intensity for the three cases.
It appears that the overages at higher $k$ present, sometimes considerably, deviations. This can be easily understood if recalling the cross-phases of Figs. \ref{crossph-1}, \ref{crossph-3} and \ref{crossph-2}, where analogous behavior was found. As previously mentioned, this breakdown has little effect on the on the total quasi-linear transport, since the most relevant part of fluxes is carried by the low $k$ scales. Recalling the formulation of Eq. \eqref{QLval7}, this is due to the weak weight of the nonlinear saturation spectrum at the high $k$ (Fig. \ref{mQL-spectra}). The apparent breakdown of the mQL approach at high $k$ could support the guess of a nonlinear wave-particle resonance broadening represented by the additional $k^2D$ term \cite{dupree68,waltz83}. In this direction, it is useful to note that the gyrokinetic code GENE has been used to show that the $E\times B$ nonlinearity can be well represented by a (purely real) linear damping in the form $k_y^2D$ for a cold ion TEM turbulence \cite{merz08}. The really large jump in the overages for $k_{\theta}\rho_s>0.5$ in the GA-ITG-TEM case can be associate with the transition from the dominance of the ion (ITG) to the electron (TEM) branch at higher-$k$. The ITG-TEM case with its inward (pinched) particle flow, i.e. $D_e<0$, seems to have the most tenuous validity according to the mQL approach and this point will be deepened in the following.\\
\indent Turning to the validation of the full frequency spectrum quasi-linear approach (fQL approach) for the GA-ITG-TEM case, it is important to firstly verify the plasma tracers method itself. Indeed, when properly retaining the nonlinear terms in the gyrokinetic equations governing the dynamics of ion and electron tracers, turbulent diffusivities are found to be identical between the main plasma species and the tracers\footnote{In particular, for the GA-ITG-TEM case, the results of the GYRO simulation are: $\chi_{i,tr}/\chi_{i,main}=12.0/12.0$, $\chi_{e,tr}/\chi_{e,main}=3.22/3.28$, $D_{e,tr}/D_{e,main}=-2.01/-2.0$. Some small difference in the electron diffusivities are expected, since the tracer electrons (and also both main and tracer ions) are evolved according to an explicit numerical method, whereas the main electrons use a mixed explicit-implicit scheme (for fast parallel motion)}.\\
Table \ref{fQL-table} provides instead the overages for the fQL approach by deleting the nonlinearity in the tracer species.
\begin{table}[!h]
	\begin{center} \begin{tabular}{|c|c|c|c|c|}
			\hline
			{ } & $\chi_{i,tr}^{fQL}/\chi_{i,main}^{nl}$ & $\chi_{e,tr}^{fQL}/\chi_{e,main}^{nl}$ & $D_{e,tr}^{fQL}/D_{e,main}^{nl}$ & $D_{i,tr}^{fQL}/D_{i,main}^{nl}$ \\
			\hline ITG-TEM & 16.7/11.3=\textbf{1.48} & 3.66/3.17=\textbf{1.15} & -2.7/-1.90=\textbf{1.43} & -5.1/-1.9 \\ \hline
	\end{tabular} \end{center}
	\caption{\label{fQL-table}Overages for the one-step tracer tests of the fQL approach as ratios of quasi-linear to nonlinear; the apparent discrepancy with respect to Table \ref{mQL-table} for the main species diffusivities is due to slightly different grids and time averages. Diffusivities are in Gyro-Bohm units.}
\end{table}
As expected, the tracer particle flow is not ambipolar: the ion tracer particle diffusivity is two times larger than the electron tracer one. Since electron non-adiabaticity controls particle flow, we consider only the electron particle flow overage of 1.43, which is a value very close to the ion energy overage. In this case, the electron energy transport presents the lowest quasi-linear over nonlinear ratio, with an overage factor which is close to unity. As already mentioned, it is reasonable to expect that quasi-linear over nonlinear overages obtained according to the fQL (Table \ref{fQL-table}) approach are smaller than those from the mQL approach (Table \ref{mQL-table}), since the nonlinear spectral intensity is also in part distributed over the all sub-dominant, modes which should have smaller quasi-linear weights.

\indent The validation of the quasi-linear response does not only rely on the structure of the quasi-linear over nonlinear overage across the $k$ wave-number and the transport channels. Even more importantly for the application to quasi-linear models, these overages should also be reasonably constant over a wide variation of the plasma parameters. In Figure 3, the GA-ITG-TEM case is reconsidered over a wide range of temperatures gradients $6.0<R/L_{Ti,e}<27.0$, using both the mQL and the fQL approach.
\begin{figure}[!htbp]
  \begin{center}
    \leavevmode
      \includegraphics[width=12 cm]{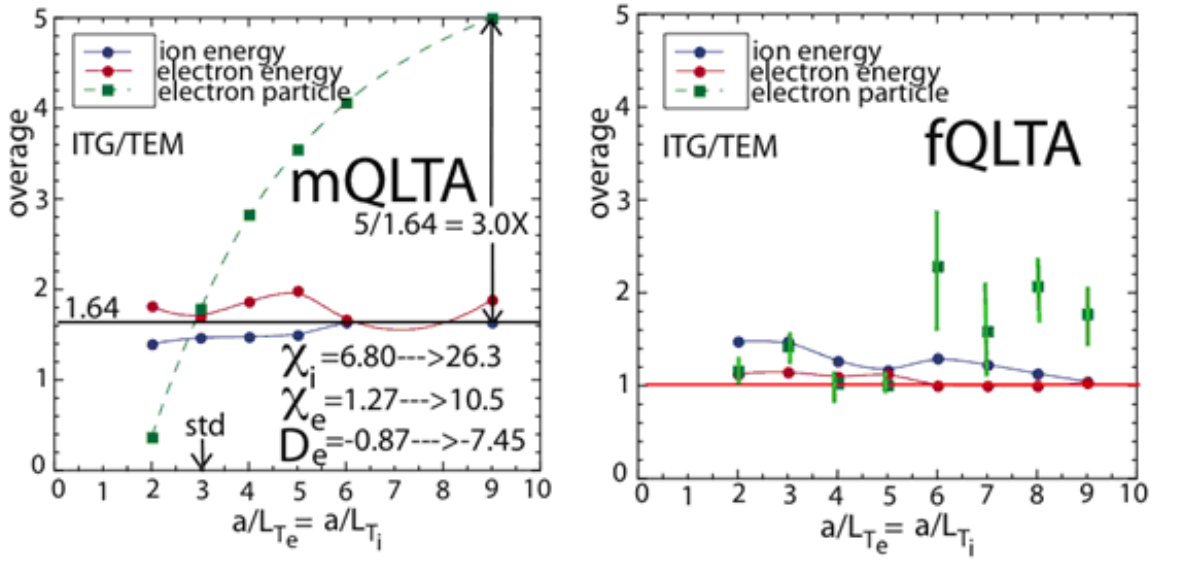}
    \caption{Quasi-linear over nonlinear overages across a temperature gradient scan based on the GA-ITG-TEM standard case, using both the mQL and the fQL approaches.}
    \label{QL-ITG-TEM-scan}
  \end{center}
\end{figure}
The results of Fig. \ref{QL-ITG-TEM-scan} show that the quasi-linear over nonlinear overages in both ion and electron energy channels remain nearly constant, even though the overall transport level increases by 4-8 times. As expected, the fQL approach leads to lower average quasi-linear over nonlinear ratio, about 1.3 versus 1.6 from the mQL approach.\\
According to the mQL approach, the overage for the particle transport, characterized by a strongly inward flow at high gradients ($D_e/\chi_{GB}$ up to -7.45), considerably breaks away from the overages on the energy transport. The mQL particle transport overage fails with an underflow at $a/L_T=2.0$ ($R/L_T=6.0$). This latter breakdown is likely due to the lack of sub-dominant modes within the mQL approach. More importantly, the particle transport overage shows a monotonic increase with higher temperature gradients, with an over-prediction with respect to the nonlinear factor by a factor up to 3. It is worth noting that this failure of the quasi-linear particle transport with strong turbulence drives, is consistent with the previous findings on the cross-phases $\delta\phi-\delta n$ shown in Fig. \ref{crossph-3}, where the linear phase shifts significantly overestimate the nonlinear ones. The key point that could not be verified through the study of the cross-phases is checking if a proper accounting of the linear sub-dominant modes can fix this over-prediction and get closer to the nonlinear predictions.
\indent The latter point can now be investigated thanks to the fQL approach. At high gradients, the quasi-linear particle transport over-prediction appears less pronounced (because of the accounting of sub-dominant modes), but still present, with quasi-linear over nonlinear ratios 2 times bigger than those of the energy transport. Hence, the quasi-linear over-prediction observed for high gradients, $R/L_T>\approx 15$, has necessarily a different origin, and it can not be related to the accounting of sub-dominant modes. This failure is in fact equally evident using both the mQL and the fQL approaches, where the latter one correctly retain the whole linear modes structures.\\
\indent These results are then in favor of recognizing the failure of the quasi-linear response when trying to reproduce the nonlinear particle transport of strongly inward flows. At this regard, it is worth noting that realistic tokamak plasma conditions foresee rather weak core particle source, so that the typical operating point has a tiny or null net particle flow in the core. A strong outflow down the density gradient takes place instead at the edge, balancing gas feed and wall recycling. The strong inward particle pinch reported in Fig. \ref{QL-ITG-TEM-scan} for $R/L_T>\approx 13$, is mostly due to ion unstable modes, can then appear as a rather pathological condition.\\ 
For this reason, another temperature gradient scan based on a modified GA-TEM2 case is explored. Referring to the parameters reported in Table \ref{GA-ITG-TEMtable}, GA-TEM2 is modified considering $R/L_{Ti}=3.0\Rightarrow0.3$. In this case, the strong density gradient results into a turbulence largely dominated by electron TEM modes while the reduced ion temperature gradient suppresses the ion ITG branch. The particle transport is hence characterized by outward flows across the whole scan of the electron temperature gradient $3.0\le R/L_{Te}\le 27.0$. Fig. \ref{QL-TEM2-scan} reports the results using the mQL approach for the quasi-linear expectations. Here, the quasi-linear/nonlinear overages appear relatively constant for both the ion and electron energy but most of all for the particle transport, even at high temperature gradients (compare with Fig. \ref{QL-ITG-TEM-scan}). 
\begin{figure}[!htbp]
  \begin{center}
    \leavevmode
      \includegraphics[width=7 cm]{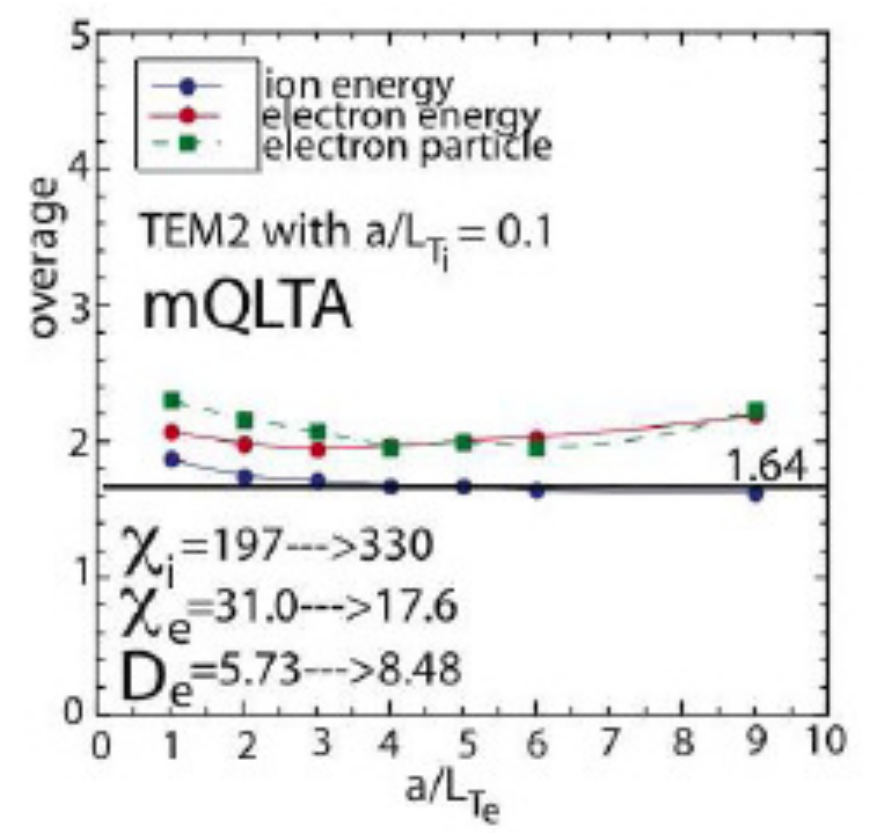}
    \caption{Quasi-linear over nonlinear overages across an electron temperature gradient scan based on a modified GA-TEM2 standard case, using both the mQL approache.}
    \label{QL-TEM2-scan}
  \end{center}
\end{figure}
The results of Fig. \ref{QL-TEM2-scan} are then in favor of supporting the guess that the quasi-linear response may not appropriate for capturing the nonlinear physics responsible for strong inward particle flux, while it appears still reliable for more usual moderate particle flows.\\

\section{Summary}

\indent The most relevant findings of the present section are here briefly summarized:
\begin{enumerate}
	\item The validation of the hypothesis of the quasi-linear response is systematically and quantitatively studied on comprehensive nonlinear and linear gyrokinetic simulations of tokamak micro-turbulence. Three different methods are defined and applied: (a) characteristic turbulence times, or Kubo-like numbers, (b) cross-phases of the fluctuating quantities, (c) transport weights.
	\item The Kubo-like numbers evaluated across a wide temperature gradient scan on tokamak relevant conditions, fully satisfy the quasi-linear ordering of the characteristic turbulence times. The effective auto-correlation time is found to be smaller than the eddy turnover time, a condition that should ensure that the particles are not trapped in the field pattern.
	\item The phase shifts between the transported quantities and the fluctuating potential highlights a good agreement between the nonlinear and linear (most unstable mode) regimes for relevant tokamak parameters. Nevertheless, the cross-phases alone can not be used to quantitatively infer the validity of the quasi-linear response. A major caveat concerns in fact the accounting of sub-dominant modes.
	\item The properly defined transport weights are finally able to quantitatively address the validation of the quasi-linear response. The latter one presents a typical over-prediction with respect to the nonlinear transport levels. Most importantly this overall factor is shown to be reasonably constant across (a) the different transport channels (particle, ion and electron energy), (b) the variation of the plasma parameters on tokamak relevant conditions. Nevertheless, the quasi-linear particle transport appears to significantly deviate from the nonlinear expectations when dealing with strong inward flows.
\end{enumerate}

\chapter{Improving the underlying hypotheses of the model}
\label{cap4-improving-QL}

\indent In the previous chapter, the gyrokinetic quasi-linear transport modeling has been investigated firstly in terms of its formal derivation and actual formulation applied to tokamak plasmas. Secondly, the hypothesis of quasi-linear response has been studied and validated by mean of systematic comparisons with comprehensive nonlinear gyrokinetic simulations. This chapter is instead dedicated to the other fundamental part contributing to the quasi-linear transport, i.e. the model of the nonlinear potential saturation, recalling the generic expression of the quasi-linear flux Eq. \ref{QLval7}. Once again we stress the crucial importance of a clear distinction between the latter issue and the one examined in the previous chapter. Since the intrinsic nature of the quasi-linear theory can not provide any information on the nonlinear saturation process, the choices and the resulting implications adopted for modeling the saturated spectrum deserve careful dedicated analysis.\\

\indent \textsl{The general aim of this chapter is the elaboration of a nonlinear saturation spectrum model, which can integrate the information carried by the quasi-linear response in order to produce reliable estimates of the turbulent fluxes.} \\

\indent Following from these preliminary remarks, the tools needed to investigate this issue has to be defined. In this work, we propose a novel physics oriented approach, where the information is carried in parallel from two different sources:
\begin{enumerate}
	\item The properties of the nonlinear saturation process are studied by comprehensive nonlinear gyrokinetic simulations. As the latter ones represent the most advanced tool actually available to gain quantitative information on the tokamak turbulence, the use of nonlinear gyrokinetic simulations is essential in order to model the saturation mechanisms.
	\item The experimental fluctuation measurements are a primal information about the real nonlinear processes that are at play in tokamak plasmas. The analysis of these data is of great importance in order to validate and improve the hypotheses adopted in the model for the saturation spectrum.
\end{enumerate}
Both the nonlinear simulations and the measurements must equally contribute to the improvement of the potential saturation choices, which otherwise would remain completely arbitrary. Conversely, the integration of the experimental and the numerical results allows relevant advances in the understanding of the underlying physics and therefore justifies the choices that will be adopted in QuaLiKiz.\\
\indent The approach here just briefly described, is a unique feature proposed within the present work. In fact, in most of the existing quasi-linear tokamak transport models the quasi-linear response is coupled to generic mixing length saturation rules, without a clear explanation of the underlying choices. More recent quasi-linear transport models, like TGLF, have more carefully treated the issue of the nonlinear saturation hypotheses. Nevertheless, the strategy adopted by TGLF is different from the one here proposed. In the case of TGLF in fact, the quasi-linear gyro-fluid response is coupled to an artificial nonlinear saturation intensity\footnote{called V-norm, further details on this point can be found in \cite{kinsey08}}, which proves superior when fitting the resulting TGLF total fluxes with a large database of GYRO nonlinear simulations.\\
Conversely, in the present work we do not rely on a numerical database best fit approach, but the fluctuation measurements are used to physically interpret and validate the expectations provided by the nonlinear gyrokinetic simulations. Once we trust these first principle numerical results, they can be used to powerfully improve the choices retained in the saturation model of a quasi-linear transport model.\\
\indent The chapter is organized as follows. The first part addresses the issue of the validation of the numerical tools, i.e. the nonlinear gyrokinetic simulations, against the turbulence measurements done in the Tore Supra tokamak. More in particular, after a brief introduction about the experimental techniques used to diagnose the tokamak plasma fluctuations, detailed analysis is devoted to the spectral structure of the density fluctuation in terms of both the wave-number $k$ and the frequency $\omega$ dependences. Consequently, the choices concerning the saturated potential adopted in QuaLiKiz are discussed. Here, both the spectral structure (in the $k$ and $\omega$ spaces) and the saturation rules adopted to weight the different contributions to the total transport are treated.

\section{Validating the nonlinear predictions against measurements}

\indent The understanding of turbulent fluctuations in the core of tokamak plasmas, causing a degradation of the confinement, is a crucial issue in view of future fusion reactors. Global empirical scaling laws based on the existing experiments \cite{yushmanov90,kaye97} are often used to extrapolate the performance of the next generation devices like ITER and DEMO. However, predictive capabilities should rely on first principle models retaining comprehensive physics. Hence, the development of validated predictive codes is an essential task for ensuring the success of the present and the future fusion experiments.\\
\indent While the general tokamak plasma instabilities framework has an history of intensive analytical and computational efforts from more than 30 years, the recent times are characterized by the following very relevant advances:
\begin{itemize}
	\item Numerical codes matured to the point to realistically include what it is believed to be the essential complexity of tokamak physics, allowing to provide quantitative information about the plasma turbulence and the associated transport.
	\item Great progresses have been achieved in developing reliable diagnostics, able to accurately measure the properties of the tokamak turbulence in wide range of parameters and experimental conditions.
	\item The computing power has significantly raised, especially thanks to the massive use of parallel high performance computing, allowing to perform highly sophisticated simulations on tractable time scales.
\end{itemize}
\indent The issue of the nonlinear codes validation has very recently produced a great amount of research. Some of the most relevant efforts have been done across several tokamaks worldwide, namely on DIII-D \cite{holland08,white08}, Alcator C-mod \cite{llin09,llin09b} and Tore Supra \cite{casati09}.\\
\indent In order to truly validate a tokamak turbulence model, two different levels of information have to be accurately predicted. The first one refers to the detailed spectral structure of the turbulence (lower-order quantities), i.e. both the spatial and time spectral shapes and amplitudes of the different fluctuation fields; the second level instead deals with the macroscopic observables (higher-order quantities), such as the total turbulent flows and fluctuation amplitudes, that integrate the previous spectral quantities. This approach has also been summarized by a practical strategy towards the validation in the fusion research, identifying \cite{terry08} the following two key concepts:
\begin{enumerate}
	\item Primacy Hierarchy: \textit{Ranking of a measurable quantity in terms of the extent to which other effects integrate to set the  value of the quantity. Assesses ability of measurement to discriminate between different non-validated models}.
	\item Validation Metric: \textit{A formula for objectively quantifying a comparison between a simulation result and experimental data. The metric may take into account errors and uncertainties in both sets of data as well as other factors such as the primacy of the quantities being compared}.
\end{enumerate}
\indent In this section, nonlinear gyrokinetic simulations are quantitatively compared to turbulence measurements on the Tore Supra tokamak. The relevance of this analysis relies on the fact that the high-order scalar observables are coherently verified also through the investigation of the lower-order spectral quantities. With respect to the previous works, the present study is qualified by the simultaneous validation of (1) the total heat transport coefficient, (2) rms values of the density fluctuations $\delta n$, (3) $k_{\theta}$ and (4) $k_r$ wave-number $\delta n$ spectra and (5) the frequency $\delta n$ spectra.

\subsection{Turbulent fluctuation measurements}

\begin{figure}[!htbp]
  \begin{center}
    \leavevmode
      \includegraphics[width=6 cm]{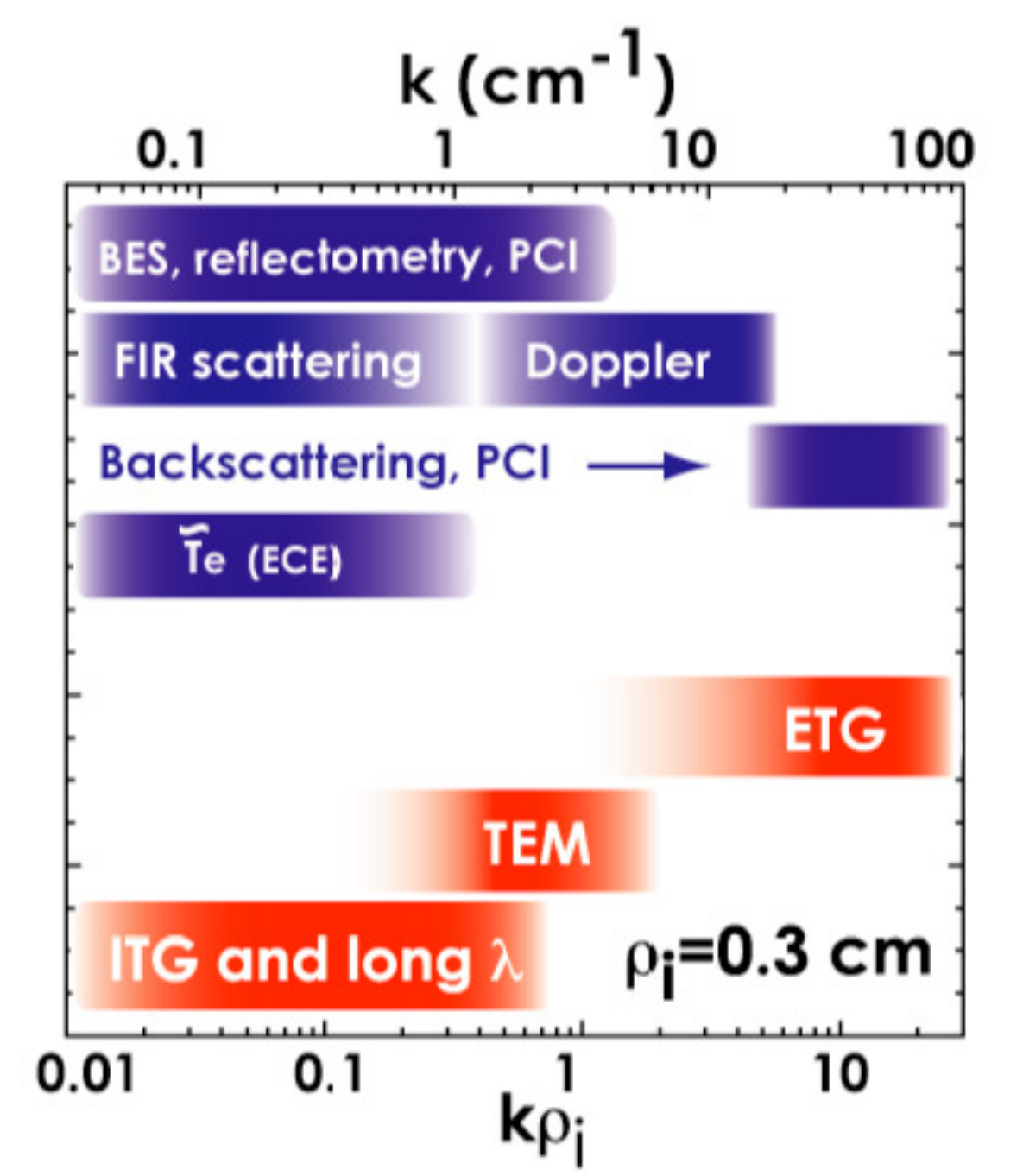}
    \caption{Picture summarizing some actual tokamak fluctuation measurement techniques, comparing the diagnostic typical spectral ranges to those of the ITG, TEM and ETG instabilities.}
    \label{Turb-diag}
  \end{center}
\end{figure}

\begin{figure}[!htbp]
  \begin{center}
    \leavevmode
      \includegraphics[width=12 cm]{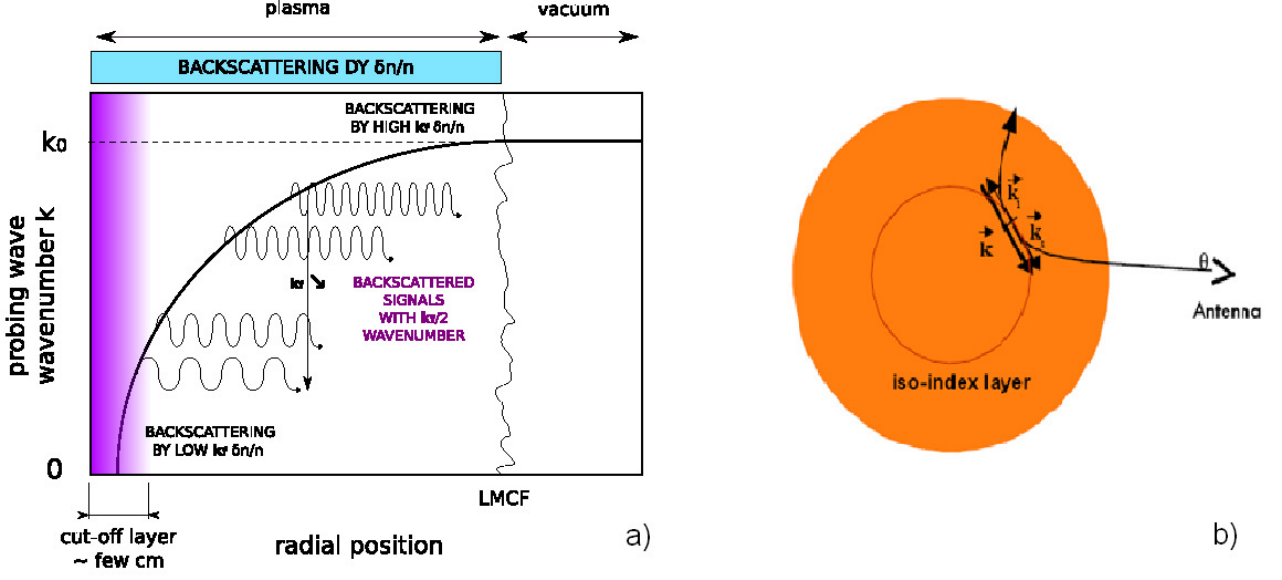}
    \caption{Schematic representation of (a) the perpendicular incidence reflectometry and (b) the Doppler reflectometry.}
    \label{Refl-pict}
  \end{center}
\end{figure}

\indent Several techniques have been developed in order to investigate the spectral structure and the main features of the density and temperature fluctuations in tokamak plasmas. An exhaustive review of these diagnostic systems is largely beyond the scope of this work. In Fig. \ref{Turb-diag}, we just remind some of the actual most important techniques, comparing their general spectral range in $k_{\perp}$ to the typical spatial scales of the tokamak instabilities.
The diagnostics appearing in Fig. \ref{Turb-diag} are (details can be found in the cited references): the Beam Emission Spectroscopy BES \cite{mckee07}, the reflectometry (Doppler \cite{hennequin06}, fast-sweeping and fixed frequency \cite{clairet01,gerbaud06}), the Far Infrared FIR \cite{brower87} and laser scatterings \cite{truc92,hennequin04}, the Correlation Electron Cyclotron Emission CECE \cite{white08}.\\
\indent The TORE SUPRA tokamak is particularly well suited to study the local spectral structure of the density fluctuations in the low and medium range $k_{\perp} \rho_s<2$: it is in fact equipped with complementary microwave diagnostics, fast-sweeping~\cite{clairet01} and Doppler~\cite{hennequin06} reflectometers. Though the geometrical arrangement and measurement techniques are different, these two methods are both based on the detection of the field backscattered on density fluctuations, whose wave number matches the Bragg rule $k_{f}=-2k_{i}$, where $k_{i}$ is the local probing wave-vector. Synthetic schemes of both the reflectometry systems are presented in Fig. \ref{Refl-pict}.\\
\indent In the fast-sweeping system, the probing wave is launched in the equatorial plane, with a $k_{i}$ in radial direction. The fast sweeping of the probing frequency (from $50$ to $110~\rm{GHz}$ in $20 \mu\rm{s}$) allows sensitive measurements of the density profile $n(r)$. This acquisition time is sufficiently fast and at high repetition rate, in order to extract the density fluctuation profile $\delta n(r)$. The latter is obtained from the integration of local spectra $S\left[\delta n/n\right]\left(k_r,r\right)$ between $1<k_{r}<10~\rm{cm^{-1}}$. The primal quantities measured by the reflectometer are the phase fluctuations, which are linked to the plasma density fluctuation through a transfer function: an iterative process of comparisons between full-wave simulations and experimental measurement is used (for more details see Ref.~\cite{gerbaud06}). The local $k_{r}$ spectrum can then be obtained using a sliding spatial Fourier transform of $\delta n/n$. The typical size of the spatial window allows probing the spectrum between $1\lesssim k_{r}\lesssim20~\rm{cm^{-1}}$, with a resolution on the spatial localization of $\Delta r/a\approx0.04$ ($a$ is the plasma minor radius). Note that in this method, fluctuations with poloidal wave-numbers up to $k_{\theta}\approx10~\rm{cm^{-1}}$ are also included in the signal due to finite beam size. This has to be taken into account when comparing with theory or simulations. The probability density function of $\delta n/n\left(r,t\right)$ shows a log-normal distribution law for all radial positions. Then time averaged $\delta n/n\left(r\right)$ values are obtained over a statistics of 1000 measurements. The computation of transfer function uses a process based on 100 normally distributed random samples, implying thus a 10\% uncertainty on its determination. The uncertainties coming from the sensitivity to the temperature and density are also accounted. Finally, the uncertainties on $\delta n/n\left(r\right)$ are estimated by quadrature addiction, with a typical level of 15 \%.\\
\indent The Doppler reflectometry technique shares with the standard reflectometry the advantages of good spatial and temporal resolution, easy access and low cost. Differently from the fast-sweeping reflectometry, in the case of the Doppler reflectometry system, the probing beam is launched in oblique incidence with respect to the cut-off layer. The scattering process is mostly localised at the cut-off, whose position is set by the (fixed) frequency of the beam~\cite{hennequin06}. In the case of O-mode beam polarisation, the selected wavenumber at the cut-off is mainly poloidal. The rms value of the signal is directly proportional to the power spectrum of density fluctuations at the selected wave-number: the wave-number spectrum is then obtained by varying the antenna tilt angle. This allows us to probe the spatial domain $0.5\lesssim r/a\lesssim0.95$ and the range of wave-numbers $4\lesssim k_{\theta}\lesssim15~\rm{cm^{-1}}$, where only very small $k_{r}\lesssim1~\rm{cm^{-1}}$ are included in the signal.\\

\noindent \textbf{Experimental set-up}\\

\begin{figure}[!htbp]
  \begin{center}
    \leavevmode
      \includegraphics[width=7 cm]{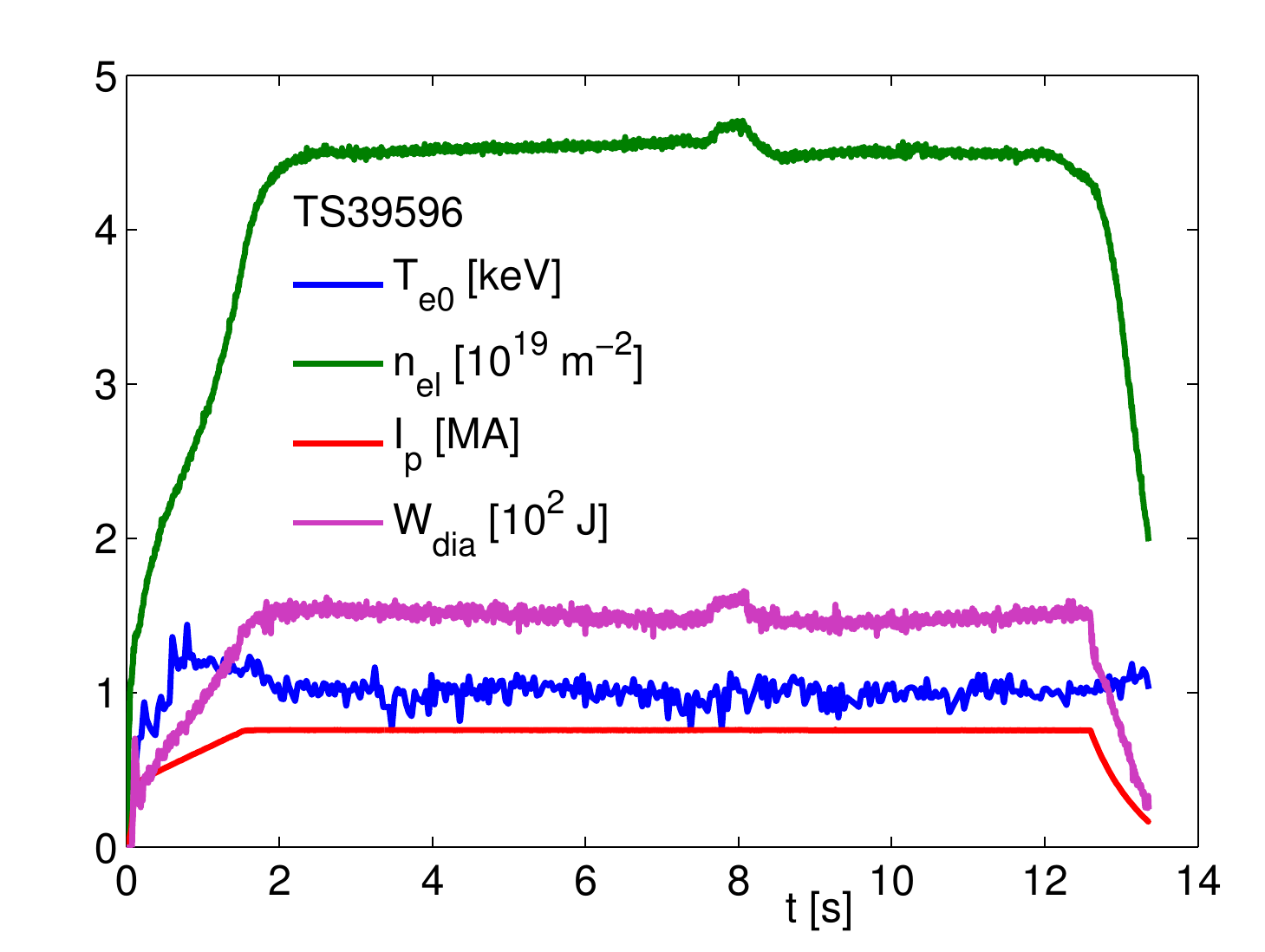}
    \caption{Time evolution of the plasma current, central electron temperature, line averaged density and diamagnetic energy content for the discharge TS39596.}
    \label{TS39596-tprf}
  \end{center}
\end{figure}

\indent One Tore Supra L-mode ohmic discharge, TS39596, has been chosen as target for the validation of the nonlinear simulations against the turbulence measurements. TS39596 is a Tore Supra representative standard shot, characterized by high reliability and repeatability. This simple discharge in fact does not differ from many other ohmic shots and it presents a large time interval with steady plasma profiles and no external momentum input. The time evolution of the principal plasma parameters is shown in Fig. \ref{TS39596-tprf}.
The plasma parameters characterizing TS39596 are moreover summarized in Table \ref{TableTS39596gl}.\\
\begin{table} 
		\begin{center} \begin{tabular}{|c|c|c|c|c|c|c|} \hline
			$B_T$ & $R_0/a$ & $I_p$ & $T_{e,0.5}$ & $T_{i,0.5}$ & $Z_{eff}$ & $n_{e0,l}$ \\ \hline
		  2.4 T & $\frac{2.38~\rm{m}}{0.73~\rm{m}}=3.25$ & 0.78 MA & 0.6 keV & 0.67 keV & 1.6 & 4.5$\cdot10^{19}~\rm{m^{-2}}$ \\ \hline
		  \end{tabular} 	\end{center} 
	\caption{\label{TableTS39596gl}Main plasma parameters for the discharge TS39596; electron and ion temperatures are referred to $r/a=0.5$}
\end{table} 
\indent The set of diagnostics installed on Tore Supra and here used to measure the plasma quantities is here briefly described. The radial profiles of the electron density are measured by the fast-sweeping reflectometers \cite{clairet01} for magnetic fields $B_T \ge 2.82~\rm{T}$, since the localization of the reflectometry cut-off layer, and hence the measurements, depends on the values of the plasma $B_T$. Since TS39596 has $B_T=2.40~\rm{T}$, the electron density profile is obtained by an Abel inversion of the line density measurements made by the interferometry. The resolution on the radial location of the measurements is $5~\rm{cm}$ when using the interferometry chords, and $1~\rm{cm}$ when using the reflectometry. The uncertainty on the density values is estimated at 10\%.\\
The Radial electron temperature profiles are measured by both Thomson scattering and electron cyclotronic emission ECE \cite{segui05}. The radial location resolution of the measurements is $10~\rm{cm}$ for the Thomson scattering, it is while close to $1~\rm{cm}$ when using the ECE. The measurements uncertainties are estimated at around 15\% for the Thomson scattering and 7\% for the ECE.\\
The ion temperatures are measured by the charge exchange diagnostic \cite{hess02}. Eight chords are available up to normalized radii of $r/a = 0.7$. The experimental uncertainty is around 20\% for the core region and 15\% in the gradients zone. The radial resolution varies between $2$ and $6~\rm{cm}$. High quality temperature and density profiles are crucial in order to minimize the resulting uncertainties on the gradient lengths, which are derived from radial derivatives of these quantities.\\
The global effective ion charge $Z_{eff}$ is measured by Brehmstralhung emission. The estimate on the radiative power given through the Matthews law by the line integrated measurements, is consistently checked with the direct radiative power measurements. The $Z_{eff}$ radial profile can not be obtained in this case, thus a flat $Z_{eff}$ profile is assumed. Simulations done with the CRONOS code \cite{basiuk03} using the measured $Z_{eff}$ have verified that the current diffusion is compatible with the measured flux consumption.\\
Finally, the safety factor $q\left(r\right)$ profile is also estimated using the CRONOS runs constrained by 10 polarimetry angles measurements. The current diffusion reproduces both the measured flux consumption and the internal value of inductance.\\
\indent The discharge TS39596 has been analyzed by mean of nonlinear gyrokinetic simulations on 4 different radial locations, namely $r/a=0.4$, $r/a=0.5$, $r/a=0.6$ and $r/a=0.7$. The experimental parameters characterizing the plasma at each location are reported in Table \ref{TableTS39596lc}.
\begin{table}
	\begin{center} \begin{tabular}{|c|c|c|c|c|c|c|c|c|c|} \hline \hline
			$r/a$ & $R/L_{Ti}$ & $R/L_{Te}$ & $R/L_{n}$ & $q$ & $s$ & $T_i/T_e$ & $\rho_*$ & $\beta$ & $\nu_{ei}$ \\
			\hline
		  0.4 & 5.34 & 9.72 & 2.6 & 1.28 & 0.55 & 1.0 & 0.0023 & 0.42\% & 0.41 \\ \hline
		  0.5 & 6.80 & 7.77 & 2.2 & 1.48 & 0.72 & 1.1 & 0.0020 & 0.31\% & 0.63 \\ \hline
		  0.6 & 7.19 & 7.64 & 2.6 & 1.71 & 0.95 & 1.1 & 0.0018 & 0.22\% & 0.90 \\ \hline
		  0.7 & 10.53 & 13.38 & 4.5 & 2.04 & 1.33 & 1.2 & 0.0016 & 0.17\% & 1.46 \\ \hline \hline
	 \end{tabular} \end{center} 
	\caption{\label{TableTS39596lc}Experimental parameters for the discharge TS39596 at different radial locations; $\rho_*=\rho_s/a$, $\nu_{ei}$ frequencies are in units of $c_s/a$ ($c_s$ is the ion-sound speed) and $s$ is the magnetic shear.}
\end{table}
The dimensionless parameters appearing in the Table \ref{TableTS39596lc} are particularly worth of note. These are $\rho_*$, the ratio of the the ion-sound Larmor radius and the plasma minor radius, $\beta$, the ratio of the plasma kinetic and magnetic pressure, and $\nu_*$, the electron-ion collision frequency normalized to the electron bounce frequency. The last ones are generally recognized as relevant parameters for fusion plasmas, so that global scalings of the energy confinement time have been proposed with crucial dependence on these three dimensionless parameters. Despite the low value of the magnetic field, the discharge TS39596 exhibit $\rho_*$ dimensionless parameters which are not so far from the values expected for ITER. At the same time, the collisionality in TS39596 is very high, while in ITER very low $\nu_*$ values are foreseen. In fact, adopting the following definition, $\nu_*=\nu_{ei}\sqrt{\frac{m_e}{T_e}}\frac{qR_0}{\epsilon^{3/2}}$, where $\epsilon=r/R_0$ is the local inverse aspect ratio, for TS39596 we obtain $\nu_*=0.66,\:0.82,\:1.05,\:1.60$ respectively for $r/a=0.4,\:0.5,\:0.6,\:0.7$.\\

\noindent \textbf{Local nonlinear gyrokinetic simulations}\\

\indent Local nonlinear gyrokinetic simulations on TS39596 are performed using the GYRO code. The latter one has in fact the capability to treat several among the key physical mechanisms needed to gain a quantitatively relevant information on the tokamak turbulence. These nonlinear simulations are performed in the local (flux-tube) limit of vanishing $\rho_*$, considering the 4 radial locations $r/a=0.4,\:0.5,\:0.6,\:0.7$. The GYRO simulations make use of all the experimental parameters reported in Table \ref{TableTS39596lc} without any modification. A summary of the numerical grid adopted is here given.
\begin{itemize}
	\item Electromagnetic fluctuations are retained.
	\item Collisions are treated trough the electron-ion pitch-angle scattering operator.
	\item Realistic magnetic geometry via the Miller equilibrium is adopted (i.e. finite aspect ratio effects are considered).
	\item Gyro-kinetic ions and drift-kinetic electrons are retained (i.e. electron FLR effects are neglected).
	\item $Z_{eff}=1$, hence $n_e=n_i$ considering the real mass ratio $\sqrt{m_i/m_e}=60$.
	\item Box size in the perpendicular directions $\left[L_x/\rho_s,L_y/\rho_s\right]=\left[122,122\right]$ with a radial resolution $\Delta x/\rho_s=0.30$ (i.e. 400 radial grid points).
	\item 128-point velocity space discretization per spatial cell: 8 pitch angles, 8 energies and 2 signs of velocity.
	\item 12 grid points in the parallel direction, 18 in the gyroaverage and 5 in the radial derivative.
	\item The simulation at $r/a=0.7$ use 32 complex toroidal harmonics resolving $0.0<k_{\theta}\rho_s<1.59$ scales, while for $r/a=0.4,\:0.5,\:0.6$, the range $0.0<k_{\theta}\rho_s<1.0$ is solved with 20 Fourier modes.
	\item Statistical averages on the saturated nonlinear state are performed on a typical physical equivalent time of 2.6 ms.
\end{itemize}

Nonlinear GYRO simulations do achieve statistically steady states of turbulence through dissipative upwind advection schemes in the real (not velocity) space. All the simulations presented in this thesis work verify the condition that both the flux and the fluctuation levels are well converged with respect to the velocity-space resolution at fixed spatial resolution. The velocity space resolution here employed (128 points for spatial grid point) is the standard one for typical GYRO simulations: the convergence with respect to this standard velocity-space resolution is systematically investigated in \cite{candy03jcp}. This aims to check that there is no missing velocity-space structure that would affect either transport levels or entropy production. Indeed this kind of approach is based on the statement that spatial upwind numerical schemes, hence finite numerical dissipation, provide a physically meaningful procedure to achieve turbulent steady states. Obviously the strategy followed by the GYRO code is not the only one possibility; other solutions deal with the issue of the dissipation looking for the direct implementation of a physical collision operator, like the recent proposals by \cite{abel08,barnes09}.\\
\indent It is important to briefly discuss the two main physical limitations of this kind of simulations, namely the local approach and the lack of electron spatial scales. This first issue is not expected to be a severe limit. Even if the argument is still an open subject of research, it is widely recognized that for such small values of $\rho_*\approx2\cdot 10^{-3}$, the local treatment of the turbulence appears as a very good approximation. This has been also demonstrated by the success in recovering the breaking from the gyro-Bohm (corresponding to the local limit of small $\rho_*$) to the Bohm transport scaling using nonlinear gyrokinetic simulations \cite{candy03}. On the other hand, the fact that the electron spatial scales, i.e. $k_{\theta}\rho_s\gg1.0$ are here neglected, deserves some additional care. The main reason for this simplification is the extremely high computational cost required by the coherent treatment of this multi-scale nonlinear problem. The problem is still presently an open issue. A first example of massive nonlinear gyrokinetic simulations of coupled ITG-TEM-ETG turbulence \cite{waltz07}  indicated small relative contributions to the total transport level coming from the high-$k$ scales. More recently instead, in Ref. \cite{goerler08}, it has been shown that in the case of reduced low-$k$ linear drives, a scale separation between electron and ion thermal transport is observed, originating a non negligible transport contribution at the small electron scales. Nevertheless, both these example do not adopt the real ion/electron mass ratio, because of computational cost constraints. Therefore, the present analysis focuses only on the spatial scales smaller or comparable to the ion Larmor radius, i.e. $k_{\theta}\rho_s\le1.0$, where the low $k$ ranges are well resolved by the nonlinear simulations and the spectral quantities provide a quantitatively relevant information.\\

\noindent \textbf{Results on the higher-order quantities}\\

\begin{figure}[!htbp]
  \begin{center}
    \leavevmode
      \includegraphics[width=8 cm]{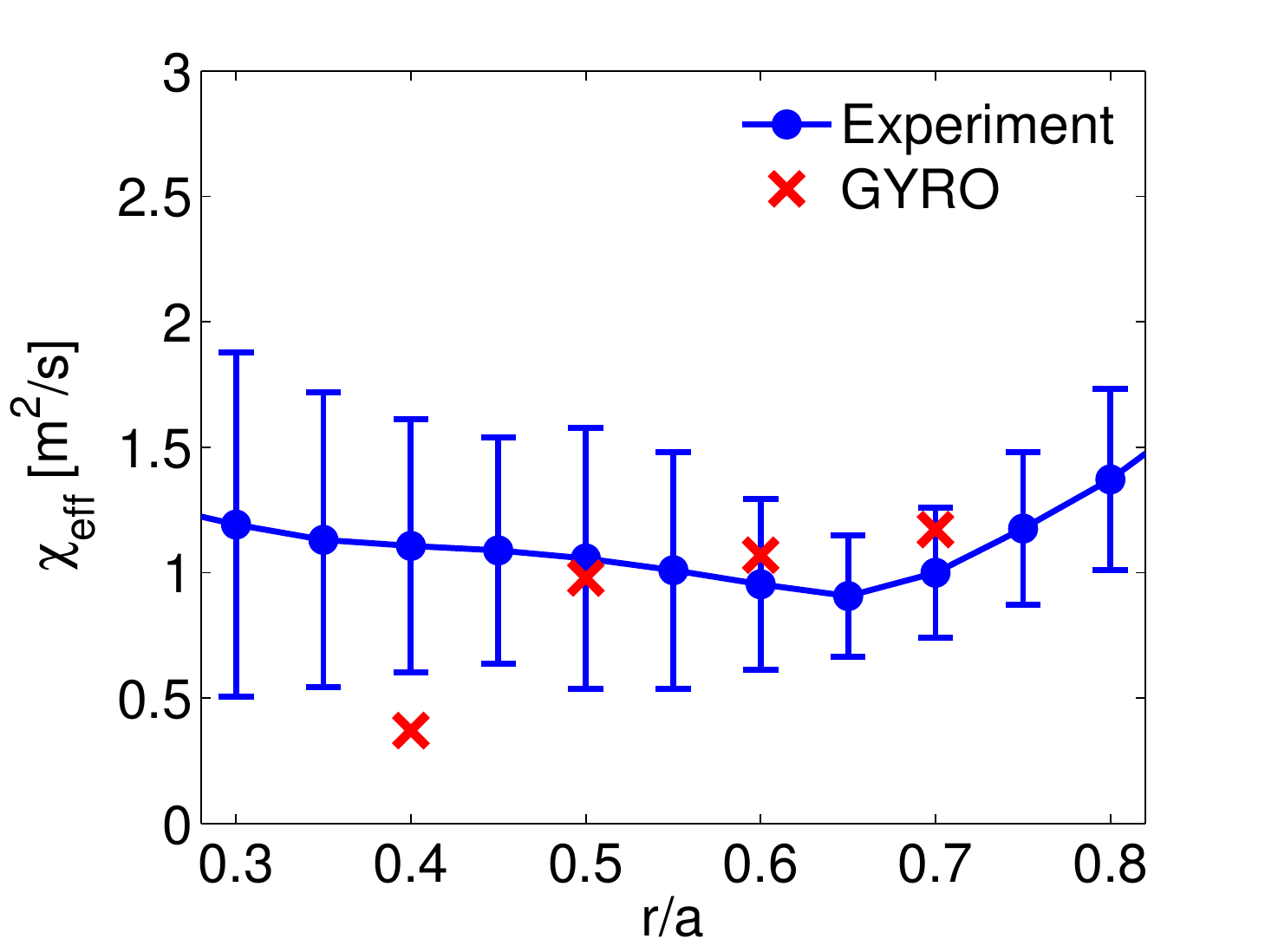}
    \caption{Radial profile of the experimental effective heat diffusivity and comparison with the GYRO predictions.}
    \label{TS39596-chieffprf}
  \end{center}
\end{figure}

\begin{figure}[!htbp]
  \begin{center}
    \leavevmode
      \includegraphics[width=8 cm]{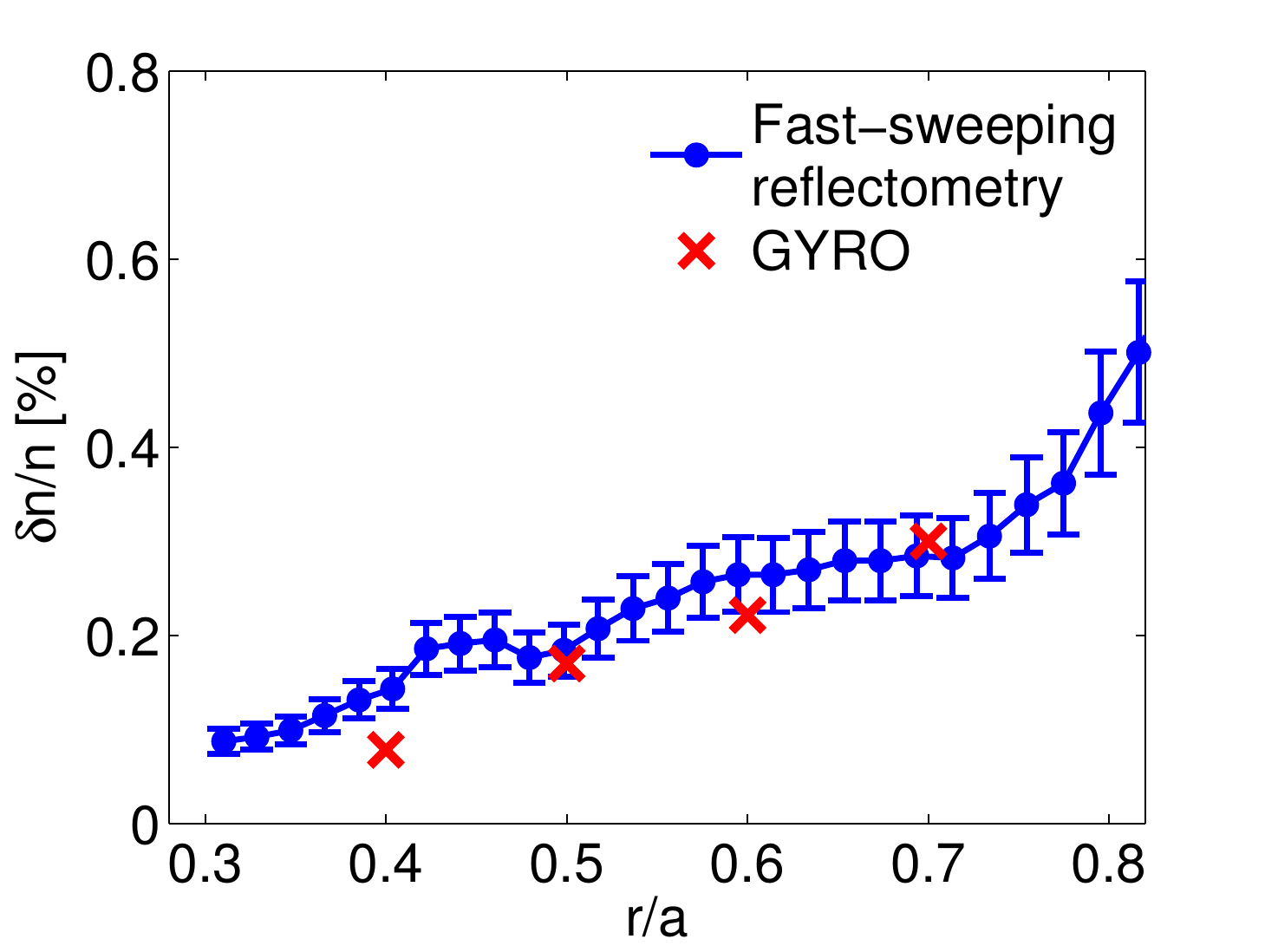}
    \caption{Radial profile of the experimental RMS $\delta n/n$ and comparison with the GYRO predictions.}
    \label{TS39596-deltanprf}
  \end{center}
\end{figure}

\indent The primal experimental quantities that have to be compared to the expectations by the nonlinear gyrokinetic simulations are the higher-order scalar quantities. Here, we focus on both the total heat transport coefficients and the RMS values of the density fluctuations.\\
The total effective heat diffusivity $\chi_{eff}$ is experimentally obtained from a power balance analysis, performed with an interpretative CRONOS run, with the following definition:
\begin{eqnarray}  \chi_{eff}=-\frac{q_e+q_i}{n_e\left(\nabla_r T_e+\nabla_r T_i\right)}
  	\label{cap4eq1} \end{eqnarray}
where $q_e$ and $q_i$ are the electron and ion heat fluxes respectively.\\ 
The experimental uncertainty is estimated taking into account the time evolution of the profiles during $1~\rm{s}$. Unfortunately, it is not possible to have a quantitatively reliable information on the separate contribution from the electron and ion heat transport; in the case of high collisional plasmas like TS39596 in fact, that would imply a detailed knowledge on the coupling terms between the two species.\\
\indent It is worth noting that in the case of steady plasma conditions, as verified for TS39596, the plasma profiles accommodate to null particle flows in the core, because no source is present in that region. For this reason, in absence of convection terms, the heat flux calculated by the CRONOS interpretative run effectively coincides with the energy flux. Conversely, the gyrokinetic simulations do not solve the physical problem at fixed flux,  whereas the temperature and density gradients are the quantities that are imposed. A certain particle flux is then calculated by the code and has to be appropriately treated. At this scope, the following relation is used, that links the energy and the heat fluxes for a generic species $s$:
\begin{eqnarray}  q_s=Q_s-\frac{3}{2}T_s\Gamma_s
  	\label{cap4eq2} \end{eqnarray}
In Eq. \eqref{cap4eq2}, $Q_s$ and $\Gamma_s$ are respectively the energy and particle fluxes predicted by GYRO, averaged over a flux-surface and a nonlinear saturation time interval (see Appendix \ref{appx2-GYRO-GYSELA} for details).\\
\indent The experimental results are compared with the expectations from the GYRO simulations in Fig. \ref{TS39596-chieffprf}.
A very good agreement between the numerical expectations and the experimental profile of $\chi_{eff}$ within the error bars is achieved. The exception of the radial point corresponding to $r/a=0.4$ is due to marginal turbulence found by the simulation on the experimental parameters. Nevertheless, the stiffness to $\nabla_r T_{i,e}$ (whose experimental uncertainty is about $\pm30$\%) inherent in the transport problem~\cite{candy03}, may induce to find an agreement with the experimental findings if varying the normalized gradients within the uncertainty levels. This argument has in fact been demonstrated already in several studies \cite{candy03,llin09,llin09b}. This kind of considerations should suggests that a reliable validation of the turbulence model can not be limited to the comparison of the scalar $\chi_{eff}$ quantity. The approach here adopted relies instead on a different strategy for the validation of the nonlinear simulations, which are constrainted to the simoultaneous comparison with several scalar and spectral measurements.\\
\indent The radial profile of the RMS density fluctuations is provided by the fast-sweeping reflectometry and is shown in Fig. \ref{TS39596-deltanprf}.
In order to validate the predictions by the nonlinear simulations, it is essential to properly reconstruct the same quantity that is measured by the diagnostics. Recalling the previous description of the fast-sweeping reflectometry system, the following relation is applied to the GYRO results:
\begin{eqnarray}  \left.\frac{\delta n}{n} \right|_{RMS} = \left\{ \int^{10cm^{-1}}_{0cm^{-1}}dk_{\theta} \int^{10cm^{-1}}_{1cm^{-1}}dk_r \left|\frac{\delta n}{n}\left(k_{\theta},k_r\right)\right|^2 \right\}^{1/2}
  	\label{cap4eq3} \end{eqnarray}
In Eq. \eqref{cap4eq3}, $\delta n/n\left(k_{\theta},k_r\right)$ refers to the physical density fluctuations predicted by the GYRO simulations computed at the outboard midplane, according to the diagnostic's line of sight. For simplicity of notation, the time average over the nonlinear saturation phase is skipped in the equation. The results shown in Fig. \ref{TS39596-deltanprf} reports a remarkable quantitative agreement with the experimental radial $\delta n/n$ profile within the error bars. Also the slight increase of the density fluctuations towards the external radii found by the diagnostic is matched by the GYRO simulations.\\

\subsection{The wave-number spectrum}
\noindent \textbf{Results on the lower-order quantities}\\

\begin{figure}[!htbp]
  \begin{center}
    \leavevmode
      \includegraphics[width=12 cm]{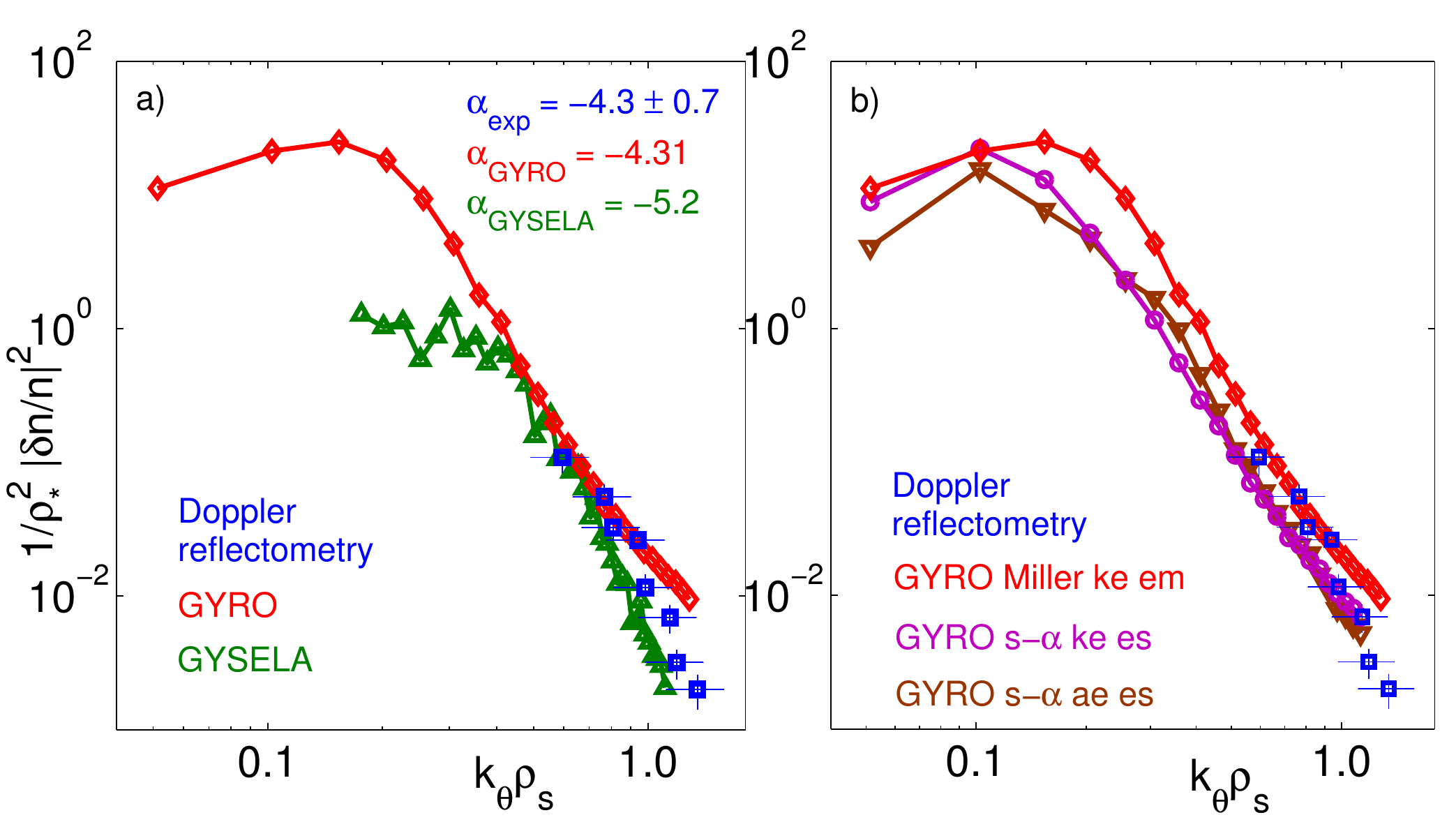}
    \caption{a) Experimental density fluctuation $k_{\theta}$ spectrum at $r/a=0.7$ from the Doppler reflectometry (squares), and comparison with the GYRO (diamonds) and GYSELA (triangles) predictions; b) Comparison between the Doppler $k_\theta$ spectrum and GYRO simulations retaining different physics, namely with Miller geometry - kinetic electrons - electromagnetic fluctuations (\textit{full physics}, diamonds), with circular $s-\alpha$ geometry - kinetic electrons - electrostatic fluctuations (circles) and with circular $s-\alpha$ geometry - adiabatic electrons - electrostatic fluctuations (down-triangles).}
    \label{TS39596-kthetasp}
  \end{center}
\end{figure}

\begin{figure}[!htbp]
  \begin{center}
    \leavevmode
      \includegraphics[width=8 cm]{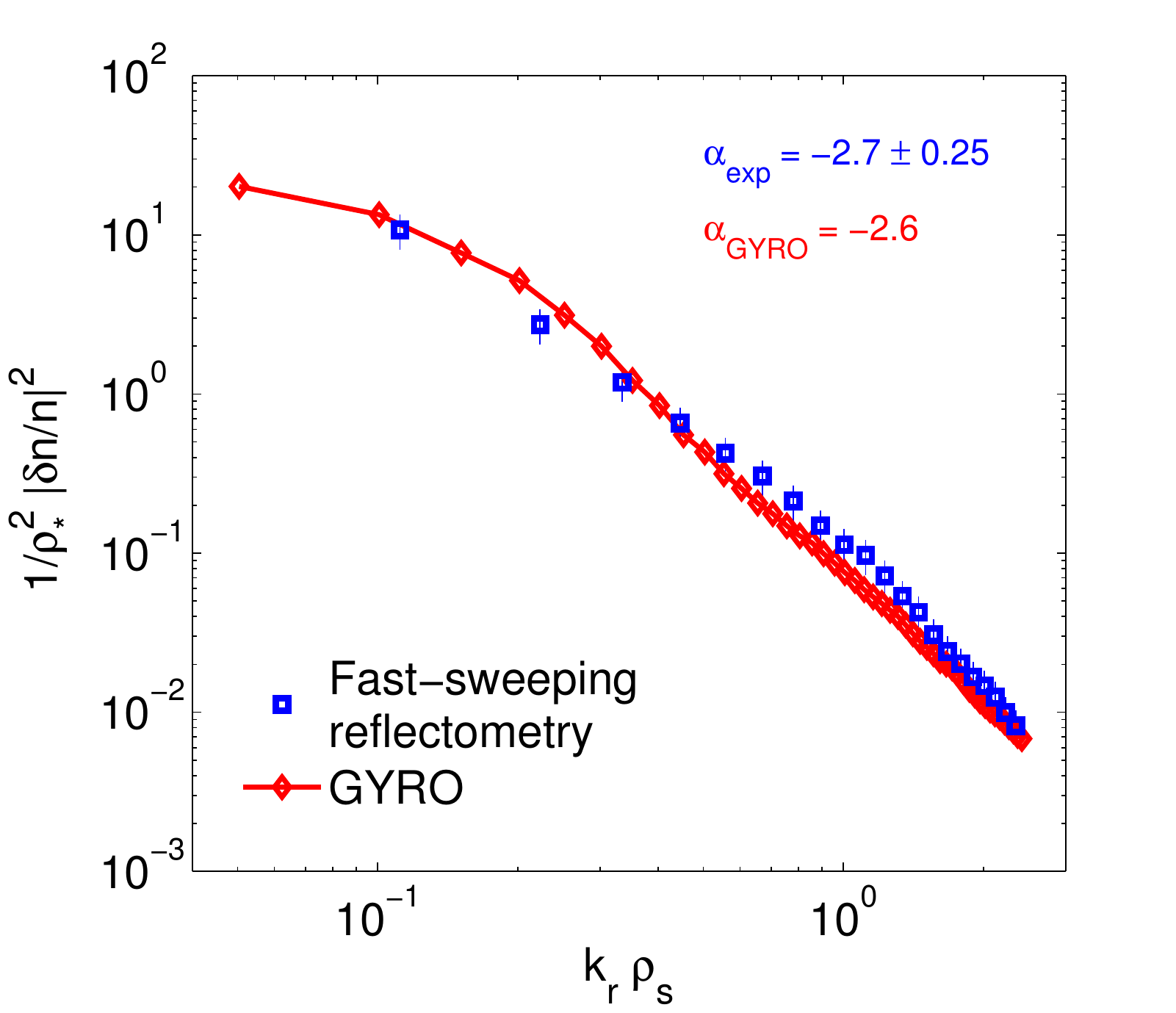}
    \caption{Experimental density fluctuation $k_r$ spectrum from the fast-sweeping reflectometry at $r/a=0.7$ , and comparison with the GYRO predictions.}
    \label{TS39596-krsp}
  \end{center}
\end{figure}

\indent The local nonlinear GYRO simulations have been shown to simultaneously reproduce the radial profiles of both the total effective heat diffusivity $\chi_{eff}$ and the RMS density fluctuations $\delta n/n$. In order to further verify the effectiveness of the turbulence predictions by the nonlinear gyrokinetic code, it is important to examine the spectral structure of the fluctuations, possibly in both the perpendicular directions to the magnetic field.\\
\indent As already discussed, the large amount of free energy in tokamak plasmas originates a wide variety of unstable modes. Due to the intense magnetic field, those are anisotropic: the convective cells exhibit much smaller parallel (to the magnetic field) wave vectors than transverse ones, $k_\parallel \ll k_\perp$, such that tokamak turbulence is quasi two-dimensional~\cite{hasegawa83}. The amount of turbulence anisotropy in the perpendicular plane is a particularly relevant issue for tokamak applications: any asymmetry favoring the formation of radially elongated structures could in fact enhance the cross-field anomalous transport, with consequent loss of confinement. Analysis of the fluctuation spectral power density in the transverse wave-number space $(k_\theta,k_r)$ (poloidal  and radial wave-vector, respectively) allows to characterize the turbulence structure and gives insight into its dynamics in terms of flow of the turbulent energy $E(k)$ at the scale $1/k$. Following Kraichnan~\cite{kraichnan67}, the injected energy in two-dimensional fluid turbulence is expected to be nonlinearly distributed through an inertial range, leading to a direct enstrophy cascade $E(k)=k^{-3}$ down to the dissipative length, and an inverse cascade towards larges scales, with $E(k)=k^{-5/3}$. Nevertheless in tokamak turbulence, three-dimensional effects, multi-field dynamics and energy injection at disparate spatial scales may originate significant departures from this simplified picture, thus requiring dedicated nonlinear simulations. On most experiments~\cite{ritz87,devynck93,brower87,hennequin04} the density fluctuation wave-number spectrum $S(k_{\perp})=|\delta n(k_{\perp})/n|^2$ shows a decay $S(k_{\perp})\sim k_{\perp}^{\alpha}$ with $\alpha=-3.5\pm0.5$, whilst assuming an adiabatic electron response, Kraichnan dual cascade would predict $\alpha=-14/3$ and $\alpha=-6$, respectively for the inverse and forward ranges. Kraichnan 2D dual cascade foresees in fact the energy density scalings $E\left(k\right)\sim k^{-5/3}$ and $E\left(k\right)\sim k^{-3}$. In 2D, the energy in an infinitesimal shell of thickness $dk$ can be written as $E\left(k\right)\sim\int\left(\frac{1}{2}mv_k^2\right)kdk$. Since $v_k\sim i\frac{k\delta\phi_k}{B}$, one finds $E\left(k\right)\sim k^3\left|\delta\phi_k\right|^2$. The scalings for the fluctuating potential are then $\left|\delta\phi_k\right|^2\sim k^{-14/3}$ and $\left|\delta\phi_k\right|^2\sim k^{-6}$. These scalings are only here presented to give a reference that applies to the more common 2D fluid turbulence, which anyway cannot be applied to the case of tokamak plasmas.\\
\indent Among the 4 different radial positions considered by the GYRO simulations (see Figs. \ref{TS39596-chieffprf} and \ref{TS39596-deltanprf}), here we focus on the local nonlinear simulation at $r/a=0.7$. The latter is used to validate the gyrokinetic predictions against the measurements of the bidimensional $\left(k_\theta,k_r\right)$ density fluctuation spectra. Because of the low value of the magnetic field in the discharge TS39596 in fact, the data from the fast-sweeping and from the Doppler reflectometry are simultaneously available only for $r/a\ge0.7$.\\
The density fluctuations $k_{\theta}$ spectrum at $r/a=0.7$ from the Doppler reflectometry is presented in Fig.~\ref{TS39596-kthetasp}. Focusing on the region $k_{\theta}\rho_s\le1.0$, the experimental data exhibit a power law decay with spectral index $\alpha_{\theta}=-4.3\pm0.7$. Possible transition towards steeper slopes at $k_{\theta} \rho_s>1$ as reported in Ref.~\cite{hennequin04} is not addressed here. In order to correctly reproduce the quantiy measured by the Doppler measurements, the following relation is adopted:
\begin{eqnarray}  \left.S\left(k_\theta\right)\right|_{Doppler}=\int^{1cm^{-1}}_{0cm^{-1}}dk_r \left|\frac{\delta n}{n}\left(k_\theta,k_r\right)\right|^2
  	\label{cap4eq4} \end{eqnarray}
which is used on the density fluctuations simulated by GYRO at the outboard midplane, according to the diagnostic's line of sight. The simulation gives a spectral index of $\alpha_{\theta}=-4.3$ for $0.4<k_{\theta} \rho_s<1.0$, in remarkable agreement with the reflectometry data within experimental uncertainty.\\
\indent An additional insight on the $k_\theta$ spectrum of the turbulent density fluctuations has been gained also by mean of global gyrokinetic simulations performed with the semi-Lagrangian code GYSELA. The latter one retains in fact the radial variations of the background gradients, conversely to the GYRO local simulations reported here above. Since the equilibrium profile is self-consistently evolved in such a code, the equilibrium gradients decay with time due to turbulent transport. Long enough simulation runs allow the system to reach a new equilibrium, characterized by well defined averaged profiles which are controlled by both the prescribed temperatures at the radial boundaries and the turbulent transport level. The main additional difference with respect to the GYRO simulations is the adiabatic assumption for the electron response, such that density and electrostatic potential fluctuations are equal, i.e. $\delta n/n=e\delta\phi/T_e$. This approximation can be though justified in this case due to the high levels of collisionality, with a consequent reduction of the non-adiabatic electron response coming from the trapped electron modes. The local Tore Supra parameters of Table \ref{TableTS39596lc} have been matched in the global simulations at $r/a=0.7$, but the normalized gyroradius was increased up to $\rho_*=8\cdot10^{-3}$ because of limited numerical resources. It should be noticed, however, that such a mismatch is likely not to impact the results provided the turbulence exhibits a gyroBohm scaling, as expected at these low $\rho_*$ values.\\
A full torus simulation on $0.5<r/a<0.9$ has been run with GYSELA using the grid points resolution $[r,\theta,\varphi,v_{\parallel},\mu]=[256,256,128,64,8]$ ($\theta$ and $\varphi$ are the poloidal and toroidal angles, $v_{\parallel}$ the parallel velocity and $\mu$ the adiabatic invariant). GYRO makes use of a field-aligned coordinate system, while GYSELA operates with the poloidal $\theta$ and toroidal $\varphi$ angles. For this reason, specific relations on the different Fourier expansions of the two codes, have to be adopted in order to compute the same spectra at $\theta=0$ needed for the comparison with the measurements. These details are reported in Appendix \ref{appx2-GYRO-GYSELA}.\\
\indent The GYSELA $k_{\theta}$ fluctuation spectrum is also shown in Fig.~\ref{TS39596-kthetasp}, giving a spectral exponent $\alpha_{\theta}=-5.2$ for $0.4<k_{\theta} \rho_s<1.0$. 
The purpose for which the GYSELA results are here shown is not the same as for GYRO. The aim of the GYRO simulations was to demonstrate that several measurements can be quantitatively and simultaneously matched when retaining as much physics as possible in the underlying model: the two main limits in the case of GYRO are the local approach (justified by the low $\rho_*\approx2\cdot10^{-3}$) and the cutoff on the small electron scales (whose impact on the fluxes should be sub-dominant and moreover these small electron scales are not resolved by the reflectometers in this discharge). The actual capabilities of GYSELA do not allow such a multi-level quantitative comparison with the experimental data. The reason why the GYSELA $\left|\delta\phi_{k\theta}\right|^2$ spectrum is here shown follows from the hypothesis that the tokamak turbulence wave-number spectrum is dominated by the $E\times B$ nonlinear convection terms, providing a rather general shape for the $k_\theta$ spectrum. Even if the similarity of the GYRO and GYSELA spectra for $k_\theta\rho_s>0.4$ can go in this direction, of course this statement still remains an hypothesis. 
The difference between the GYRO and GYSELA spectra at low $k$ is mostly due to the not small differences in the underlying model. Due to the intrinsic global nature of GYSELA (hence numerical cost), the $\rho_*$ used in the simulation has been increased by a factor of 5 with respect to the experimental value ($\rho_*=1.6\rightarrow8\cdot10^{-3}$): large sheared flows in the range of the geodesic acoustic modes are damping the large scale fluctuations, hence flattening the spectrum. For this reason and the lack of the electron physics in the model, the transport resulting from the GYSELA simulation cannot be compared to the expectations by GYRO.\\
\indent The latter argument could rise the following question: which is the essential physics that has to be retained in the gyrokinetic simulations in order to match the measurements within the experimental uncertainty? This point is investigated again by mean of GYRO simulations, retaining different levels of complexity in the underlying physical model. Fig. \ref{TS39596-kthetasp}(b) reports the results. The first GYRO case deals with the \textit{full physics} simulation, i.e. using a proper Miller magnetic equilibrium coupled to the treatment of kinetic electrons and electromagnetic fluctuations. At lower complexity, two other simulations are performed, both adopting the simplified $s-\alpha$ geometry: hence a second one still includes the kinetic electrons, while the last one retains only adiabatic electrons. It is worth noting that in principle, this latter case should correspond to the local ($\rho_*\rightarrow0$) limit of the global GYSELA simulation.\\
The results shown in Fig. \ref{TS39596-kthetasp}(b) highlight that, even if differences are present, the three GYRO spectra retaining different physics, still match the Doppler measurements within the experimental uncertainty in the range $0.4\le k_\theta \rho_s\le1.0$. The power spectral indexes varies from the $\alpha_\theta=-4.3$ for the \textit{full physics} simulation to $\alpha_\theta=-5.1$ in the case of the $s-\alpha$ adiabatic electron case: as already seen by GYSELA, this confirms the expected result that neglecting the non-adiabatic electron response produces steeper spectral slopes. Moreover, the adiabatic electrons simulations (both GYRO and GYSELA) lead to an apparent better agreement on the high-$k$ part of the spectrum; nevertheless this effect is only a fortuitous effect. It appears that the most relevant differences when changing the physical complexity in the GYRO simulations, are present at the low-$k$ scales $k_\theta \rho_s\le0.4$. These changes have a non negligible impact on the total transport level, which is in fact mostly carried by the scales $k_\theta \rho_s\approx0.2$. The simplified simulations evidently underestimate the total level of turbulence with respect to the \textit{full physics} case, which has instead already demonstrated to correctly match the experimental values of the heat flux (Fig. \ref{TS39596-chieffprf}). On the other hand, the turbulence diagnostic has not access to the $k_\theta \rho_s\le0.4$ scales; hence in this case, the Doppler reflectometry alone can not be used to discriminate between the different physical effects retained in the simulations.\\
\indent The local $k_r$ density fluctuation spectrum is obtained by the fast-sweeping reflectometry; the measurements at the same radial position $r/a=0.7$ for TS39596 are shown in Fig.~\ref{TS39596-krsp}.
The measurements still exhibit a power law decay with a spectral exponent $\alpha_{r}=-2.7\pm0.25$ for scales corresponding to $0.4<k_r\rho_s<2.0$. This spectral quantity is reconstructed through the relation:
\begin{eqnarray}  \left.S\left(k_r\right)\right|_{Fast-sw}=\int^{10cm^{-1}}_{0cm^{-1}}dk_\theta \left|\frac{\delta n}{n}\left(k_\theta,k_r\right)\right|^2
  	\label{cap4eq5} \end{eqnarray}
which is applied to the GYRO $\delta n$ predictions at the outboard midplane. A very good agreement with the fast-sweeping reflectometry data is achieved, both in the magnitude and the slope of the relative fluctuation level, covering also the larger spatial scales up to $k_{r,min}\approx1~\rm{cm^{-1}}$.\\

\noindent \textbf{Discussion on the $\delta n$ $k$ spectra}\\

\indent Recalling the results on the wave-number density spectra just presented in Figs. \ref{TS39596-kthetasp} and \ref{TS39596-krsp}, it clearly appears that at the same radial location $r/a=0.7$, the two reflectometers provide different fluctuation spectral exponents in the perpendicular plane. This discrepancy is above the experimental uncertainties, showing in particular $\alpha_{\theta}=-4.3\pm 0.7$ by the Doppler reflectometry while $\alpha_{r}=-2.7\pm 0.25$ according to the fast-sweeping reflectometry, both in the range of $k_{\perp}\rho_s\ge0.4$. Such a discrepancy could initially suggest a highly anisotropic turbulence, favoring the formation of radially elongated structures. The nonlinear simulations can be used in this case as a powerful tool for clarifying this relevant issue. The GYRO results motivate in fact a revised interpretation of the experimental evidence: the two dissimilar exponents may be simply ascribed to intrinsic instrumental effects.\\
\begin{figure}[!htbp]
  \begin{center}
    \leavevmode
      \includegraphics[width=8 cm]{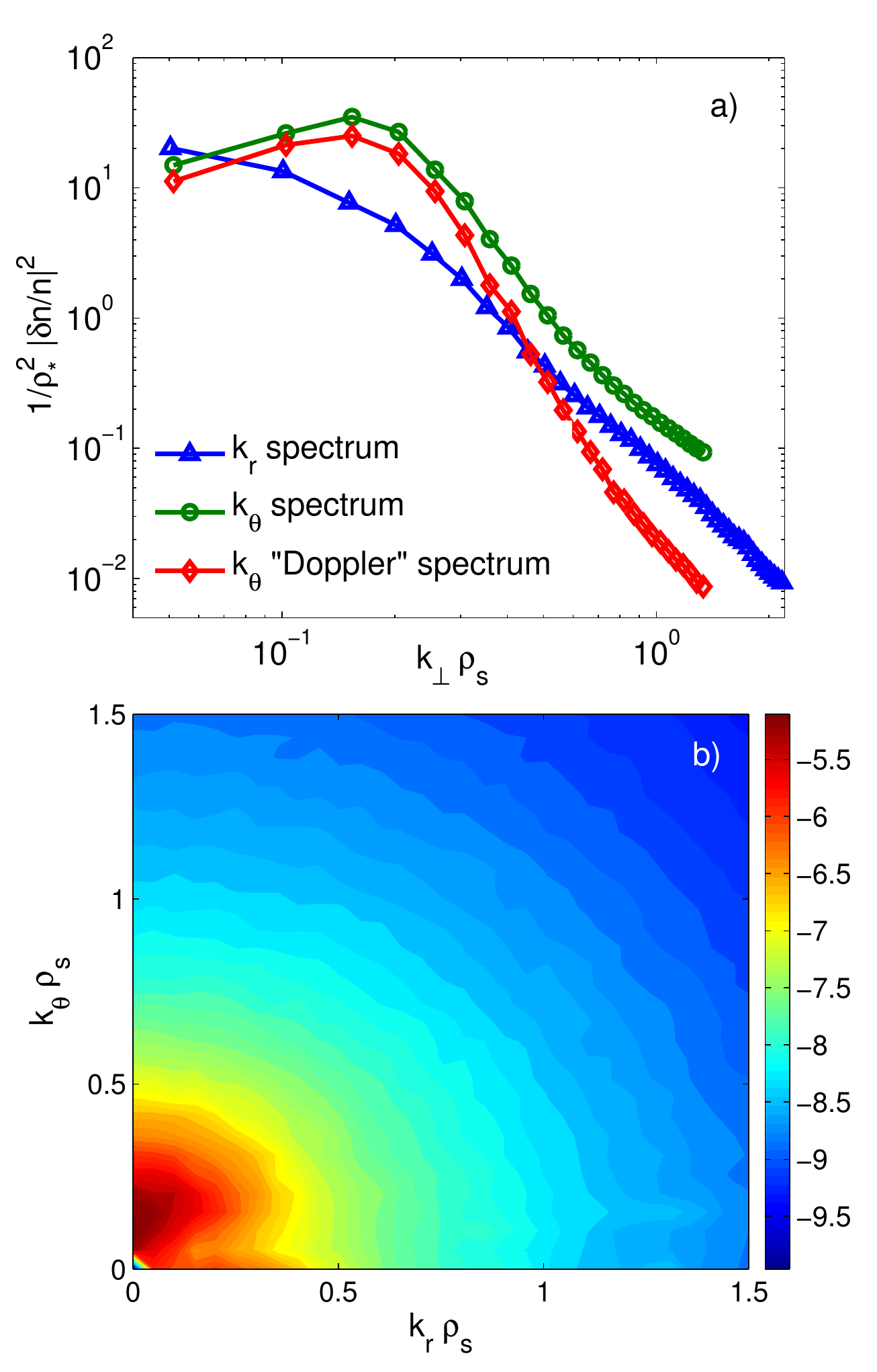}
    \caption{(a) $|\delta n/n|^2$ spectra from the GYRO simulation at $r/a=0.7$, and the impact of reconstructing the Doppler reflectometry instrumental response on the $k_{\theta}$ spectrum. (b) Contour plot of the GYRO $log_{10}|\delta n(k_r,k_{\theta})/n|^2$.}
    \label{TS39596-k2Dsp}
  \end{center}
\end{figure} 
\indent In the case of the fast-sweeping $k_r$ spectrum, the contributions of medium-low $k_{\theta}\le10~\rm{cm^{-1}}$ wave-numbers are retained, as it appears from Eq. \eqref{cap4eq5}. On the other hand, with the Doppler reflectometry, the $k_{\theta}$ spectrum selects only very low radial wave-numbers $k_{r}\le1~\rm{cm^{-1}}$, as explicated by Eq. \eqref{cap4eq4} \cite{hennequin06}. The density fluctuation spectra predicted by the GYRO simulations reflect this asymmetry. More in particular, as it is shown in Fig. \ref{TS39596-k2Dsp}, the $k_{\theta}$ spectral exponents given by GYRO clearly exhibit a difference when integrating over the diagnostic Doppler range $0<k_r<1~\rm{cm^{-1}}$ rather than accounting for all the radial wave-numbers. In the first case in fact, $\alpha_{\theta}=-4.3$ is obtained, the results already shown in Fig. \ref{TS39596-kthetasp} and in agreement with the measurements. In the second case, the spectral $\alpha_{\theta}\approx -2.9$ is produced.\\
The numerical predictions observe that the strong anisotropy carried by the peak in the $k_{\theta}$ axis significantly affects the $k_{\perp}\rho_s<0.4$ ranges, while the asymmetry appears weaker, but still present, at smaller spatial scales~\cite{waltz07,ritz87}. The iso-level contours of $|\delta n/n(k_r,k_{\theta})|^2$ in the perpendicular plane computed by GYRO and reported in Fig.~\ref{TS39596-k2Dsp}(b) confirm this kind of expected picture, identifying a linearly driven turbulence anisotropy around the $k_{\perp} \rho_s\approx0.2$ scales, which are not accessible by our Doppler reflectometry system.\\

\subsection{The frequency spectrum}

\indent Firstly, it is of interest to compare some quantities concerning the frequency of the turbulent modes provided by the numerical simulation on the nonlinear and the linear regimes. The $k_\theta$-dependent frequency spectrum computed by the nonlinear gyrokinetic simulation provides a relevant information on the dominant nonlinear frequency $\omega_k^{nl}$ and broadening $\Delta\omega_k^{nl}$ at a given wave-number. Both these quantities can be compared respectively to the linear frequency $\omega_k^{lin}$ and growth rate $\gamma_k^{lin}$. This kind of study is directly useful to validate and improve the choices assumed in the quasi-linear transport models. As discussed in the paragraph \ref{sec-quasilinear-new-model}, QuaLiKiz keeps the linear frequencies $\omega_k^{lin}$ in the calculation of the turbulent fluxes, while it assumes a frequency broadening with the widths equal to the linear growth rates, $\Delta\omega_k=\gamma_k$.\\
\begin{figure}[!htbp]
  \begin{center}
    \leavevmode
      \includegraphics[width=12 cm]{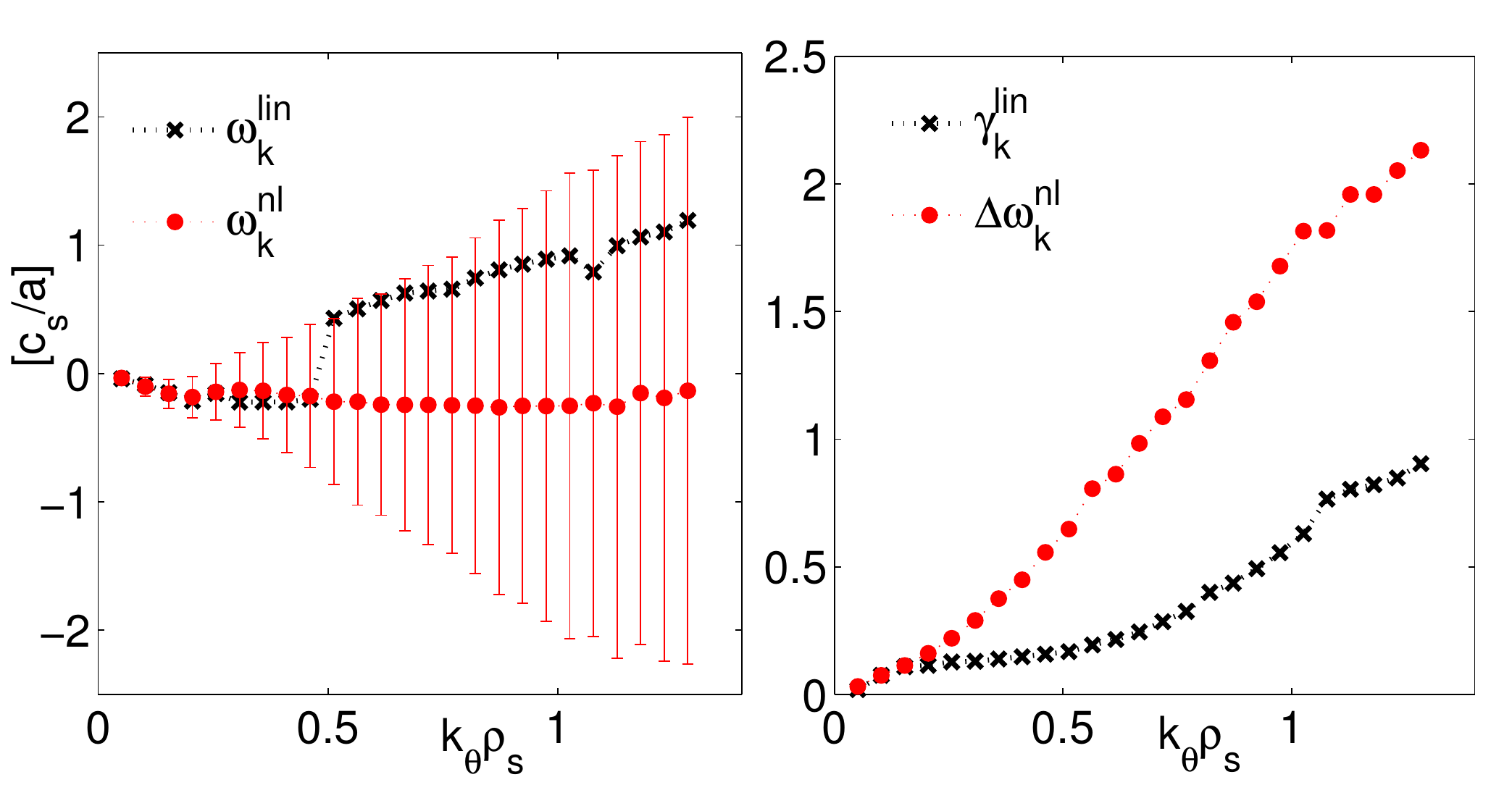}
    \caption{(a) $k_\theta$ spectrum of the frequencies of the linear most unstable mode and of the nonlinear frequencies (the bars indicate the statistical variance from the mean value). (b) $k_\theta$ spectrum of the linear growth rates and of the nonlinear frequency widths. The results refer to nonlinear and linear (most unstable mode) simulations usuing GYRO on TS39596 at $r/a=0.7$.}
    \label{TS39596-frqlnlsp}
  \end{center}
\end{figure}
\indent The results reported in Fig. \ref{TS39596-frqlnlsp} show that for $k_\theta\rho_s<0.5$, where most part of the transport is driven, the mean nonlinear frequencies\footnote{The mean nonlinear frequencies are computed from the simulation as the ones corresponding to the maximal amplitude of the frequency spectrum at a given wave-number $k_\theta$; each value has an associated statistical variance.} are in good agreement with the linear ones of the most unstable mode. For $k_\theta\rho_s>0.5$, the linearly most unstable mode jumps to the electron diamagnetic direction with increasing $\left|\omega_k^{lin}\right|$, while nonlinearly $\omega_k^{nl}$ tends to zero, with a non negligible dispersion. On the other hand, the nonlinear frequency broadenings show a departure from the growth rates of the linear most unstable mode. The numerical results illustrate that the nonlinear decorrelations operate stronger than the linear ones, hence $\Delta\omega_k^{nl}>\gamma_k^{lin}$. The evidence of Fig. \ref{TS39596-frqlnlsp} is then consistent with the previous evaluation of the Kubo-like numbers from nonlinear gyrokinetic simulations (in paragraph \ref{sec-turbulence-times}, Fig. \ref{Gyro-Kubo}). This point certainly deserves more accurate study in the future; interestingly, even if this observation is at odd with the actual choices of QuaLiKiz, this seems not to dramatically affect the total turbulent fluxes, which will be shown in the following of this work to correctly reproduce the nonlinear expectations in a number of cases.\\

\indent From the experimental point of view, the investigation of the frequency spectra of the density fluctuations, is not a very recent subject of research. Relevant works documented in the literature are for example Refs. \cite{mazzuccato82,antar99,hennequin06}. Most part of the theoretical efforts on this subject are related to the statistical theories of turbulence, like the DIA \cite{waltz83,krommes02}. On the other hand, only very recently it has been possible to investigate the frequency spectra also by mean of comprehensive nonlinear simulations. Indeed, long enough simulations are needed for such an analysis. Ref. \cite{holland08} provides a first important example; nevertheless in this latter work, where the BES technique is used, the turbulence measurements are not resolved in $k_\theta$. Here we report some preliminary results about the $k_\theta$ resolved frequency spectra and the comparison between measurements and nonlinear local GYRO simulations.\\
\indent The Doppler reflectometry technique measures the instantaneous spatial Fourier analysis of the density fluctuations, $\delta n\left(\mathbf{k},t\right)=\int_V d\mathbf{x}\:n\left(\mathbf{x},t\right)e^{i\mathbf{k}\cdot\mathbf{x}}$. In the last paragraph we have discussed the information on the density fluctuations related to the wave-vector $\mathbf{k}$, providing the $k_\theta$ fluctuation spectrum. On the other hand, thanks to the good time resolution of the diagnostic, a simple Fourier analysis on the same scattered signal can provide localized measurements of the $k_\theta$ resolved $\delta n$ frequency spectra.\\
The Doppler shifted component corresponding to the scattering close to the cut-off layer, typically dominates these frequency spectra (or at least it can be clearly separated from any spurious signal that appears mainly at zero frequency). The parasitic components at $f=0$ can be due to the direct reflection of the beam, or of the refracted beam on inner elements of the chamber, or again to a backscattering all along the beam path. In most cases, however, the frequency spectrum clearly exhibits the Doppler backscattered component without any $f = 0$ spurious component, as detailed in Ref. \cite{hennequin06}, revealing the good selectivity of the diagnostic.\\

\indent The interest in studying the fluctuation frequency spectrum for a given toroidal wave-number is strictly linked to the issue of the quasi-linear modeling. The argument has been already introduced in the paragraph \ref{sec-quasilinear-new-model}. The statistical DIA theory for example, foresees a spectrum which presents a Lorentzian shaped broadening: this is also the choice adopted in our quasi-linear transport model QuaLiKiz. From the experimental point of view, the Doppler measurements are generally not compatible with a such Lorentzian shape, neither with a Gaussian one, as detailed in Ref. \cite{hennequin06}. Therefore, a different approach has been developed for the interpretation of the experimental fluctuation frequency spectra, that will be here described.\\
\indent Since the $\delta n$ frequency spectrum corresponds to the fluctuation time-correlation function in the Fourier space, a classical statistical argument can be introduced as a very simple model. The approach closely recalls the discussion of the paragraph \ref{sec-quasilinear-ordering}, where a Lagrangian time correlation function $C_v\left(\tau\right)$ has been introduced in Eq. \eqref{QLeq11}. Very simply, the mean square value of the displacement of a particle with a velocity $v$ in the turbulent field, $\left\langle\Delta^2 \right\rangle$, can be classically evaluated for times either long or short with respect to the Lagrangian velocity correlation time $\tau_L$. This would lead to two limits, respectively $\left\langle\Delta^2 \right\rangle=2D\tau=u^2\tau_L\tau$ and $\left\langle\Delta^2 \right\rangle=u^2\tau^2$, where $D$ is a diffusion coefficient and $u^2=\left\langle v^2 \right\rangle$. It clearly appears that the first expression corresponds in fact to a diffusive behavior, while the second to a convective one.
The former considerations can be applied to a simplified model which describes a general transition from the convective to the diffusive behavior, leading to the expression:
\begin{eqnarray}  \left\langle\Delta^2 \right\rangle=u^2\tau_L^2\left(\frac{\tau}{\tau_L}-1+e^{-\frac{\tau}{\tau_L}}\right)
  	\label{cap4eq5b} \end{eqnarray}
which correctly recovers the previous limits.\\
An expression for the correlation function of the scattered field emitted by the Doppler reflectometer's antenna has to be derived. This has been done in \cite{gresillon92, gervais94}, giving the following expression:
\begin{eqnarray}
   C_\mathbf{k}\left(\tau\right) = \left\langle \sum_{i,j}e^{i\mathbf{k}\cdot\left(\mathbf{r}_i\left(t\right)-\mathbf{r}_j\left(t+\tau\right)\right)} \right\rangle_t \rightarrow S\left(\mathbf{k}\right)F\left(\mathbf{k},\tau\right)
\end{eqnarray}
i.e. the product of the static form factor $S\left(\mathbf{k}\right)$ by the characteristic function of the probability of turbulent displacement $F\left(\mathbf{k},\tau\right)$. The first one is a static factor which defines the $k_\theta$ spectrum previously considered, while the second one corresponds to the probability of the displacements. The latter one can be rewritten when assuming a given statistics for the distribution of these turbulent displacements (here denoted by $\boldsymbol{\Delta}_\tau$); under the (rather strong) hypothesis of normal statistics one has:
\begin{eqnarray}
   F\left(\mathbf{k},\tau\right) =  \left\langle e^{i\mathbf{k}\cdot\boldsymbol{\Delta}_\tau} \right\rangle = 
    \int P\left(\boldsymbol{\Delta}|\tau\right) e^{i\mathbf{k}\cdot\boldsymbol{\Delta}}d\boldsymbol{\Delta}
    \rightarrow  e^{ik^2\left\langle \Delta^2\right\rangle/2} 
\end{eqnarray}
Here the game is to find an explicit expression for the mean squared turbulent displacement $\left\langle \Delta^2\right\rangle$. A general expression follows from the classical statistical approach of the Lagrangian dynamics of a particle with the velocity $v$ \cite{monin73}:
\begin{eqnarray}
   \left\langle \Delta^2 \right\rangle = 2 u^2 \int_0^\tau \left(\tau-s\right) C_v \left(s\right) ds  \qquad \mathrm{with} \qquad
     C_v\left(s\right) =  \left\langle v\left(0\right)v\left(s\right)\right\rangle/u^2 
\end{eqnarray}
where $C_v$ is the Lagrangian velocity correlation function and $u^2=\left\langle v^2\right\rangle$ is the variance of the velocity. A Lagrangian velocity correlation time $\tau_L$ is then given by $\tau_L=\int_0^\infty C_v\left(\tau\right)d\tau$.
Under these hypotheses and using Eq. \eqref{cap4eq5b}, the signal correlation function $\bar{C}\left(\mathbf{k},\tau\right)$ results proportional to the following expression:
\begin{eqnarray}  \bar{C}\left(\mathbf{k},\tau\right) \propto \left\langle e^{i\mathbf{k}\cdot\boldsymbol{\Delta}_\tau} \right\rangle = e^{-k^2\left\langle\Delta^2\right\rangle/2} = e^{-k^2u^2\tau_L^2\left(\frac{\tau}{\tau_L}-1+e^{-\tau/\tau_L}\right)}
  	\label{cap4eq7} \end{eqnarray}
In \eqref{cap4eq7}, the passage between the second and the third term has been possible thanks to the hypothesis of Gaussian statistics for the turbulent displacements; a rigorous justification can be found in \cite{krommes02}, making use of the mathematical theory of cumulants. In this work, Eq. \eqref{cap4eq7} is here referred as the model test T function.\\
\indent In the Fourier space, Eq. \eqref{cap4eq7} corresponds to a frequency spectrum at a given wave-number $k$, which can not be analytically calculated. Nevertheless, it is particularly useful to examine the frequency spectral shape resulting from the Fourier transform of the Eq. \ref{cap4eq7} in two relevant limits formerly cited:
\begin{enumerate}
	\item $\tau\gg\tau_L$: \textsl{Diffusive limit}. The diffusive behavior leads to a Lorentzian broadening of the frequency spectrum. This was the expected result, since the similar argument of a diffusive random-walk of the particles in the turbulent field, produces the Lorentzian spectrum in the quasi-linear theory framework. Consequently, the wave-number $k$-dependence of the spectral widths $\rm{FWHM}_k=2\Delta\omega_k$ follows $\rm{FWHM}_k \propto k^{2}$.
	\begin{eqnarray} \bar{C}\left(k,\tau\right) \approx e^{-k^2u^2\tau\tau_L} \qquad \Leftrightarrow \qquad S_k\left(\omega\right) \approx \textrm{Lorentzian} \qquad \Leftrightarrow \qquad \Delta\omega_k \propto k^2
  	\nonumber \end{eqnarray} 
	\item $\tau\ll\tau_L$: \textsl{Convective limit}. The convective component originates a Gaussian shape for the frequency spectrum. Differently from the Lorentzian broadening, this behavior results in a $k$-dependence for the frequency widths of the type $\rm{FWHM}_k \propto k^{1}$.
	\begin{eqnarray} \bar{C}\left(k,\tau\right) \approx e^{-k^2u^2\tau^2} \qquad \Leftrightarrow \qquad S_k\left(\omega\right) \approx \textrm{Gaussian} \qquad \Leftrightarrow \qquad \Delta\omega_k \propto k^1
  	\nonumber \end{eqnarray} 
\end{enumerate}
Nevertheless, it has to be stressed that the previous expectations on the $k$-dependence of the frequency broadenings are given in the hypothesis that both $u^2$ and $\tau_L$ do not depend on the wave-number $k$. Even if this could appear a strong limit, when assuming dominant $\bar{u}^2$ and $\bar{\tau}_L$ across the $k$-spectrum, the previous scalings are meaningful to indicate dynamics dominated by diffusive or convective behaviors.\\
The frequency spectrum corresponding to the whole T model, i.e. the Fourier transform of Eq. \eqref{cap4eq7}, presents a spectral shape which is intermediate between a Lorentzian and a Gaussian. Hence in this case, it is reasonable to expect that the $k$-dependence of the frequency broadenings obeys to $\Delta\omega \propto k^{\alpha\left(k\right)}$, where the exponent has an intrinsic $k$ dependence $\alpha=\alpha\left(k\right)$, and $1<\alpha\left(k\right)<2$. Under the same former hypothesis of $u^2$ and $\tau_L$ independent of $k$, the spectral exponent $\alpha$ can be numerically studied from the Fourier analysis of Eq. \eqref{cap4eq7}, recognizing a transition between the diffusive and the convective regimes. The properties of this simple T model can then be summarized as:
\begin{eqnarray}  \bar{C}\left(k,\tau\right) \approx e^{-k^2u^2\tau_L^2\left(\frac{\tau}{\tau_L}-1+e^{\frac{\tau}{\tau_L}}\right)} \qquad \Leftrightarrow \qquad S_k\left(\omega\right) \approx \textrm{intermediate} \qquad \Leftrightarrow \qquad \Delta\omega_k \propto k^{\alpha\left(k\right)}
  	\nonumber \end{eqnarray}
\indent The study of the fluctuation frequency spectra is here addressed comparing the Doppler reflectometry measurements to the nonlinear gyrokinetic simulations using GYRO. At this scope, the following relation is adopted to reproduce the experimental signals:
\begin{eqnarray}  \left. S_k\left(\omega\right)\right|_{Doppler} = \left\langle \left|\mathcal{F}\left\{\frac{\delta n_k}{n}\left(r,t\right)\right\}\right|^2 \right\rangle_r
  	\label{cap4eq6} \end{eqnarray}
where $\mathcal{F}$ stands for a time Fourier transform. The latter expression is applied to the local nonlinear GYRO simulations done on the discharge TS39596. The time Fourier transform is performed over a sufficiently long interval of the nonlinear saturation regime predicted by the code. The density fluctuations are computed at the outboard midplane, according to the diagnostic line of sight, as already described in the previous paragraphs. The wave-number $k$ here refers to the poloidal wave-number $k_\theta$ selected by the Doppler reflectometry. Finally, since the GYRO simulations are done in the local limit, hence without any variation of the radial profiles, the average over the radial coordinate $\left\langle\ldots \right\rangle_r$ is simply useful to cumulate statistics on the numerical predictions. This study refers to the same Tore Supra discharge already described, TS39596; also, the same radial position used for the $k$ spectra analysis, $r/a=0.7$, is considered. We refer to Tables \ref{TableTS39596gl} and \ref{TableTS39596lc} for the experimental parameters.\\
\begin{figure}[!htbp]
  \begin{center}
    \leavevmode
      \includegraphics[width=13 cm]{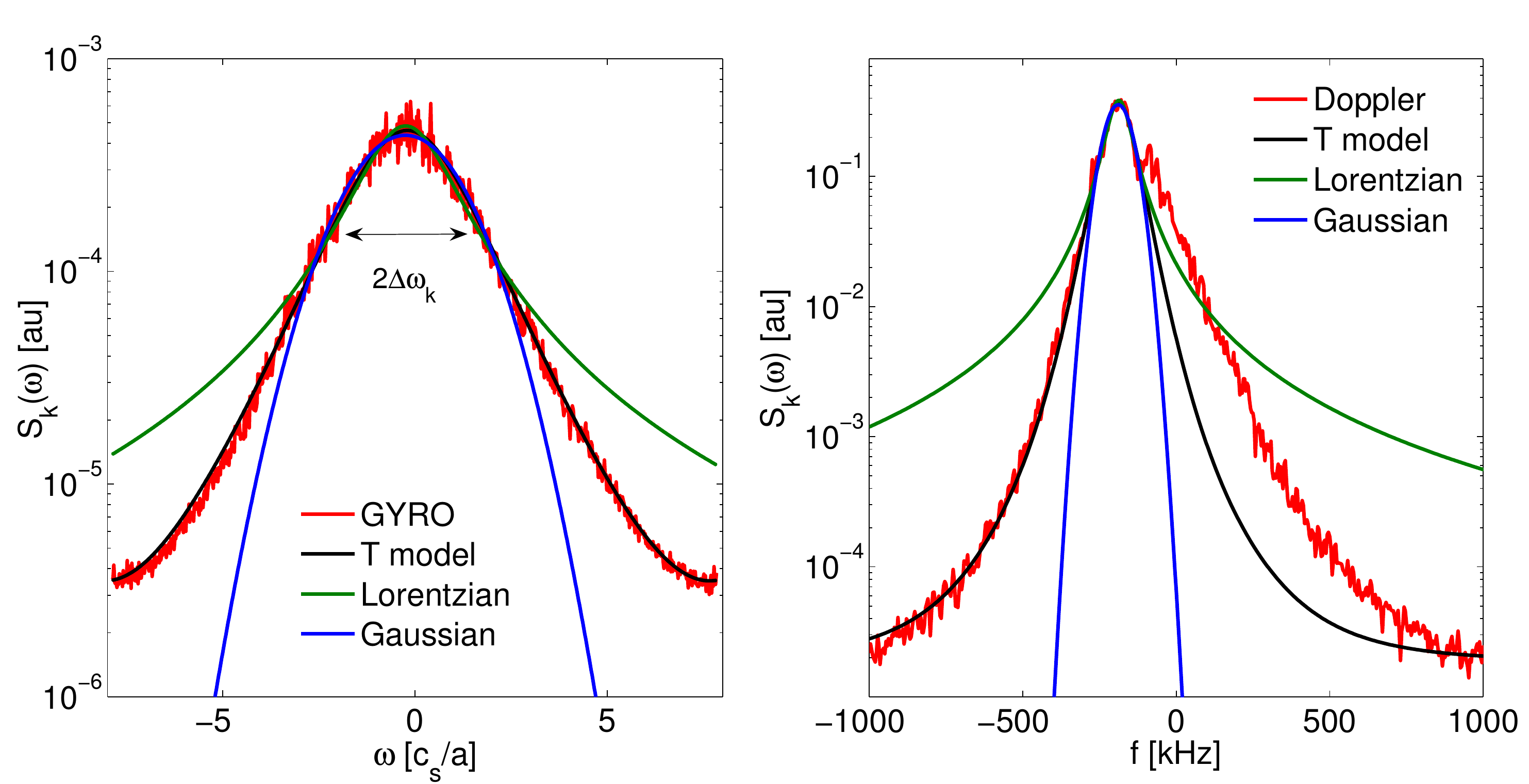} \\
    \caption{(a) Frequency spectrum of $\delta n_{\bar{k}}$ (where $\bar{k}_\theta\rho_s=0.82$) from the GYRO simulation of TS39596 at $r/a=0.7$ and comparison with different spectral shapes. (b) Experimental frequency spectrum from the Doppler reflectometry on the same discharge and radial position; best fits from Gaussian, Lorenzian and T model spectral shapes are also shown.}
    \label{TS39596-frfitsp}
  \end{center}
\end{figure}
\indent The gyrokinetic nonlinear simulations are here used for the first time in order to quantitatively investigate the issue of the spectral shape of the frequency spectrum and to compare it to the experimental evidences. In Fig. \ref{TS39596-frfitsp}, the frequency spectrum of the density fluctuations at a given wave-number, computed by the GYRO simulation according to Eq. \eqref{cap4eq6}, is shown.
The GYRO density fluctuation frequency spectrum is compared with a Lorentzian and a Gaussian spectral shapes. It clearly appears that both functions fail in reproducing the broadening observed in the nonlinear simulations. Nevertheless, it is worth noting that the Lorentzian and the Gaussian shapes can still be useful to obtain reasonable estimates on the values of the FWHM of the spectra.\\
\indent The T model described by Eq. \eqref{cap4eq7} is moreover applied to the analysis of the GYRO results. Since there is not an analytical expression for the Fourier transform of Eq. \eqref{cap4eq7}, a nonlinear bisquare regression has been applied in the direct space, i.e. to the time correlation function of the density fluctuations predicted by GYRO; hence an inverse FFT is used to compare the frequency spectrum of the T model to the one directly obtain by the simulation. Surprisingly, the T model shows an extremely good agreement in reproducing the GYRO frequency spectral shape (a factor $R^2=0.9996$ is obtained from the nonlinear regression).\\  
This result is even more relevant when compared to the experimental evidences. Until now, the T model is often used as practical solution for the fitting of the experimental frequency spectra of the Doppler reflectometry \cite{hennequin06}, but without a clear understanding of the reasons of its superiority with respect to the Lorentzian and the Gaussian shapes. In Fig. \ref{TS39596-frfitsp}(b), the experimental $\delta n$ frequency spectrum at a given wave-number $k_\theta$ selected by the Doppler reflectometry, is shown for the same discharge TS39596 at $r/a=0.7$. The asymmetry seen on the positive side of the spectrum is most probably due to spurious components coming from the backscattered signal all along the beam path. The experimental spectral shape exhibits analogous properties with respect to the GYRO results, with a frequency broadening intermediate between a Gaussian and a Lorentzian function, but conversely well fitted by the simple T model. The present study provides then a very good validation of the methods used in the analysis of the Doppler reflectometry.\\ 
\begin{figure}[!htbp]
  \begin{center}
    \leavevmode
      \includegraphics[width=8 cm]{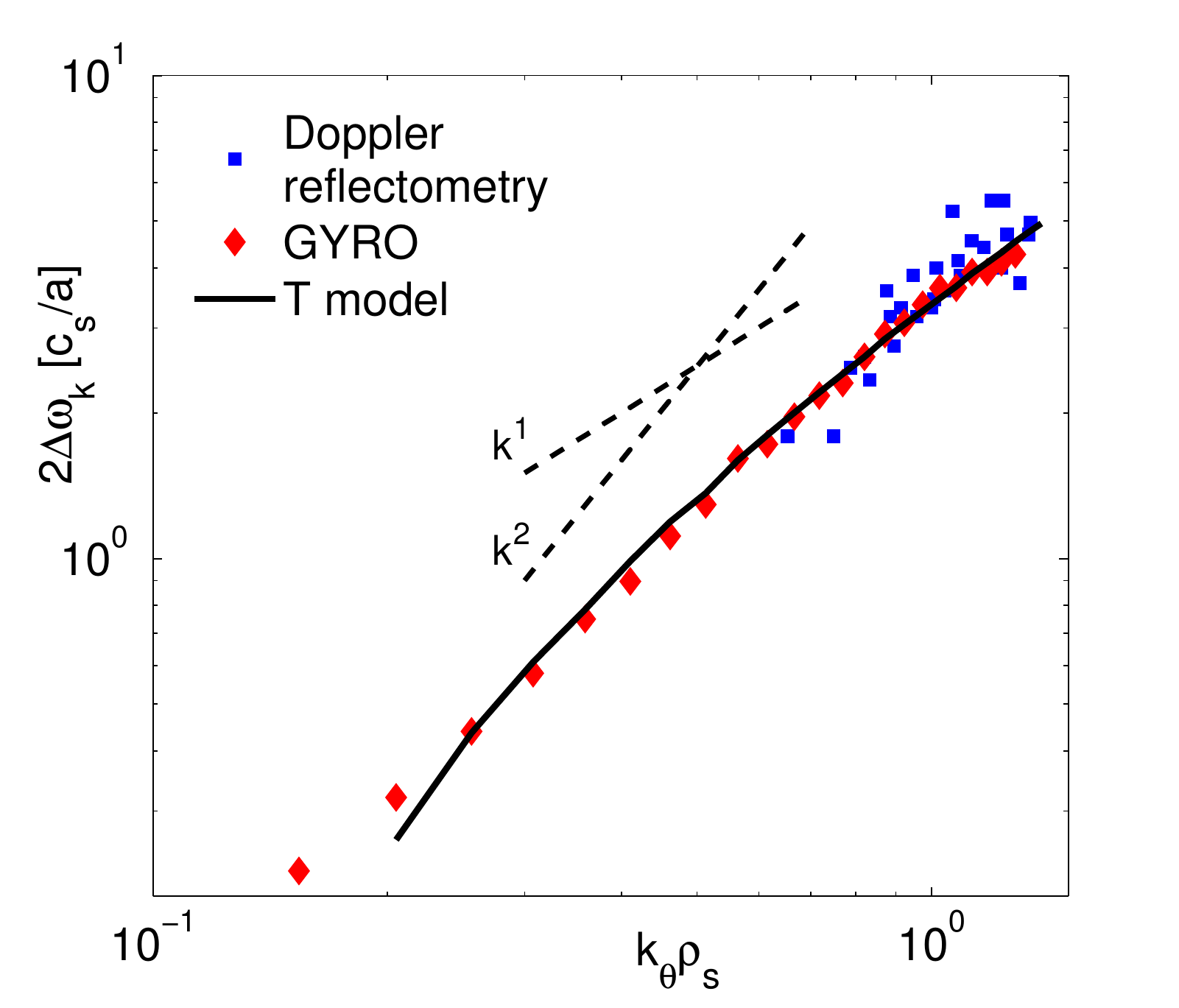}
    \caption{$k_\theta$ dependence of the FWHM relative to the $\delta n$ frequency spectra on TS39596 at $r/a=0.7$: the measurements by the Doppler reflectometry are compared to the expectations by GYRO and from the T model.}
    \label{TS39596-frsp}
  \end{center}
\end{figure}
\indent Following from these results, the consequent crucial test for the validation of the nonlinear simulations, is finally the quantitative comparison of the $k$-dependence of the frequency broadenings. Previous experimental studies addressed this issue using a $\rm{CO}_{\rm{2}}$ laser back-scattering technique for the analysis of the density fluctuations. The findings revealed that the frequency widths $\Delta f_k$ increase at higher $k$ according to $\Delta f_k \propto k^{\alpha}$, where $1.1<\alpha<1.5$. More precisely, a transition of the spectral exponent was observed across $k$: $\alpha\approx1.5$ was found at low $k$, while decreasing for large $k$, $\alpha\approx1.1$.\\
On the discharge TS39596, the radially localized Doppler measurements at $r/a=0.7$ allow reliable evaluation of the FWHM of the $\delta n$ frequency spectra associated to each wave-number $k_\theta$. These are quantitatively compared to the FHWM computed by the GYRO simulation: the results are shown in Fig. \ref{TS39596-frsp}.\\
\indent Firstly, the GYRO simulation shows a remarkably good quantitative agreement with the experimental spectral widths from the Doppler reflectometry. The experimental data are characterized by a non negligible dispersion, but they can be fitted by a power law whose mean spectral exponent is $\alpha_{Doppler}=1.4\pm0.4$. The predictions coming from the simple T model are in remarkable agreement with both the measurements and the nonlinear gyrokinetic simulation. More in particular, the T model $k^{\alpha\left(k\right)}$ dependence for the broadenings exhibits a slight decrease of the spectral exponent $\alpha\left(k\right)$ at larger $k$, as seen by the nonlinear simulation. These evidences provide a relevant additional information with respect to the former result of Fig. \ref{TS39596-frfitsp}, obtained at a given wave-number $k_\theta$. Here in fact, the whole $k$-dependence of the fluctuation frequency spectra is consistent with the outcomes of the T model.\\

\noindent \textbf{Summary}\\

\indent In summary, the turbulent dynamics studied by mean of both density fluctuation measurements with the Doppler reflectometry and nonlinear gyrokinetic simulations, appears in good quantitative agreement with the following main hypotheses of the simple T model:
\begin{enumerate}
	\item In the nonlinear saturation phase, the particle displacements in the turbulent field $\left\langle\Delta^2\right\rangle$ are well described by a simple transition between diffusive and convective behaviors, described by Eq. \eqref{cap4eq5b}. Individually, neither the first nor the second mechanism are able to reproduce the observed frequency spectral shape. 
	\item The distribution of the particle turbulent displacements appears to be consistent with the hypothesis of a normal Gaussian statistics.
	\item Assuming dominant $u^2$ and $\tau_L$ as fitting parameters independent of the wave-number $k$, the $k^{\alpha\left(k\right)}$ dependence of the frequency broadenings is correctly reproduced for a wide spectrum of wave-numbers $k$.
\end{enumerate}
Despite the success of this very simple model in reproducing both the Doppler measurements and the nonlinear gyrokinetic simulations, the issue of the fluctuation frequency spectra certainly deserves additional study in the future. In particular, the analysis with more sophisticated theoretical models, namely the advanced statistical turbulence theories, is highly desirable for relevant advances in this research area.\\

\section{Modeling of the nonlinear saturation process}

\indent In the last section we have treated the validation of the nonlinear gyrokinetic simulations against turbulence measurements, concerning both scalar, i.e. rms $\delta n/n\left(r\right)$ and $\chi_{eff}\left(r\right)$, and spectral quantities, i.e. $\delta n/n\left(k_\theta,k_r\right)$ and $\delta n/n\left(k,\omega\right)$. The aim of the present section is then using the information carried by these nonlinear simulations in order to produce accurate hypotheses for the saturation model required by the quasi-linear modeling. In order words the question is: how to model the saturation spectral intensity $\left\langle \left|\delta \phi_{k}\right|^2\right\rangle$? \\
\indent It is important to clarify that the answer to the latter question can not be immediately derived from the direct comparison between the turbulence measurements and the nonlinear simulations illustrated in the previous section. The reason is simply due to the fact that the measured quantities do not coincide with what is required by the quasi-linear transport modeling. In fact, recalling the quasi-linear expressions Eqs. \eqref{QLpartflux}-\eqref{QLenergyflux}, it appears that (1) the spectral saturation intensity in the quasi-linear formulation refers to the potential fluctuations, and not to the density fluctuations measured by the diagnostics\footnote{The approximation $\delta n/n\approx e\delta \phi/T_e$ is in fact rigoursly valid only if neglecting the non-adiabatic electron response}, (2) $\left|\delta \phi_{k}\right|^2$ has to be calculated according to a proper flux-surface average, and not at the given position in the poloidal cross section where the diagnostics measure the fluctuations (typically on the low field side).\\
\indent The model of the saturation intensity $\left\langle \left|\delta \phi_{k}\right|^2\right\rangle$ used in QuaLiKiz can be written according to the expression:
\begin{eqnarray}  \left\langle \left|\delta \phi_{k}\right|^2\right\rangle = \mathcal{S}_{k}\sum_j L_{k,j}
  	\label{cap4eq9} \end{eqnarray}
with the explicit summation over the $j$ linear unstable modes. The expression \eqref{cap4eq9} is composed of two different parts:
\begin{enumerate}
	\item The first term $\mathcal{S}_{k}$, is the spectral functional shape in the $k$ space, which is assumed the same for each unstable mode.
	\item The second part $L_{k,j}$, concerns instead the saturation level that weights the $j$ unstable solution at each wave-number $k$.
\end{enumerate}
After a brief discussion on the issue of the frequency spectrum in the actual version of QuaLiKiz, the following two paragraphs will be dedicated respectively to these two issues.\\

\subsection{Choices on the frequency spectrum}

\indent At this point it is important to mention that the issue of the frequency spectrum, just discussed in the previous paragraph, does not explicitly appear in the expression \eqref{cap4eq9}. In fact in QuaLiKiz, as shown in the paragraph \ref{sec-quasilinear-new-model}, the choice of a particular frequency spectrum for the saturated potential is included in the formulation of the quasi-linear response, through the additional imaginary contribution in the resonant denominator (Eqs. \eqref{QLeq22}-\eqref{QLeq23}). Namely, in QuaLiKiz, the frequency spectrum is assumed with a Lorentzian shape characterized by broadening equal to linear growth rate, i.e. a choice which is not consistent with the evidences of the nonlinear gyrokinetic simulations, as shown by Figs. \ref{TS39596-frfitsp}-\ref{TS39596-frqlnlsp}. Therefore this point clearly deserves dedicated further studies and improvements with respect to the present formulation in QuaLiKiz have to be done. Nevertheless interestingly, as it will be shown in the next chapter, the actual choices on the frequency spectrum provide a good agreement on the total turbulent fluxes by QuaLiKiz with the expectations by nonlinear gyrokinetic simulations.\\ 

\subsection{Spectral structure of the saturated potential}
\label{sec-spectrum-sat-model}

\begin{figure}[!htbp]
  \begin{center}
    \leavevmode
      \includegraphics[width=8 cm]{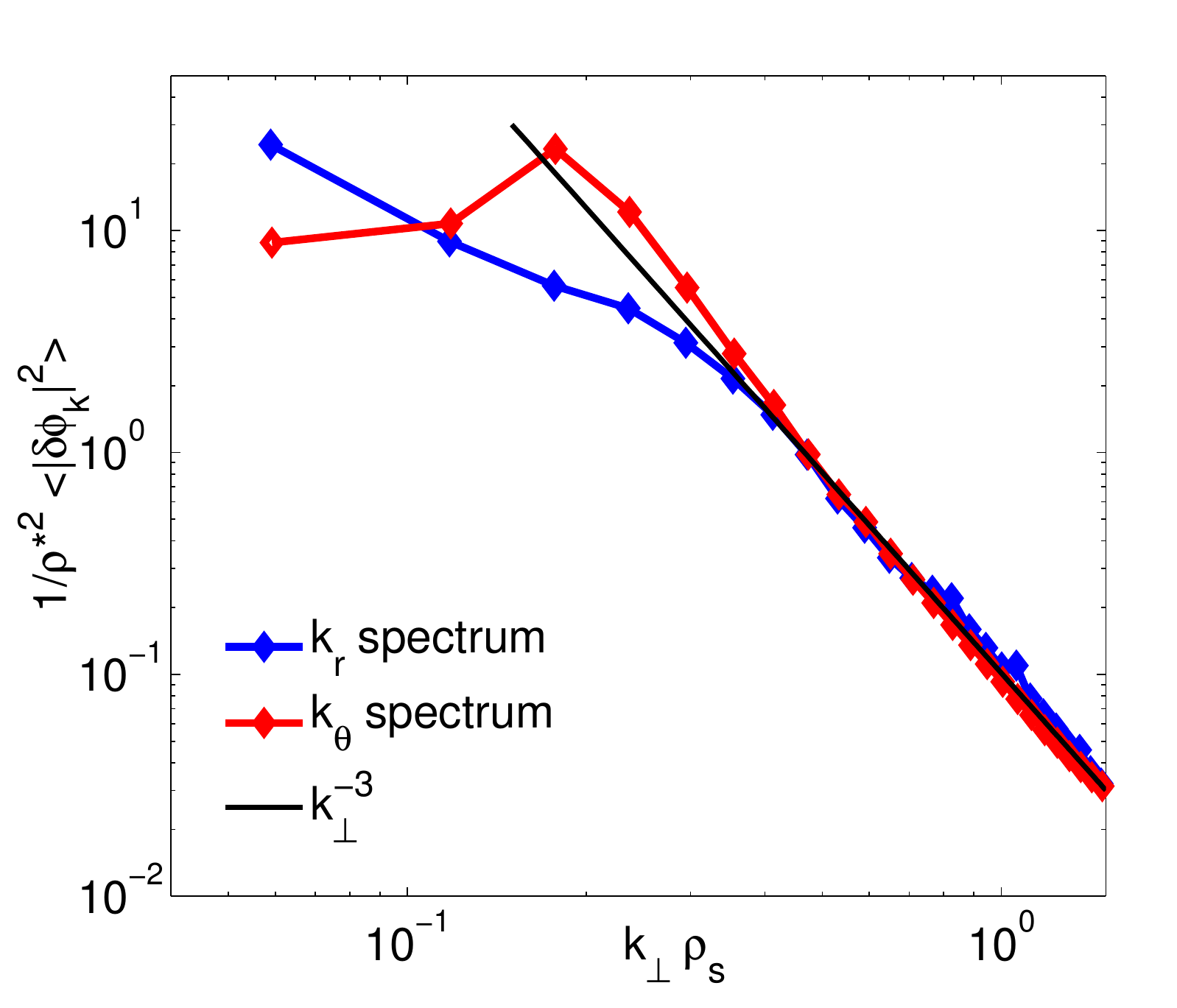}
    \caption{$k_\theta$ and $k_r$ spectra of the fluctuating potential from the GYRO simulation of TS39596 at $r/a=0.7$.}
    \label{TS39596-kspQL}
  \end{center}
\end{figure}

\indent Local nonlinear gyrokinetic simulations are again used in order to determine the $k$ spectral shape of the saturated potential required by the quasi-linear model. It is important to remind that nonlinearly, the $k$ spectrum of the saturated potential can not be separated into distinct contributions corresponding to the linear modes $j$, as already pointed out in the paragraph \ref{sec-transport-weights}. For this reason in QuaLiKiz, the $k$ saturation spectral shape is assumed identical for all the linear unstable modes $j$. This point remains an arbitrary hypothesis that could be object of further studies.\\
\indent The quantity that has to be computed from the simulations is $\left\langle \left|\delta\phi\left(k_\theta\right)\right|^2\right\rangle$, where $\left\langle \ldots\right\rangle$ corresponds here to the time and flux-surface averages. The result from the GYRO simulation of TS39596 at $r/a=0.7$ are reported in Fig. \ref{TS39596-kspQL}.\\
Even if for the purposes of the quasi-linear modeling only the dependence of the saturated potential on the toroidal wave-number $n$, hence the $k_\theta$ spectrum, is important, also the $k_r$ spectrum is shown in Fig. \ref{TS39596-kspQL} for comparison. It appears that the power law $k_\perp\rho_s^{-3}$ fits very well the nonlinear results in the range $0.2<k_\theta\rho_s<1.0$, i.e. for the $k_\theta$ scales above the peak corresponding to $k_{\theta,max}^{nl}$.\\
\indent Interestingly, the $k\rho_s^{-3}$ cascade found by the nonlinear gyrokinetic simulation of Fig. \ref{TS39596-kspQL} is well reproduced also by a recently proposed analytical model for the $k$ spectrum \cite{gurcan09}. This work deals with the minimal hypotheses of homogeneous and isotropic fluctuations, dealing with a weak turbulence theory, in particular the nearly adiabatic limit of the Hasegawa-Wakatani model \cite{hasegawa83}. In the limit of dominant nonlocal disparate scales interaction through the large scale flows (zonal flows), the following analytical expectation is derived:
\begin{eqnarray}  \left\langle \left|\delta \phi_{k}\right|^2\right\rangle = \left\langle \left|\delta n_{k}\right|^2\right\rangle \propto \frac{k\rho_s^{-3}}{\left(1+k\rho_s^{2}\right)^2}
  	\label{cap4eq10} \end{eqnarray}
One of the most original feature of this model is the presence of a change of slope in the nonlinear $k$ cascade around $k\rho_s\approx1$, leading to a steeper spectral exponent for the smaller spatial scales $k\rho_s>1.0$. Even if this point has been found in nice agreement with laser back-scattering turbulence measurements \cite{hennequin04}, it will not be examined here, since the higher $k$ components of the spectrum $k\rho_s>1.0$, are not resolved by our nonlinear simulations, because of the computational cost. On the other hand, in the range $0.2<k\rho_s<1.0$, both the analytical model and the nonlinear simulations coherently find the spectral shape $k_\perp\rho_s^{-3}$.\\
\indent In the light of these results, in QuaLiKiz, the $k$ saturation spectral shape $\mathcal{S}_{k}$ is chosen according to the following relations:
\begin{eqnarray}  \mathcal{S}_{k} \propto k\rho_s^{3} \qquad k\rho_s<k_{\theta,max}^{nl}\rho_s \nonumber \\
     \mathcal{S}_{k} \propto k\rho_s^{-3} \qquad k\rho_s>k_{\theta,max}^{nl}\rho_s
  	\label{cap4eq11} \end{eqnarray}
The symmetry of the spectrum model around $k_{\theta,max}^{nl}\rho_s$ described by \eqref{cap4eq11}, is justified by both experimental results using the beam emission spectroscopy \cite{mckee01} and nonlinear simulations \cite{parker93}.\\

\subsection{Saturation rules}

\indent In order to completely define the saturated $k$ spectral shape $\mathcal{S}_k$ adopted in the quasi-linear calculation, a choice for the value of the nonlinear peak $k_{\theta,max}^{nl}$ appearing in Eq. \eqref{cap4eq11} has to be made. Generally, the nonlinear spectrum peaks in the $k_\theta$ axis for values $k_\theta\rho_s\approx0.2$ (see Fig. \ref{TS39596-kspQL}), i.e. lower than the typical linear growth rate maximum of the most unstable mode $k_\theta\rho_s\approx0.4$. This feature has been confirmed by both experimental evidences \cite{mckee01} and numerical simulations \cite{dannert05}. Moreover, a relevant dependence of the value of $k_{\theta,max}^{nl}$ on the safety factor $q$ is observed in nonlinear simulations \cite{dannert05,hirose05}. In QuaLiKiz, a simple mixing length argument, as introduced in the paragraph \ref{sec-quasilinear-ordering}, Eq. \eqref{QLeq14}, is used with two different goals: firstly, to produce a pertinent choice of $k_{\theta,max}^{nl}$, and secondly to provide the saturation level weighting the contributions from the different linear unstable modes.\\
\indent The value of $k_{\theta,max}^{nl}$ is chosen such that the mixing length diffusivity factor is maximum, according to:
\begin{eqnarray}  \textrm{max}\left(D_{eff}\right) = \left. \frac{\gamma}{\left\langle k_\perp^2\right\rangle} \right|_{k_{\theta,max}^{nl}}
  	\label{cap4eq12} \end{eqnarray}
In Eq. \eqref{cap4eq12}, both the growth rate and the $\left\langle k_\perp^2\right\rangle$ mode structure refers to the linear most unstable mode, so that the resulting $k_{\theta,max}^{nl}$ unambiguously defines a saturation shape $\mathcal{S}_k$ which is kept identical for all the linear unstable modes.\\
In QuaLiKiz, $\left\langle k_\perp^2\right\rangle$ is chosen according to the form originally proposed by \cite{jenko05,dannert05,angioni05}, valid for strongly ballooned modes and adding the impact of the MHD parameter $\alpha$:
\begin{eqnarray}  \left\langle k_\perp^2\right\rangle = k_\theta^2\left[1+\left(s-\alpha\right)^2\left\langle \theta^2\right\rangle\right]
  	\label{cap4eq13} \end{eqnarray}
where
\begin{eqnarray}  \left\langle \theta^2\right\rangle = \frac{\int\theta^2\left|\delta\phi_k^{lin}\left(\theta\right)\right|^2d\theta}{\int\left|\delta\phi_k^{lin}\left(\theta\right)\right|^2d\theta}
  	\label{cap4eq14} \end{eqnarray}
In Eq. \eqref{cap4eq14}, the $\theta$ ballooning structure of the linear eigenmodes is involved; in QuaLiKiz a trial Gaussian eigenfuction is used, such that $\delta\phi^{lin}\left(\theta\right)\propto\sqrt{w}e^{-\theta^2w^2/\left(2d^2\right)}$, where $w$ is the mode width, solution of the fluid limit, and $d$ is the distance between two resonant surfaces. In that case, Eq. \eqref{cap4eq14} then reduces to:
\begin{eqnarray}  \left\langle \theta^2\right\rangle = \frac{2d^2\Gamma\left(3/4\right)}{w^2\Gamma\left(1/4\right)}
  	\label{cap4eq15} \end{eqnarray}
where $\Gamma$ is the mathematical Gamma function.\\
\indent The functional shape of the saturation $k$ spectrum $\mathcal{S}_k$ of Eq. \eqref{cap4eq11} can be specified with the former hypotheses derived from a mixing length argument. Finally, the last choice refers then to the absolute value of $\mathcal{S}_k$. It is reasonable to expect that the linear unstable eigenmodes contribute in a different way to the total turbulent flux, according to the dominant or sub-dominant nature of the mode. Hence, the saturation model adopted in the quasi-linear transport modeling has then to account for weighting factors on the different linear unstable roots. In the actual version of QuaLiKiz, each unstable mode $j$ is weighted through the mixing length hypothesis using its corresponding growth rate and mode structure. In other words, recalling Eq. \eqref{cap4eq9}, we have:
\begin{eqnarray}  L_{k,j} = \frac{\gamma_{k,j}}{\left\langle k_{\perp,j}^2\right\rangle}
  	\label{cap4eq16} \end{eqnarray}
This choice naturally implies that the largest contribution to the total turbulent flux is carried by the leading linear mode at each wave-number $k_\theta$, while the sub-dominant solutions bring smaller terms. Nevertheless, these sub-dominant terms can have a crucial role in conditions where two modes are competing with comparable growth rates, while retaining only the most unstable solution would imply a failure with respect to the nonlinear expectation for the total turbulent flux.\\
The weighting factor adopted in QuaLiKiz expressed by Eq. \eqref{cap4eq16} is completely equivalent to the choice adopted in the TGLF model, demonstrating that this rule is able to achieve a good agreement with the predictions by nonlinear gyrokinetic simulations. Analogous good agreement will be treated in the next chapter.\\

\noindent \textbf{Summary}\\

\indent Finally, the fundamental choices on the saturation model adopted in QuaLiKiz are here summarized:
\begin{enumerate}
	\item A general functional shape for the $k$ saturation spectrum $\mathcal{S}_k$ is identified, characterized by a nonlinear cascade with a spectral exponent $\alpha=-3$ and symmetric around the nonlinear peak $k_{\theta,max}^{nl}$. The validity of this model is coherently supported by nonlinear gyrokinetic simulations, turbulence measurements and simplified analytical models.
	\item The choice of a Lorentzian frequency spectrum with a finite broadening equal to the linear growth rate at each wave-number $k$ is assumed in the formulation of the quasi-linear response. This hypothesis does not agree with the results inferred by both nonlinear simulations and turbulence measurements. Actually, further improvements on this point can be investigated in future work.
	\item A mixing length argument for the effective turbulent diffusivity is used to derive both the nonlinear peaking $k_{\theta,max}^{nl}$ in the model of the $k$ spectrum, and the weighting factors for the different linear unstable modes.
\end{enumerate}

\chapter{Operating the quasi-linear model}
\label{cap-model-operation}

\indent In the previous chapters, both the validity of the gyrokinetic quasi-linear response and the improvement of the model of the saturated potential have been accurately discussed. Finally, in this last chapter, the two main parts of the quasi-linear transport model are put together, in order to produce the estimates of the total turbulent fluxes of energy and particles. Two levels of validation will be here discussed.\\
\indent The first one is the direct comparison between the quasi-linear turbulent fluxes and the predictions by the nonlinear gyrokinetic simulations. The first section will be then devoted to the verification of the quasi-linear QuaLiKiz versus the nonlinear GYRO expectations. Several parametric scans are presented for tokamak relevant conditions, allowing to investigate in which conditions the quasi-linear approximation is able to track the nonlinear results.\\
\indent The second level of validation concerns the application of the quasi-linear transport model to realistic tokamak scenarios. The aim is here gaining interpretative and predictive capabilities on the actual experimental observations. This step necessarily implies that the quasi-linear model is coupled within an integrated transport solver. The latter integrates the quasi-linear turbulent fluxes with the sources of particles and energy and the reconstruction of the magnetic equilibrium, solving the time evolution of the plasma profiles.\\
\indent The second section will be dedicated to the coupling between QuaLiKiz and the integrated transport code CRONOS, while the third section deals with some first analysis of the experimental results.

\section{Validation against the nonlinear predictions}

\subsection{Parametric scans}
\label{parametricscans}

\indent In this paragraph the quasi-linear turbulent fluxes computed by the actual version of QuaLiKiz are compared to local nonlinear gyrokinetic simulations using GYRO. GYRO has been run using the standard numerical resolution summarized in the paragraph \ref{sec-turbulence-times}. The fundamental quantities that are compared are the effective energy and particle diffusivities, defined according to the relations $\Gamma_s=-D_s\nabla_r n_s$ and $q_s=-n_s\chi_s\nabla_r T_s$ for each species $s$. Additional details on the calculation of the total turbulent flux by the nonlinear GYRO simulations are given in Appendix \ref{appx2-GYRO-GYSELA}.\\
\indent The quasi-linear fluxes computed by QuaLiKiz follows from the Eqs. \eqref{QLpartflux}-\eqref{QLenergyflux}, using the model for the saturated potential described in the previous chapter \ref{cap4-improving-QL}. Finally, a single arbitrary constant $C_0$ is used to renormalize the quasi-linear results: in other words, $C_0$ is the factor responsible of rescaling the quasi-linear over-prediction with respect to the nonlinear results, as discussed in the paragraph \ref{sec-transport-weights}. Practically for QuaLiKiz, $C_0=1.6^{-1}$ has been chosen in order to renormalize the value of the quasi-linear ion energy flux to the GYRO nonlinear prediction for the GA-standard case. This value is fixed and applies to all the transport channels (energy and particle) and to all the plasma parameters used in the simulations.\\

\begin{figure}[!htbp]
  \begin{center}
    \leavevmode
      \includegraphics[width=7 cm]{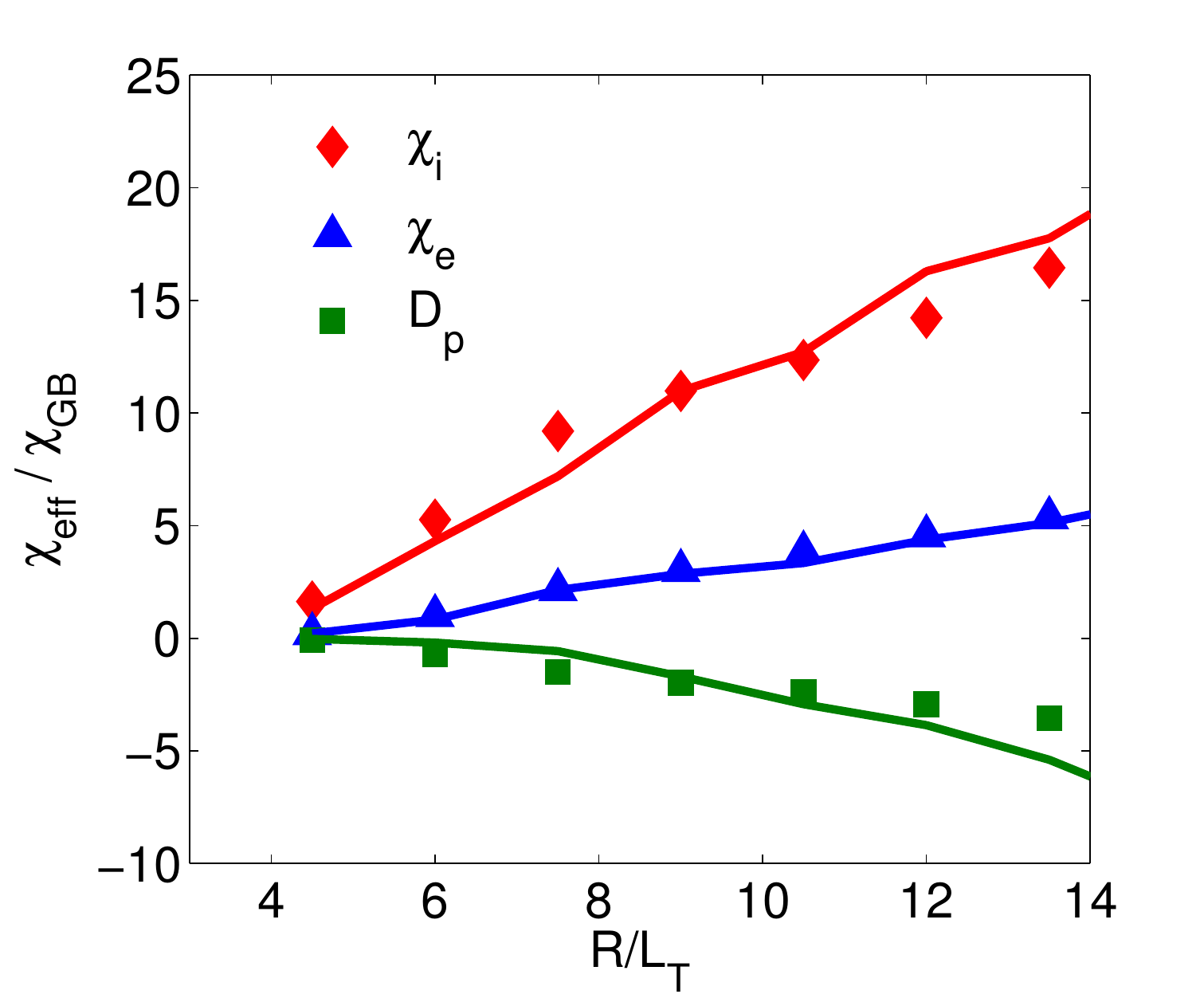}
    \caption{Ion energy, electron energy and particle effective diffusivities from GYRO (points) and QuaLiKiz (lines) for the $R/L_T$ scan based on the GA-standard case.}
    \label{QLK-scan1}
  \end{center}
\end{figure}
\indent In the first scan, both the ion and electron temperature gradients are simultaneously varied on a wide range: $4.5<R/L_{Ti}=R/L_{Te}<13.5$. The effective energy and particle diffusivities are expressed in GyroBohm units $\chi_{GB}=\rho_s^2c_s/a$. Across the whole scan, the ion and electron energy and particle fluxes computed by QuaLiKiz and GYRO agree within 15\%. Both the ratio between the transport channels and the parametric dependence are well captured by the quasi-linear approach.\\
\indent Focusing on the $R/L_T=9.0$ point of the scan presented in Fig. \ref{QLK-scan1}, i.e. the GA-standard case, a first preliminary test is done exploring the coupled ITG-TEM-ETG transport. Nowadays, this kind of exercise is still challenging beacause of the extremely high computational resources needed to run nonlinear gyrokinetic simulations retaining disparate spatial scales. Here we refer to the results by the first coupled ITG-TEM-ETG gyrokinetic nonlinear simulations reported in \cite{waltz07}, using the GYRO code. In this work, both the medium-large ion and the small electron spatial scales up to $k_\theta\rho_e\approx1$ are coherently solved by the nonlinear simulations, but using a reduced mass ratio $\sqrt{m_i/m_e}=20$ instead of $\sqrt{m_i/m_e}=60$.\\
From the point of view of the quasi-linear modeling with QuaLiKiz, the actual hypotheses described in the previous chapters are kept without any modification. In particular it is worth to stress that the spectral exponent of the $k$ saturation spectrum $\mathcal{S}_k$ is kept $\alpha=-3$, while there are experimental and theoretical insights suggesting a steepening of the spectral slope for sub-ion spatial scales $k_\theta\rho_i>1$, as discussed in the paragraph \ref{sec-spectrum-sat-model}.
Referring to the GA-standard case, QuaLiKiz predicts that the ETG contribution (i.e. for $k_\theta\rho_s>1$) to the total electron energy flux is 11\%, in very good agreement with the value of 10\% obtained by the massive GYRO simulation.\\
As important point it has to be noted that the GA-standard case does not contain any additional $E\times B$ shearing rate coming from an external radial electric field. The latter one has been found in Ref. \cite{waltz07}, to be very efficient in quenching the low $k$ ITG-TEM transport, while leaving almost unaffected the ETG contribution. This evidence has suggested \cite{kinsey08} that in certain experimental conditions, with high $E\times B$ shearing rates, the ETG driven transport can still become dominant, since the ITG-TEM contributions are strongly reduced.\\

\indent A second example is a direct application to an experimental collisionality $\nu_*$ scan realized on Tore Supra plasmas. This has been realized as a dimensionless scaling experiment, since both the other two relevant dimensionless parameters, $\rho_s$ and $\beta$ are kept constant across the different discharges, thanks to a coherent variation of the magnetic field and the temperatures, so that only the collisionality is changed. The main plasma parameters are here summarized in Table \ref{TScoll-table}.\\
\begin{table}[!h]
	\begin{center} \begin{tabular}{|c|c|c|c|c|c|c|c|c|c|c|} \hline
			$R_0/a$ & $r/a$ & $R/L_{Ti}$ & $R/L_{Te}$ & $R/L_{n}$ & $q$ & $s$ & $T_i/T_e$ & $Z_{eff}$ & $\rho_*$ & $\beta$ \\
			\hline
		  3.25 & 0.5 & 8.0 & 6.5 & 2.5 & 1.48 & 0.72 & 1.0 & 1.0 & 0.002 & 0 \\ \hline
	\end{tabular} \end{center}
	\caption{\label{TScoll-table}Plasma parameters of the Tore Supra dimensionless collisionality scan at $r/a=0.5$. With respect to the experimental values, only $\beta$ is artificially set to $0$ in the GYRO simulations.}
\end{table}
\begin{figure}[!htbp]
  \begin{center}
    \leavevmode
      \includegraphics[width=7 cm]{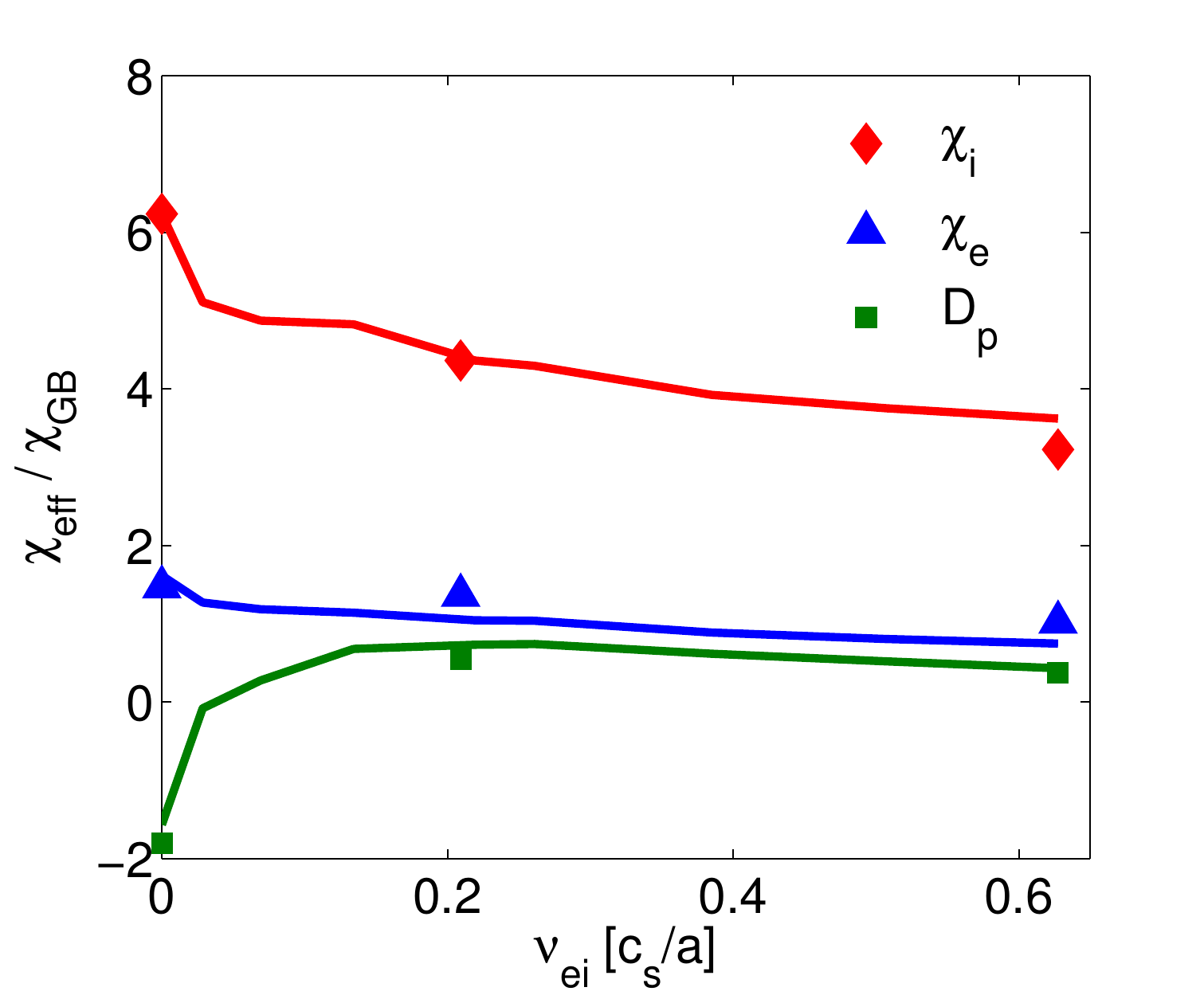}
    \caption{Ion energy, electron energy and particle effective diffusivities from GYRO (points) and QuaLiKiz (lines) for the collisionality scan based on the Tore Supra discharges.}
    \label{QLK-scan2}
  \end{center}
\end{figure}
\indent The $\nu_*$ scaling of transport is a crucial test for quasi-linear models. Two effects are potentially at play with opposite consequences on the total flux levels. The collisional damping of sheared flows would in fact result into an increase of the turbulence level at higher collisionality; hence the fluxes would be enhanced, as pointed in Ref. \cite{falchetto04}. This effect is not taken into account in QuaLiKiz, which does not include sheared flows. Conversely, the linear collisional TEM damping would act in the opposite direction, reducing the linear instability drive as therefore the total turbulent flux at higher collisionality.\\ 
For the plasma parameters here explored, the GYRO simulations find a visible reduction of the transport level when increasing the collisionality, as shown in Fig. \ref{QLK-scan2}. Therefore the linear TEM damping is found to be dominant in these nonlinear simulations with respect to the collisional drag of sheared flows. Finally, Fig. \ref{QLK-scan2} demonstrates that, for experimental values of collisionality, the quasi-linear modeling by QuaLiKiz is able to well reproduce the nonlinear diffusivities predicted by GYRO simulations, performed with pitch-angle scattering operators on both electrons and ions. The coupled dynamics between ion and electron non-adiabatic responses is crucial for both GYRO and QuaLiKiz. In particular, the particle flux reverses direction as $\nu_*$ increases, as already noticed in Ref. \cite{angioni05}.\\

\begin{figure}[!htbp]
  \begin{center}
    \leavevmode
      \includegraphics[width=7 cm]{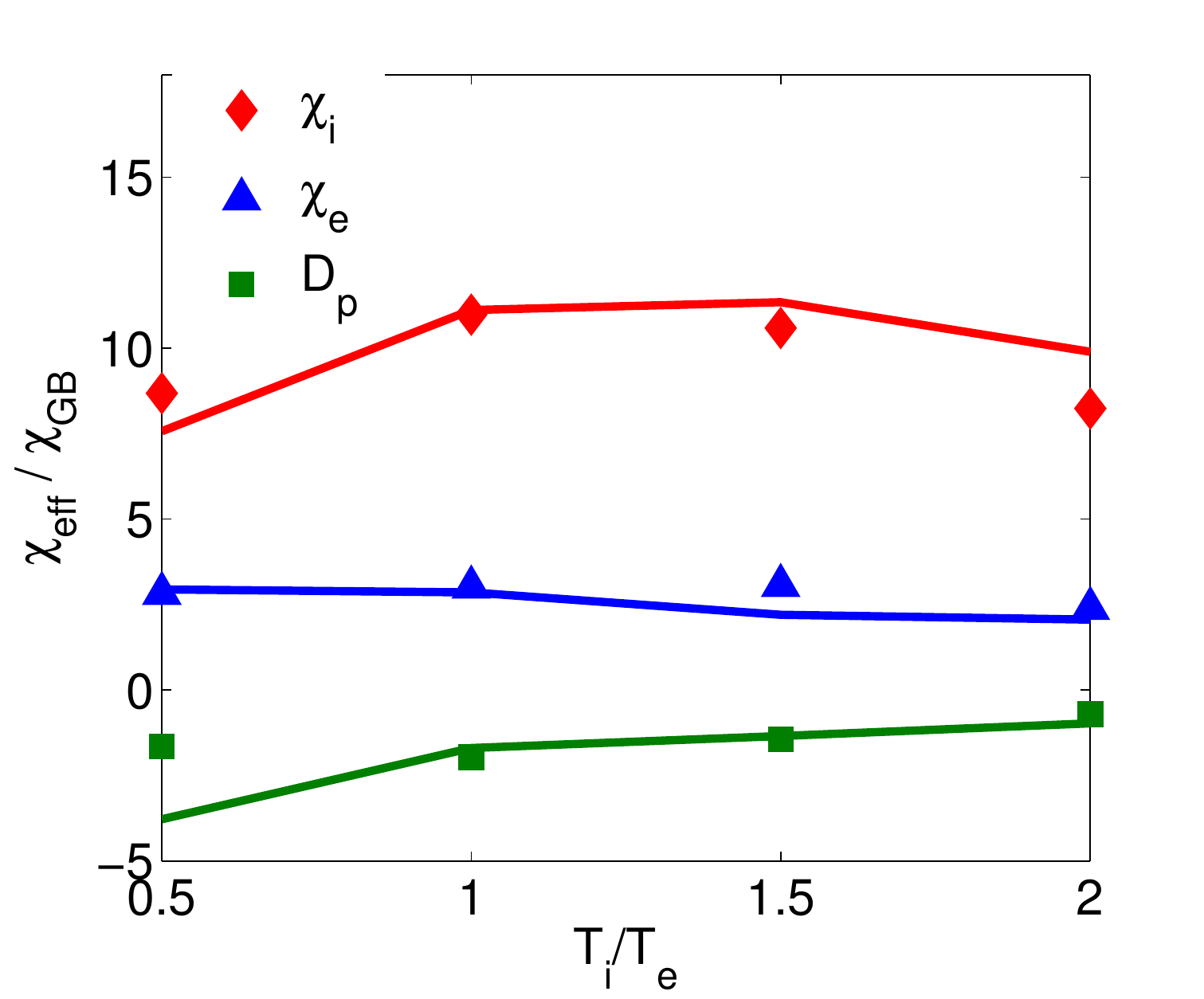}
    \caption{Ion energy, electron energy and particle effective diffusivities from GYRO (points) and QuaLiKiz (lines) for the $T_i/T_e$ scan based on the GA-standard case.}
    \label{QLK-scan3}
  \end{center}
\end{figure}

\indent The third scan illustrates a variation of the ratio $T_i/T_e$ performed on the GA-standard case. Both the nonlinear and the quasi-linear results are reported in Fig. \ref{QLK-scan3}.\\
Some discrepancy between the quasi-linear particle flux predicted by QuaLiKiz and the nonlinear result by GYRO is found for $T_i/T_e=0.5$. For this point of the scan in fact, ITG are expected to be dominant, because of the proportionality of the linear ITG threshold with $T_i/T_e$. The disagreement between the quasi-linear and the nonlinear particle flux is then coherent with the previous results, that indicated the weakness of the quasi-linear modeling when dealing with inward ITG particle flows (see paragraph \ref{sec-transport-weights}). Actually this aspect appears as one of the most significant failures of quasi-linear modeling.\\

\begin{figure}[!htbp]
  \begin{center}
    \leavevmode
      \includegraphics[width=7 cm]{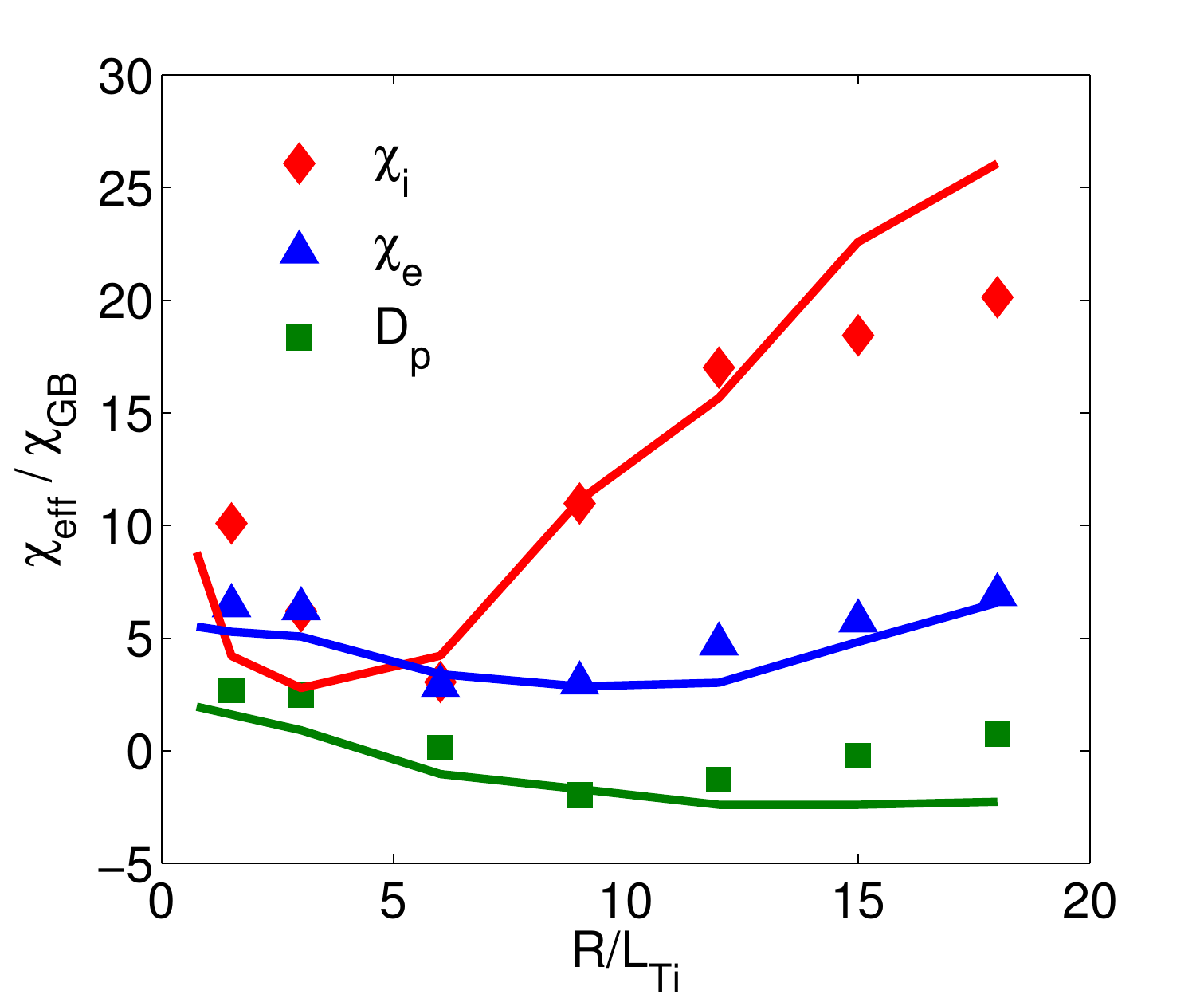}
    \caption{Ion energy, electron energy and particle effective diffusivities from GYRO (points) and QuaLiKiz (lines) for the TEM to ITG transition based on the GA-standard case, fixing $R/L_{Te}=9.0$ and varying $R/L_{Ti}$.}
    \label{QLK-scan4}
  \end{center}
\end{figure}

\indent Fig. \ref{QLK-scan4} illustrates the TEM to ITG transition on the GA-standard case, realized keeping fixed $R/L_{Te}=9.0$ and varying only the ion gradient $R/L_{Ti}$.\\
The electron energy fluxes are well matched; discrepancies are instead observed on the particle fluxes for strong ITG turbulence and for the ion energy flux for $R/L_{Ti}<4$ and $R/L_{Ti}>13$. At $R/L_{Ti}<4$, TEM become the dominant unstable modes, while the marginal conditions for ITG turbulence could be responsible for this quasi-linear failure on the ion energy flux. The quasi-linear model TGLF for example, adopts a modified mixing length rule that is proportional to $\gamma+\gamma^{1.5}$ instead of the simple linear relation with $\gamma$. On the other hand, above $R/L_{Ti}=13$, the QuaLiKiz overestimations can be ascribed to a more pronounced effect of zonal flows in the nonlinear saturation of the ion energy transport for ITG dominated turbulence. Hence, the impact of both sheared flows and marginal turbulence on the quasi-linear transport modeling deserves dedicated additional analysis.\\

\begin{figure}[!htbp]
  \begin{center}
    \leavevmode
      \includegraphics[width=13 cm]{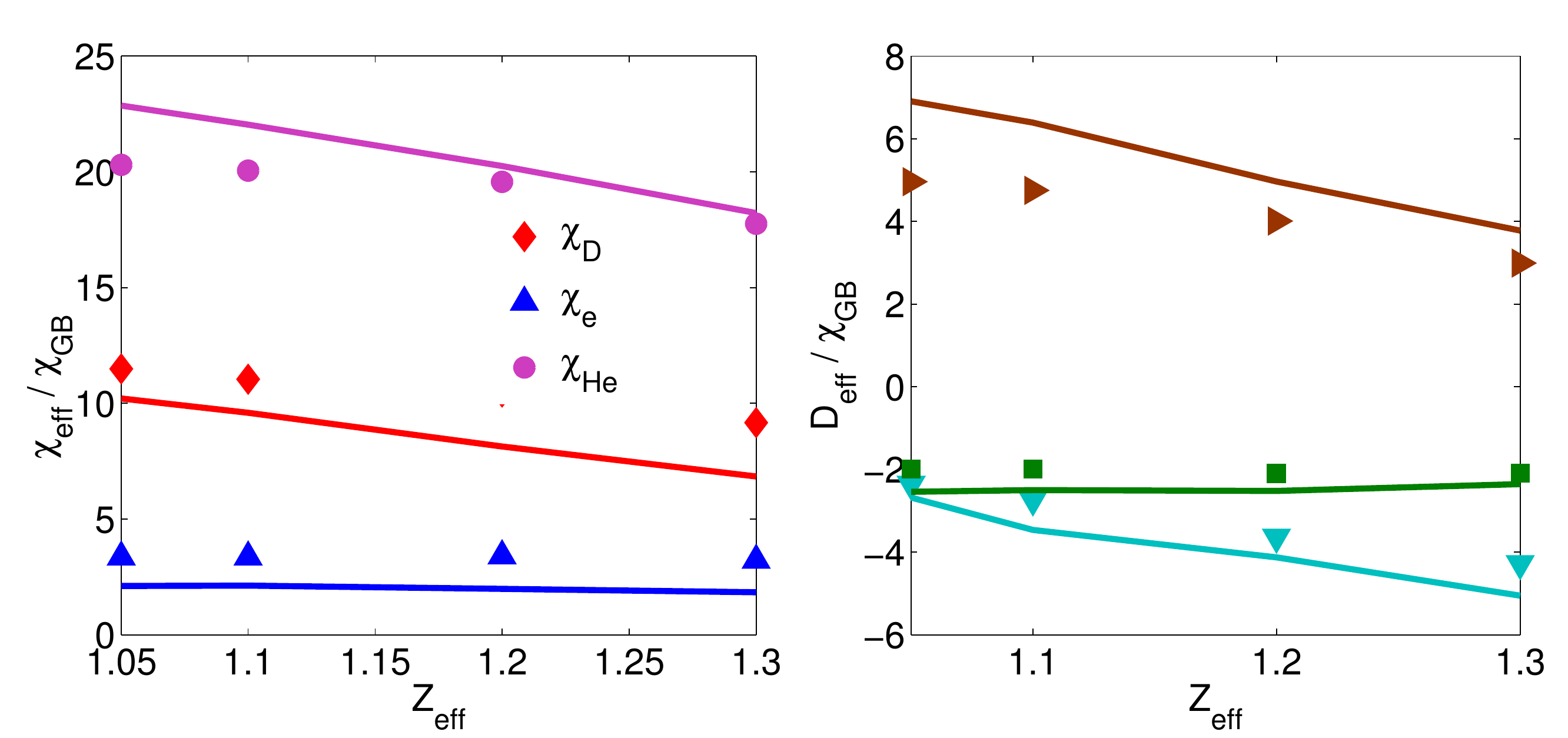}
    \caption{Effective energy and particle diffusivities from GYRO (points) and QuaLiKiz (lines) for a dilution scan, based the on GA-standard case considering $D$ ions and electrons plasma with $He$ impurity.}
    \label{QLK-scan5}
  \end{center}
\end{figure}

\indent Fig. \ref{QLK-scan5} presents a dilution scan operated on plasmas with $D$ main ions, electrons and $He$ impurity on the GA-standard case. Moreover it is assumed that: $T_{He}=T_i$, $R/L_{THe}=R/L_{Ti}$ and $R/L_{nHe}=R/L_n$. For both energy and particle transport of all three species, the quasi-linear predictions by QuaLiKiz are shown to rather well agree with GYRO simulations realized with full nonlinear gyrokinetic treatment of the three plasma species.\\

\section{Coupling with an integrated transport solver}

\indent This section is dedicated to the coupling of the gyrokinetic quasi-linear transport model QuaLiKiz into a system of time dependent fluid transport equations for the evolution of the plasma profiles. Starting from a general formulation, the discussion will be particularly referred to the application within the integrated transport code CRONOS.


\subsection{The time dependent transport equations}

\indent The definition of the fundamental quantities which are evolved in the transport equation directly follows from the velocity space integration of the kinetic Fokker-Planck equation for the species $s$, which describes the evolution of the distribution function $f_s\left(\mathbf{x},\mathbf{v},t\right)$. The velocity moments from the $\rm{0}^{\rm{th}}$ to the $\rm{3}^{\rm{rd}}$ order are respectively the density, the particle flow, the stress tensor (pressure) and the energy flux:
\begin{eqnarray}
	n=\int d^3v\:f \qquad \boldsymbol{\Gamma}=n\mathbf{u}=\int d^3v\:\mathbf{v}f \label{cap5eq1}
\end{eqnarray}
\begin{eqnarray}
	\overline{\overline{P}}=p\overline{\overline{I}}+\overline{\overline{\Pi}}=m\int d^3v\:\mathbf{v}\times\mathbf{v}f \qquad \mathbf{Q}=\int d^3v\:\frac{1}{2}mv^2\mathbf{v}f \label{cap5eq2}
\end{eqnarray}
where the scalar pressure $p$ is defined from the stress tensor as $Tr\left(\overline{\overline{P}}\right)/3$. Another important $\rm{3}^{\rm{rd}}$ velocity moment is the heat flux, i.e. the energy flux in the frame of the moving fluid. Two alternative definitions can be found in literature \cite{hinton76}:
\begin{eqnarray}
	\mathbf{q}=\int d^3v\:\frac{1}{2}m\left|\mathbf{v}-\mathbf{u}\right|^2\left(\mathbf{v}-\mathbf{u}\right)f	\qquad \mathbf{q}'=\int d^3v\:\frac{1}{2}mv^2\left(\mathbf{v}-\mathbf{u}\right)f		 \label{cap5eq3}
\end{eqnarray}
Using the hypothesis that the distribution function $f$ can be decomposed into a Maxwellian component plus an expansion based on the parameter $\epsilon=\rho/L$, i.e. the ratio between the ion gyroradius and the typical macroscopic transport length, at the $\rm{0}^{\rm{th}}$ order in $\epsilon$ Eqs. \eqref{cap5eq3} are linked to the other fluid moments according to:
\begin{eqnarray}
	\mathbf{q}=\mathbf{Q}-\frac{5}{2}p\mathbf{u}	\qquad \mathbf{q}'=\mathbf{Q}-\frac{3}{2}p\mathbf{u}		 \label{cap5eq4}
\end{eqnarray}
\indent At this point it is particularly relevant to make a link with the quantities that are computed by QuaLiKiz. The latter one in fact deals only with the fluctuations of the distribution function $\delta f$ with respect to a local Maxwellian equilibrium. Recalling the QuaLiKiz expressions \eqref{QLpartflux}-\eqref{QLenergyflux} for the particle and energy turbulent flows, these corresponds to the quasi-linear estimates of the following quantities:
\begin{eqnarray}
	\Gamma^{ql}\approxeq\int d^3v\:\mathbf{v}\delta f\cdot\nabla\rho	\qquad Q^{ql}\approxeq\int d^3v\:\frac{1}{2}mv^2\mathbf{v}\delta f\cdot\nabla\rho		 \label{cap5eq4b}
\end{eqnarray}
where $\rho$ represents the perpendicular cross-field radial coordinate. Eqs. \eqref{cap5eq4b} have to be compared to cross-field components of the flows expressed by Eqs. \eqref{cap5eq1}-\eqref{cap5eq2}. It appears that the these two approaches are compatible only if assuming that the neoclassical and the turbulent transport are additive. The collisional neoclassical transport can in fact not be captured by the $\delta f$ approach used in the quasi-linear modeling. In other terms, a possible influence of the turbulence on the evaluation of the neoclassical terms is here neglected. This point represents still nowadays an open subject of research and the validity of the hypothesis of additivity remains to be addressed.\\
\indent The fundamental time dependent transport equations stem from the conservation of particles and energy in the system. These conservation equations can be derived taken respectively the $\rm{0}^{\rm{th}}$ and the $\rm{2}^{\rm{nd}}$ velocity moments of the Fokker-Planck equation. The conservation of particles is then expressed by:
\begin{eqnarray}
	\frac{\partial n}{\partial t}+\nabla\cdot\left(n\mathbf{u}\right)=S		 \label{cap5eq5}
\end{eqnarray}
where the RHS represents a source term, while the equation for the energy conservation reads:
\begin{eqnarray}
	\frac{3}{2}\frac{\partial p}{\partial t}+\nabla\cdot\mathbf{Q}=Q_c+\mathbf{u}\cdot\left(\mathbf{F}+en\mathbf{E}\right)		 \label{cap5eq6}
\end{eqnarray}
In Eq. \eqref{cap5eq6}, the two terms $Q_c$ and $\mathbf{F}$ result from the integration of the collision operator $C\left(f\right)$, defining the collisional energy exchange $Q_c=\int d^3v\: \frac{m}{2}\left|\mathbf{v}-\mathbf{u}\right|^2C\left(f\right)$ and the friction force $\mathbf{F}=m\int d^3v\:\mathbf{v}\times\mathbf{v}\:C\left(f\right)$. Collisions have to respect the conservation of momentum and energy, so that it can be written:
\begin{eqnarray}
	\sum_s\mathbf{F}_s=0 \qquad \sum_s Q_{c,s}+\mathbf{u}_s\cdot\mathbf{F}_s=0		 \label{cap5eq7}
\end{eqnarray}
\indent The main goal of the integrated transport modeling codes is to solve the flux-surface averaged version of the local continuity equations \eqref{cap5eq5}-\eqref{cap5eq6} for the particles and the heat flux. The key point of this kind of approach is the reduction of the dimensionality of the problem: the original 6D plasma dynamics (3 dimensions in the real space and 3 dimensions in the velocity space) is reduced to a 1D approach. The first reduction, lowering by a factor of 3 the dimensionality, corresponds to the $d^3v$ velocity integration of the Fokker-Planck equation (fluid approximation). A further reduction of 2 dimensions is linked to the fact that, at the first order in the expansion parameter $\epsilon$, the thermodynamical plasma quantities such as density, temperature and pressure are constant over the flux surfaces of an axis-symmetric magnetic equilibrium. The latter one is usually numerically solved in a 2D space, but the transport processes can be simplified to a 1D approach, which deals only with the radial cross-field transport. This procedure requires then a proper flux-surface average\footnote{The system solved by the integrated transport codes can also be referred as a 1.5D problem, since an axis-symmetric magnetic equilibrium is solved in a 2D space, but the transport processes are mapped into a 1D grid.}.\\
\indent The average of a scalar quantity $z$ over a given magnetic surface can be expressed considering an infinitesimal volume around the flux surface:
\begin{eqnarray}
	\left\langle z\right\rangle=\frac{\partial}{\partial V}\int_V d^3x\:z=\frac{1}{V'} \int_S dS\frac{z}{\left|\nabla\rho\right|} \qquad \qquad V'\equiv\frac{\partial V}{\partial \rho}		 \label{cap5eq8}
\end{eqnarray}
In the case of the continuity equation for the electrons, this can be obtained from the flux-surface average of Eq. \eqref{cap5eq5}. The cross-field electron flow can be written as $\Gamma_e=n_e\left(\mathbf{u}_e-\mathbf{u}_\rho\right)\cdot\nabla\rho$, where $\mathbf{u}_\rho$ represents the possible velocity of the flux-surface labeled by the coordinate $\rho$ in the laboratory frame. The continuity equation equation then reads:
\begin{eqnarray}
	\frac{\partial}{\partial t}\left(V' n_e\right)+\frac{\partial}{\partial\rho}\left(V'\left\langle \Gamma_e\right\rangle\right)=V'\left\langle S_e\right\rangle		 \label{cap5eq9}
\end{eqnarray}
Actually, the equation which is effectively solved by most of the integrated transport codes, including CRONOS, prefers to use as unknown, a modified form the particle flux, expressed by $\Gamma_e^*=\frac{\left\langle \Gamma_e\right\rangle}{\left\langle \left|\nabla\rho\right|^2\right\rangle}$.\\
\indent The issue of the magnetic equilibrium geometry deserves some additional care. Firstly, the transport equations solved by CRONOS, like Eq. \eqref{cap5eq9}, employ flux-surface averaged quantities, while the quasi-linear turbulent fluxes computed by QuaLiKiz do not depend on the poloidal angle. Therefore, the metric factor $\left\langle\left|\nabla\rho\right|\right\rangle$ has to be introduced in order to account for the quantities averaged across a surface of constant $\rho$. This affects both the profile gradients provided as input by CRONOS to QuaLiKiz and the resulting quasi-linear estimates for the turbulent flows: referring for example to the case of the particle flux, one has:
\begin{eqnarray}
	\left\langle \nabla_\rho n\right\rangle=\partial_\rho n\left\langle\left|\nabla\rho\right|\right\rangle	\qquad \mathrm{and} \qquad \left\langle \Gamma^{ql}\right\rangle=\Gamma^{ql}\left\langle\left|\nabla\rho\right|\right\rangle	 \label{cap5eq9b}
\end{eqnarray}
Secondly, as here above explained, the CRONOS code has to possibility to treat a generally shaped axis-symmetric magnetic equilibrium, while QuaLiKiz solves the linear gyrokinetic dispersion relation on a simplified $s-\alpha$ circular equilibrium. Hence, the direct impact of the shape of the magnetic geometry on the characteristics of the turbulence is not captured by our quasi-linear model. Recently, this aspect has been object of several studies using nonlinear gyrokinetic simulations \cite{kinsey07,angelino09}. Both global PIC and local Eulerian simulations recognized a stabilizing effect of the elongation on the ITG-TEM turbulence, in terms of a reduction of both the linear growth rates and the nonlinear saturation level. Even if these effects are presently completely neglected in QuaLiKiz, several strategies can be explored in the future in order to operate an heuristic correction. The TGLF model for example, uses an adjusted factor for the total quasi-linear flows, which allows to achieve the best fit with a large database of local GYRO simulations run with a shaped Miller magnetic equilibrium. Otherwise, in Ref. \cite{angelino09}, it has been proposed an effective normalized gradient length (obtained by dividing the real gradients by the square root of the elongation), which is shown to correctly reproduce the linear growth rates in presence of finite elongation.\\
\indent The solution of the continuity equation Eq. \eqref{cap5eq9} requires some cautions. Pinch terms are very important for the the particle transport, since they usually balance the diffusive terms to produce a net flux close to zero. For this reason in CRONOS, the particle flux appearing in Eq. \eqref{cap5eq9} is explicitly split into two separate diffusive and convective terms, according to:
\begin{eqnarray}             \left\langle\Gamma_e\right\rangle=\Gamma_e^*\left\langle\left|\nabla\rho\right|^2\right\rangle=-D_e\left\langle\left|\nabla\rho\right|^2\right\rangle\partial_\rho n_e+V_e\left\langle\left|\nabla\rho\right|^2\right\rangle n_e		 \label{cap5eq10}
\end{eqnarray}
The latter expression has to be compared to the quasi-linear flow calculated by QuaLiKiz, that can be computed separating the diffusive and convective components, and it can be written as:
\begin{eqnarray}             \Gamma_e^{ql}=\frac{\left\langle\Gamma_e^{ql}\right\rangle}{\left\langle\left|\nabla\rho\right|\right\rangle}=-D_e^{ql}\left\langle\nabla_\rho n_e\right\rangle+\left(V_{th}^{ql}+V_{c}^{ql}\right)n_e  		 \label{cap5eq11}
\end{eqnarray}
where $V_{th}$ and $V_{c}$ are respectively the thermo-diffusion and the compressibility components of the convective velocity. Finally, recalling Eqs. \eqref{cap5eq9b}, the QuaLiKiz coefficients that are used in CRONOS in order to solve the continuity equation \eqref{cap5eq9} are\footnote{For the diffusion coefficient, the approximation $\left\langle\left|\nabla\rho\right|^2\right\rangle \approx \left\langle\left|\nabla\rho\right|\right\rangle^2$ is used.}:
\begin{eqnarray}             
       D_e=D_e^{ql} \qquad \mathrm{and} \qquad V_e=\left(V_{th}^{ql}+V_{c}^{ql}\right)\frac{\left\langle\left|\nabla\rho\right|\right\rangle}{\left\langle\left|\nabla\rho\right|^2\right\rangle} 		 \label{cap5eq12}
\end{eqnarray}

\indent The flux-surface averaged version of the electron heat transport equation follows from the general expression for the energy conservation Eq. \eqref{cap5eq6}. Recalling that CRONOS uses the conventional definition for the heat flux, i.e. $\mathbf{q}$ in Eqs. \eqref{cap5eq3}-\eqref{cap5eq4}, after several algebra (see \cite{imbeaux09} for details) one obtains the equation for the radial cross-field electron heat transport $q_e=\mathbf{q}_e\cdot\nabla\rho$:
\begin{align}             
       & \frac{3}{2}\frac{1}{V'^{\frac{2}{3}}}\frac{\partial}{\partial t}\left(\left\langle p_e \right\rangle V'^{\frac{5}{3}}\right) + 
		 \frac{\partial}{\partial\rho}\left[V'\left\langle q_e+\frac{5}{2}T_e\Gamma_e\right\rangle\right] = \nonumber \\
		 & \qquad -V'\left\langle\mathbf{u}_\rho\cdot\nabla p_e\right\rangle -V'\left\langle Q_{c,i}\right\rangle -V'\left\langle\mathbf{u}_i\cdot\left(\mathbf{F}_i+Z_in_i\mathbf{E}\right)\right\rangle + V'\left\langle\mathbf{J}\cdot\mathbf{E}\right\rangle + V'\left\langle S_e\right\rangle 
 \label{cap5eq13}
\end{align}
It is worth noting that both the latter heat transport equation \eqref{cap5eq13}, and the previous continuity equation Eq. \eqref{cap5eq9}, are derived retaining a finite velocity of the flux-surface in the laboratory frame $\mathbf{u}_\rho$, while QuaLiKiz computes the turbulent fluxes only relative to the magnetic surface , i.e. to the velocity $\mathbf{u}-\mathbf{u}_\rho$. That implies that the total electron heat flux defined as $\mathbf{q}_e=\mathbf{Q}_e-\frac{5}{2}p\mathbf{u}_e$, following from Eq. \eqref{cap5eq4}, has to be approximated because of the missing $\mathbf{u}_\rho$ term in the QuaLiKiz prediction. More precisely, the flux-surface averaged quantity appearing in the second term of Eq. \eqref{cap5eq13} can be re-written as:
\begin{eqnarray}             
       \left\langle q_e+\frac{5}{2}T_e\Gamma_e\right\rangle = \left\langle\left[\mathbf{q}_e+\frac{5}{2}p_e\left(\mathbf{u}_e-\mathbf{u}_\rho\right)\right]\cdot\nabla\rho\right\rangle	= 
       \left\langle\left(\mathbf{Q}_e-\frac{5}{2}p_e\mathbf{u}_\rho\right)\cdot\nabla\rho\right\rangle	 \label{cap5eq14}
\end{eqnarray}
The approximation can then be easily expressed when neglecting the $\left\langle\frac{5}{2}p_e\mathbf{u}_\rho\cdot\nabla\rho\right\rangle$ term, i.e. this term reduces to the total energy flux $\left\langle q_e+\frac{5}{2}T_e\Gamma_e\right\rangle\rightarrow\left\langle\mathbf{Q}_e\cdot\nabla\rho\right\rangle$ Even if this point represents a limit of the actual formulation, in most practical cases, the corrections due to the movement of the magnetic surfaces are negligible.\\
Likewise for the continuity equation, CRONOS solves Eq. \eqref{cap5eq14} in terms of the variable $q_e^*=\frac{q_e}{\left\langle\left|\nabla\rho\right|^2\right\rangle}$. On the other hand, the flux-averaged energy flux computed by QuaLiKiz reads as $\left\langle Q_e^{ql}\right\rangle=Q_e^{ql}\left\langle\left|\nabla\rho\right|\right\rangle$. Using the just mentioned approximation, the QuaLiKiz term which enters in the CRONOS electron heat transport equation is the following\footnote{Practically, CRONOS solves the Eq. \eqref{cap5eq13} splitting the diffusive and convective terms, in order to avoid injecting a negative diffusion coefficient, analogously to the continuity equation for the particles. On the other hand, such explicit splitting is not done for the energy transport inside Qualikiz, in order to speed up the calculations and because the energy flux is almost always outward, i.e. it corresponds to $\chi_{eff}>0$. For this reason the QuaLiKiz energy flux is initially stored as a purely diffusive effective transport coefficient, while a purely numerical $(D,V)$ splitting is done in the CRONOS solver in the case of of inward heat flux.}:
\begin{eqnarray}                    q_e^*+\frac{5}{2}T_e\Gamma_e^*=Q_e^{ql}\frac{\left\langle\left|\nabla\rho\right|\right\rangle}{\left\langle\left|\nabla\rho\right|^2\right\rangle}	 \label{cap5eq15}
\end{eqnarray}
\indent A brief description of the RHS terms of the Eq. \eqref{cap5eq13} is here given. The first one represents an apparent energy flux due to a finite velocity $\mathbf{u}_\rho$ of the flux-surfaces. As previously observed, this additional component is usually negligible in comparison to the other source terms, and in any case it is not retained by QuaLiKiz. The second one refers to the non-negligible collisional energy exchange from the electrons to the ions. This terms is calculated by CRONOS according to the derivation of Ref. \cite{hinton76}. The third term formally represents the friction forces between the electron and the ion species: this small correction is treated in CRONOS using the neoclassical theory. The last two terms correspond to the heating sources, respectively to the $\mathbf{J}\cdot\mathbf{E}$ ohmic heating induced by the plasma current, and to external heating sources, like the RF power and the Neutral Beam Injection.\\

\indent The equation that governs the cross-field ion heat transport $q_i=\mathbf{q}_i\cdot\nabla\rho$ is completely analogous to the electron one. It reads:
\begin{align}             
       & \frac{3}{2}\frac{1}{V'^{\frac{2}{3}}}\frac{\partial}{\partial t}\left(\left\langle p_i \right\rangle V'^{\frac{5}{3}}\right) + 
		 \frac{\partial}{\partial\rho}\left[V'\left\langle q_i+\frac{5}{2}T_i\Gamma_i\right\rangle\right] = \nonumber \\
		 & \qquad -V'\left\langle\mathbf{u}_\rho\cdot\nabla p_i\right\rangle +V'\left\langle Q_{c,i}\right\rangle +V'\left\langle\mathbf{u}_i\cdot\left(\mathbf{F}_i+Z_in_i\mathbf{E}\right)\right\rangle + V'\left\langle S_i\right\rangle 
 \label{cap5eq16}
\end{align}
The whole discussion following from the previous electron heat equation can be readily applied here. It is worth noting that the second and the third term of the RHS are identical to the Eq. \eqref{cap5eq13}, but with opposite sign.\\
CRONOS only solves the Eq. \eqref{cap5eq16} for the total ion pressure, assuming the same temperature for all the $s$ ion species, i.e. $p_{i,tot}=T_i\sum_s n_s$. Conversely, QuaLiKiz can presently account for electrons, ions and an additional ion species, computing separate particle and energy fluxes. With analogous notation with the previous one, the QuaLiKiz estimations that enter in the CRONOS total ion heat transport equation are finally:
\begin{eqnarray}                    q_{i,tot}^*+\frac{5}{2}T_i\Gamma_{i,tot}^*=\left(Q_i^{ql}+Q_{imp}^{ql}\right)\frac{\left\langle\left|\nabla\rho\right|\right\rangle}{\left\langle\left|\nabla\rho\right|^2\right\rangle}	 \label{cap5eq17}
\end{eqnarray}

\section{Application to the experiments}

\subsection{Preliminary results}

\begin{figure}[!htbp]
  \begin{center}
    \leavevmode
      \includegraphics[width=8 cm]{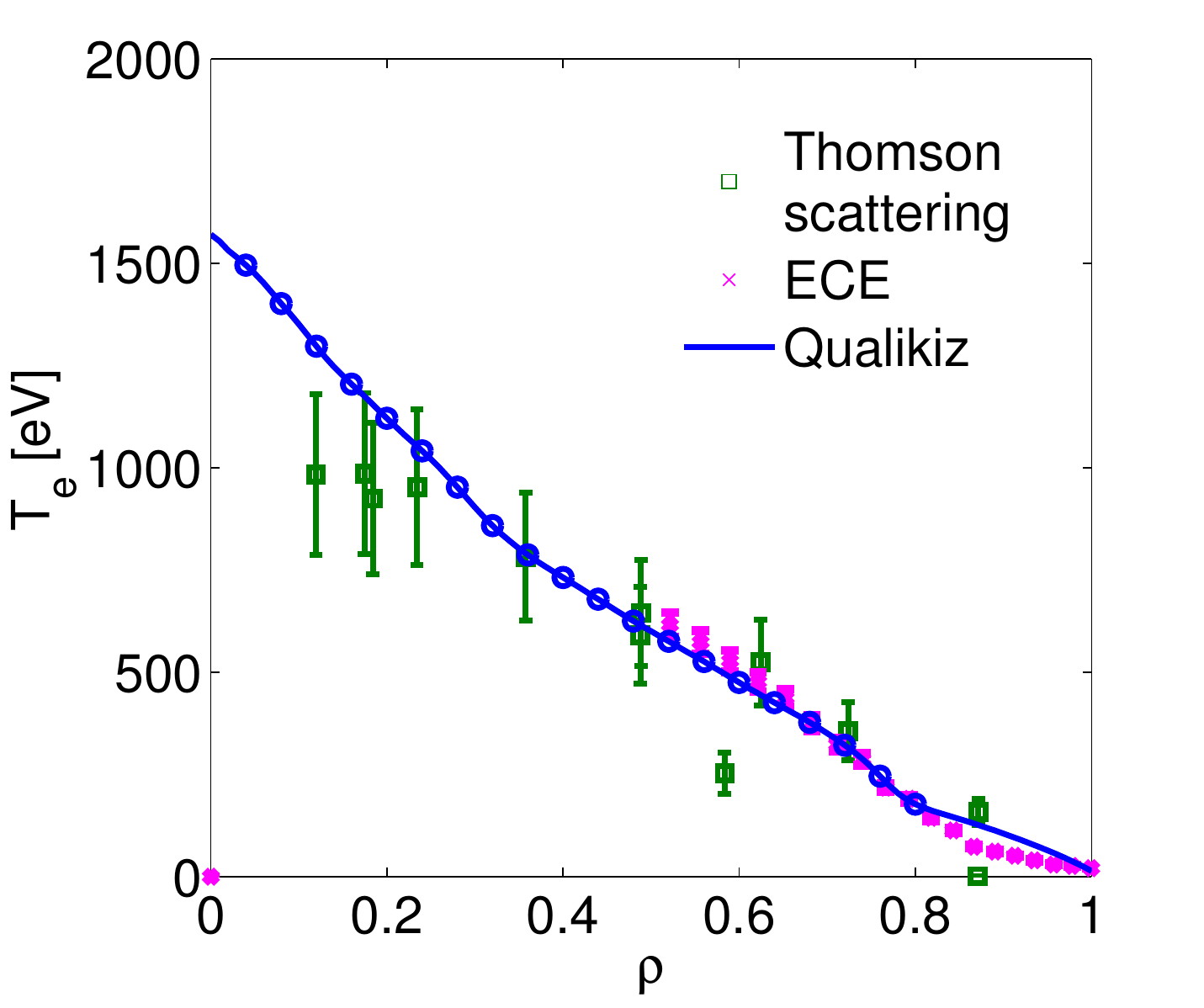}
    \caption{First example of the application of QuaLiKiz integrated into the CRONOS code for the predictive analysis of the discharge TS39596 at $t=7.668~\rm{s}$. In this case only $T_e$ and $T_i$ are evolved using the experimental value of $Z_{eff}=1.6$. The simulation achieves a steady state solution on a time $\Delta t\approx1\tau_E$. This plot shows the radial $T_e$ profiles predicted by QuaLiKiz compared to the experimental data from both Electron Cyclotron Emission and Thomson Scattering diagnostics.}
    \label{QLK-Te-TS39596}
  \end{center}
\end{figure}

\begin{figure}[!htbp]
  \begin{center}
    \leavevmode
      \includegraphics[width=8 cm]{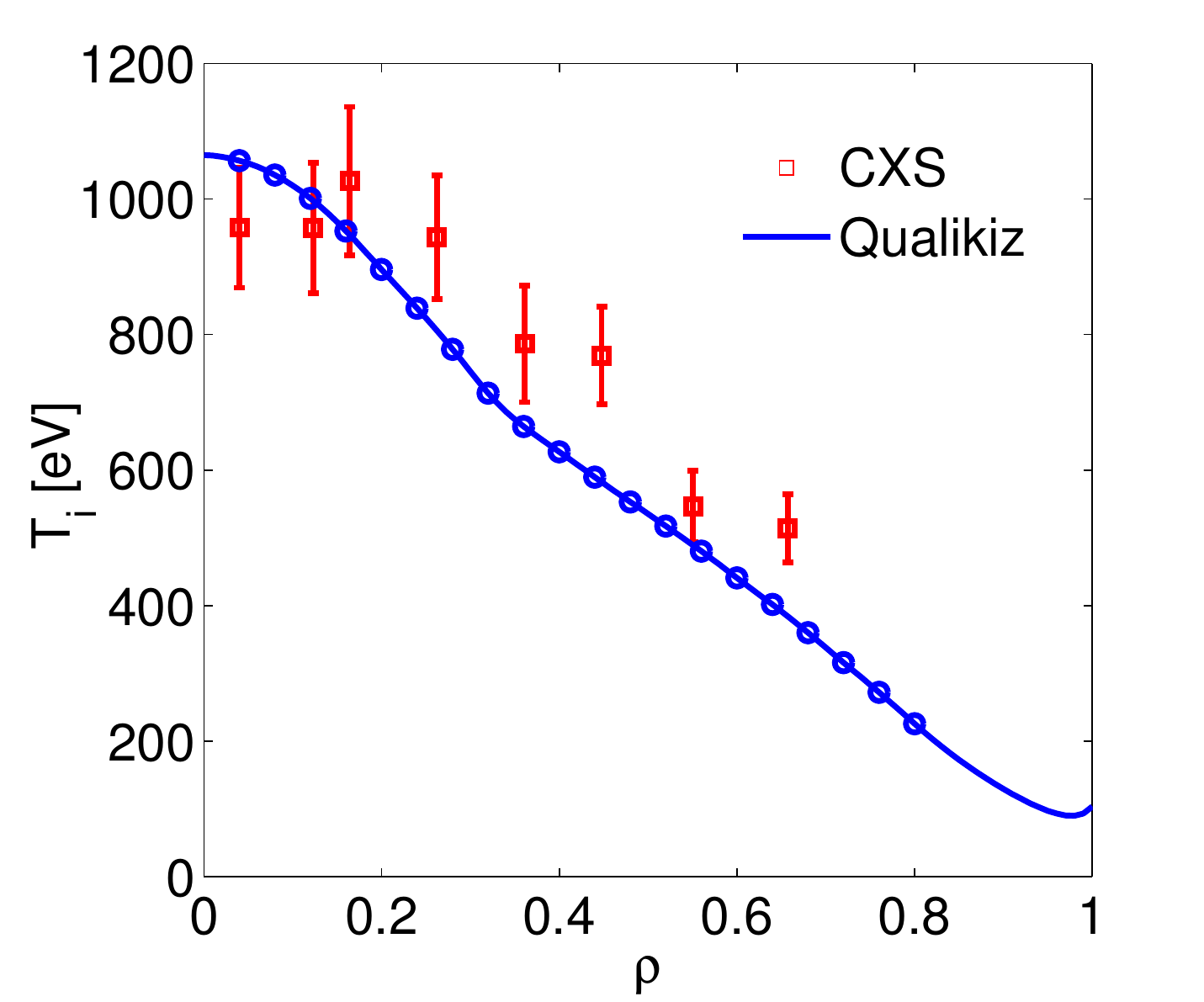}
    \caption{Radial $T_i$ profiles predicted by QuaLiKiz compared to the experimental data from the Charge Exchange diagnostic.}
    \label{QLK-Ti-TS39596}
  \end{center}
\end{figure}

\begin{figure}[!htbp]
  \begin{center}
    \leavevmode
      \includegraphics[width=8 cm]{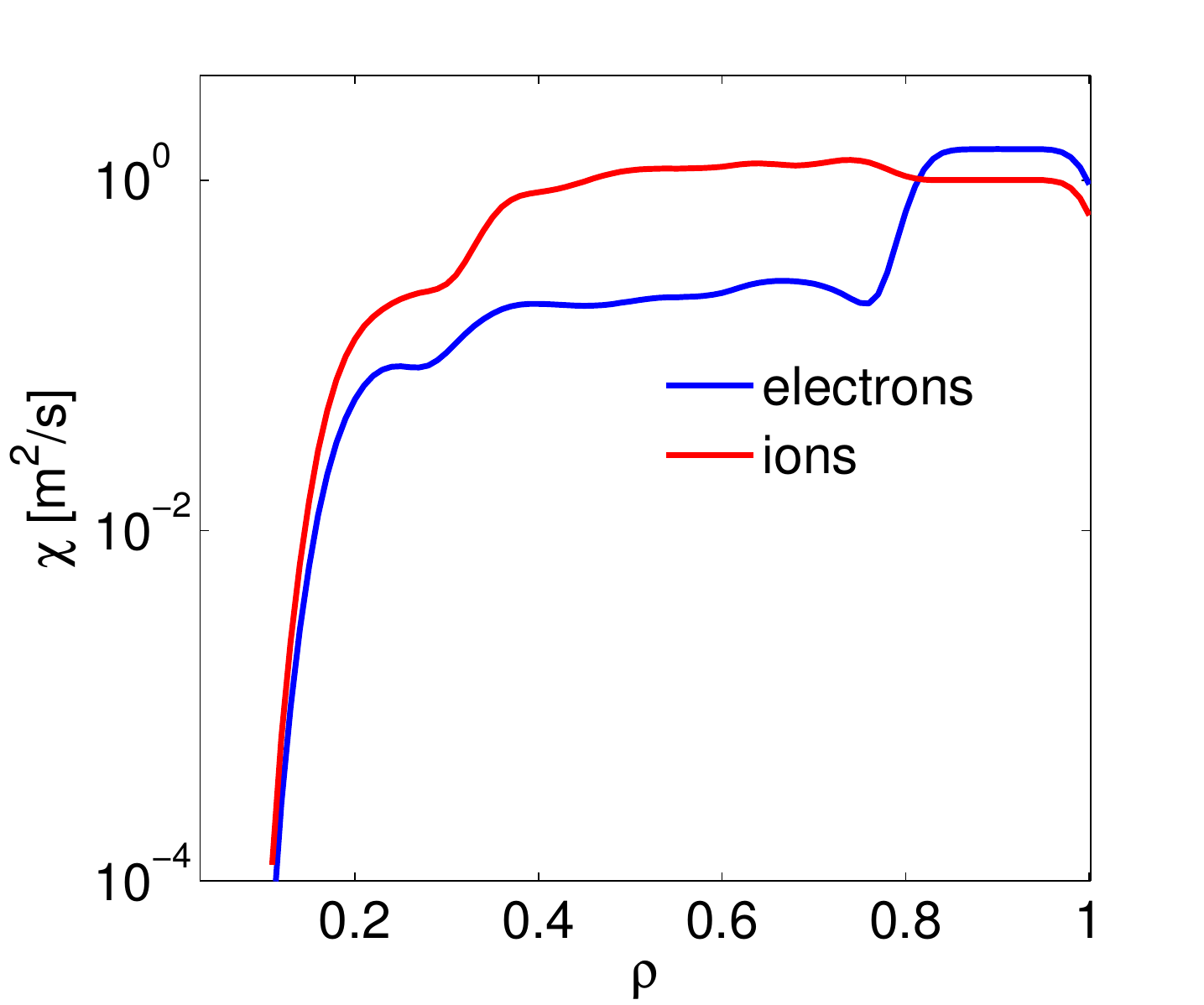}
    \caption{Radial profiles of the ion and electron energy diffusivities predicted by QuaLiKiz.}
    \label{QLK-chi-TS39596}
  \end{center}
\end{figure}

\indent The present version of QuaLiKiz is being coupled and tested within the CRONOS code. Unfortunately, only preliminary results are here presented, since at the time of the redaction of the thesis this is still an ongoing work, which certainly deserves much more attention. An important task that is currently being realized is due to the fact that for each iteration of the transport solver, QuaLiKiz has to be run at each position of the radial grid. For this reason, in order to perform time-dependent simulations within a reasonable time, the numerical routine associated to QuaLiKiz has to be parallelized.\\
\indent The very first interest of running QuaLiKiz within an integrated transport solver is to analyze a tokamak discharge at a given time corresponding to quasi steady-state conditions. The main goal is the following: the radial profile of the effective heat and particle diffusivities, as well of the temperatures and the density obtained from the interpretative analysis, should be matched by the predictive transport simulation using QuaLiKiz, provided that the heat and particle sources are constant in time. It is expected that, by the effect of the QuaLiKiz transport coefficients, the temperature and density profiles and their relative gradients undergo to a time evolution, which should converge to the experimental values within the experimental uncertainty. More in particular, due to the absence of particle fueling in the center of the plasma, the steady-state condition corresponds to a null particle flow in the core.\\
\indent A standard Tore Supra discharge, TS39596, already discussed in Chapter \ref{cap4-improving-QL} and whose experimental parameters are summarized in Tables \ref{TableTS39596gl} and \ref{TableTS39596lc}, has been chosen as a first example to be analyzed with QuaLiKiz. The simulation has been run keeping a fixed density profile, while evolving both the ion and electron temperatures retaining the real $Z_{eff}=1.6$. The transport coefficients predicted by QuaLiKiz are applied in the normalized radial domain $0<\rho<0.8$; for the outer plasma region corresponding to $\rho>0.8$, an arbitrary profile of diffusivity is assumed, providing suitable boundary conditions for the temperature profiles. The results presented in Figs. \ref{QLK-Te-TS39596} and \ref{QLK-Ti-TS39596} show the comparison between the $T_e$ and $T_i$ profiles obtained by the QuaLiKiz simulation and the measurements from the diagnostics. The CRONOS simulation has been run for a time $t\approx\tau_E$, where physically $\tau_E\approx200~\rm{ms}$, such that these temperature profiles reproduce a steady state on a macroscopic transport time scale. A reasonable agreement between the QuaLiKiz predictions and the measurements is obtained for both $T_e$ and $T_i$. A careful examination of the ion temperature profile of Fig. \ref{QLK-Ti-TS39596} reveals that the boundary condition for $\chi_i$ assumed at $\rho=0.8$ is not optimal, as highlighted by the comparison of the reconstructed $T_i$ profile with the Charge Exchange data. This can warn about the actual sensitivity of the prediction with respect to the boundary conditions.\\
On the other hand, the peaking of the temperature profiles in the inner region ($\rho<0.3$) is clarified when looking at the heat diffusivities calculated by the quasi-linear model and shown in Fig. \ref{QLK-chi-TS39596}. For $\rho<0.3$ in fact, the model finds no turbulent transport since the local plasma parameters are below the linear mode thresholds: only the neoclassical contributions are left, hence a \textit{barrier-like} effect appears in the temperature profiles. This point can be regarded as a typical consequence of the hypothesis of local transport. This feature is commonly shared among all the local transport simulations and, more importantly, it is not linked to the quasi-linear approximation, since a nonlinear local approach would lead to a similar result (see for example Ref. \cite{candy09}). A solution to this general problem is beyond the scope this thesis work and represents one of the important challenges that have to be addressed to progress in the understanding and the prediction of the tokamak turbulent transport.\\


\chapter{Conclusion}

The development of a gyro-kinetic quasi-linear transport model for tokamak plasmas, ultimately designed to provide physically comprehensive predictions of the thermodynamic relevant quantities, is a task that required tight links among theoretical, experimental and numerical studies. The framework of QuaLiKiz, which operates a reduction of complexity on the nonlinear self-organizing plasma dynamics, allows in fact multiple validations of the current understanding of the tokamak micro-turbulence.\\
\indent The main outcomes of this thesis derive from the fundamental steps involved by the formulation of the quasi-linear transport model, namely: (1) the verification of the quasi-linear response against the numerically computed nonlinear one, (2) the improvement of the saturation model through an accurate validation of the nonlinear codes against the turbulence measurements, (3) the integration of the quasi-linear model within an integrated transport solver.\\
\indent The kinetic description of the plasma response is believed to be crucial to correctly capture the essential physics at play in the tokamak turbulent transport. The $T_i/T_e$ dependence of the ion ITG and electron TEM linear instability threshold is chosen as relevant example. Both the analytical calculations and the gyro-kinetic simulations recognize, for the first time in the case of the electron modes, that the temperatures ratio dependence of these thresholds is intrinsically linked with the kinetic wave-particle resonance. Hence, the gyro-kinetic formulation is preferred in QuaLiKiz, with respect to an arbitrary closure of a fluid or gyro-fluid model.\\
\indent The system of equations that defines the quasi-linear expectations for the turbulent fluxes of energy and particle is presented, as it is implemented in QuaLiKiz. At the level of this formulation, two possible resonance broadenings are identified. The first one is related to the kinetic wave-particle resonance and its finite broadening is necessary to fulfill the ambipolarity of the particle fluxes. The second one is linked to an intrinsic frequency spectral shape of the fluctuation potential around the real frequency of the unstable mode. The validity of the hypothesis of the quasi-linear response is for the first time systematically investigated against fully nonlinear gyrokinetic simulations, apart from any choice on the structure of the saturated potential. Three different levels are examined. Firstly, the comparison of the characteristic turbulence times allows to derive an estimation of a Kubo number. Secondly, the phase relations between the transported quantities and the fluctuating potential are studied in the linear as well in the nonlinear phase. Finally, the overall quasi-linear over nonlinear ratio for the transport of both the energy and particle channels is analyzed. The main outcome is that, with an appropriate re-normalization, the quasi-linear approximation is able of reasonably recovering the nonlinear fluxes over a wide range of plasma parameters, within a mean rms error of 13\%. This result surprisingly extends also to conditions where the high instability drive would have suggested strong turbulence conditions that could not be captured by the quasi-linear physics. Nevertheless, the quasi-linear particle flux can deviate from the nonlinear expecations when dealing with strong inward flows driven by ion modes. Fortunately, such conditions very rarely appear in the experiments.\\
\indent The model for the fluctuating saturated potential assumed in QuaLiKiz results from the validation of the numerical and theoretical predictions against the measurements. The nonlinear gyrokinetic simulations, using the GYRO code, are here shown for the first time to quantitatively reproduce complementary measurements from a standard Tore Supra ohmic discharge, namely: the radial profiles of the heat diffusivity $\chi_{\textrm{eff}}$ and of the rms $\delta n/n$, as well the local $k_{\theta}$ and $k_r$ spectra. When properly reconstructing the instrumental response of the diagnostic system, the simulations help moreover to revise the apparent experimental turbulence anisotropy at scales comparable to the ion gyroradius. Also, the similarity of the fluctuation $k_{\theta}$ power spectra between the measurements and both local and global, using the GYSELA code, gyrokinetic simulations, suggests a rather general character of the tokamak turbulence wave-number spectrum. At the scales of the order of the ion gyroradius, the power law scaling $\left|\delta n_k\right|^2\approx\left|\delta\phi_k\right|^2\propto k^{-3}$, found on many other devices, is also recovered by an analytical shell model for the drift-wave turbulence. Simple theoretical considerations are again successful when applied to the experimental and the numerical frequency spectra of the fluctuations. An intermediate statistics between diffusive and convective behaviors for the particle displacements in the turbulent field, is found to agree with the experiments and the nonlinear simulations on both the frequency spectral shape and the $k_{\theta}$ dependence of the spectral broadenings.\\
\indent Finally, using only one renormalization constant, the total quasi-linear fluxes predicted by QuaLiKiz are compared the a wide number of nonlinear gyrokinetic simulations. When coupling the choices for the saturated potential with the quasi-linear response, the QuaLiKiz fluxes are shown to agree with the nonlinear predictions for the energy transport in the ion and electron channels, as well as for particle fluxes for a wide range of tokamak relevant plasma parameters. QuaLiKiz is now coupled to the integrated transport platform CRONOS; its first applications to the experiment are encouraging but demand further investigation from both the quasi-linear and nonlinear modeling. 

\begin{center} $\therefore$ \\ \end{center} 

\indent A number of challenging issues still remains to be tackled. In fact, as it has been demonstrated in this thesis work, developing a reduced transport model allows to explore a very wide range of theoretical, experimental and numerical issues of interest for the nuclear fusion research. A first priority is certainly linked to further assess the effectiveness of QuaLiKiz in predicting the tokamak confinement properties, solving the time dependent transport problem within CRONOS. A partial list of other potential topics of study and improvement are: i) The choices for the saturated potential deserve additional comparisons with nonlinear simulations and experimental measurements dealing with different plasma scenarios. ii) The way of accounting for the subdominant unstable modes in the quasi-linear formulation can be refined. iii) The domain in which the quasi-linear approximation fails (particle flux, marginal conditions, strong ITG turbulence, onset of zonal/large scale flows, etc.) should be more precisely understood, especially in view of the experimental plasma conditions. iv) Further improvements to the quasi-linear model could be addressed, namely accounting for the $E\times B$ shear stabilization effects and for shaped plasma geometries. v) The quasi-linear model can be extended to deal with the transport of momentum, in order to perform fully predictive studies. vi) The numerical optimization within the CRONOS transport solver will contribute to provide a fast and reliable tool for tokamak plasma studies.

\appendix
\chapter{The $\textrm{T}_\textrm{i}/\textrm{T}_\textrm{e}$ dependence of linear ITG-TEM thresholds}
\label{appx1-TiTe}

\noindent \textbf{The fluid approach}\\

\begin{figure}[!htbp]
  \begin{center}
    \leavevmode
      \includegraphics[width=7 cm]{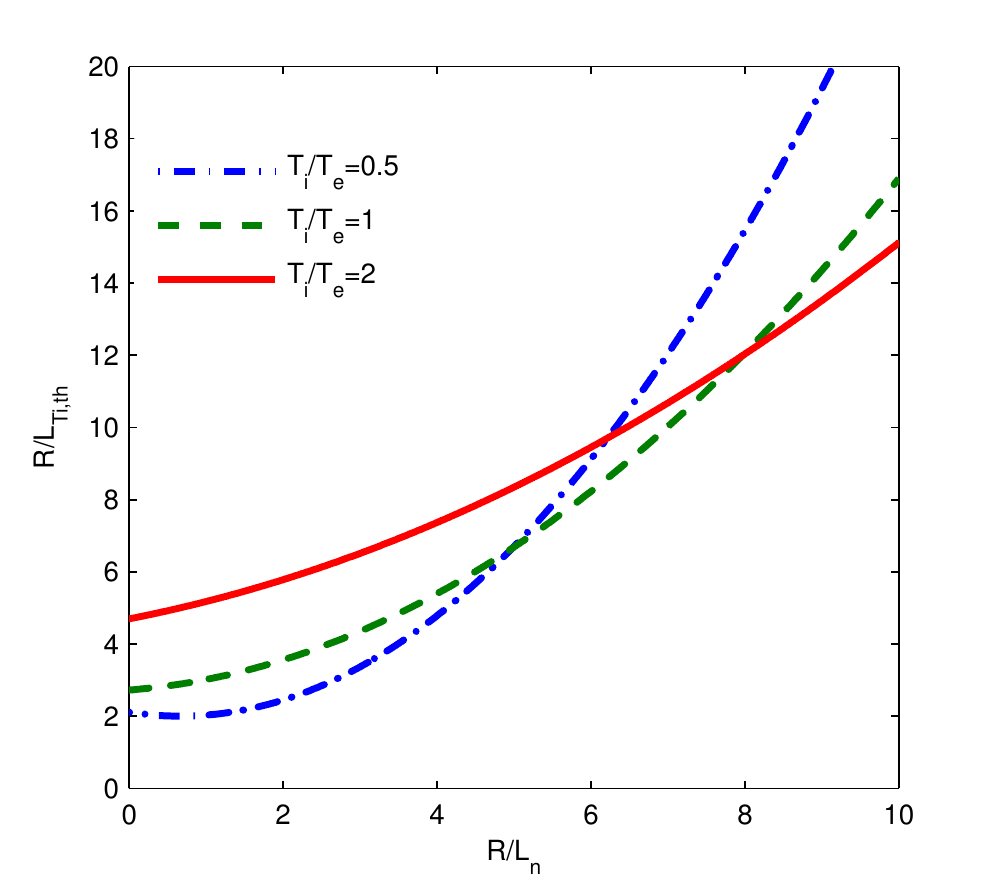}
    \caption{ITG threshold vs $R/L_n$ derived according to the Weiland model (Eq. \eqref{appeq5}) for different values of the ratio $T_i/T_e$.}
    \label{figTiTe1}
  \end{center}
\end{figure}

\indent Advanced fluid theories have been widely used for deriving analytical threshold expressions of electrostatic unstable modes. A significant reference is the Weiland reactive two-fluids model \cite{weiland89,weiland00}, which accounts for the influence and the interactions of convection, compression, and thermalization of the plasma species.
Within the Weiland fluid approach, equations for ion and electron continuity, trapped electron and ion energy are considered. The hierarchy of fluid equations has necessarily to be truncated by a closure, which actually represents the wave-particle resonance. The Weiland model adopts the so called Righi–Leduc or diamagnetic heat flow closure following by that of Braginskii \cite{braginskii65}. In the first approximation, the parallel ion motion can be neglected; this is reasonable if the fastest growing mode fulfills $\left(k\rho_i\right)^2\approx0.1$ \cite{weiland00}, where $k$ is the mode wave vector and $\rho_i$ is the ion Larmor radius. The electron density perturbation $\delta n_e$ is written as:
\begin{eqnarray}  \frac{\delta n_e}{n_e}=f_t\frac{\delta n_{et}}{n_{et}}+\left(1-f_t\right)\frac{\delta n_{ep}}{n_{ep}}
  \label{appeq1} \end{eqnarray}
where index \textit{t} and \textit{p} stand, respectively, for trapped and passing particles. Passing electrons are allowed to reach a Boltzmann distribution, i.e., they are assumed adiabatic according to $\delta n_{ep}/n_{ep}=e\delta\phi/T_e$. These assumptions are coupled with the quasi-neutrality condition (assuming a single hydrogenous ion species):
\begin{eqnarray}  \delta n_{i} = \delta n_{ep} + \delta n_{et}
  \label{appeq2} \end{eqnarray}
The linear part of the fluid equations leads to the following dispersion relation for a two-stream instability \cite{weiland89,nordman89,weiland00}, where finite Larmor radius FLR effects have been neglected:
\begin{align}  &\frac{1}{N_i}\left[\omega\left(\omega_{ne}^*-\omega_{DTe}\right) + 
\left(\omega_{Ti}^*-\frac{7}{3}\omega_{ne}^*+\frac{5}{3}\omega_{DTe}\right)\omega_{DTi}\right] = \nonumber \\
&\qquad \frac{f_t}{N_e} \left[\omega\left(\omega_{ne}^*-\omega_{DTe}\right) + 
\left(\omega_{Te}^*-\frac{7}{3}\omega_{ne}^*+\frac{5}{3}\omega_{DTe}\right)\omega_{DTe}\right] + \left(1-f_t\right)
  \label{appeq3} \end{align}
with the resonant denominator for the species $j$ in the form:
\begin{eqnarray}  N_j = \omega^2-\frac{10}{3}\omega\omega_{DTj}+\frac{5}{3}\omega_{DTj}^2
  \label{appeq4} \end{eqnarray}
A first unstable branch is achieved with the condition $N_i<N_e$ that corresponds to a mode propagating in the ion magnetic drift direction (ITG modes), while for $N_i>N_e$ the mode is rotating in the electron drift one (TE modes).\\
\indent Within the Weiland model these modes are thought to decouple when $N_i \ll N_e$ (pure ITG mode) or $N_i \gg N_e$ (pure TEM); for each of these cases, the full dispersion relation splits into two quadratic equations. Imposing a null imaginary part to the solution $\omega=\omega_r+i\gamma$ provides the instability threshold condition. In case of ITG modes this procedure leads to the following threshold expressed:
\begin{eqnarray}  \left.\frac{R}{L_{Ti}}\right|_{th} = \frac{R}{L_{n}}\left(\frac{2}{3}-\frac{1}{2}\frac{T_e}{T_i}+\frac{1}{8}\frac{T_e}{T_i}\frac{R}{L_{n}}\right)+\frac{1}{2}\frac{T_e}{T_i}+\frac{20}{9}\frac{T_i}{T_e}
  \label{appeq5} \end{eqnarray}
where the main $T_i/T_e$ proportionality is carried by the last term. In the advanced fluid Weiland model, this ITG $T_i/T_e$ threshold dependence is carried by the choice on the resonant denominator, restoring the same feature highlighted by the kinetic approach in \ref{sec-TiTeexample}. A plot for this threshold behavior versus $R/L_n$ is presented in Fig. \ref{figTiTe1} for different ratios $T_i/T_e$. For most common values of normalized density gradients, i.e. $R/L_n<5$, the ITG threshold raises when increasing $T_i/T_e$. Conversely, for $R/L_n>5$, the opposite $T_e/T_i$ scaling is predicted by Eq. \eqref{appeq5}.
\indent Following an analogous procedure, the Weiland fluid model provides an analytical expression for the $R/L_{Te}$ TEM linear threshold:
\begin{eqnarray}  \left.\frac{R}{L_{Te}}\right|_{th} = \frac{20}{9K_t} + \frac{2}{3}\frac{R}{L_{n}} + \frac{K_t}{2}\left(1-\frac{R}{2L_{n}}^2\right)
  \label{appeq6} \end{eqnarray}
No term dependent on the temperature ratio is present within this formulation, contrary to the ITG case Eq. \eqref{appeq5}.\\ 

\noindent \textbf{The $\nabla_r n/n$ TEM threshold in the limit of zero temperature gradients}\\

\indent The very simplified case of $\nabla_r T_e=\nabla_r T_i=0$ is addressed here, since trapped electron modes can be destabilized by the only presence of density gradients. Even if these are not relevant conditions for tokamak plasmas, the present analysis is useful in order to enlarge predictive capabilities on wider parametric space.\\
Dealing with zero temperature gradients implies that the linear kinetic dispersion relation \eqref{kintrape1} can be greatly simplified, since $\omega_{Te}^*=\omega_{Ti}^*=0$. In this case it is not necessary to isolate the imaginary resonant contribution, which would lead to $\omega_{n}^*=0$ (see Eq. \eqref{kintrape2}), where no energy dependence is present. The crude second order fluid expansion based on the condition $\omega_{De}/\omega \ll 1$ is then allowed here for a mode in the electron diamagnetic direction; the non-negligible ion response has been treated through a second order fluid expansion too.
\begin{figure}[!htbp]
  \begin{center}
    \leavevmode
      \includegraphics[width=7 cm]{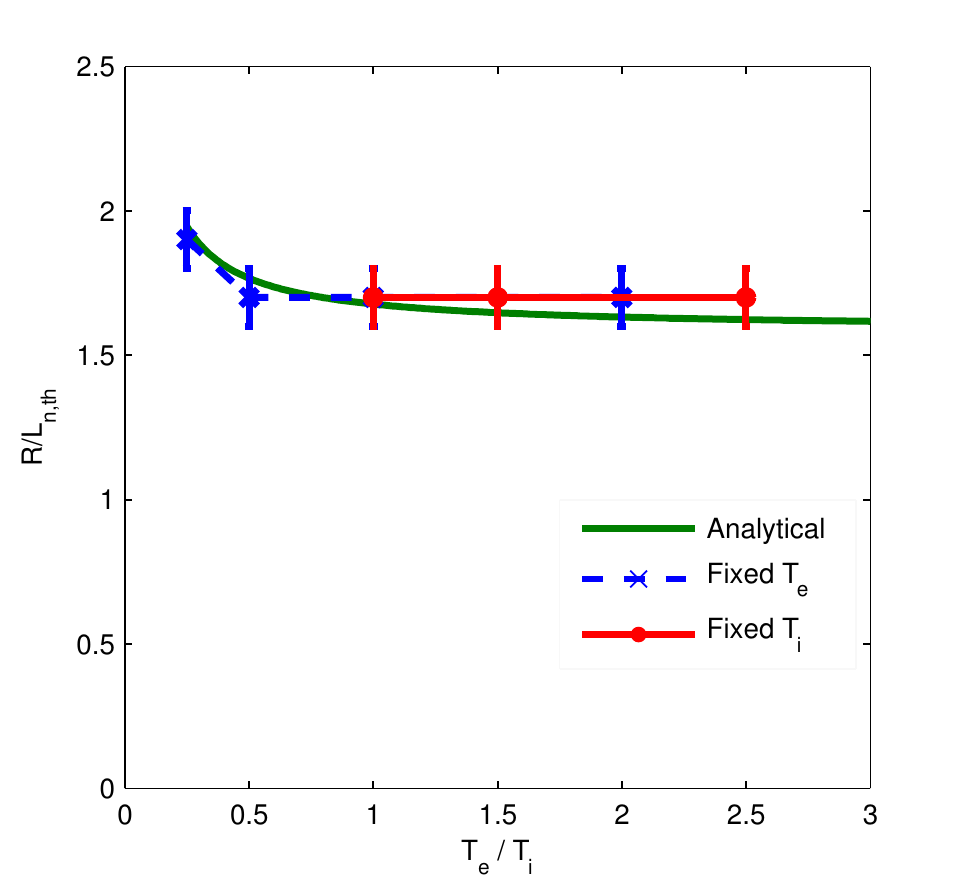}
    \caption{$T_e/T_i$ dependence of the $R/L_n$ TEM instability threshold calculated with Kinezero at $R/L_{Ti}=R/L_{Te}=0$ and at fixed $T_i$ and $T_e$. Analytical predictions following from a fluid expansion are also plotted.}
    \label{figTiTe4}
  \end{center}
\end{figure}
The critical $R/L_n$ value above which unstable modes set in can only be numerically derived from the quadratic dispersion
relation:
\begin{equation} \begin{split} 1-f_t \left\langle 1- \frac{\omega_{ne}^*}{\omega} + \frac{\omega_{DTe}}{\omega}\mathcal{E} - \frac{\omega_{ne}^*\omega_{DTe}}{\omega^2}\mathcal{E} + \frac{\omega_{DTe}^2}{\omega^2}\mathcal{E}^2 \right\rangle_{\mathcal{E},\lambda} \\
+ \frac{T_e}{T_i} \left\langle \frac{\omega_{ni}^*}{\omega} + \frac{\omega_{DTi}}{\omega}\mathcal{E} - \frac{\omega_{ni}^*\omega_{DTi}}{\omega^2}\mathcal{E} + \frac{\omega_{DTi}^2}{\omega^2}\mathcal{E}^2 \right\rangle_{\mathcal{E},\lambda} = 0
	\label{kintrape4} \end{split} \end{equation}
\indent This $T_e/T_i$ dependence of $R/L_{n,th}$ has been compared with Kinezero simulations in Fig. \ref{figTiTe4}, choosing the set of plasma parameters defined by the Table \ref{TableTiTe} and $R/L_{Ti}=R/L_{Te}=0$. The fluid expansion foresees a weak decrease of the $R/L_n$ TEM threshold for higher $T_e/T_i$. On the other hand, linear gyrokinetic simulations identify a temperature ratio impact in agreement with analytical predictions only for low values of $T_e/T_i$. When considering conditions of $\nabla_r T_e=\nabla_r T_i=0$, the crude fluid expansion based on nonresonant effects can then be regarded as an adequate treatment, capable of qualitatively reproducing the numerical features found by linear gyrokinetic simulations. Moreover, in the absence of temperature gradients, the ion response contribution has a weak impact of the ratio $T_e/T_i$ on the $R/L_n$ TEM threshold.\\

\noindent \textbf{Derivation of a consistent electron (ion) response within a kinetic approach}\\

The linear kinetic dispersion relation written in Eq. \eqref{kineqtrape} can describe both ion and trapped electron modes. The analytical approach here presented refers to a single unstable branch in the ion or in the electron diamagnetic direction. The key point is evaluating the response of the opposite species with respect to the one giving the modes direction. In this appendix we examine in detail the case of electron modes; ion modes relations will not be here explicitly rewritten since they can be analogously obtained simply reversing the role of the species and accounting for the passing ions.\\
\indent As already explained in paragraph \ref{sec-TiTeexample}, imposing a null imaginary part coming from the kinetic resonance provides fundamental conditions for the modes frequency (Eq. \eqref{kintrape1b} for electron modes, Eq. \eqref{kinitglin1} for ion ones) that can be rewritten for electron modes as:
\begin{eqnarray}  \frac{\omega}{\omega_{DTe}} = \frac{\frac{3}{2}\frac{\omega_{Te}^*}{\omega_{DTe}} - \frac{\omega_{ne}^*}{\omega_{DTe}} } {\frac{\omega_{Te}^*}{\omega_{DTe}}}
  \label{appeq7} \end{eqnarray}
The term proportional to $R/L_n$ can be eventually set to zero when considering strict flat density limit $\omega_{ne}^*=0$. It is worth noting that the role of the density gradient inside expression \eqref{appeq7} limits the validity of this kind of approach entirely based on the kinetic resonance, because the modes frequency is not allowed to change sign. Kinetic effects will then be dominant on the threshold behavior until the conditions 
\begin{eqnarray} \frac{R}{L_n} <\approx \frac{3}{2} \left.\frac{R}{L_T}\right|_{th} \qquad
   \textrm{and} \qquad \left.\frac{R}{L_T}\right|_{th} >\approx 2
  \label{appeq8} \end{eqnarray}
are fulfilled; beyond these limits the hypothesis of modes frequency close to the resonance is not valid and this procedure is not applicable anymore.\\
\indent Since the modes frequency is assumed to obey to Eq. \eqref{appeq7}, it is possible to evaluate the ion response at the same frequency; ion resonant effects are in fact excluded because of the opposite sign of $\omega$. On the other hand, the real part of Eq. \eqref{kineqtrape} leads to:
\begin{eqnarray}  \left\langle \frac{\omega_{Te}^*}{\omega_{DTe}} \right\rangle_{\lambda} = \frac{1}{f_t} 
  \left( 1+\aleph\frac{T_e}{T_i} \right)
  \label{appeq9} \end{eqnarray}
The parameter $\aleph$ represents in fact the ion response in both the adiabatic and non-adiabatic components; it can be written as:
\begin{align}  \aleph &= 1-   \left\langle \frac{\omega-\omega_{i}^*}{\omega-\omega_{Di}} \right\rangle_{\mathcal{E},\lambda}     \nonumber \\
  &= 1- \left\langle \frac{ \frac{\omega}{\omega_{DTe}}+\frac{T_i}{T_e}\left[\frac{\omega_{ne}^*}{\omega_{DTe}}+\frac{\omega_{Ti}^*}{\omega_{DTi}}\left(\mathcal{E}-\frac{3}{2}\right)\right]}  {\frac{\omega}{\omega_{DTe}}+\frac{T_i}{T_e}\mathcal{E}} \right\rangle_{\mathcal{E},\lambda}
  \label{appeq10} \end{align}
The term proportional to $R/L_{Ti}$ can be eventually set to zero in the absence of ion temperature gradients, $\omega_{Ti}^*=0$. Physically, Eq. \eqref{appeq10} expresses the ion response in its adiabatic and non-adiabatic contributions driven by ion temperature and density gradients. The only unknown is the normalized frequency $\omega/\omega_{DTe}$; within our approach, the latter one is assumed to obey to Eq. \eqref{appeq8}.\\
\indent Apart from $R/L_n$, Eq. \eqref{appeq8} appears depending on the electron mode threshold itself through the term $\omega_{Te}^*/\omega_{DTe}$, which has to be evaluated according to Eq. \eqref{appeq10}; the unknown threshold value and the ion response result then intrinsically linked one to each other. Both Eqs. \eqref{appeq8} and \eqref{appeq9} can be substituted in Eq. \eqref{appeq10}, obtaining an implicit integral expression for $\aleph$. Auto-consistent solutions for $\aleph$ can be finally numerically found, depending on the parameters $T_i/T_e$, $R/L_n$, $R/L_{Ti}$, and $f_t$ externally imposed.\\
Within the already mentioned limits, this kind of procedure is then rigorously valid; the only approximations regard having neglected FLR effects and the simplification of the $\lambda$-integration.\\
\indent Self-consistent calculations of Eqs. \eqref{appeq8},\eqref{appeq9},\eqref{appeq10}, considering $\omega_{Ti}^*=\omega_{ne}^*=0$, have been carried out for deriving the $T_e/T_i$ dependence of the pure TEM threshold in strict flat density limit: the result is shown in Fig. \ref{figTiTe2}. Retaining $\omega_{Ti}^*,\omega_{ne}^*\ne0$ has instead lead to the analytical expectations shown in Fig. \ref{figTiTe3}. An analogous set of coupled equations as Eqs. \eqref{appeq8},\eqref{appeq9},\eqref{appeq10} can be easily written when considering ion modes. With $\omega_{Te}^*,\omega_{ni}^*\ne0$ this has lead to the results shown in Fig. \ref{figTiTe6}.

\chapter{Brief notes on the nonlinear gyrokinetic codes}
\label{appx2-GYRO-GYSELA}

\noindent \textbf{About the GYRO code}\\

\indent GYRO is a nonlinear tokamak micro-turbulence code. Developed at General Atomics since 1999, GYRO is presently one of the most advanced and comprehensive tool available for investigating the tokamak turbulence transport. GYRO uses a fixed Eulerian grid to solve the 5-D gyrokinetic-Maxwell equations using a $\delta f$ scheme. Despite the complexity of the nonlinear problem which requires massive parallel computing, the main feature of this code relies on its flexible operation, with the capability to treat:
\begin{itemize}
	\item A local (otherwise called flux-tube) or global radial domain, in a full or partial torus.
	\item Generally shaped or simple circular $s-\alpha$ magnetic geometry.
	\item Full nonlinear gyrokinetic treatment of ions, electrons and impurity species (both trapped and passing domains).
	\item Electrostatic or electromagnetic fluctuations.
	\item Electron-ion and ion-ion pitch-angle scattering operators.
\end{itemize}
\indent The right-handed, field-aligned coordinate system $(\psi,\theta,\zeta)$ together with the Clebsch representation \cite{kruskal58} for the magnetic field is used. This has been briefly introduced in paragraph \ref{sec-geometry}, see in particular Eq. \eqref{Bfalign}, in the simplified case of concentric circular flux surfaces. Keeping the magnetic field representation $\mathbf{B}=\nabla\eta \times \nabla\psi$, more generally, the angle $\zeta$ can be written in terms of the toroidal angle $\varphi$ as:
\begin{eqnarray}
	\zeta=\varphi + \nu\left(\psi,\theta\right) \label{app2eq1}
\end{eqnarray}
In these coordinates, the formal Jacobian is :
\begin{eqnarray} \mathcal{J}_{\psi}=\frac{1}{\nabla\psi\times\nabla\theta\cdot\nabla\zeta} = \frac{1}{\nabla\psi\times\nabla\theta\cdot\nabla\varphi}=\left(\frac{\partial \psi}{\partial r}\right)^{-1}\mathcal{J}_r
	 \label{app2eq2}
\end{eqnarray}
The advantage of this representation is the capability of treating a general shaped magnetic equilibrium, solution of the Grad-Shafranov equation, and not only circular flux surfaces. Recalling the usual expression for the magnetic field from Eq. \eqref{Bfield}, $\nu\left(\psi,\theta\right)$ is expressed by an integral relation:
\begin{eqnarray} \nu\left(\psi,\theta\right) = -I\left(\psi\right)\int_0^{\theta}\mathcal{J}_{\psi}\left|\nabla\varphi\right|^2d\theta
	 \label{app2eq3}
\end{eqnarray}
While for a general Grad-Shafranov equilibrium, $\nu\left(\psi,\theta\right)$ can be only numerically solved, in the simplified case of concentric circular $s-\alpha$ geometry, we recover the more usual result $\nu\left(\psi,\theta\right)=-q\left(\psi\right)\theta$ and therefore $\zeta=\varphi-q\left(\psi\right)\theta$.\\
Finally in this framework, the following flux-surface operator acting on the general quantity $z$ is defined:
\begin{eqnarray} \mathcal{F} z = \frac{\displaystyle{\int_0^{2\pi}d\zeta\int_0^{2\pi}d\theta\mathcal{J}_{\psi}z}}{\displaystyle{\int_0^{2\pi}d\zeta\int_0^{2\pi}d\theta\mathcal{J}_{\psi}}}
	 \label{app2eq3b}
\end{eqnarray}
\indent The $\delta f$ expansion refers to the single particle distribution function $f_s$, which is written as a sum of an equilibrium part and fluctuating terms:
\begin{eqnarray} f_s\left(\mathbf{x},E,\mu,t\right) = f_{0s}\left(\mathbf{x},E\right) + \delta f_s\left(\mathbf{x},E,\mu,t\right)
	 \label{app2eq4}
\end{eqnarray}
where $\mathbf{x}=\mathbf{R}+\boldsymbol{\rho}$ is the particle position, $\boldsymbol{\rho}$ is the gyroradius vector and $\mathbf{R}$ is the guiding-center position. The equilibrium function $f_{0s}$ is assumed to be a Maxwellian.\\
In GYRO, all the perturbed quantities are expended as Fourier series in $\zeta$. For example, the normalized electrostatic potential $\delta \hat{\phi}=e\delta\phi / T_s$ is written as:
\begin{eqnarray} \delta\hat{\phi}\left(r,\theta,\zeta\right) = \sum_{j=0}^{N_n-1} \delta\phi_n\left(r,\theta\right)e^{in\zeta}
\qquad n=j\Delta n
	 \label{app2eq5}
\end{eqnarray}
Contrarily to other nonlinear codes, which use different discretization schemes, GYRO operates through a direct discretization of the quantities $z\left(r,\theta\right)$. From the numerical point of view, GYRO always solves the discrete linear or nonlinear gyrokinetic and Poisson/Amp\'ere equations in $\left(n,r,\theta\right)$, plus the time advance. One of the perpendicular coordinates is then the radial one $r$ (otherwise referred as $x$); the other perpendicular direction (usually referred as $y$) is not a coordinate in GYRO, which uses instead the Fourier index n, conjugate to the Clebsch angle $\zeta=\varphi+\nu\left(r,\theta\right)$.\\
\indent Considering a physical function represented by the real field $z\left(r,\theta,\zeta\right)$, a $\theta$ periodicity condition can be written as:
\begin{eqnarray} z\left(r,0,\varphi+\nu\left(r,0\right)\right) = z\left(r,2\pi,\varphi+\nu\left(r,2\pi\right)\right)
	 \label{app2eq6}
\end{eqnarray}
So, even if the physical field $z$ is $2\pi$-periodic in $\theta$, the Fourier coefficients of the expansion \eqref{app2eq5} are not periodic, while they satisfy the phase condition:
\begin{eqnarray} z_n\left(r,0\right) = e^{2\pi inq\left(r\right)} z_n\left(r,2\pi\right)
	 \label{app2eq7}
\end{eqnarray}
Moreover, since $z$ is real, the Fourier coefficients satisfy the relation $z_n^*=z_{-n}$. The spectral form of Eq. \eqref{app2eq5} is then $2\pi/\Delta n$-periodic in $\zeta$ (or in $\varphi$) at fixed $\left(r,\theta\right)$. A direct consequence of the representation \eqref{app2eq5} allows to easily map the physical quantities at the outboard midplane in terms of the coordinates $\left(r,\varphi\right)$:
\begin{eqnarray} z\left(r,\theta=0,\varphi,t\right) = \sum_{n=-N_n}^{N_n-1}z_n\left(r,\theta=0,t\right)e^{-in\varphi}
	 \label{app2eq8}
\end{eqnarray}
\indent The perturbed distribution function $\delta f_s$ solved by GYRO is given by (Gaussian CGS units are here used):
\begin{eqnarray} \delta f_s\left(\mathbf{x},E,\mu,t\right) = -\frac{e_s}{T_s}f_{0s} 
\left[ \delta\phi\left(\mathbf{x},t\right) -\mathcal{G}^s\delta\phi\left(\mathbf{R},t\right)+\frac{v_{\parallel}}{c}\mathcal{G}^s\delta A_{\parallel}\left(\mathbf{R},t\right) \right] + 
h^s\left(\mathbf{R},E,\mu,t\right)
	 \label{app2eq9}
\end{eqnarray}
Using the following notation
\begin{align}  \tilde{h}_s &= \frac{h_s\left(\mathcal{E},\lambda\right)}{f_{0s}\left(\mathcal{E}\right)} \\
   \tilde{U}_s &= \frac{e}{T_s}\left[\delta\phi-\frac{v_{\parallel s}\left(\mathcal{E},\lambda\right)}{c}\delta A_{\parallel}\right] \\
   \tilde{\mathbf{v}}_U &= c\frac{T_e}{e}\frac{-\nabla\tilde{U}\times\mathbf{b}}{B} \\
   \tilde{H}_s &= \tilde{h}_s + \frac{e_s\hat{n}_{0s}}{\hat{T}_s}\mathcal{G}^s\tilde{U}_s \\
    \omega_* &= c_s\left[\frac{1}{L_{ns}}+\frac{1}{L_{Ts}}\left(\mathcal{E}-\frac{3}{2}\right)\right]k_y\rho_s 
    \end{align}
where $\mathcal{G}$ is the gyroaverage operator, the gyrokinetic equation used in GYRO for a generic species $s$ can be synthetically written as:
\begin{eqnarray} \frac{\partial\tilde{h}_s}{\partial t} + \underbrace{ \tilde{\mathbf{v}}_{Us}\cdot\nabla\tilde{h}_s }_1 + \underbrace{ i\hat{n}_{0s}\omega_{*s}\mathcal{G}^s\tilde{U}_s }_2 = \underbrace{ -v_{\parallel s}\nabla_{\parallel}\tilde{H}_s }_3 - \underbrace{ \mathbf{v}_{Ds}\cdot\nabla\tilde{H}_s }_4 + \underbrace{ C\left(\tilde{h}_s\right) }_5
	 \label{app2eq10}
\end{eqnarray}
The driver perturbed field $\mathcal{G}^s\tilde{U}_s$ is given by the Poisson-Amp\'ere equation: the perturbed charge and current densities determining $\mathcal{G}^s\tilde{U}_s$ enter as velocity space integrals over $\tilde{h}_s$, which is proportional to $\hat{n}_{0s}=n_{0s}/n_{0e}$, the density of the species relative to the electron density. In the gyrokinetic Eq. \eqref{app2eq10}, the terms $1$ and $2$ are respectively the nonlinear and linear generalized $E\times B$ drift motions, including the magnetic fluctuations. The terms $3$ and $4$ represent instead the parallel and curvature drift motions, while the term $5$ accounts for the collisions.\\
\indent Finally, the particle and energy flux for each species are defined as:
\begin{align} \Gamma_s\left(r\right) &= \mathcal{F}\int dv^3\delta f_s\left(\mathbf{x}\right)\left(\frac{1}{B}\mathbf{b}\times\nabla U\right)\cdot\nabla r \label{app2eq11} \\
	 Q_s\left(r\right) &= \mathcal{F}\int dv^3 E \delta f_s\left(\mathbf{x}\right)\left(\frac{1}{B}\mathbf{b}\times\nabla U\right)\cdot\nabla r \label{app2eq12}
\end{align}

\noindent \textbf{Tracer and quasi-linear transport}\\

\indent A species is here qualified as \textit{tracer} when $\hat{n}_{0s}\ll1$. The following statements are derived and can be verified by GYRO simulations:
\begin{enumerate}
	\item The tracer species have a negligible contribution to the fields through the Poisson-Amp\`ere equation; hence, the background plasma turbulence and transport is unaffected by the presence of the tracers.
	\item If the tracer species have identical mass and charge to the main plasma species, and all the terms of the gyrokinetic equation \eqref{app2eq10} for tracers are kept, then the tracer particle and energy fluxes are identical to those of the main plasma species.
	\item The tracer particle and energy fluxes, unlike the main plasma species ones, are independent of the tracer gradients $1/L_{n,tr}$ and $1/L_{T,tr}$ as long as the tracer $\eta_{tr}=L_{n,tr}/L_{T,tr}$ is held fixed. The tracer transport follows in fact a Fick's law, i.e. linear with the gradients only at fixed $\eta_{tr}$.
\end{enumerate}
\indent The full frequency spectrum quasi-linear approach (fQL) described in the paragraph \ref{sec-transport-weights} refers to GYRO tracer (test-particle) simulations. Apart from the main plasma species, ion and electrons, two additional tracers are retained, with identical masses and charges to the main ions and electrons. Conversely, these tracers are characterized by negligible densities, i.e. $\hat{n}_{0e,tr}=\hat{n}_{0i,tr}=10^{-7}$. Moreover, in order to deal with quasi-linear test-particle transport, the nonlinear term $1$ in the gyrokinetic equation \eqref{app2eq10} governing both the tracers is dropped. \\
As highlighted in the paragraph \ref{sec-transport-weights}, this method should allow to obtain the \textit{best} quasi-linear transport estimate. The main advantages of the present fQL approach are in fact: (1) the whole structure of the linear modes (and not only the linear most unstable and ballooning mode) is retained within the calculation, (2) the nonlinear saturation of the fluctuating potential is completely and self-consistently imposed by the main plasma species. On the other hand, the main drawback of the method is that the particle flux ambipolarity is not respected anymore.\\
\begin{figure}[!htbp]
  \begin{center}
    \leavevmode
      \includegraphics[width=7 cm]{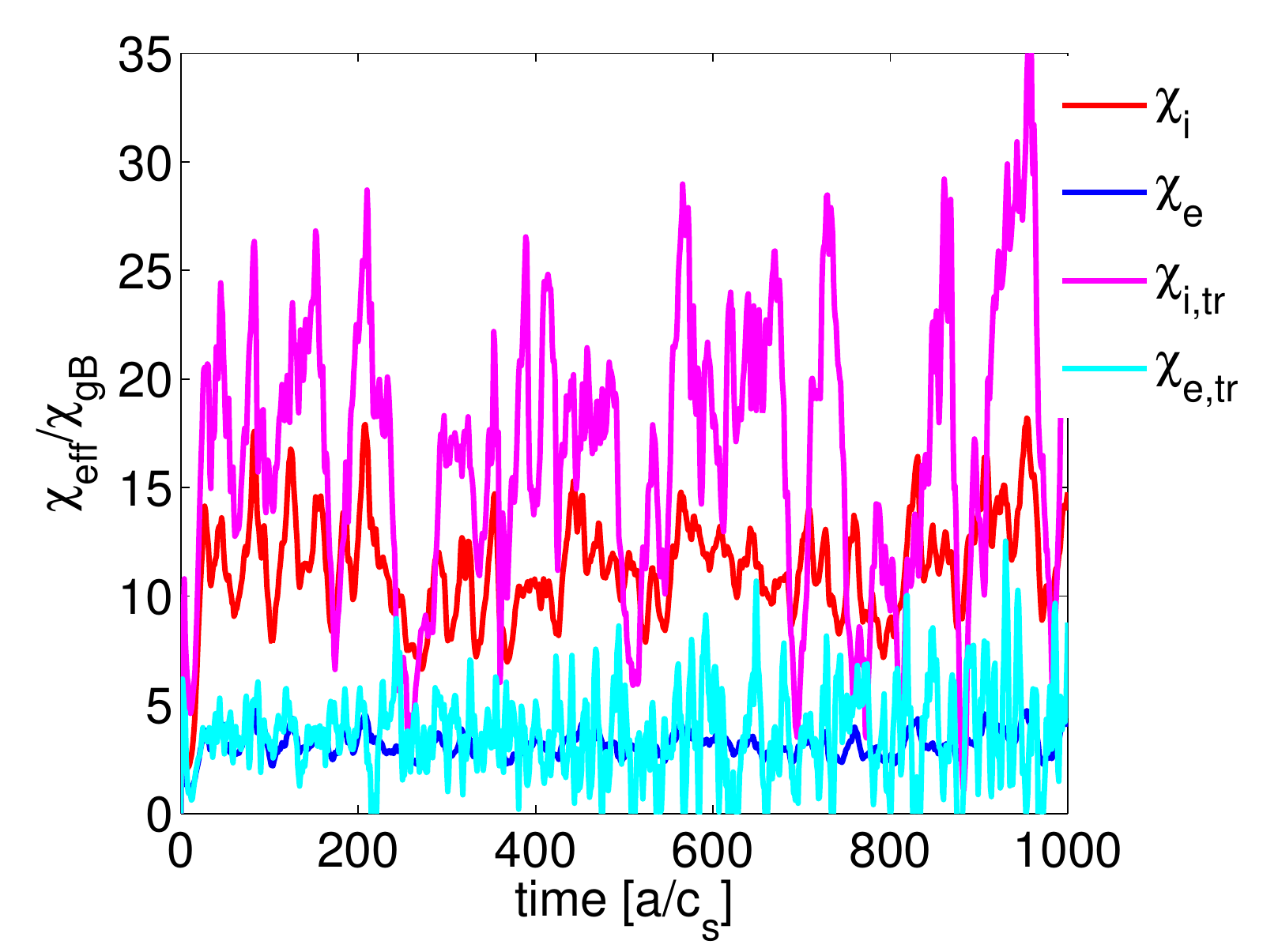}
    \caption{GYRO simulation on the GA-ITG-TEM standard case using the quasi-linear test-particle method: the time evolution of the energy diffusivities is shown for both the main plasma and the quasi-linear tracer species.}
    \label{GYRO-tr-tevol}
  \end{center}
\end{figure}
\indent Apart from the numerical tracers experiments, an alternative approach can be used to investigate the quasi-linear transport. For a standard ions and electrons plasmas, recalling the gyrokinetic equation Eq. \eqref{app2eq10}, only the linear $E\times B$ drive expressed by the term $1$ can be artificially suppressed in a nonlinear simulation for a single toroidal wave-number. Consequently, there will be no linearly unstable mode in the deleted wave number but still the nonlinear dynamics. Fig. \ref{GYRO-quasinl-tr} reports the results of this exercise with the GA-ITG-TEM standard case, where the linear drive has been suppressed for the wave-numbers corresponding to $k_{\theta}\rho_s=0.1$ and $k_{\theta}\rho_s=0.6$. As shown in the figure, some transport appears also where the linear drive is deleted, due to purely nonlinear couplings in the $k$ space. It is worth noting that in this case, the more relevant purely nonlinear contribution in Fig. \ref{GYRO-quasinl-tr} refers to the particle transport, i.e. the channel where the quasi-linear response appears more feeble (see paragraph \ref{sec-transport-weights}). Analogous demonstration of an high-$k$ transport nonlinearly driven by low-$k$ scales has been investigated in Ref. \cite{waltz07}.
\begin{figure}[!htbp]
  \begin{center}
    \leavevmode
      \includegraphics[width=13 cm]{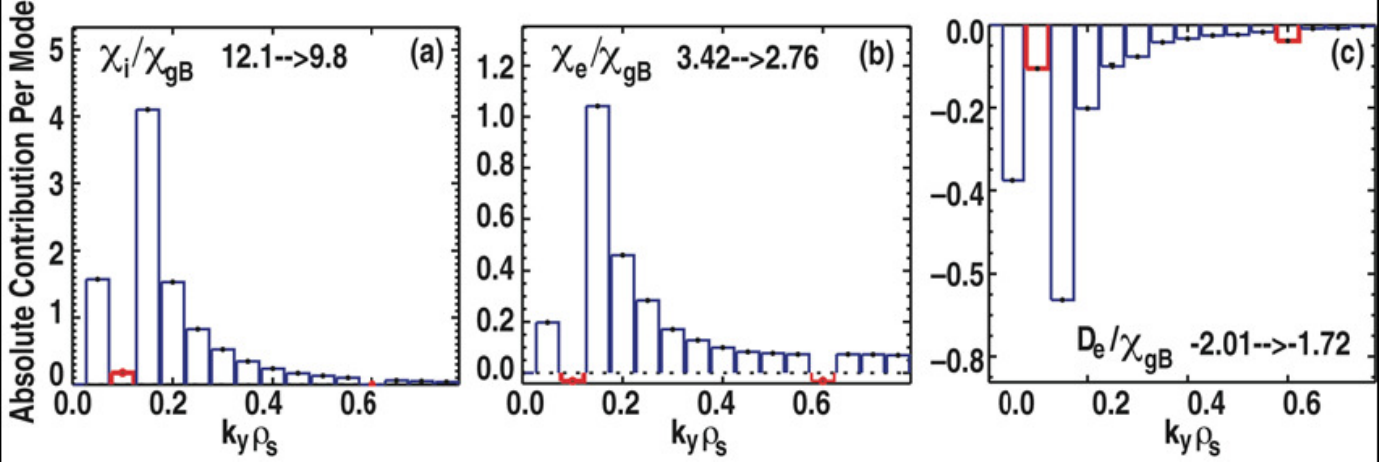}
    \caption{GA-ITG-TEM standard case ion energy (a), electron energy (b), and particle (c) contribution per mode diffusivity versus wave number with the linear $E\times B$ drive at $k_{\theta}\rho_s=0.10$ and $k_{\theta}\rho_s=0.60$ deleted hence linearly stable. Purely nonlinearly induced transport in highlighted in red. The total transport is somewhat reduced.}
    \label{GYRO-quasinl-tr}
  \end{center}
\end{figure}

\noindent \textbf{Coordinates systems}\\

\indent As previously mentioned, GYRO use a field aligned coordinate systems and it does not employ an expansion in poloidal harmonics. Since several nonlinear tokamak micro-turbulence codes adopt different coordinates, it is of interest to detail some basic relations useful in order to make comparisons between different representations. In particular, here we refer to the coordinate system used in the GYSELA code. The latter one operates in fact with the coordinates $\left(r,\theta,\varphi\right)$ (respectively the radial coordinate, the poloidal and the toroidal angles): conversely to GYRO, the fields are expanded in Fourier series in both $\theta$ and $\varphi$.\\
\indent In GYSELA, the expansion for a generic field $z$ is written as:
\begin{eqnarray} z^{GS}\left(r,\theta,\varphi\right) = \sum_{n,m} z^{GS}_{nm}e^{i\left(n\varphi+m\theta\right)}
	 \label{app2eq13}
\end{eqnarray}
which has to be compared to the analogous expression for GYRO. The superscripts \textit{GR} and \textit{GS} are here introduced to label GYRO and GYSELA notations respectively. The following relations link the GYSELA representation with the GYRO one within a simplified circular magnetic $s-\alpha$ equilibrium, where the relation $\zeta=\varphi+\nu\left(r,\theta\right)$ reduces to the analytically tractable form $\zeta=\varphi-q\left(r\right)\theta$. A fundamental relation is then:
\begin{eqnarray} \sum_{n} z^{GR}_{n}\left(r,\theta\right)e^{in\zeta} = \sum_{n,m} z^{GS}_{nm}\left(r\right)e^{i\left(n\varphi+m\theta\right)}
	 \label{app2eq14}
\end{eqnarray}
hence it follows:
\begin{align} z^{GR}_{n}\left(r,\theta\right) &= \sum_{m} z^{GS}_{nm}\left(r\right)e^{i\left[m+nq\left(r\right)\right]\theta} \\
  z^{GS}_{nm}\left(r\right) &= \int_{-\pi}^{\pi}\frac{d\theta}{2\pi} z^{GR}_{n}\left(r,\theta\right)e^{-i\left[m+nq\left(r\right)\right]\theta}
	 \label{app2eq15}
\end{align}
which can be practically used to map one representation into another one.\\
\indent When dealing with quantities that are defined through the flux-surface average defined by Eq. \eqref{app2eq3b}, it appears that:
\begin{align} Q_n^{GR}\left(r\right) &= \int_{-\pi}^{\pi}\frac{d\theta}{2\pi}z_n^{GR}\left(r,\theta\right)v_n^{*GR}\left(r,\theta\right) \\
 &= \int_{-\pi}^{\pi}\frac{d\theta}{2\pi} \sum_{m,m'}z_{nm}^{GS}\left(r\right)v_{nm'}^{*GR}\left(r\right)e^{i\left(m-m'\right)\theta} = \sum_m z_{nm}^{GS}\left(r\right)v_{nm}^{*GR}\left(r\right)
	 \label{app2eq16}
\end{align}
The latter relation is useful when defining for example a flux contribution coming from a certain toroidal wave-number $n$. It also immediately leads to the definition of a $n$-dependent flux-surface averaged rms quantity, i.e.
\begin{eqnarray}   \left.z_n^{GR}\left(r\right)\right|_{rms, flxav} = \left\{\sum_m\left|z_{nm}^{GS}\left(r\right)\right|^2\right\}^{1/2}
	 \label{app2eq17}
\end{eqnarray}

\chapter{The case of quasi-linear pure ITG turbulence}
\label{appx3-ITG-breeakdown}

\begin{figure}[!htbp]
  \begin{center}
    \leavevmode
      \includegraphics[width=7 cm]{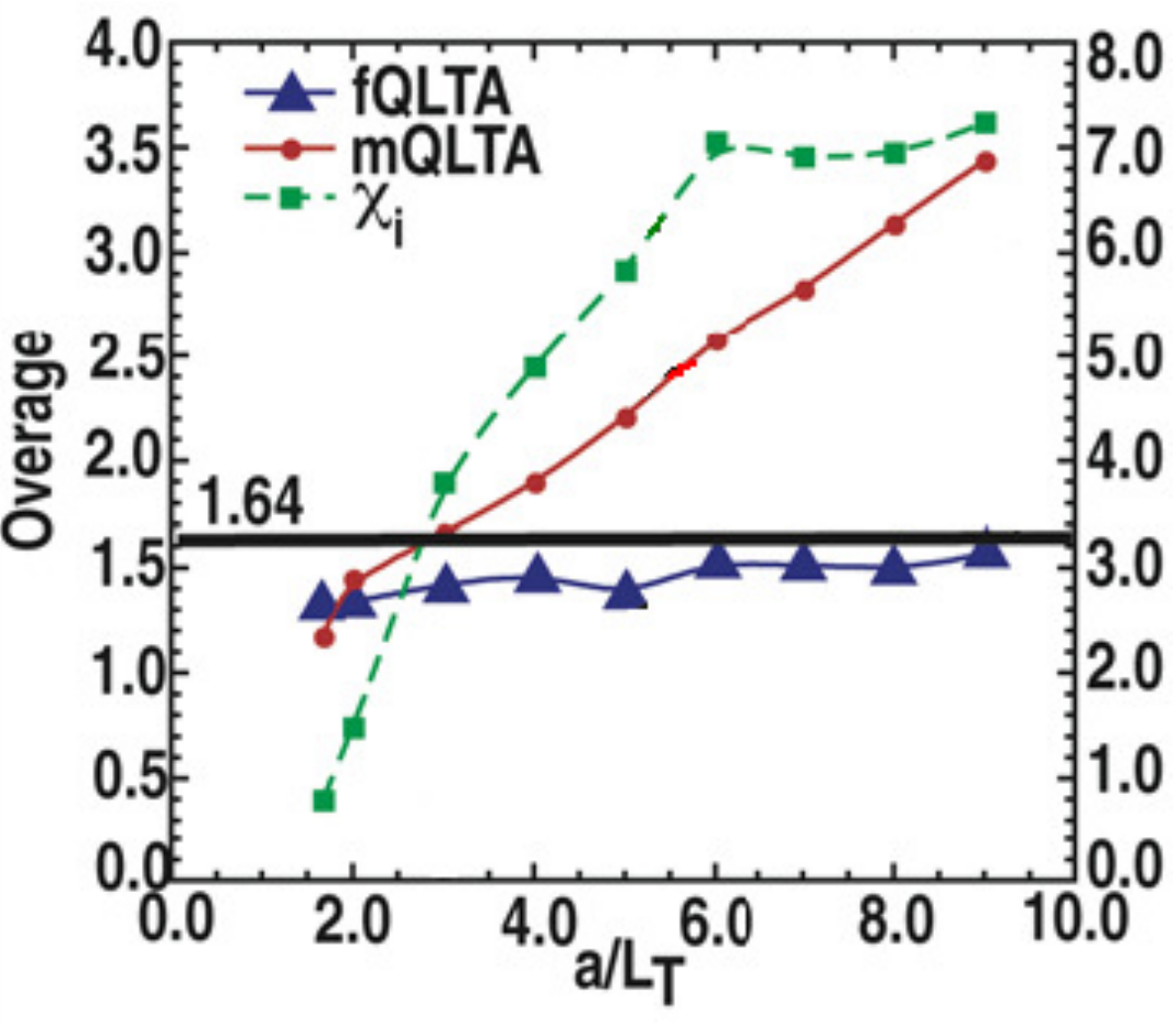}
    \caption{Quasi-linear over nonlinear overages across a ion temperature gradient scan based for pure ion ITG turbulence, using both the mQL and the fQL approaches. $\chi_i$ are also given in terms of $\chi_{GB}$ units.}
    \label{GYRO-ql-ITG}
  \end{center}
\end{figure}

\indent Even if realistic application of the transport models to tokamak plasmas demands that the non-adiabatic electron physics is retained, it is still relevant to test the validity of the quasi-linear approximation in the case of pure ion ITG turbulence, i.e. considering adiabatic electrons. This has been done again using the GYRO code in the local limit, adopting the same parameters and numerical resolution summarized in Table \ref{GA-std-table}, except for the electron response which is forced to be completely adiabatic.\\
\indent As previously done, the validity of the quasi-linear response is tested with GYRO through both the approaches mQL, i.e. retaining only the linear leading mode, and fQL, i.e. the full frequency spectrum approach (see paragraph \ref{sec-transport-weights}). A wide scan of the normalized ion temperature gradient $1.5<a/L_{Ti}<9.0$ is performed and the results are reported in Fig. \ref{GYRO-ql-ITG}. A clear breakdown of the mQL approximation is found for the ion energy flow driven in the case of pure ITG, going far from threshold to very high turbulence levels. Fig. \ref{GYRO-ql-ITG} shows that quasi-linear over nonlinear ratio (overage) computed according to the mQL approach increases up by a factor of 2.1 from the overage of 1.64 at the reference value of $a/L_{Ti}=3.0$.Surprisingly, the quasi-linear over nonlinear ratio computed with the fQL approach does not exhibit a similar feature, while it stays reasonably constant across the whole scan.\\
\begin{figure}[!htbp]
  \begin{center}
    \leavevmode
      \includegraphics[width=13.5 cm]{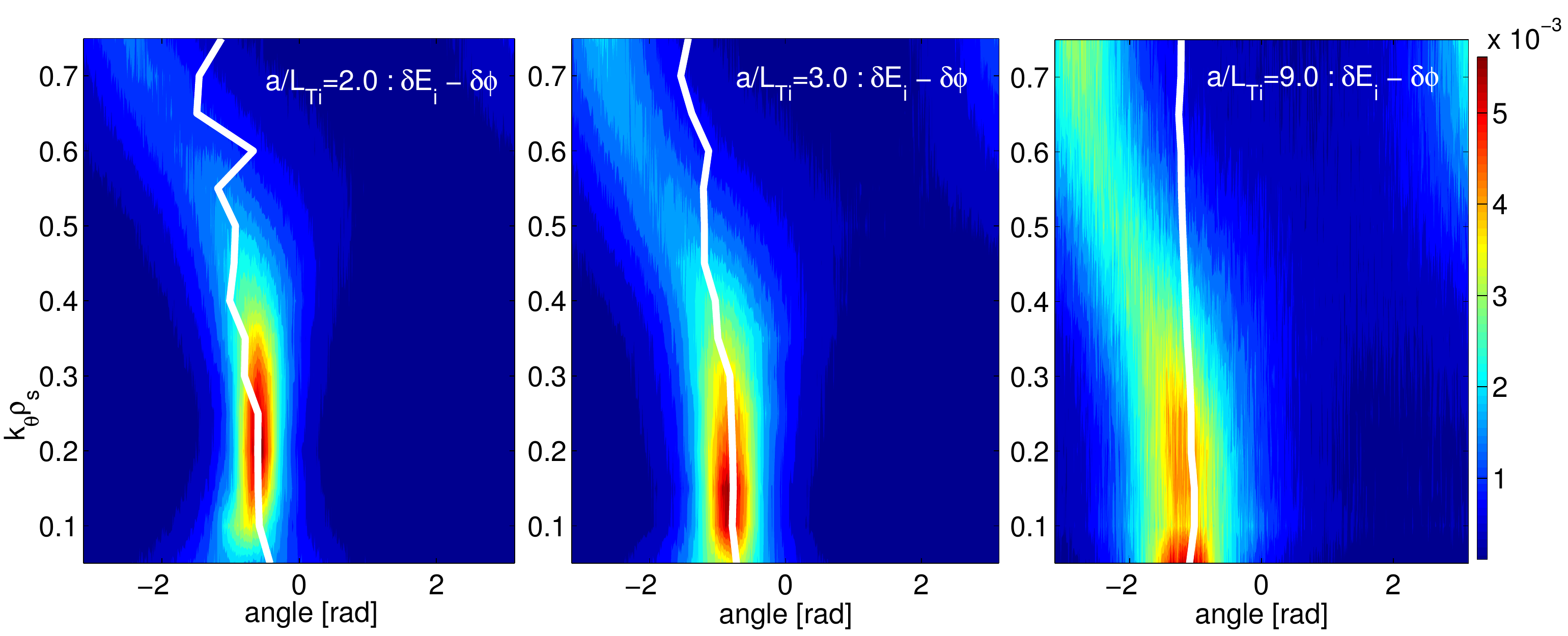}
    \caption{PDF of the nonlinear cross-phases and the linear cross-phase of the most unstable mode (white line): a) $\delta E_i-\delta\phi$ at $a/L_{Ti}=2.0$, b) $\delta E_i-\delta\phi$ at $a/L_{Ti}=3.0$ and c) $\delta E_i-\delta\phi$ at $a/L_{Ti}=9.0$, from local GYRO simulations of pure ITG turbulence with adiabatic electrons.}
    \label{GYRO-ql-ITG-crph}
  \end{center}
\end{figure}
Relevant insight can be gained when studying the cross-phase relations, reported in Fig. \ref{GYRO-ql-ITG-crph}, comparing the phase angles of the nonlinear saturation regime and those of the linear most unstable mode. Still surprisingly, the linear cross-phases relative to the leading linear mode accurately track the nonlinear ones, even for the highest turbulence levels. The results of Fig. \ref{GYRO-ql-ITG-crph} are then a clear example highlighting that the information of the quasi-linear response is not completely carried by the cross-phases. In the case of pure ITG turbulence in fact, the failure of the mQL approximation reported in Fig. \ref{GYRO-ql-ITG}, is linked to the quasi-linear relative amplitudes $\left|\delta E_i\right|/\left|\delta\phi\right|$, which originate an over-estimate of the total turbulent energy flux. A relevant question at this point is then: why the fQL approach does not lead to a similar failure? A possible explanation is here below proposed.\\
\indent The effects coming from the toroidicity are very strong for pure ITG turbulence. The guess is that a relevant physical mechanism linked to this point is missing in the mQL approach, which is instead properly retained with the fQL approximation. It has been already discussed that mQL deals with only the linear leading mode: more in particular this always refers to a ballooning normal mode, i.e. a mode centered around a fixed angle $\bar{\theta}_0$ (this angle is usually, but not necessarily, $\bar{\theta}_0=0$). Normally, other less unstable modes at different ballooning angles are active and contributing to the nonlinear saturation. Previously, it has been shown how in the mQL approach, the neglect of these less ballooning and less unstable modes can account for the different quasi-linear over nonlinear ratios seen between the mQL and the fQL approximations for coupled ITG-TEM turbulence, as reported in Fig. \ref{QL-ITG-TEM-scan}.\\
\indent This argument can be made more explicit recalling the general expression for the quasi-linear turbulent flux \eqref{QLval7}. A more general relation can in fact be written as:
\begin{eqnarray}  \textrm{QL-flux} \propto \sum_{k,j,\theta_0} \textrm{QL-weight}_{k,j,\theta_0} \otimes \textrm{Spectral-intensity}_{k,j,\theta_0}
	\label{app3eq1} \end{eqnarray}
where also the convolution over the $\theta_0$ ballooning angles is retained. The mQL approach results from the Eq. \eqref{app3eq1} in the limit of both $j\rightarrow\bar{j}$ and $\theta_0\rightarrow\bar{\theta}_0$, meaning respectively the most unstable $\bar{j}$ mode centered at the $\bar{\theta}_0$ ballooning angle. On the other hand, the fQL approximation correctly retains the quasi-linear convolution over $j$ and $\theta_0$.\\ 
Hence, a possible origin of the failure of the mQL approximation for pure ITG turbulence is that the quasi-linear weight is computed only at a fixed ballooning angle $\bar{\theta}_0$. Since the nonlinear spectral intensity can not be separated into distinguishable contributions in $j$ and $\theta_0$, this implies that the saturation potential spectrum is entirely applied to the single ballooning mode $\bar{\theta}_0$ instead of the proper convolution of Eq. \eqref{app3eq1}.\\
\indent Finally it can be concluded that the validity of the quasi-linear approximation has been successfully tested in the case of pure ITG turbulence across a wide scan of the ion temperature gradient, using the fQL approach (Fig. \ref{GYRO-ql-ITG}). Nevertheless it has to be stressed that this requires accounting for the whole spectrum of the $\theta_0$ ballooning linear modes.

\bibliographystyle{plain}
\bibliography{phdref}

\end{document}